\documentclass[twoside,12pt]{article}
\usepackage{epsfig}
\usepackage{rotating}
\usepackage{xspace}
\usepackage{color}
\def\Journal#1#2#3#4{{#1} {#2} (#3) #4 }

\def\etal{{\it et al.}}
\def\coll{Collaboration}
\def\colls{Collaborations}

\newcommand{\be}{\begin{equation}}
\newcommand{\ee}{\end{equation}}
\newcommand{\bea}{\begin{eqnarray}}
\newcommand{\eea}{\end{eqnarray}}

\newcommand{\xgobs}{\ensuremath{x_{\gamma}^{obs}\,}}
\newcommand{\xgobsm}{x_{\gamma}^{obs}}
\newcommand\mct{\mbox{$M_c$}}
\newcommand\mcto{\mbox{$M_c^{\rm opt}$}}
\newcommand\mbt{\mbox{$M_b$}}

\newcommand{\ftb}{$F_2^{b\bar{b}}\, $}
\newcommand{\ftq}{$F_2^{Q\bar{Q}}\, $}

\newcommand{\flq}{$F_L^{Q\bar{Q}}\, $}

\newcommand{\fzq}{$F_3^{Q\bar{Q}}\, $}
\newcommand{\redc}{$\sigma_{\rm red}^{c\bar{c}}\, $}

\newcommand{\redq}{$\sigma_{\rm red}^{Q\bar{Q}}\, $}
\newcommand{\ft}{$F_2\, $}
\newcommand{\redi}{$\sigma_{\rm red}\, $}

\newcommand{\dstarp}{$D^{*+}\,$}
\newcommand{\dst}{$D^{*}\,$}

\newcommand{\dzero}{$D^{0}\,$}

\newcommand{\dplusp}{$D^{+}\,$}

\newcommand{\epem}{$e^+e^-\,$}
\newcommand{\Rud}{$R_{u/d}\,$}
\newcommand{\Pdv}{$P^d_{V}\,$}
\newcommand{\gs}{$\gamma_s\,$}
\newcommand{\bmp}[2]{\begin{minipage}[#1]{#2}}
\newcommand{\emp}{\end{minipage}}
\newcommand{\donec}{$D_1(2420)^+\,$}
\newcommand{\donez}{$D_1(2420)^0\,$}

\newcommand{\donecz}{$D_1(2420)^{0,+}\,$}
\newcommand{\dtwocz}{$D_2^*(2460)^{0,+}\,$}
\newcommand{\dsone}{$D_{s1}(2536)^{+}\,$}
\newcommand{\ptrel}{$p_T^{\rm rel}\,$}
\newcommand{\herai}{\mbox{HERA~I}\xspace}
\newcommand{\heraii}{\mbox{HERA~II}\xspace}
\newcommand{\mvtx}{m_{\mathrm{vtx}}}
\newcommand{\calo}{\mathcal{O}}
\newcommand{\gp}{\gamma p}
\newcommand{\etjet}{E^{\mathrm{jet}}_{T}}
\newcommand{\etajet}{\eta^{\mathrm{jet}}}

\newcommand{\rnge}{\hbox{$\,\textnormal{--}\,$}}
\newcommand{\fig}[1]{Fig.~\ref{fig:#1}}

\newcommand{\Fig}[1]{Figure~\ref{fig:#1}}

\newcommand{\tab}[1]{Table~\ref{tab:#1}}
\newcommand{\Tab}[1]{Table~\ref{tab:#1}}
\newcommand{\Taband}[2]{Tables~\ref{tab:#1} and~\ref{tab:#2}}
\newcommand{\Sect}[1]{Section~\ref{sect:#1}}
\newcommand{\Sectand}[2]{Sections~\ref{sect:#1} and~\ref{sect:#2}}
\newcommand{\eq}[1]{Eq.~(\ref{eq:#1})}
\newcommand{\Eq}[1]{Equation~(\ref{eq:#1})}
\newcommand{\pcite}[1]{{\protect\cite{#1}}}
\newcommand{\mum}{\,\mu\textnormal{m}}
\newcommand{\mm}{\,\textnormal{mm}}
\newcommand{\cm}{\,\textnormal{cm}}
\newcommand{\m}{\,\textnormal{m}}
\newcommand{\km}{\,\textnormal{km}}

\newcommand{\pbi}{\,\textnormal{pb}^{-1}}
\newcommand{\fbi}{\,\textnormal{fb}^{-1}}

\newcommand{\fb}{\,\textnormal{fb}}
\newcommand{\Hz}{\,\textnormal{Hz}}
\newcommand{\MHz}{\,\textnormal{MHz}}
\newcommand{\eVdist}{\kern-0.06667em}

\newcommand{\Gev}{{\textnormal{Ge}\eVdist\textnormal{V\/}}}

\newcommand{\mev}{{\,\textnormal{Me}\eVdist\textnormal{V\/}}}
\newcommand{\gev}{{\,\textnormal{Ge}\eVdist\textnormal{V\/}}}
\newcommand{\tev}{{\,\textnormal{Te}\eVdist\textnormal{V\/}}}

\newcommand {\lapprox}
   {\raisebox{-0.7ex}{$\stackrel {\textstyle<}{\sim}$}}
\def\gsim{\,\lower.25ex\hbox{$\scriptstyle\sim$}\kern-1.30ex%
\raise 0.55ex\hbox{$\scriptstyle >$}\,}
\def\lsim{\,\lower.25ex\hbox{$\scriptstyle\sim$}\kern-1.30ex%
\raise 0.55ex\hbox{$\scriptstyle <$}\,}

\graphicspath{{figures/}}

\begin{document}

\title{
{\small \tt DESY-15-085}\hspace*{15cm} \\
 \vspace{1cm} Charm, Beauty and Top at HERA}
\author{O.~Behnke, A.~Geiser, M.~Lisovyi$^{*}$,
\\
DESY, Hamburg, Germany \\
{\small $^{*}$ now at Physikalisches Institut, Universit\"at Heidelberg, Heidelberg, Germany} \\
}

\maketitle

\begin{abstract} Results on open charm and beauty production and on the search 
for top production in high-energy electron-proton collisions at HERA are 
reviewed. 
This includes a discussion of relevant theoretical aspects, 
a summary of the available measurements and measurement techniques, 
and their impact on improved understanding of QCD and its parameters, 
such as parton density functions and charm- and beauty-quark masses. 
The impact of these results on measurements at the LHC and elsewhere 
is also addressed.
\end{abstract}

\tableofcontents

\newpage

% Introduction
\newpage
\section{Introduction}

\begin{figure}[h]
     \hspace{0.6 cm}
\centering
\includegraphics[width=0.28\linewidth]{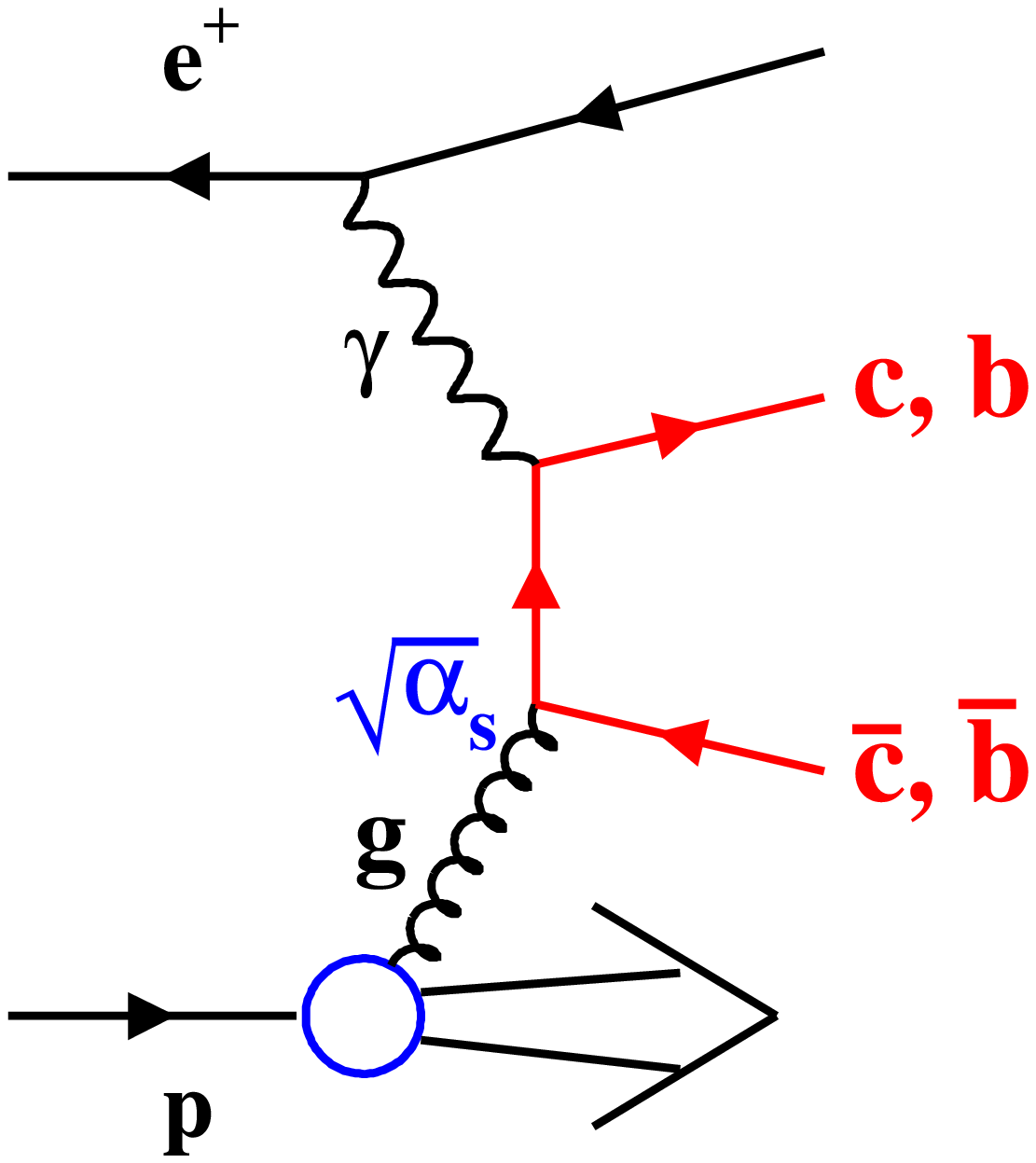}
\hspace{3cm}
\includegraphics[width=0.28\linewidth]{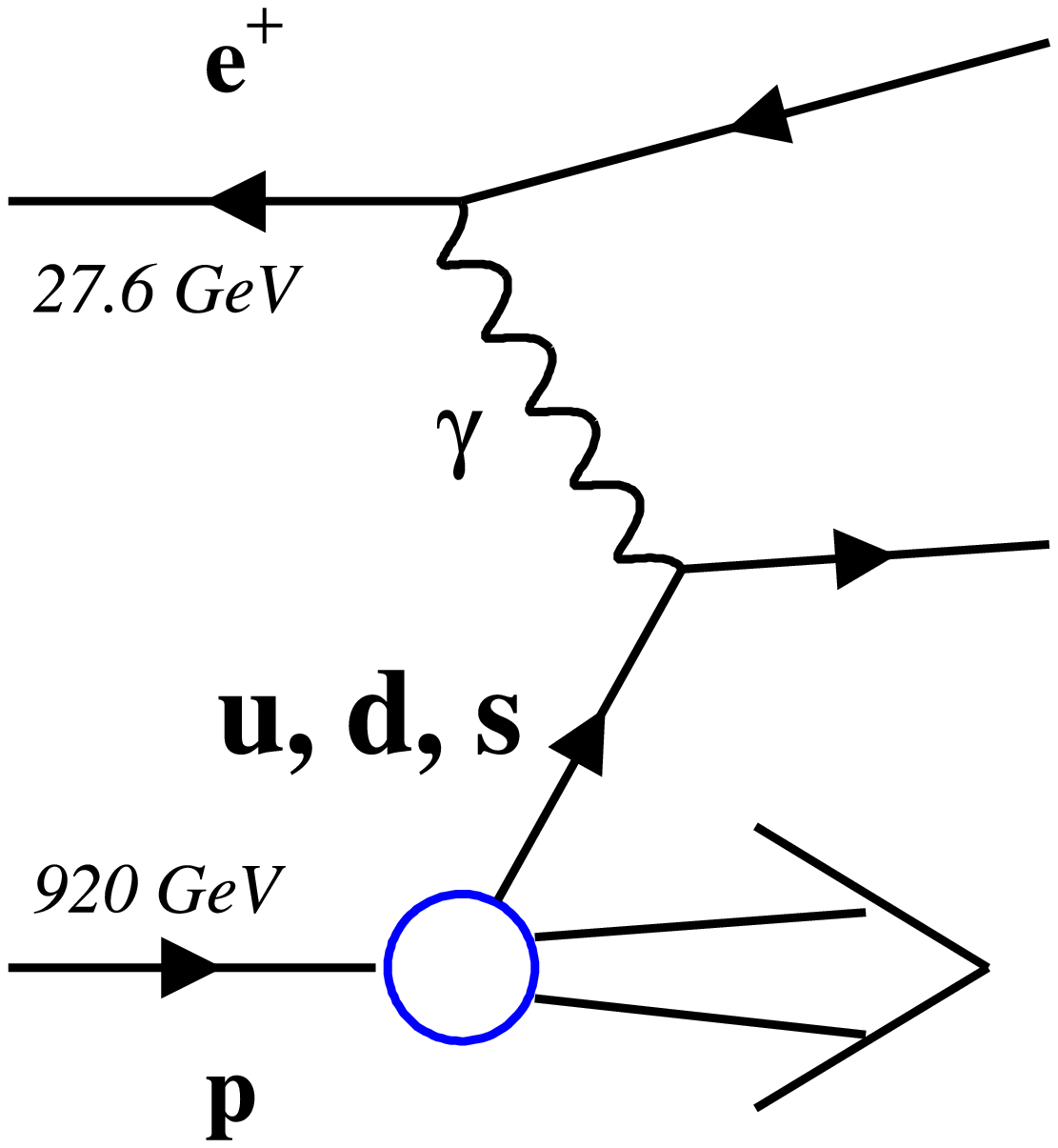}
\\
\vspace{-5cm}
(a) \hspace{8cm} (b) \hspace{6cm}
\vspace{5cm}
\caption{ (a) The dominant production process for
charm and beauty quarks in $ep$ collisions
at HERA, the boson-gluon fusion (BGF) reaction.
(b) The simplest Quark-Parton-Model diagram for deeply inelastic scattering on a light quark. 
}
\label{fig:intro1}
\end{figure}

HERA was the first and so far only high energy electron\footnote{Throughout this document, the term ``electron'' includes positrons, 
unless explicitly stated otherwise.}-proton collider.
The production of heavy-quark final states in deeply inelastic scattering (DIS)
and photoproduction ($\gamma p$) from $ep$ interactions at HERA 
(\fig{intro1}) 
originally was \cite{HERAHQ} and still is (this review)
one of the main topics of interest of HERA-related physics, and of
Quantum Chromodynamics (QCD) in general.

%
%%%%%%%%%%%%%%%%%%%%%%%%%%%%%%%%%%%%%%%%%%%%%%%%%%%%%%%%%%%%%%%%%%%%%%%
%
% BGF process and QPM scattering
%
%
%
%
%/////////////////////////////////////////////////////////////////////
%

A quark is defined to be ``heavy'' if its mass is significantly 
larger than the QCD scale parameter $\Lambda_{QCD} \sim 250$~MeV.
The heavy quarks 
kinematically accessible at HERA are the charm and beauty quarks, 
which are the main topic of this review.
At the time of the proposal of the HERA collider and experiments in the
1980's \cite{HERAprop}, a search for the top quark was one of the
major goals \cite{HERAtop}. 
This influenced parts of the detector design: if at all, top 
quarks would be produced boosted into the proton direction, and top-quark 
mass reconstruction from hadronic final states would profit from an excellent
hadronic energy resolution.  
As we know today, top-quark pair production was out of the 
kinematic reach of the HERA collider. Single top-quark production is
kinematically possible, but strongly suppressed by Standard Model couplings.
This allows the search for non-Standard Model top-production processes which 
will be covered in Section \ref{sect:top}.

Charm production at HERA, in particular in deeply inelastic scattering, was 
realised from very early on to be of particular interest for the understanding of QCD \cite{HERAHQ,HERAHQ91}. 
Up to one third of the HERA cross section is 
expected to originate from processes with charm quarks 
in the final state: assuming ``democratic'' 
contributions from all quark flavours, which is a reasonable assumption 
at very high momentum transfers, this fraction $f(c)$ can be approximated
by the ratio of photon couplings in Fig. \ref{fig:intro1}, which
are proportional to the square of the charges $Q_q,~q=u,d,s,c,b$ of the 
kinematically accessible quark flavours:
\begin{equation} \label{eq:cfrac}
f(c) \sim \frac{Q_c^2}{Q_d^2+Q_u^2+Q_s^2+Q_c^2+Q_b^2}
%        = \frac{\frac{4}{9}}{\frac{1+4+1+4+1}{9}}
        = \frac{4}{11} \simeq 0.36, 
\end{equation}
while a similar approximation for beauty yields 
$ f(b) \sim \frac{1}{11} \simeq 0.09 $.
In general, the impact of beauty on inclusive cross sections at HERA is thus 
smaller than the impact of charm.

\begin{figure}[h]
\centering
\includegraphics[width=0.3\linewidth]{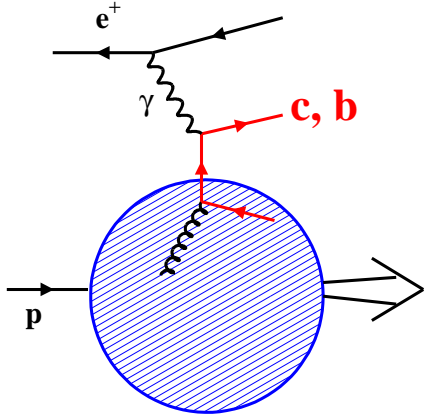}
\caption{Quark-parton-model view of heavy flavour production
in $ep$ collisions at HERA.
}
\label{fig:intro6}
\end{figure}

At momentum transfers large enough for these approximations to be meaningful, 
charm and beauty can be treated as an integral part
of the ``quark-antiquark sea'' inside the proton (Fig. \ref{fig:intro6}), 
similar to the light quarks 
in Fig. \ref{fig:intro1}b, originating from the initial state splitting
of virtual gluons. Since the proton has no net charm and beauty flavour number,
charm and beauty quarks
in the proton can only arise in pairs of quarks and anti-quarks 
(Fig. \ref{fig:intro1}(a) and shaded part of Fig. \ref{fig:intro6}). 
However, due 
to the large charm- and beauty-quark masses of about $1.5\gev$ and $5\gev$, 
respectively, such a pair is considerably
heavier than the mass of the proton. From purely kinematic considerations,
it can thus not exist as a ``permanent'' contribution to the proton in the 
low-energy limit. Considerations of so-called ``intrinsic charm'' 
\cite{intrinsic} have been 
challenging this simple point of view. Since there is no evidence 
for such a contribution from HERA data \cite{intrlimit}, 
this will not be pursued further
in this review. Thus, charm and beauty ``in the proton'', as depicted in Fig. \ref{fig:intro6}, are always
considered to be virtual, and to arise as fluctuations from the perturbative 
splitting of gluons inside the proton. This establishes heavy quark production
as a primary probe of the gluon content of the proton. 

%%%%%%%%%%%%%%%%%%%%%%%%%%%%%%%%%%%%%%%%%%%%%%%%%%%%%%%%%
% 
% massless scheme diagram(s) 
%
%
%
%/////////////////////////////////////////////////////////
%

\begin{figure}[htb]
%     \hspace{0.6 cm}
\centering
%\hspace{-4mm}
\includegraphics[width=0.4\linewidth]{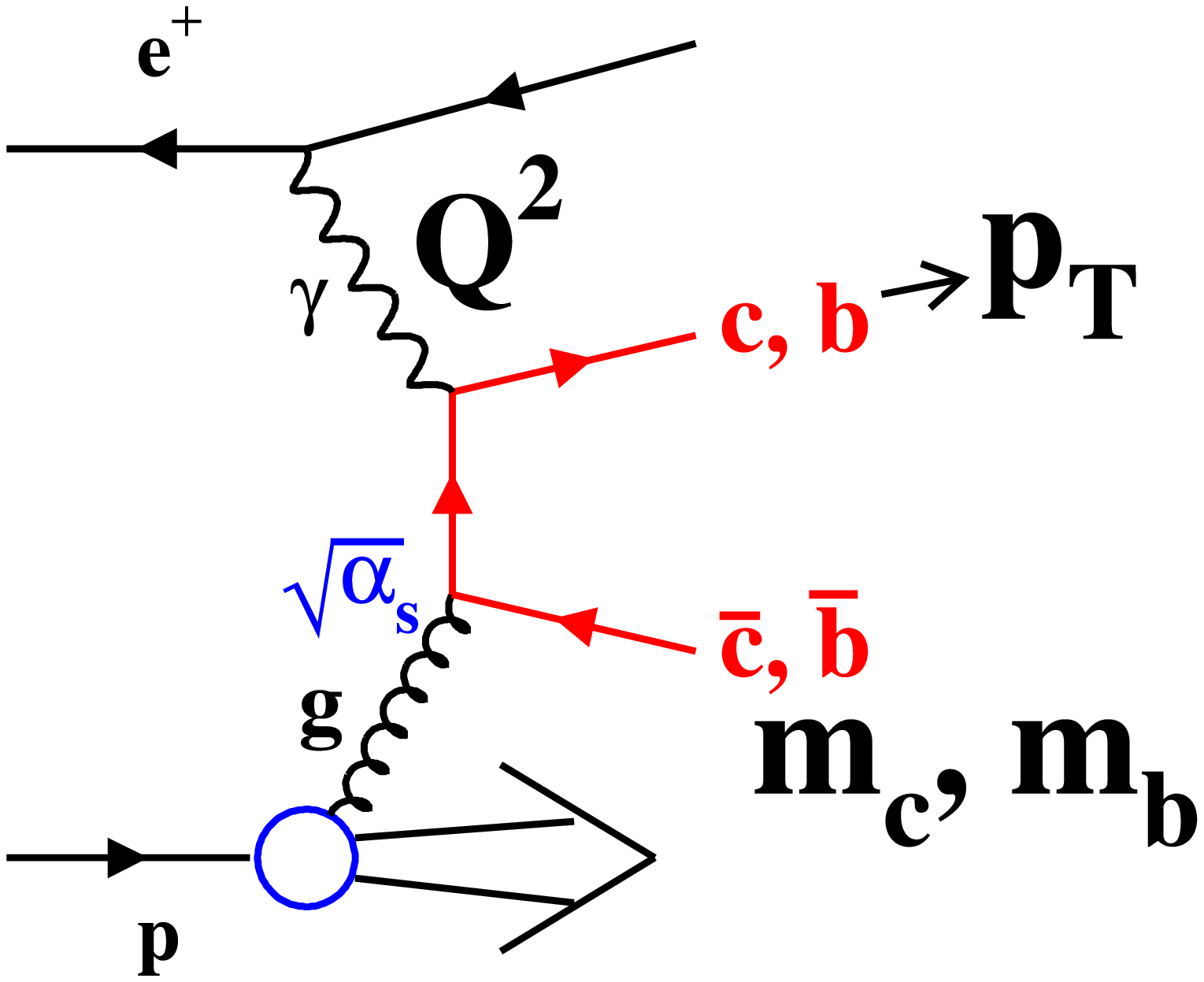}
\caption{Possible hard scales in the boson-gluon-fusion 
process. % {Behnkereview}. 
For the explanation of the symbols see text.}
\label{fig:intro5}
\end{figure}

Different approaches to the theoretical treatment of charm- and 
beauty-quark production at HERA are discussed in Section \ref{sect:theory}. 
All these treatments crucially make use 
of the fact that the heavy-quark mass acts as a kinematic cut-off parameter
in most of the QCD processes in which heavy quarks occur. 
Furthermore, the fact that
the heavy-quark mass is ``large'' compared to the QCD scale 
$\Lambda_{QCD} \sim 0.25$ GeV allows the usage of this mass as a 
``hard scale'' in QCD perturbation theory, appropriately taking into account 
quark mass effects in perturbative calculations (Fig. \ref{fig:intro5}). 
On the other hand, the ``smallness'' in particular of the charm-quark mass 
with respect to other scales appearing in the perturbative 
expansion, such as the virtuality of the photon, $Q^2$, or the transverse 
momentum of a jet or a quark, $p_T$, 
can give rise to potentially large logarithmic corrections, e.g. of the form 
\begin{equation}
\sim [\alpha_s \ln(p^2_{T}/m_Q^2)]^n \quad \mbox{or} 
\quad
\sim [\alpha_s \ln(Q^2/m_Q^2)]^n
\label{eq:intro1}
\end{equation}
where $n$ is the order of the logarithmic expansion, $\alpha_s$ is the 
strong coupling constant, and $m_{Q,~Q=c,b}$ is the heavy-quark mass.
The size and treatment of these corrections is one of the issues to be 
investigated.

%
%%%%%%%%%%%%%%%%%%%%%%%%%%%%%%%%%%%%%%%%%%%%%%%%%%%%%%%%%%%%%%%%
%
% Hard scales in BGF process
%
%
%//////////////////////////////////////////////////////////////////////
%

During the lifetime of the HERA collider, of order $10^9$ charm and $10^7$
beauty events should have been produced in the H1 and ZEUS detectors, 
of which $\calo(10\%)$
have been recorded to tape via both inclusive and dedicated triggers.
HERA can thus truly be considered to be a charm factory. 
Furthermore, the fact that charm- and beauty-quark production at HERA can be 
studied essentially over its complete kinematic range, from the $c\bar c$- or 
$b\bar b$-mass threshold up to squared momentum
transfers of order $1000\gev^2$, offers the opportunity to treat HERA as 
a ``QCD laboratory'' to test the different possible theoretical approaches
to heavy-quark production against experimental data.
Many such tests are presented in 
Sections \ref{sect:charmphoto} and \ref{sect:cbDIS}.

In particular, the charm-production measurements can be used to constrain 
important QCD parameters, such as the charm-quark mass and its running,
and has important consequences for the determination of other parameters like 
the QCD strong coupling constant, $\alpha_s$. The measurements can also
be used to constrain the charm fragmentation parameters, 
to constrain the flavour composition of virtual quarks in the proton,
and to determine or cross-check the gluon distribution inside the proton.
Such measurements and results are discussed in \Sect{QCD}.    
Finally, the HERA charm results have a significant impact on 
measurements and theoretical predictions for many QCD-related
processes at hadron colliders, such as the LHC. For instance, the 
resulting constraints on the flavour composition of quarks in the proton 
reduce the uncertainties of the 
LHC $W$- and $Z$-production cross sections, and the constraints on the gluon 
content of the proton are an important ingredient for the determination 
of the Higgs Yukawa coupling to top quarks from the dominant 
gluon-gluon-fusion Higgs-production process via an intermediate
top quark loop.
Such cross-correlations are also discussed 
in detail in \Sect{QCD}.

Beauty production at HERA (Sections \ref{sect:beautyphoto} and 
\ref{sect:cbDIS}) offers further complementary 
insight into the theoretical intricacies of heavy-flavour production 
in QCD. Due to its higher mass ($m_b \sim 5\gev$) and a correspondingly 
smaller value of the strong coupling constant, its perturbative QCD behaviour 
is somewhat better than the one of charm. However, the beauty mass 
remains non-negligible over essentially the full accessible phase space of 
HERA, and a large fraction of the cross section is close to the kinematic
$b\bar b$-mass threshold. This offers a particularly sensitive
handle on the treatment of mass effects in QCD but also
requires a particularly careful treatment of these mass effects in order 
to obtain reliable predictions.
Finally, the coupling of the photon to $b$ quarks is four times smaller 
than the coupling to charm quarks (Eq. (\ref{eq:cfrac})), 
and the higher $b$-quark mass yields a 
strong kinematic suppression. Therefore in practice, depending on the region 
of phase space probed, the $b$-production 
cross section at HERA is about $1\rnge2$ orders of magnitude smaller than the 
cross section 
for charm production. This makes separation from the background and 
accumulation of a significant amount of statistics experimentally much 
more challenging.
Also, the experimental analyses of beauty are often not fully separable 
from those of charm production. One of the highlights is the measurement of the 
beauty-quark mass (Section \ref{sect:QCD}).
Others are the measurement of the total beauty-production cross section 
at HERA (Section \ref{sect:beautyphoto}), and the potential impact of HERA
measurements on $b$-quark-initiated production processes at the LHC
(Section \ref{sect:QCD}). 

Last but not least, most of the results presented depend on a good 
understanding of the performance of the HERA machine and the HERA detectors,
as well as on mastering the heavy-flavour detection techniques.
Unfortunately, only a small fraction of the original data can make it through
the various event filtering and reconstruction stages.  
These aspects will be addressed in Sections \ref{sect:experiments} and 
\ref{sect:Tagging}.

Except for the shortest ones, each section will start with a brief
introduction and 
close with a summary, such that a reader less interested in the 
details may decide to skip the reading of the more detailed parts of the 
section. 

Some of the material in this review has been adapted from an earlier 
unpublished review \cite{Behnkereview} of one of the authors.
Further complementary information, in particular on charmonium and 
bottomonium production or diffractive charm production, which are not covered 
by this review, is available 
elsewhere \cite{Meyer:2005au, Quarkonium, Kramer:2001hh,Wolf, Wing}.
The broader context of other physics topics can be explored in a more 
general review on collider physics at HERA \cite{Klein}.

\newpage

% HQ theory
\section{Theory of heavy-flavour production at HERA}
\label{sect:theory}

This section describes the different theoretical approaches to charm 
and beauty cross-section predictions, which will be needed later in the 
discussion of
\begin{itemize} 
\item the Monte-Carlo (MC) based acceptance corrections for the data sets 
used to obtain cross sections;
\item the extrapolation of different measurements to a 
common phase space, such that they can be compared or combined;
\item the comparison of different theory predictions to the measured
cross sections;
\item the parton-density fits including the heavy-flavour data; 
\item the fits of the charm and beauty masses and their running.
\end{itemize}
Since there is a large overlap between the theoretical approaches for these 
different purposes they will be discusssed
in a common framework in the following.  

\subsection{HERA kinematic variables and phase space}
\label{sect:kinvar}
\begin{figure}[h]
     \hspace{0.6 cm}
\centering
\includegraphics[width=0.4\linewidth]{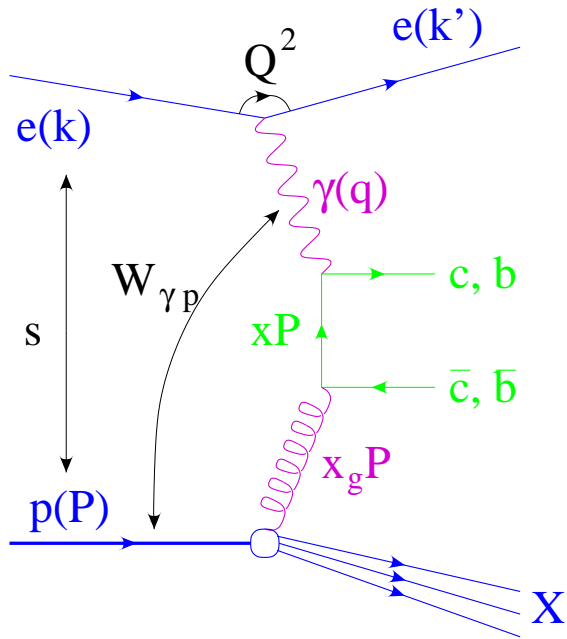}
\caption{
Illustration of event kinematic variables
for $ep$ scattering at HERA with heavy-quark production
via the boson-gluon-fusion process.}
\vspace{-4mm}
\label{fig:t1}
\end{figure}

The measurements of
heavy-quark production at HERA have
been restricted, for statistical reasons, to neutral current events 
(exchange of a neutral boson) and to the kinematic
region of the negative four-momentum transfer squared 
$Q^2\ \lapprox\ 2000\gev^2$, where 
photon exchange dominates and $Z^0$ exchange can be neglected.
Figure~\ref{fig:t1} illustrates the 
event kinematic variables for $ep$ 
scattering with heavy-quark production 
via the boson (i.e. photon) gluon fusion process 
(see also Fig. \ref{fig:intro1}).
%
%%%%%%%%%%%%%%%%%%%%%%%%%%%%%%%%%%%%%%%%%%%%%%%%%%%%%%%%%%%%%%%%%%%%%%
%
% Intro of kinematic variables
%
%
%
%/////////////////////////////////////////////////////////////////
%
%

The four-momenta of the incoming electron $k$, the 
outgoing electron $k'$ and the proton $P$ can be used to define
the following Lorentz-invariant variables:
\begin{eqnarray} \label{eq:kine}
s    &  = &  (k+P)^2 \\
Q^2  &  = &  -q^2 = -(k-k')^2 \\
x    &  = & \frac{Q^2}{2P\cdot q}\\
y    &  = & \frac{P \cdot q}{P \cdot k} \\
W_{\gamma p}^2 &  = & (P+q)^2 
\end{eqnarray}
Here $\sqrt{s}$ is the centre-of-mass energy of the $ep$ system
and $Q^2$ is the photon virtuality.
$W_{\gamma p}$ is the centre-of-mass 
energy of the $\gamma^{(*)} p$ system.
In the simple Quark Parton Model \cite{QPM} (QPM) the
Bjorken scaling variable $x$ describes the proton momentum fraction carried by
the scattered parton (Figs. \ref{fig:intro1}(b) and \ref{fig:intro6}). 
The inelasticity, $y$, gives the fraction of the electron energy
taken by the photon in the proton rest frame.
Only three of these five kinematic variables are independent.
Neglecting the masses of the electron and the proton the following
relations between these quantities hold:
\begin{eqnarray} \label{eq:kine2}
Q^2 & = & s \cdot x \cdot y \label{eq:Qxys}\\
W_{\gamma p}^2 & = &  y \cdot s - Q^2 \label{eq:WysQ}
\end{eqnarray}
In the full QCD case this picture becomes more complicated, as illustrated in Fig. \ref{fig:t1},
where the proton momentum fraction $x_g$ carried by the gluon does not coincide any longer with
Bjorken $x$. However, Eqs. (\ref{eq:kine}) -- (\ref{eq:WysQ}) remain mathematically valid.

The $ep$ scattering events are classified by the photon virtuality
$Q^2$. 
The regime of small $Q^2\approx 0\;\mbox{GeV}^2$ 
is called  photoproduction ($\gamma p$)
and the regime $Q^2\gsim 1\;\mbox{GeV}^2$ is called
Deeply Inelastic Scattering (DIS).

More details on inclusive DIS results and proton structure can be found 
elsewhere \cite{DISreview,Bluemlein,DeRoeck,Engelen,Cooper}.

\subsection{Perturbative QCD calculations}
\label{sect:perturbative}

\begin{figure}
\centering
\includegraphics[width=0.5\linewidth]{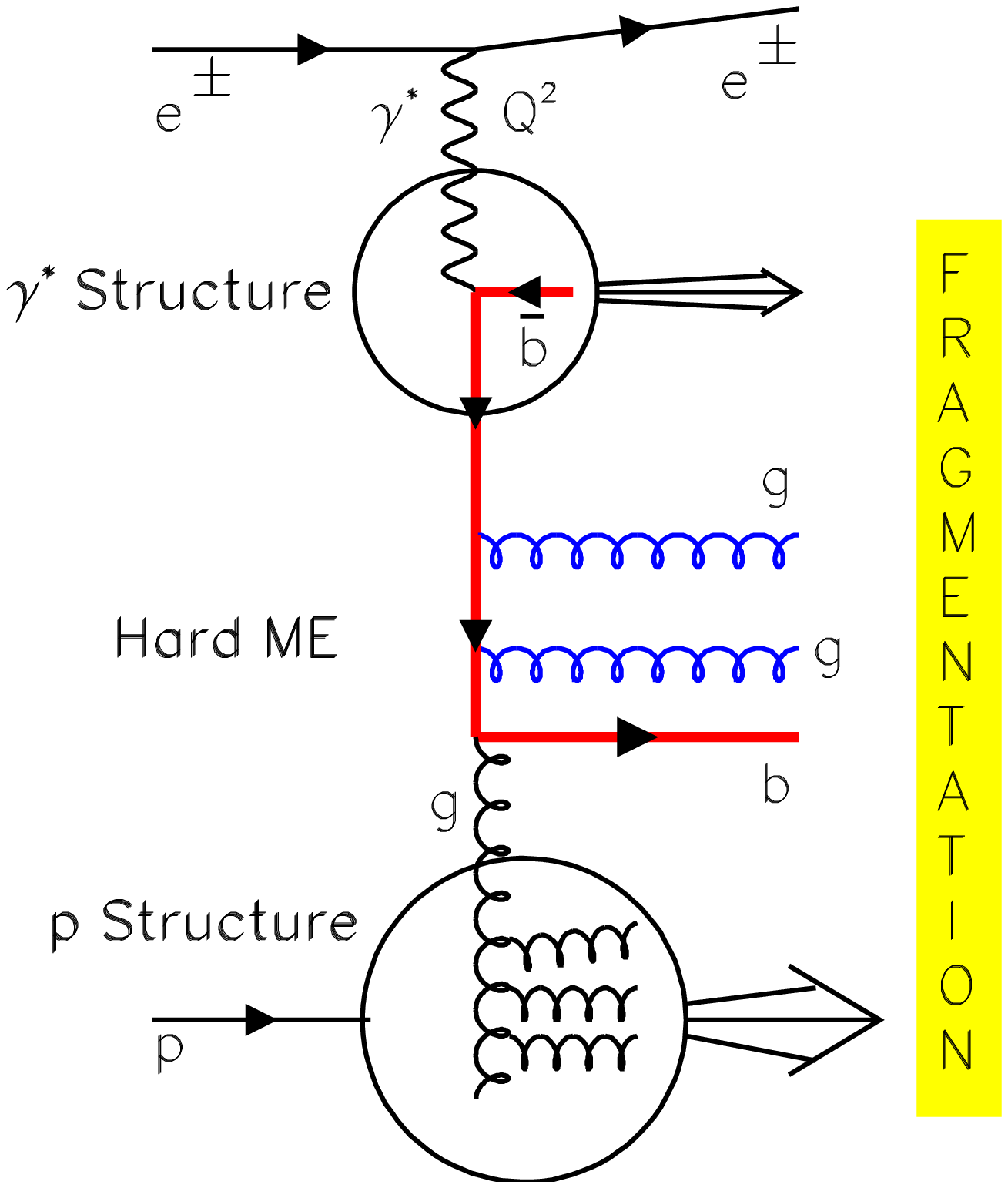}
\caption{Example for the factorisation of heavy-flavour production in QCD 
into proton structure, photon structure, hard matrix element and
fragmentation.} 
\label{fig:t2}
\end{figure}

In fixed-order perturbative QCD (pQCD) the calculation of any parton-level 
cross section in 
$ep$, $\gamma p$, $\bar pp$ or $pp$ collisions can be expressed as 
\begin{equation}
\label{eq:factorization}
\sigma(ab) = \int dx_a dx_b f^a_{p_a}(x_a,\mu_a) f^b_{p_b}(x_b,\mu_F)
\hat \sigma_{p_a p_b}(x_a P_a,x_b P_b,\mu_a,\mu_F,\alpha_s(\mu_R))
\end{equation}
where $a = e, \gamma, \bar p$ or $p$ is one incoming beam particle, 
and the other, $b$, is a proton. $p_a$ is a ``parton'' taken from $a$,
e.g. the electron or photon itself (DIS and real photons), a slightly virtual 
photon radiated from the electron (photoproduction), or a gluon or 
quark from the structure 
of a real photon or (anti)proton. $p_b$ is a parton taken from the proton, 
i.e. a gluon or quark.
$x_a$ and $x_b$ represent the respective momentum fractions of these partons 
with respect to their ``parent'' momenta $P_a$ and $P_b$. Note that these 
correspond to Bjorken $x$ and $y$
as defined in the previous section\footnote{and using the improved 
Weizs\"acker-Williams approximation \cite{Weizsaecker} in the case of $y$.} 
only
in the case of the quark-parton-model approximation to deeply inelastic 
$ep$ scattering, while they have a different meaning in other cases.  
For example, in Fig. \ref{fig:t1}, the quantity $x_g$ (rather than $x$) 
corresponds to $x_b$ 
as defined in Eq. (\ref{eq:factorization}), while $x_a = y$ in the 
photoproduction interpretation, 
and $x_a=1$ in the hard DIS interpretation.
$f^a_{p_a}$ and 
$f^b_{p_b}$ are the probability density functions, or parton density functions 
(PDFs), which give the probability, e.g. in the $f_{p_b}^b$ case, to find a parton of type $p_b$ with 
momentum fraction $x_b$ in a proton.
$\hat \sigma_{p_a p_b}$ represents the cross section for the partonic 
hard scattering reaction. This is sometimes split into the so-called 
hard process, i.e. the part of the reaction with the highest momentum 
transfer, and so-called initial state (i.e. occurring before the hard process)
or final state (i.e. occurring after the hard process) radiation
(Fig. \ref{fig:t2}).
Part of the initial state radiation can also be absorbed into the parton 
density definition.
Three energy scales $\mu_a$, $\mu_F$ and $\mu_R$ appear in the expansion
given by Eq. (\ref{eq:factorization}), 
which is also called factorisation,
because the cross section is separated into semi-independent factors.

%
%%%%%%%%%%%%%%%%%%%%%%%%%%%%%%%%%%%%%%%%%%%%%%%%%%%%%%%%%%
%
% Factorisation
%
%
%
%/////////////////////////////////////////////////////////
%
%

The renormalisation scale $\mu_R$ determines the scale at which the value of 
the strong coupling constant $\alpha_s$ is evaluated, i.e. it is the 
reference point around which the perturbative Taylor expansion of 
QCD matrix element (ME) calculations is performed.
If the expansion is done to all orders in $\alpha_s$, the result does 
not depend on the choice of this scale. After truncation of the series to 
finite order, the neglected higher-order corrections arising from the 
contribution of a particular subprocess are minimised 
if this scale is chosen to be close to the physical scale of the 
momentum transfer in this subprocess. 
Since at high enough perturbative order there are 
always different subprocesses with differing physical scales (see e.g. Figs.
\ref{fig:intro5}, \ref{fig:t2}), no single scale 
choice can universally cover all such scales. The
variation of the cross section with respect to a variation of the 
renormalisation
scale is used to estimate the uncertainty due to the finite-order truncation 
of this perturbative series. Some further aspects concerning the choice of this
scale are discussed in \Sect{scale}.

The factorisation scale $\mu_F=\mu_b$ determines at which scale the proton 
PDFs are evaluated. By default, any initial state radiation 
(lower blob in Fig. \ref{fig:t2}) with a momentum transfer smaller than the 
factorisation scale will be absorbed into the (usually collinear) PDF 
definition. In contrast, any 
initial state radiation with a momentum transfer larger than this scale,
and all final state radiation down to the fragmentation scale 
(see \Sect{frag}), will be considered
as part of the matrix element, with correct (noncollinear) kinematics.
On one hand, the choice of a lower factorisation scale therefore gives a 
more detailed 
description of the initial state radiation kinematics at a given order. 
On the other hand, the explicit treatment of initial state QCD radiation in the matrix 
element ``uses up'' a power of $\alpha_s$ that would otherwise have been available
for a real radiation elsewhere in the process, or for a virtual correction.
This effectively reduces the order of the calculation with respect to the case
where the same radiation is absorbed into the PDF definition, and therefore 
reduces the overall accuracy of the calculation. Empirically, choosing a 
factorisation scale equal to or at least similar to the renormalisation
scale has been found to be a good compromise. 

The third scale, $\mu_a$, is conceptually the same as $\mu_F$ in the 
$\bar p$, $p$ and resolved $\gamma$ cases, and therefore taken to be equal to it, 
while in the electron and direct photon case 
it is the scale at which the electromagnetic coupling $\alpha$ is evaluated for the electromagnetic 
part of the matrix element
(see \Sect{QED}).

\subsection{Heavy-quark production at HERA in ``leading order''}
\label{sect:leading}

In general, the terminology ``leading order'' (LO), next-to-leading order (NLO), etc. for a perturbative QCD expansion  is not unique.
It can either refer to a specific power of the strong coupling constant $\alpha_s$ or to a specific number of loops in the perturbative
expansion of the matrix elements and/or parton splitting functions contributing to a given process. In order to be precise, this 
additional information thus needs to be quoted explicitly.  

\begin{figure}[htbp]
\setlength{\unitlength}{1cm}
\begin{picture}(14.0,5.0)
\put(1.,0.5){\epsfig{file=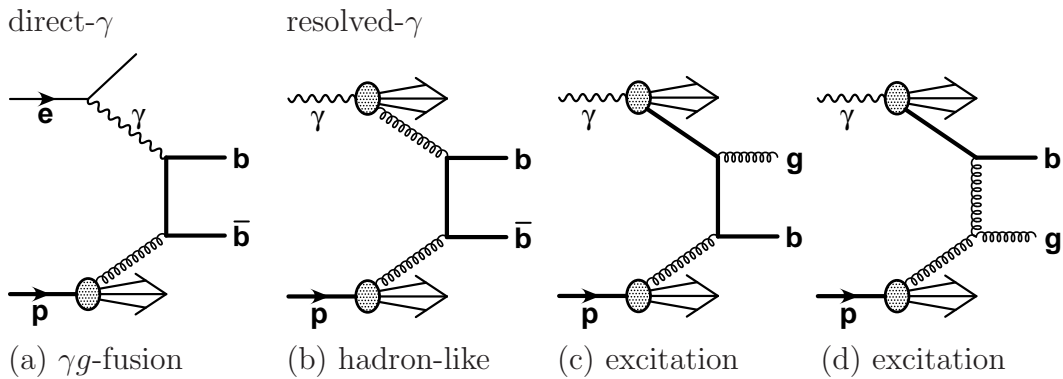,width=14cm}}
\put(1.,4.5){direct-$\gamma$}
\put(4.7,4.5){resolved-$\gamma$}
\put(1.,0.){(a) $\gamma g$-fusion}
\put(4.7,0.){(b) hadron-like}
\put(8.3,0.){(c) excitation}
\put(11.8,0.){(d) excitation}
\end{picture}
\caption{Beauty production processes in leading order (0 loop) QCD 
as implemented e.g. in PYTHIA \cite{PYTHIA}. Note that in (c) and (d) the 
``photon remnant'' (arrow arising from the photon) contains a 
$\bar b$ quark.}
\label{fig:fey1}
\end{figure}

At leading (0-loop) order  in QCD, as implemented in the form 
of tree-level $2\to2$ hard matrix elements in most Monte 
Carlo generators, charm and beauty production in $ep$ collisions 
is dominated by boson-gluon fusion (Fig. \ref{fig:fey1}(a)),
complemented by other diagrams (Figs. \ref{fig:fey1}(b-d)).
Since a $c\bar c$ or $b\bar b$ pair is being produced (collectively referred 
to as $Q\bar Q$), there is a natural lower cut-off 
$2m_Q$ for the mass of the hadronic final state.

In the so-called massless approach (Zero-Mass Variable-Flavour-Number Scheme, 
ZMVFNS), in which the heavy quark mass is set to 0 for the computation of the 
matrix elements and kinematics, this natural cut-off is replaced by an 
artificial cut-off (``flavour threshold'') at $Q^2 \sim m_Q^2$ for deeply 
inelastic scattering, or $p_T \sim m_Q$ for photoproduction.  Below this 
threshold, which is often applied at the level of the factorisation scale, 
the heavy-flavour production cross section is (unphysically) assumed to vanish. 
Above this threshold, heavy quarks are assumed to occur as massless partons
in the proton, like the $u,d,$ and $s$ quarks 
(Figs. \ref{fig:intro1}(b) and \ref{fig:intro6}). 
Except for cases in which both final state charm quarks have large 
transverse momenta $p_T^2 > \mu_F^2$, 
the gluon splitting to $Q\bar Q$ in Fig. \ref{fig:fey1}(a) is thus 
assumed to happen inside the proton, and to be part of the evolution of 
the parton density functions. The running of $\alpha_s$ is calculated
using 3 flavours ($u,d,s$) below the renormalisation scale $m_c$, 
using 4 flavours (including charm) between $m_c$ and $m_b$, and using 5 
flavours above the scale $m_b$.
This results in a quark-parton-model-like scattering of the 
electron off a heavy quark ``in the proton'' (Fig. \ref{fig:intro6}), 
defining the concept of the heavy-quark PDF. 
In this picture the leading-order process is now an $\calo(\alpha_s^0)$ process, 
while the boson-gluon-fusion graph (Fig. 1(a), with both heavy quarks at high 
$p_T$)
is treated as part of the $\calo(\alpha_s)$ next-to-leading order corrections.
This illustrates the partially ambigous meaning of terms like LO, NLO, etc.,
discussed at the start of this subsection.

Higher order corrections can be applied either by explicitly including them into the calculation 
of the matrix elements and/or splitting functions, or, if they are to be applied at tree level only,
by adding an additional so-called parton shower step. In the latter case, also referred to as leading 
order plus leading log parton shower (LO+PS), the outgoing and incoming
partons of the core ``hard'' matrix elements are evolved forward or backwards using splitting 
functions as they are applied during the PDF evolution. Most MCs (e.g. PYTHIA \cite{PYTHIA}, HERWIG \cite{HERWIG}
and RAPGAP \cite{RAPGAP}) use the standard DGLAP evolution, as implemented 
e.g. in JETSET \cite{JETSET}
for this purpose. However, in contrast to the PDF evolution, finite transverse momenta are assigned
to the partons. Some MCs (e.g. ARIADNE \cite{ARIADNE}) use a colour dipole model for this 
evolution, while others use BFKL \cite{BFKL}  or CCFM \cite{CCFM} inspired so-called $k_t$ factorisation 
(e.g. CASCADE \cite{CASCADE}).    
In the context of such MCs, the first diagram in Fig. \ref{fig:fey1} is referred to as direct production or flavour creation,
and the third and fourth are referred to as flavour excitation (in the photon).
Either the second only (e.g. PYTHIA) or 
collectively the last three (e.g. HERWIG) are being referred to as 
resolved photon processes.
The second can uniquely be referred to as a hadron-like resolved-photon process.     

For the explicit generation of heavy-flavour final states in such LO+PS MCs, the boson-gluon-fusion
diagram (Fig. 1(a)) is optionally treated using massive matrix elements, while 
for all other diagrams the massless treatment remains the only available 
option. 

\subsection{Quark-mass definition}
\label{sect:mass}

The heavy-quark masses appear in theoretical QCD calculations in several ways.
Their physical definition arises from their appearance as parameters in the 
QCD Lagrangian. The exact value of the masses depends on the renormalisation
scheme applied. In the $\overline{MS}$ scheme, the masses are defined as 
perturbative
scale-dependent running parameters ($\overline{MS}$ running mass), 
similar to the running strong coupling constant.
In the on-shell mass renormalisation scheme, the masses are defined as the 
poles of the quark propagator (pole mass), similar to the usual definition
of the lepton masses. This is also the definition which one would naively 
expect to enter phase space calculations. 
However, since quarks do not exist as free particles,
and since the definition of the propagator pole inevitably involves 
contributions from the nonperturbative region, the pole mass definition 
has an intrinsic uncertainty of order $\Lambda_{QCD}$ \cite{renormalon}.  
At next-to-leading (one loop) order in perturbation theory, the relation 
between the pole and running mass definitions can be expressed 
as \cite{PDG2014}
\begin{equation}
\label{eq:polerun}
m_Q(m_Q) = m_Q^{pole} (1-\frac{4\alpha_s(m_Q)}{3\pi}), 
\end{equation}
where the running mass has been expressed in terms of its value 
at ``its own scale''. Its scale dependence can be expressed as \cite{PDG2014}
\begin{equation}
\label{eq:run1}
m_Q(\mu) = m_Q(m_Q)(1-\frac{\alpha_s(\mu)}{\pi}\ln\frac{\mu^2}{m_Q^2}),  
\end{equation}
or alternatively as \cite{RUNdec}
\begin{equation}
\label{eq:run2}
m_Q(\mu) = m_Q(m_Q)\frac{(\frac{\alpha_s(\mu)}{\pi})^{\frac{1}{\beta_0}}}
                        {(\frac{\alpha_s(m_Q)}{\pi})^{\frac{1}{\beta_0}}},  
\end{equation}
with $\beta_0=\frac{9}{4}$. Higher order expressions can also be found in the 
quoted references.

At leading (0 loop) order, the difference between the two definitions 
vanishes. Finally, in the context of so-called massless schemes,
the ``mass'' is defined as a kinematic cutoff parameter in certain 
parts of the theory calculations.

The pole-mass definition
has been used in most QCD calculations relevant for this review.
In recent variants of the ABKM \cite{ABKMMSbar} and ACOT \cite{CTEQmass}
schemes, 
the $\overline{MS}$-running-mass definition is used instead.
The latter has the advantage of reducing the sensitivity of the cross sections 
to higher order corrections, and improving the theoretical precision 
of the mass definition \cite{ABKMMSbar}.

\subsection{The zero-mass variable-flavour-number scheme}
\label{sect:massless}

\begin{figure}[h]
\centering
\includegraphics[width=0.25\linewidth]{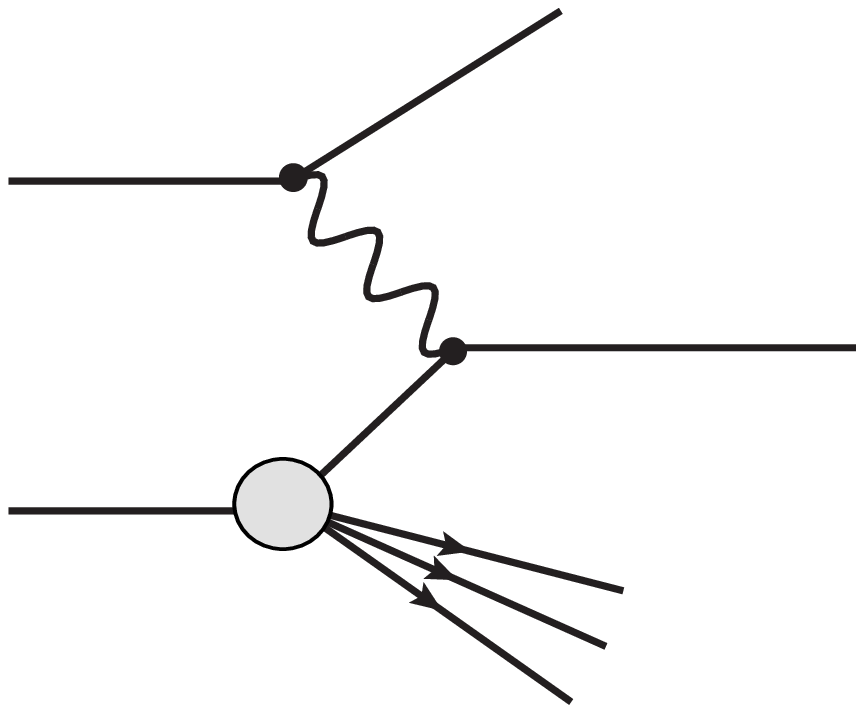}
\includegraphics[width=0.25\linewidth]{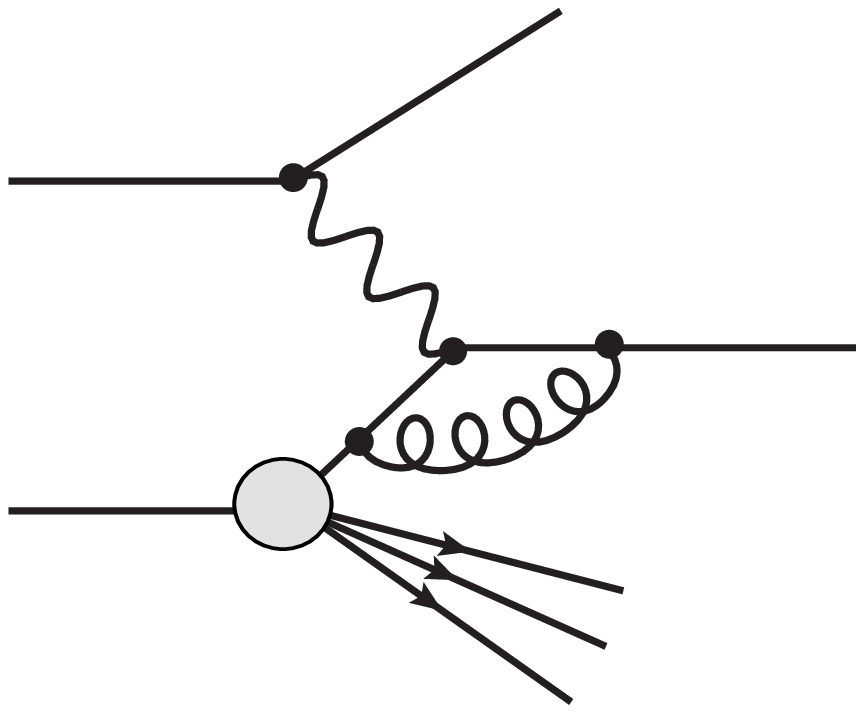}
\includegraphics[width=0.2\linewidth]{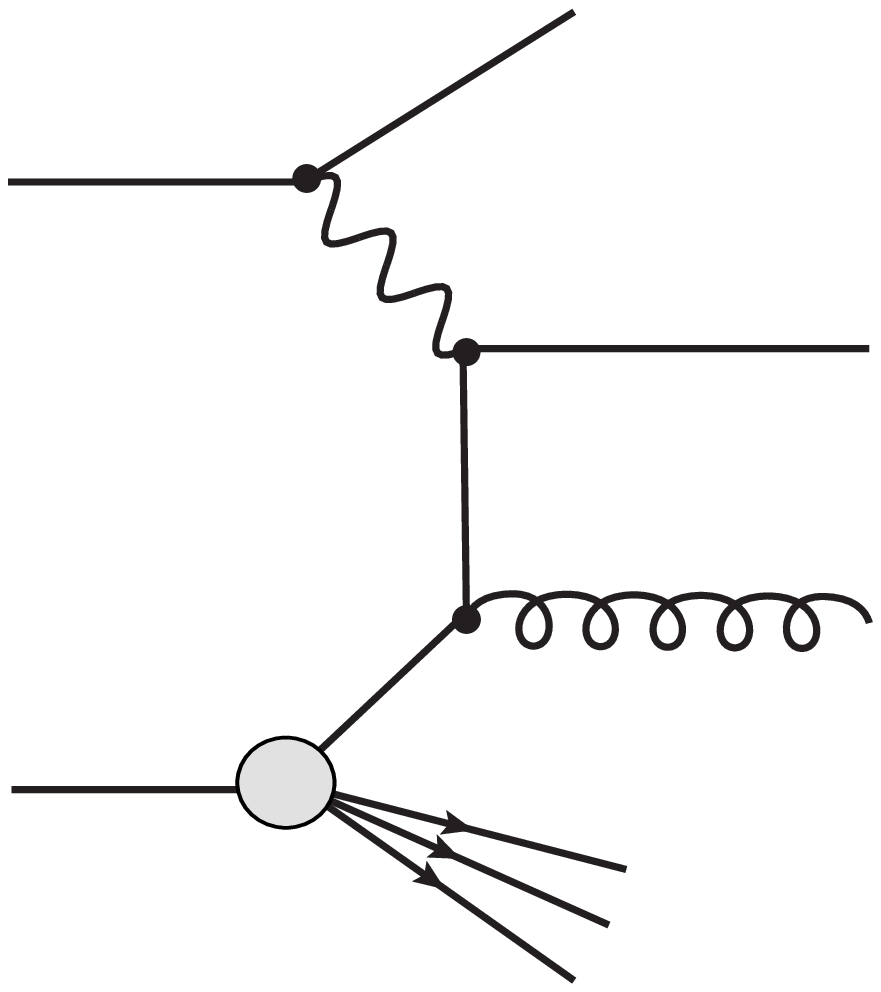}
\includegraphics[width=0.2\linewidth]{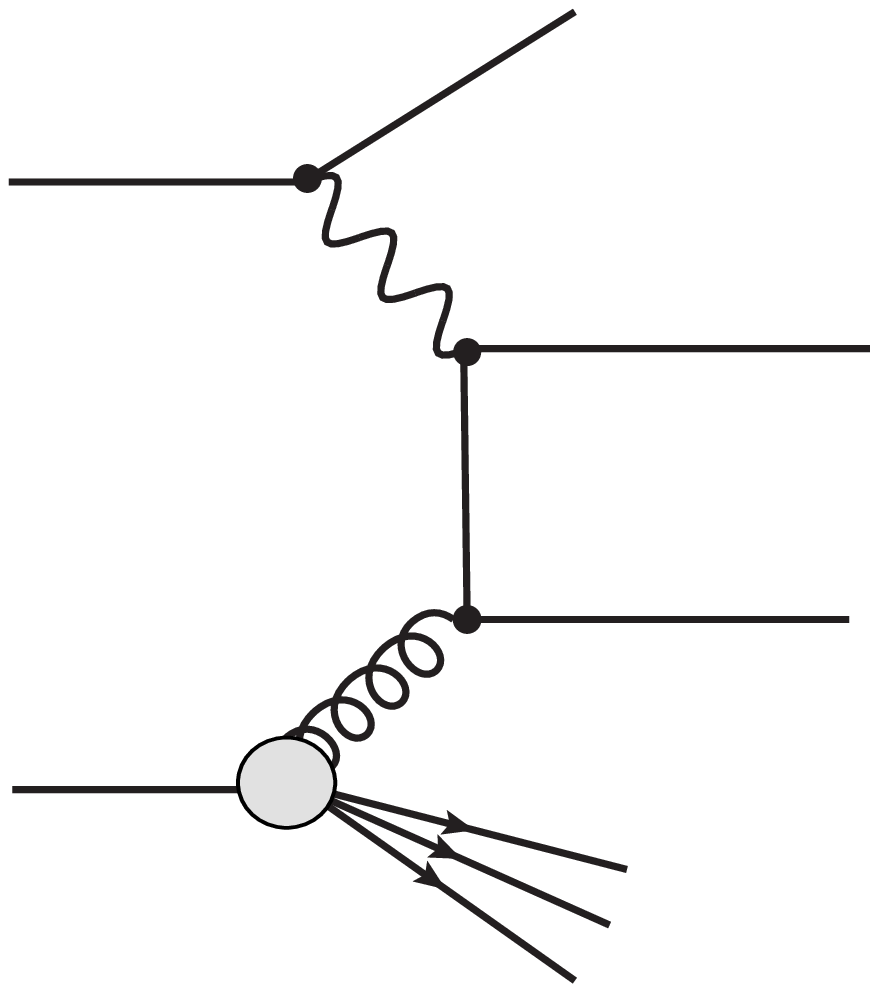}
\hspace{3cm} (a) \hspace{3.5cm} (b) \hspace{3.5cm} (c) \hspace{3.5cm} (d)
\hspace{0.5cm}
\caption{Leading order ($\calo(\alpha_s^0)$) (a) and selection of next to leading 
order ($\calo(\alpha_s^1)$) (b)--(d) 
processes for heavy flavour production in DIS in the massless 
scheme. 
For (b), only the interference term with (a) contributes at this order.}
\label{fig:intro7}
\end{figure}

In its ``NLO'' variant,
including one-loop virtual corrections (Fig. \ref{fig:intro7})(b)), 
the ZMVFNS has been used for most\footnote{
MRST98 \cite{mrst} is a notable early exception.} 
NLO variable-flavour parton-density fits up to a few years ago, such as
CTEQ6M \cite{cteq6M}, 
ZEUS-S \cite{ZEUSS}, H1 \cite{H1PDF}, 
NNPDF2.0 \cite{NNPDF20}. 

%%%%%%%%%%%%%%%%%%%%%%%%%%%%%%%%%%%%%%%%%%%%%%%%%%%%%%%%%
% 
% massless scheme diagrams 
%
%
%
%/////////////////////////////////////////////////////////
%

One of its advantages is that e.g. next-to-leading-log (NLL) resummation of 
terms proportional to $log(Q^2/m_Q^2)$ 
can be applied to all orders, avoiding
the problem that such logs could spoil the convergence of the perturbation
series at high momentum transfers. 
However, it is clear that this simplified approach can not give the correct
answer for processes near the ``flavour threshold''. This has been verified 
experimentally e.g. for the DIS case \cite{H1ZMVFNSfail} 
(\Sect{ZMVFNSfail}). Also, it was found 
that neglecting the charm mass in the cross section calculations used for the 
PDF extraction can result in 
untolerably large effects on theoretical predictions even at high 
scales, such as $W$ and $Z$ production at the LHC \cite{Tungmass}.
All more recent PDF approaches 
\cite{herapdf,HERAPDF1.5, mstw08,cteq6.6,nnpdf21,abkm,gjr} 
therefore include at least a partial explicit 
consideration of the charm mass in the matrix 
elements (\Sectand{massive}{variable}).
Nevertheless, since higher orders are more easily calculable in this scheme,
the massless approach can offer advantages e.g. in high-energy 
charm-photoproduction processes \cite{Heinrich:2004kj,massless} in which the 
consideration of an extra order of $\alpha_s$ in the final state allows a 
reduction of the theoretical uncertainty (\Sect{charmphoto}).

\subsection{The massive fixed-flavour-number scheme}
\label{sect:massive}

\begin{figure}[htbp]
\centering
\hspace{-4mm}
\includegraphics[width=0.18\linewidth]{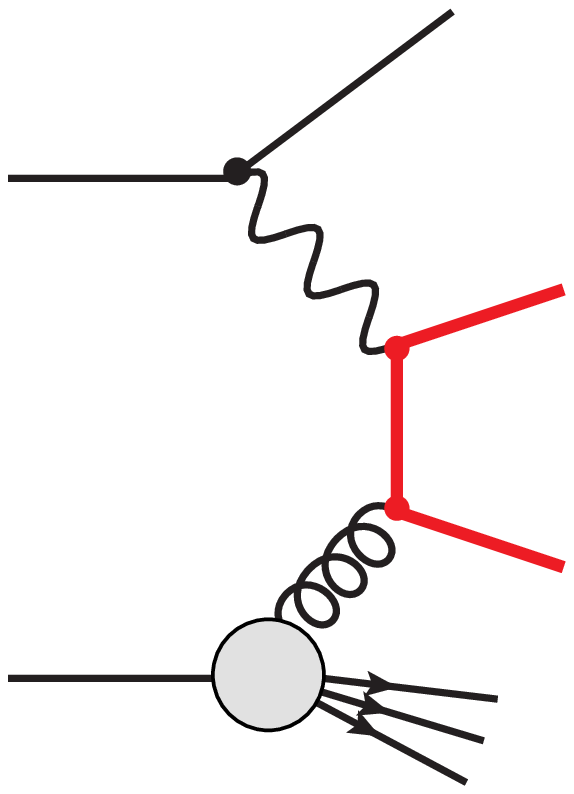}
\includegraphics[width=0.18\linewidth]{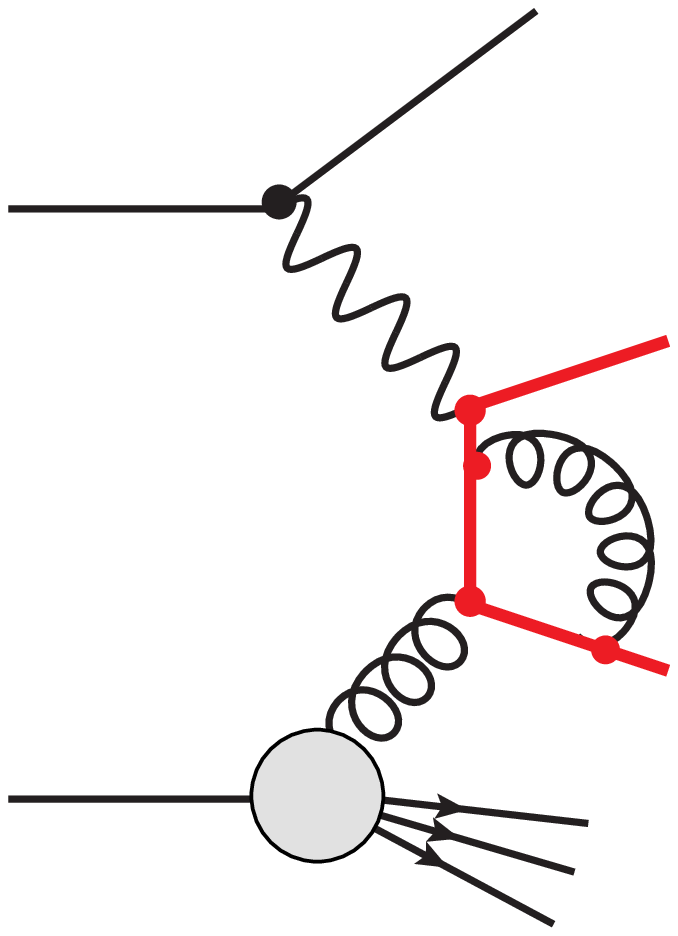}
\includegraphics[width=0.18\linewidth]{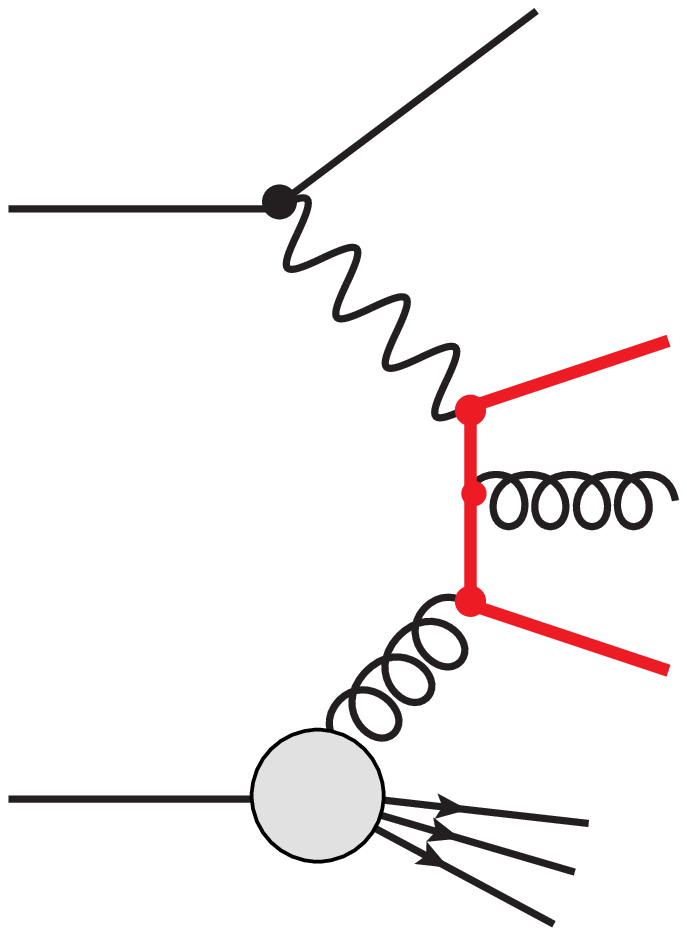}
\includegraphics[width=0.18\linewidth]{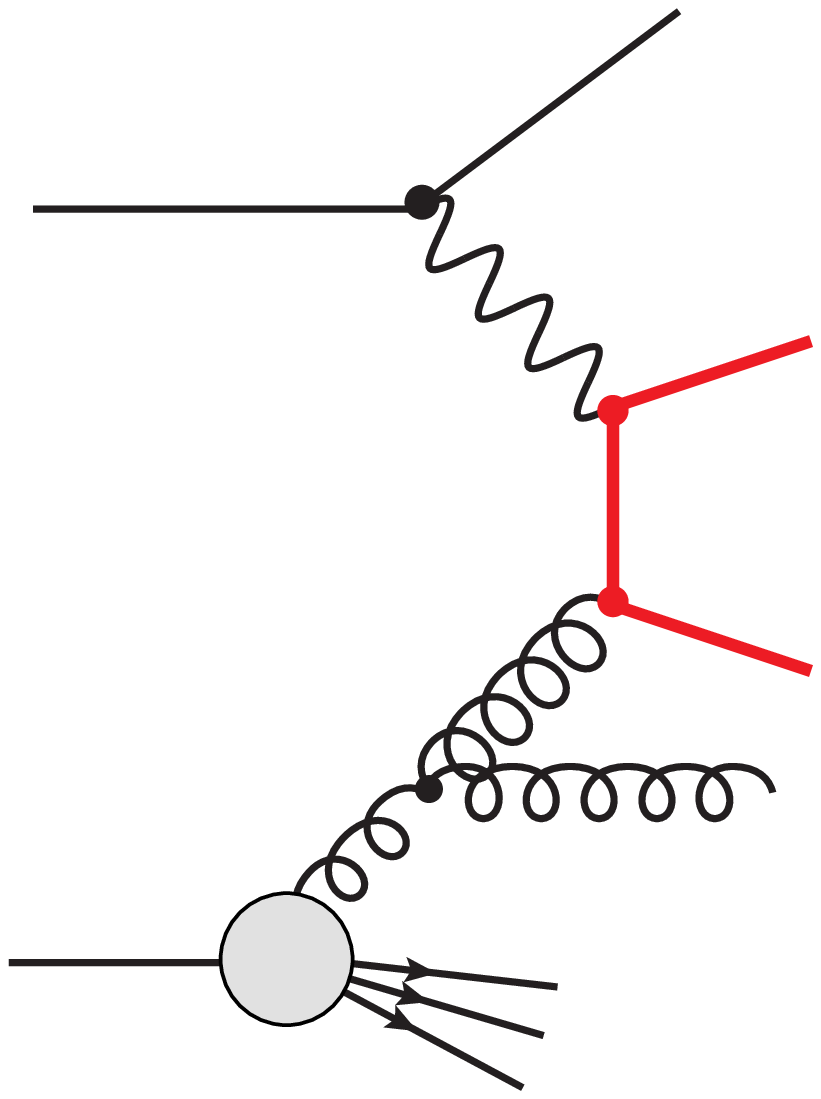}
\includegraphics[width=0.18\linewidth]{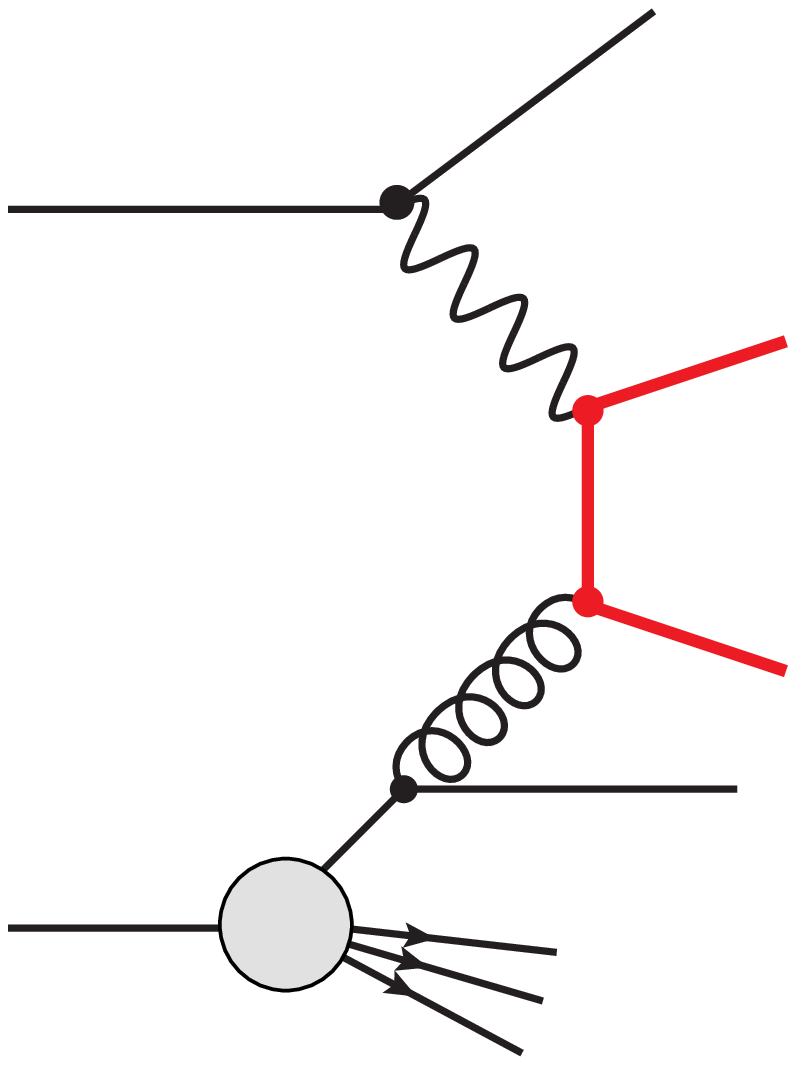}
\hspace{3cm} (a) \hspace{2.5cm} (b) \hspace{2.5cm} (c) \hspace{2.5cm} (d)
\hspace{2.5cm} (e) \hspace{0.5cm}
\caption{Leading order (a) and 
selection of next to leading order (b)--(e) 
processes for heavy flavour
production at HERA in the massive scheme.}
\label{fig:intro4}
\end{figure}

The fixed-flavour-number scheme (FFNS) treats the heavy-quark masses explicitly
and follows a rigourous quantum field 
theory ansatz. Full NLO (one loop) calculations of heavy-flavour production 
in this scheme exist for DIS
\cite{riemersma,hvqdis,mrstff3,mstw08f3,cteq5f3,ct10f3,abkm,gjr}, 
for photoproduction \cite{FMNR,FMNRtotal,Frixione:1995qc,Frixione:1997ma} 
and for hadroproduction \cite{MNR}.
Some partial NNLO (two-loop) calculations are also available \cite{abm11,abkm09msbar}.
In this scheme, heavy flavours are 
treated as massive at all scales, and never appear as an active flavour in 
the proton. In the case in which all heavy flavours are treated as massive, 
the number of light flavours in the PDFs is thus fixed to 3, and charm as well
as beauty are always produced in the matrix element (Fig. \ref{fig:intro4}). 
So-called flavour excitation 
processes (Fig. \ref{fig:fey1}(c,d)), 
which are often classified as leading order 
($\calo(\alpha_s)$) QCD in partially massless MC 
approaches \cite{PYTHIA,HERWIG,RAPGAP} of
charm or beauty production, appear as $\calo(\alpha_s^2)$
NLO corrections in the fully massive approach (Fig. \ref{fig:intro4}).

%
%%%%%%%%%%%%%%%%%%%%%%%%%%%%%%%%%%%%%%%%%%%%%%%%%%%%%%%%%%%%%%%%%%%%%%%%
%
% NLO processes in 3 flavour scheme
%
%
%//////////////////////////////////////////////////////////////////////
%

\begin{figure}[htbp]
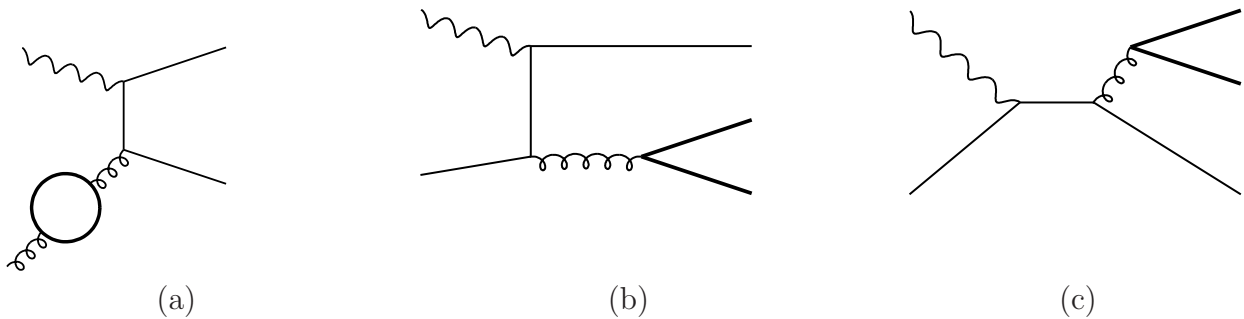

\setlength{\unitlength}{1cm}
\begin{picture}(14.0,4.2)
\put(1.,0.5){\epsfig{file=figures/bgfhqloop.epsi,height=3cm}}
\put(6.5,1.5){\epsfig{file=figures/gluonsplit.epsi,height=2.5cm}}
\put(13.,1.5){\epsfig{file=figures/gluonsplit1.epsi,height=2.5cm}}
\put(3.,0.){(a)}
\put(9,0.){(b)}
\put(15.,0.){(c)}
\end{picture}
\caption{Heavy Flavour loop correction (a) and 
gluon splitting (b,c) 
processes in the massive scheme. The thick (thin) lines indicate heavy (light)
flavours.}
\label{fig:loopsplit}
\end{figure}

There are several variants of the FFNS for heavy-flavour production in DIS
(see e.g. remarks in appendix of \cite{Kusina}). 
In one approach, here called FFNS A, 
the $\alpha_s$ evolution used together
with the 3 flavour PDFs is also 
restricted to 3 flavours. Thus, the small contribution from heavy flavour 
loops (Fig. \ref{fig:loopsplit}(a)) 
is either treated explicitly in the matrix elements 
(FFNS A) \cite{abkm}, or, somewhat incorrectly, neglected 
completely (FFNS A$^\prime$) \cite{mrstff3,cteq5f3}.
In either variant, this leads to a lower effective value of $\alpha_s$
than in the massless scheme when evolved to high reference scales, e.g.
$\alpha_s (M_Z)$. This is one of the consequences of the non-resummation of
$\log(Q^2/m_Q^2)$ terms, and is partially compensated e.g. by a conceptually 
larger gluon PDF. Despite the conceptual disadvantage of not allowing all order 
resummation of mass logarithms, this scheme yields 
very reasonable agreement
with charm and beauty data at HERA up to the highest $Q^2$ and $p_T^2$ 
(Sections \ref{sect:charmphoto}--\ref{sect:cbDIS}).
At HERA energies, the numerical differences between schemes A and A$^\prime$ 
are of order 1\%, and therefore almost negligible compared to the current data 
precision. 

%
%%%%%%%%%%%%%%%%%%%%%%%%%%%%%%%%%%%%%%%%%%%%%%%%%%%%%%%%%%%%%%%%%%%%%%%%
%
% NLO processes in 3 flavour scheme
%
%
%//////////////////////////////////////////////////////////////////////
%

In the FFNS B approach\footnote{Elsewhere \cite{qcdnum} this is 
sometimes called the mixed flavour number scheme. For a discussion see 
\cite{MST_A}\cite{Glueck}. A recent new variant of it \cite{doped} is 
referred to as the ``doped'' scheme.}, which was widely used in 
the early nineties \cite{riemersma,hvqdis,FMNR,MNR,alphavarDIS}, 
the running of $\alpha_s$ is calculated by incrementing the number of flavours 
when crossing a flavour threshold, like in the variable-flavour-number scheme. 
The class of logs corresponding to this running is thus 
resummed both in the $\alpha_s$ 
and in the PDF evolution, and the ``missing'' heavy flavour log resummation
is restricted to other cases like gluon splitting and vertex corrections in 
the matrix elements.
 This approach is possible since most of
the loop and leg corrections which diverge in the massless case, but
compensate each other to yield finite contributions, remain separately finite 
in the massive case.
They can thus be separated. 
At one-loop order, the A and B approaches differ by the way
a virtual heavy flavour correction in the BGF matrix 
element (Fig. \ref{fig:loopsplit}(a))
(which is missing in the A$^\prime$ approach) is treated. 
The FFNS B scheme conceptually yields a value
of $\alpha_s$ at high scales which is the same as the one from the 
variable-flavour approach, and is probably less sensitive to ``missing logs'' 
at very high scales than the FFNS A approach. 
Most NLO photoproduction \cite{FMNR} and 
hadroproduction \cite{MNR} calculations, as well as the electroproduction 
code HVQDIS \cite{hvqdis} have been originally designed to be used with the 
FFNS B scheme.
 
The A and B approaches both converge to the exact QCD result at infinite order
if implemented consistently.
The A$^\prime$ scheme can not converge to the exact result since heavy flavour 
loop corrections are completely missing, but, as stated earlier, the practical 
consequences at HERA energies are small. It can however serve as a useful 
ingredient to variable flavour number scheme calculations 
(Section \ref{sect:variable}).

There are also other differences.
In the ABKM approach, final state gluon 
splitting (Fig. \ref{fig:loopsplit}(b,c)) is conceptually 
treated as part of the light-flavour 
contribution, while it is treated as part of heavy flavour production in 
many others 
\cite{riemersma,hvqdis,FMNR,MNR} (see Section \ref{sect:pdf}), 
as well as in the measured cross sections, 
since it can hardly be distinguished experimentally. This difference is 
small \cite{FONLL} in most regions of phase space,
but might need to be accounted for when comparing data and theory. 

Finally, FFNS calculations in DIS are currently available in leading order 
($\calo(\alpha_s)$), or 
NLO ($\calo(\alpha_s^2)$) \cite{abkm,riemersma,hvqdis}. 
Partial NNLO (two-loop, $\calo(\alpha_s^3)$) calculations also exist, 
based on a full calculation of the 
$\calo(\alpha_s^3)log$ and the $\calo(\alpha_s^3)log^2$ terms, and the leading 
term from threshold
resummation for the $\calo(\alpha_s^3)$ constant term \cite{abm11}.
Further NNLO corrections for the high scale limit \cite{Bluemlein} have not 
yet been implemented in practice. 
Actually, both the NLO and partial NNLO ABKM heavy flavour calculations use 
the PDFs from their NNLO (two loop, $\calo(\alpha_s^2)$ in the matrix elements)
fit to the inclusive data \cite{abkm}.

Some of the differences between the calculations discussed in this section
and in the next two sections are also summarised in Table \ref{tab:theories},
using the example of reduced charm cross sections in DIS.
 
\subsection{The general-mass variable-flavour-number scheme}
\label{sect:variable}

An alternative to the fixed-flavour-number approach is given by the so-called 
general-mass variable-flavour-number schemes (GMVFNS) 
\cite{acot,TR,GMVFNSdiff,bmsn}.
In these schemes, 
charm production is treated in the FFNS approach in the low-$Q^2$ region, where 
the mass effects are largest, and in the massless approach at very high
scales, where the effect of resummation is most noticeable. 
At intermediate
scales (in practice often at all scales above the ``flavour threshold''), 
an interpolation is made between the two schemes, avoiding 
double-counting of common terms, while making a continous 
interpolation between 
differing terms. This scheme combines the advantages of the two previous 
schemes, while introducing some level of arbitrariness in the treatment 
of the interpolation.

%
%%%%%%%%%%%%%%%%%%%%%%%%%%%%%%%%%%%%%%%%%%%%%%%%%%%%%%%%%%
%
% VFNS subtraction term
%
%
\begin{figure}[htbp]
\centering
\includegraphics[width=0.8\linewidth]{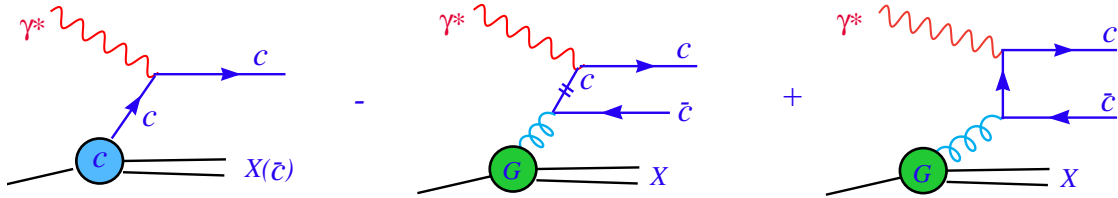}
% *** something seemed to be going wrong with the page numbering here ***
% now fixed, but just in case it reoccurs:
% tip: use the -K option with dvips
\caption{Leading order (0-loop) diagrams for charm production in DIS in the 
variable flavour number scheme: 
On the left the QPM diagram is shown, on the right the BGF diagram
and in the middle the ``subtraction diagram'' \cite{Tungfigures}.
The vertex correction loop diagram (Fig. \ref{fig:intro7}(b)), which also contributes to this order in $\alpha_s$, is not shown. 
}  
\label{fig:t4}
\end{figure}
%
%/////////////////////////////////////////////////////////
%
%

One of the most constrained schemes is the BMSN scheme \cite{bmsn} 
used by the VFNS approach of ABKM \cite{abkm09msbar}. 
At NLO, it interpolates between $\calo(\alpha_s^2)$ charm 
matrix elements in the FFNS part using the ABKM FFNS scheme, 
and $\calo(\alpha_s)$ matrix elements in the massless part. It has 
no tuneable parameters, and (currently) uses the pole-mass definition for the FFNS
part. In contrast to most other GMVFNS schemes, the switch to a larger number of flavours
should not be made at the ``flavour threshold'', but at a scale which is high enough 
that additional semi-arbitrary kinematic correction terms are not required.
In practice, the 3-flavour scheme is used for processes at HERA energies, while the 4- or 5-flavour schemes
are recommended for applications at the LHC.

\begin{sidewaystable}[p]
\centering
\linespread{1.3}
  \normalsize 
{\small  \begin{tabular}[h]{|l|l|l|c|c|c|c|c|c|c|}
    \hline
   Theory           & Scheme        & Ref.               & $F_{2(L)}$    & $m_c$      & PDF & Massive / $F_L$  & Massless $F_2$ & $\alpha_s(m_Z)$ & Scale \\
                    &               &                    &  def.        &  [GeV]     &     &
       $(Q^2\lsim m_c^2)$ \ \ \ \ \ \ & $(Q^2\gg m_c^2)$  & ($n_f=5$)  &            \\
   \hline
  MSTW08 NLO        & RT standard   &\cite{rt_std}       & $F_{2(L)}^c$  & $1.4$ (pole)       & ${\cal O}(\alpha_s^2)$ & ${\cal O}(\alpha_s^2)$ & ${\cal O}(\alpha_s)$    & $0.12108$    &$Q$ \\ 
  MSTW08 NNLO       &               &                    &              &                    & ${\cal O}(\alpha_s^3)$ & approx.-${\cal O}(\alpha_s^3)$ & ${\cal O}(\alpha_s^2)$ & $0.11707$ &     \\ 
  MSTW08 NLO (opt.) & RT optimised  &\cite{rt_opt}       &              &                    & ${\cal O}(\alpha_s^2)$ & ${\cal O}(\alpha_s^2)$ & ${\cal O}(\alpha_s)$    & $0.12108$       &     \\ 
  MSTW08 NNLO (opt.)&               &                    &              &                    & ${\cal O}(\alpha_s^3)$ & approx.-${\cal O}(\alpha_s^3)$ & ${\cal O}(\alpha_s^2)$ & $0.11707$ &     \\
  \hline 
 HERAPDF1.5 NLO     & RT standard   &\cite{HERAPDF1.5}   & $F_{2(L)}^c$  & $1.4$ (pole)       & ${\cal O}(\alpha_s^2)$ & ${\cal O}(\alpha_s^2)$ & ${\cal O}(\alpha_s)$    & $0.1176$    &$Q$ \\  
\hline 
  NNPDF2.1 FONLL A  & FONLL A       &\cite{fonllb_and_c} & n.a.         & $\sqrt{2}$         & ${\cal O}(\alpha_s^2)$ & ${\cal O}(\alpha_s)$   & ${\cal O}(\alpha_s)$    & $0.119$ & $Q$ \\
  NNPDF2.1 FONLL B  & FONLL B       &                    & $F_{2(L)}^c$  & $\sqrt{2}$ (pole)  & ${\cal O}(\alpha_s^2)$ & ${\cal O}(\alpha_s^2)\ / \ {\cal O}(\alpha_s)$ & ${\cal O}(\alpha_s)$  & &        \\
  NNPDF2.1 FONLL C  & FONLL C       &                    & $F_{2(L)}^c$  &  $\sqrt{2}$ (pole) & ${\cal O}(\alpha_s^3)$ & ${\cal O}(\alpha_s^2)$ & ${\cal O}(\alpha_s^2)$  &           &  \\
  \hline 
  CT10  NLO         & S-ACOT-$\chi$ &\cite{ct10f3}       & n.a.         & $1.3$              & ${\cal O}(\alpha_s^2)$ & ${\cal O}(\alpha_s)$   & ${\cal O}(\alpha_s)$    & $0.118$ & $\sqrt{Q^2 + m_c^2}$ \\
  CT10  NNLO        &               &\cite{ct12nnlo}  & $F_{2(L)}^{c\bar c}$ & $1.3$ (pole)    & ${\cal O}(\alpha_s^2)$ & ${\cal O}(\alpha_s^2)$ & ${\cal O}(\alpha_s^2)$  &       &        \\
  \hline
  ABKM09 NLO        & FFNS A        &\cite{abkm}  & $F_{2(L)}^{c\bar c}$  & $1.18$ ($\overline{MS}$) & ${\cal O}(\alpha_s^2)$ & ${\cal O}(\alpha_s^2)$ & - & 0.1135 & $\sqrt{Q^2+4m_c^2}$ \\
  ABKM09 NNLO       &               &               &                   &                    & ${\cal O}(\alpha_s^3)$ & approx.-${\cal O}(\alpha_s^3)$ & - &   &     \\
  \hline
  HVQDIS+ZEUS S     & FFNS B        &\cite{hvqdis} & $F_{2(L)}^c$        & $1.5$ (pole)       & ${\cal O}(\alpha_s^2)$ & ${\cal O}(\alpha_s^2)$  & - & 0.118 & $\sqrt{Q^2+4m_c^2}$ \\ 
  \hline
   \end{tabular}}
\linespread{1.0}
  \caption{Selected calculations for reduced charm cross sections in DIS from different 
theory groups as used in this review. The table
shows the heavy flavour scheme used and the corresponding reference,
the respective $F_{2(L)}$ definition (\Sect{pdf}), 
the value and type of charm mass used (\Sect{mass}), 
the order in $\alpha_S$ of the PDF part and the massive and massless
parts of the calculation (and of the massless part of $F_L$, which, 
except for FONLL B, is usually taken to be the 
same as for the massive part), the value of $\alpha_s$, the renormalisation and 
factorisation scale. The distinction between the two possible $F_{2(L)}$ 
definitions is not applicable (n.a.) for ${\cal O}(\alpha_s)$ calculations,
or in photo- or hadroproduction. Usually, the order of the PDF part is used 
to define the label LO, NLO, or NNLO.
}    
\label{tab:theories}
\end{sidewaystable}

The NLO version of the TR\footnote{also referred to as RT} 
scheme \cite{TR} combines the 
$\calo(\alpha_s^2)$ charm matrix elements in the FFNS A$^\prime$ scheme 
with the $\calo(\alpha_s)$ matrix elements of the massless scheme, 
requiring continuity of the physical observables in the threshold region.
In this case the usage of the $A^\prime$ scheme is fully appropriate, since
the missing terms will be taken care of by the massless and interpolation 
terms.  
Several variants exist for the interpolation, including the so-called
standard scheme used e.g. in MSTW08 \cite{mstw08} and 
HERAPDF1.0 \cite{herapdf}, 
and the optimised scheme preferred for more recent versions, %{MST10},
since it avoids a kink in the $Q^2$ dependence of the 
cross section \cite{rt_opt}.
Both of these variants also exist in a partial NNLO approach \cite{rt_std}, including 
approximate $\calo(\alpha_s^3)$ threshold resummation terms for the FFNS part,
and a full $\calo(\alpha_s^2)$ NNLO calculation for the massless part.
They all use the pole mass definition for the FFNS part.

The ACOT \cite{acot} scheme, used by CTEQ \cite{cteq6.6}, also exists in 
several variants. At NLO, $\calo(\alpha_s)$
(i.e. leading order) FFNS matrix elements are interpolated to $\calo(\alpha_s)$
(now NLO) massless matrix elements\footnote{The fact that similar 
matrix elements are denoted by different labels concerning their effective 
order in different context is very confusing, but unavoidable due to 
different definitions of the truncation of the perturbative QCD series.}. 
Due to the LO FFNS treatment, there 
is no difference between the pole-mass and running-mass schemes.
The interpolation is made in two variants: the S-ACOT approach \cite{SACOT}, 
and the ACOT-$\chi$ approach \cite{ACOTchi}. %, in which ***to be added***.
The NNLO variant of CT10 \cite{CT10} uses both FFNS and massless matrix 
elements at $\calo(\alpha_s^2)$, in the S-ACOT scheme.

The FONLL scheme \cite{FONLL} has 3 variants. 
The FONLL A approach, used by NNPDF2.1 \cite{nnpdf21} 
is equivalent \cite{FONLL} to the CTEQ S-ACOT approach, and uses 
$\calo(\alpha_s)$ FFNS heavy-quark matrix elements at NLO. FONLL B and C both use
$\calo(\alpha_s^2)$ FFNS heavy-quark matrix elements. FONLL B uses $\calo(\alpha_s)$
matrix elements for light quarks, like MSTW, while
FONLL C uses $\calo(\alpha_s^2)$ matrix elements for light quarks like ABKM. 
However, they differ from the latter 
in the way they treat the interpolation terms. The FONLL C scheme is also
similar \cite{FONLL} to the CTEQ S-ACOT NNLO scheme.     
A full NNLO version of the FONLL A approach also exists 
\cite{fonllb_and_c,CTEQNNLO}. 
Final state gluon splitting is not included in the charm cross-section 
predictions for any of these schemes. 

The ABM group uses the BSMN approach \cite{bmsn} to generate a GMVFNS scheme out of their
FFNS 3-, 4- and 5-flavour PDFs \cite{abkm09msbar}.

For photoproduction, a GMVFNS calculation \cite{massfrag} exists for single 
inclusive cross sections.

All GMVFNS variants use the variable-flavour approach for the running of 
$\alpha_s$. 
Although the mass is unambigously defined in the massive part of the 
calculation (usually the pole mass),
the partial arbitrariness in the treatment of the interpolation 
terms (Fig. \ref{fig:t4}) 
prevents a clean interpretation of the charm and beauty quark masses in terms 
of a single renormalisation scheme. Therefore, in contrast to the pure 
FFNS treatment, the charm mass appearing in VFNS schemes can be 
treated as an effective mass parameter \cite{HERAcharmcomb}. 
We will use the symbols $M_c$ and $M_b$ for these effective mass parameters.
Alternatively,
the presence of the interpolation terms can be included as an additional 
uncertainty on the respective mass definition \cite{CTEQmass}.

\subsection{Proton structure functions in DIS}  
\label{sect:pdf}

In analogy to the inclusive neutral current DIS cross section, the cross 
sections for heavy-quark production in DIS
can be expressed in terms of the heavy-quark contributions to the 
inclusive structure functions \cite{DeRoeck} $F_2$, $F_L$ and $F_3$, denoted 
by \ftq, \flq and \fzq ($Q=c,b$):
\begin{equation}
	\frac{d\sigma^{Q\bar{Q}}(e^{\pm} p)}{dx \, dQ^2} = 
	        \frac{2 \pi \alpha^2}{x \, Q^4} \Big ((1+(1-y)^2) \, F_2^{Q\bar{Q}} - 
	        y^2 \, F_L^{Q\bar{Q}} 
	        \mp x \, (1-(1-y)^2) \, F_3^{Q\bar{Q}}\Big ),
	\label{eq:hqstructfunct}
\end{equation}
where $\alpha$ is the electromagnetic coupling constant. 
The structure function \ftq makes the dominant contribution to the neutral current scattering 
in the kinematic regime accessible at HERA. 
\fzq contains contributions only from $\gamma Z^0$ interference and $Z^0$ exchange, 
therefore for the region $Q^{2} \ll M_{Z}^{2}$, which was studied at HERA, 
this contribution is suppressed and can be neglected. 
The longitudinal heavy-quark structure function \flq parametrises the contribution from 
coupling to the longitudinally polarised photons. 
The contribution of \flq to the \emph{ep} cross section is suppressed for $y^2 \ll 1$, 
but can be up to a few percent in the kinematic region of the heavy-quark measurements
at HERA and thus can not be neglected.

For both electron and positron beams, neglecting the  \fzq contribution, 
the reduced heavy-quark cross section, \redq, is defined as 
\begin{equation}
	\sigma_{red}^{Q\bar{Q}} (x, Q^{2}) = 
	          \frac{d\sigma^{Q\bar{Q}}(e^{\pm} p)}{dx \, dQ^2} \cdot \frac{x \, Q^4}{2 \pi \alpha^2 Y_{+}} 
	          = F_2^{Q\bar{Q}} - \frac{y^2}{Y_{+}} \, F_L^{Q\bar{Q}},
	\label{eq-sigma_reduced}
\end{equation}
where $Y_{\pm} = (1 \pm (1-y)^2)$. Thus, \redq  and \ftq only differ by a small \flq correction at high $y$~\cite{daum_fl}.

In the Quark-Parton Model, the structure functions depend on $Q^2$ only 
and can be directly related 
to the parton density functions. In the QCD case, and in particular for 
heavy flavour production, this correlation is strongly diluted, and the 
structure functions depend on both $x$ and $Q^2$.
More information on the general case can be found e.g. in 
\cite{DeRoeck,herapdf}. 

Using the example of the charm case \cite{HERAcharmcomb}, the above definition 
of $F^{c\bar c}_{2 (L)} (x,Q^2)$ (also denoted as $\tilde F_c$ \cite{FONLL} or
$F_{c,SI}$ \cite{CTEQNNLO}) is suited for 
measurements in which charm is explicitly detected. It differs 
from what is sometimes used in theoretical calculations in which
$F^c_{2 (L)} (x,Q^2)$ \cite{FONLL,rt_std,MSTWF2} is defined as the 
contribution to the inclusive $F_{2(L)}
(x,Q^2)$ in which the virtual photon couples directly to a $c$ or $\bar{c}$
quark. The latter excludes contributions from final state gluon splitting to 
a $c\bar c$ pair in events where the photon couples directly to a light quark, 
and contributions from events in which the photon is replaced by a gluon
from a hadron-like resolved photon.
As shown in table 1 of ~\cite{FONLL}, 
the gluon splitting contribution is expected to be 
small enough to allow
a reasonable comparison of the experimental results 
to theoretical predictions using this definition.
The hadron-like resolved photon contribution is expected to be heavily 
suppressed at high $Q^2$, but might not be completely negligible in the low $Q^2$ region.
From the point of view of pQCD it appears at $O(\alpha_s^3)$ and it is
neglected in all theoretical DIS calculations used in this review.

\subsection{QED corrections}
\label{sect:QED}

\begin{figure}[htbp]
\centering
\hspace{-4mm}
\includegraphics[width=0.3\linewidth]{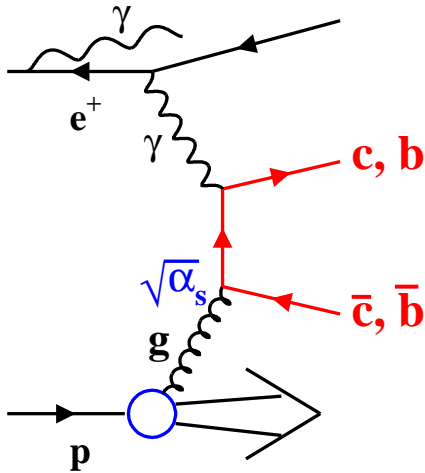}
\caption{BGF diagram with 
initial state photon radiation.}
\label{fig:photon}
\end{figure}

In addition to the different QCD schemes discussed above, predictions of 
charm production can also differ through their treatment of QED corrections.
Some of these corrections, e.g. collinear photon radiation from the initial
state electron before the hard interaction (Fig. \ref{fig:photon}) can actually
be large (of order $\alpha \ln \frac{Q^2_{max}}{m_e^2}$),
and can influence the definition of the $Q^2$, $x$ and $y$ 
variables \cite{DeRoeck}.
For photoproduction calculations, the improved Weizs\"acker-Williams 
approximation \cite{Weizsaecker}
can be used to parametrise the photon spectrum arising 
from the incoming electron.

%
%//////////////////////////////////////////////////////////////////////
%

For acceptance corrections (and partially for visible cross sections), 
predictions including full LO 
 QED radiative corrections, as implemented 
e.g. in HERACLES \cite{HERACLES}, are used.
For more sophisticated purposes, an NLO version of these corrections
is available in the HECTOR package \cite{HECTOR}.  
At the level of the DIS structure functions or $\sigma_r$ it is customary to 
translate the measured cross sections to so-called Born-level cross sections,
i.e. cross sections in which all QED corrections have been removed, to ease
comparison of the data with pure QCD predictions.
There is one potential exception: the fine structure ``constant'' $\alpha$ 
can be used in two different ways. 
\begin{itemize}
\item as a genuine atomic scale constant $\alpha = \frac{1}{137.036}$, 
      i.e. all virtual QED corrections are removed, too;
\item as running $\alpha$ in the $\overline{MS}$ scheme, i.e. the respective 
      relevant virtual corrections are kept, and a typical value for 
      HERA kinematics is then 
      $\alpha \simeq \frac{1}{134}$ \cite{alphaQED}.
      %*** remark:$\alpha(15.5\gev^2) = \frac{1}{133.557}$, 
      %corresponds to $Q^2 =6.5\gev^2$, taken from HVQDIS; differs slightly 
      %from the state-of-the-art predictions of alphaQED code 
      %$\alpha(15.5\gev^2) = \frac{1}{133.820}$***.
\end{itemize} 
The difference between these two approaches in QED has some remote 
similarity to the difference between the FFNS A and B schemes in 
perturbative QCD (Section \ref{sect:massive}), treating all quarks and leptons 
as ``heavy'' with respect to the atomic scale.  

\subsection{Fragmentation}
\label{sect:frag} 
\Eq{factorization} allows one to make predictions of heavy-quark production
with partons in the final state. 
However, cross sections are measured and reported mostly in terms 
of heavy-flavour hadrons, leptons from their decay,
or collimated jets of hadrons.
Therefore such predictions have to be 
supplemented with a fragmentation or hadronisation model. 

In analogy to \epem collisions~\cite{Nason:1993xx}, 
the factorised cross section for the production of 
a heavy-quark hadron $H$ as a function of transverse momentum $p_T^H$
can be written as:
\begin{equation}
\label{eq:FF_factorization}
\frac{d\sigma^H}{dp_T^H}(p_T^H) = \int \frac{dp_T^Q}{p_T^Q}\:
\frac{d\sigma^Q}{dp_T^Q}(p_T^Q,\mu_f) \: D_Q^H(\frac{p_T^H}{p_T^Q},\mu_f) 
\cdot f(Q \to H),
\end{equation}
where $\sigma^{Q}(x,\mu_f)$ is the production cross section for
heavy quarks (\eq{factorization}), $D_Q^H$ is the 
fragmentation function, $\mu_f$ is the fragmentation scale
and $f(Q \to H)$ in the fragmentation fraction. The latter is defined as 
the probability of the given hadron $H$ to originate from the heavy
quark $Q$. 
The fragmentation function defines the probability for 
the final-state hadron to carry the fraction $z=p_T^H / p_T^Q$
of the heavy-quark momentum.
The fragmentation function is defined similarly to the PDFs.
In the ``massless'' approximation, 
it is defined at a starting scale and has to be evolved to 
a characteristic scale $\mu_f$ of the process using perturbative QCD.
In the massive fixed flavour approach, this evolution can be conceptually
absorbed into the pole mass definition. 

Fragmentation fractions as well as the starting parametrisation
of the fragmentation function can not be calculated perturbatively.
Thus they have to be extracted from data.
Comprehensive phenomenological analyses of the charm and beauty 
fragmentation functions in \epem collisions have been 
performed~\cite{Nason:1999zj, Cacciari:2005uk, Kneesch:2007ey}.
While the QCD evolution is process-dependent, the non-perturbative
ingredients of the fragmentation model are assumed to be universal%
\footnote{The non-perturbative fragmentation function is universal 
only if it is accompanied by appropriate evolution.}.
Comparing measurements from HERA and results from \epem
colliders one can test this universality.

However, the tools that are available for $\mathcal{O}(\alpha_s^2)$ 
fixed-order calculations of 
the heavy-quark production cross sections in $ep$ collisions,
which are mostly used for exclusive final states, do not 
comprise a perturbative component of the fragmentation function.
Therefore, an ``independent'' non-perturbative 
fragmentation function $D^{\textrm{NP}}(z)$ is used in conjunction with
the parton-level cross sections. 
The parametric forms of the independent fragmentation functions most 
commonly used at HERA are due to Peterson~\cite{Peterson:1982ak}:
$$D^{\textrm{NP}}(z) \propto \frac{1}{z (1-1/z-\varepsilon/(1-z))^2},$$
Kartvelishvili~\cite{Kartvelishvili:1977pi}:
$$D^{\textrm{NP}}(z) \propto z^\alpha (1-z)$$
and to the Bowler modification of the symmetric Lund~\cite{Bowler:1981sb} 
parametrisation:
$$D^{\textrm{NP}}(z) \propto 1/z^{1+r_Q b m_Q^2} \, (1-z)^{a} \, exp(-b (m_H^2+p_T^2) / z),$$
where $\epsilon$, $\alpha$, $a$, $b$ and $r_Q$ are free parameters
that depend on the heavy-flavour hadron species and have to be extracted from 
data.
Since no QCD evolution is applied, the corresponding parameters might 
be scale- and process-dependent.

The recent GMVFNS NLO predictions for charm photoproduction~\cite{massfrag}
incorporate a perturbative fragmentation function and have been tested against 
data (\Sect{charmphoto}).

\subsection{Choice of renormalisation scale}
\label{sect:scale}

For many cross-section predictions the dominant contribution to the 
theoretical uncertainty arises from 
the variation of the renormalisation and factorisation scales by a factor 2 
around some suitably chosen default scale.
Such a variation is intended to reflect the uncertainty due to uncalculated
higher orders.
It might therefore be useful to consider some phenomenological aspects of 
these scale choices as considered in a mini-review on beauty production at 
HERA and elsewhere \cite{DIS07}, focusing in particular on the choice of the 
renormalisation scale.

Ideally, in a QCD calculation to all orders, the result of the perturbative
expansion does not depend on the choice of this scale.
In practice, a dependence arises from the truncation of the 
perturbative series. Since this is an artefact of the truncation, rather than
a physical effect, the optimal scale can not be ``measured'' from the data.
Thus, it must be obtained phenomenologically.  

Traditionally, there have been several options to choose the ``optimal'' scale,
e.g.
\begin{itemize}

%\noindent $\bullet$
\item 
The ``natural'' scale of the process. This is usually taken to be the 
transverse 
energy, $E_T$, of the jet for jet measurements, the mass, $m$, of a heavy 
particle for the total production cross section of this particle, or the 
combination $\sqrt{m^2+p_T^2}$ for differential cross sections of such a 
particle.
Often, this is the only option considered.
The choice of this natural scale is based on common sense, and on 
the hope that this will minimise the occurrance of large logs
of the kind described above, for the central hard process.
However, higher order subprocesses such as additional gluon radiation often
occur at significantly smaller scales, such that this choice might not
always be optimal.

\item
The principle of fastest apparent convergence (FAC) \cite{FAC}.
The only way to reliably evaluate uncalculated higher orders
is to actually do the higher-order calculation. Unfortunately, this is often 
not possible. Instead, one could hope that a scale choice which 
makes the 
leading-order prediction identical to the next-to-leading-order one would 
also minimise the NNLO corrections. 
This principle, which can be found in many 
QCD textbooks, can not be proven. However, recent
actual NNLO calculations might indicate that it works phenomenologically 
after all (see below).

\item
The principle of minimal sensitivity (PMS) \cite{PMS}. The idea is 
that when the derivative of the cross section with respect to the NLO scale 
variation vanishes, the NNLO corrections will presumably also be small.
Again, there is no proof that this textbook principle should work, but
actual NNLO calculations might indicate that it does (see below).    
\end{itemize}
 
\begin{figure}[tb]
\includegraphics[width=0.5\columnwidth]{figures/geiser_achim.fig5a.eps}
\hspace{1 cm}
\includegraphics[width=0.31\columnwidth]{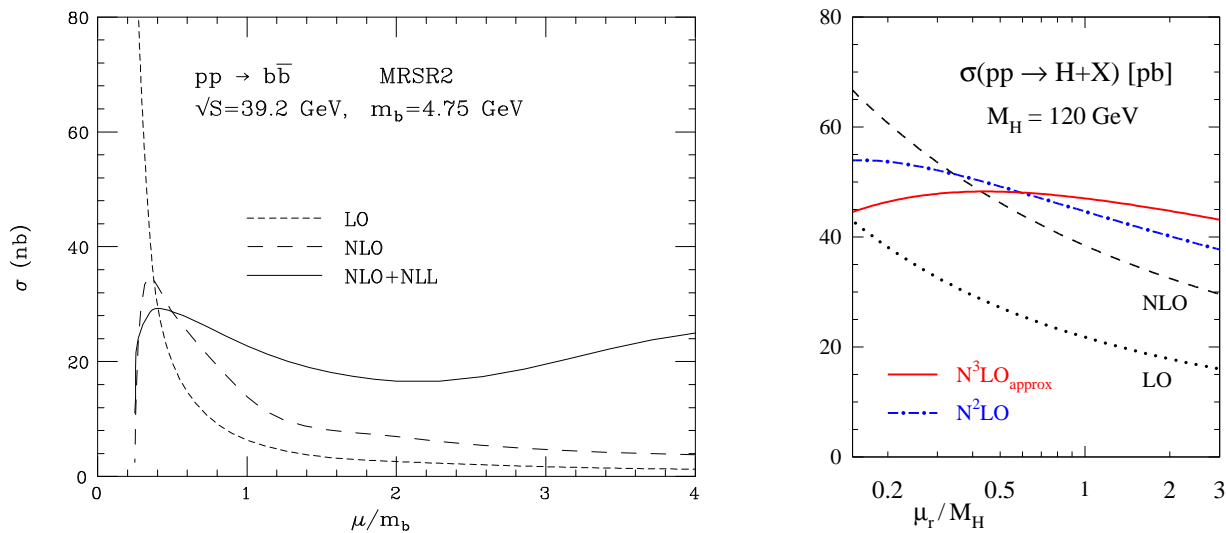}
\caption{\small
Scale dependence of the total cross section for beauty production at HERA-B 
\cite{HERABpred} (left) and for Higgs production at the LHC \cite{HiggsLHC} (right).
}
\label{fig:scalevar}
\end{figure}

To illustrate these principles, consider two examples.
First, the prediction for the total cross section for beauty production at 
HERA-B \cite{HERABpred} (Fig. \ref{fig:scalevar}). 
The natural scale for this case is the $b$-quark mass, $\mu_0 = m_b$,
and all scales are expressed as a fraction of this reference scale. 
Inspecting Fig. \ref{fig:scalevar}, one finds that both the PMS and FAC 
principles, applied to the NLO prediction and to the comparison with LO
(NLO stability), would yield an optimal scale of about half the natural scale.
The same conclusion would be obtained by using the NLO+NLL prediction, 
including resummation, and comparing it to either the LO or the NLO 
prediction (NLO+NLL stability).         
 
%%%%%%%%%%%%%%%%%%%%%%%%%%%%%%%%%%%%%%%%%%%%%%%%%%%%%%%%%%%%%%%%%%%%%%%%%%
%
% scale dependence
% 
Second, the prediction for Higgs production at the LHC \cite{HiggsLHC} (Fig. \ref{fig:scalevar}).
The reference scale is now the Higgs mass ($\mu_0 = m_H$).
However, inspecting the behaviour of the LO and NLO predictions, neither the
FAC nor the PMS principle would yield a useful result in this case, since the
two predictions do not cross, and the NLO prediction does not have a maximaum
or minimum.
This situation occurs rather frequently, and is also true for $b$ production 
at HERA. 
Fortunately, in the case of Higgs production, the NNLO and even approximate NNNLO 
predictions have actually been calculated (Fig \ref{fig:scalevar}).
Applying the FAC and PMS prescriptions to these instead (NNLO stability), again
a scale significantly lower than the default scale would be favoured.
This might indicate that choosing a scale which is smaller than the default 
one makes sense even if the FAC and PMS principles do not yield useful
values at NLO.

Beyond these examples, a more general study is needed 
to phenomenologically validate this approach.
To avoid additional complications arising from a multiple-scale problem 
caused by e.g. the scale $Q^2$ at HERA or the scale $M_Z$ at LEP, the study 
was 
limited to cross sections for photoproduction at HERA, or hadroproduction at
fixed-target energies, the Tevatron, and LHC.
The somewhat arbitrary selection of processes includes 
beauty production at the $Sp\bar pS$ \cite{UA1alpha,Frixheavy}, the Tevatron \cite{Frixheavy},
and HERA-B \cite{HERABpred},
top production at the Tevatron \cite{HERABpred,Frixheavy},
direct photon production at fixed target \cite{photon},  
$Z$ \cite{ZLHC} and Higgs \cite{HiggsLHC} production at the LHC,
jets at HERA \cite{jetPHP} and at the Tevatron \cite{TevaJet}.   
This selection is obviously not complete, and many further calculations, 
in particular NNLO calculations, have been achieved since this study 
\cite{DIS07} was originally made. However, it is not biased in the 
sense that all processes that were originally considered were included, and 
none were discarded. Clearly, a quantitative update of this study would be 
useful, but 
was not yet done.
Qualitatively, all newer predictions which the authors have been made aware of 
either confirm this conclusion, or at least do not significantly contradict 
it.

%%%%%%%%%%%%%%%%%%%%%%%%%%%%%%%%%%%%%%%%%%%%%%%%%%%%%%%%%%%%%%%%%%%%%%%%%%
%
% scale summary plot
% 
\begin{figure}[tb]
\centering
\vspace{-0.5cm}
\includegraphics[width=0.45\columnwidth]{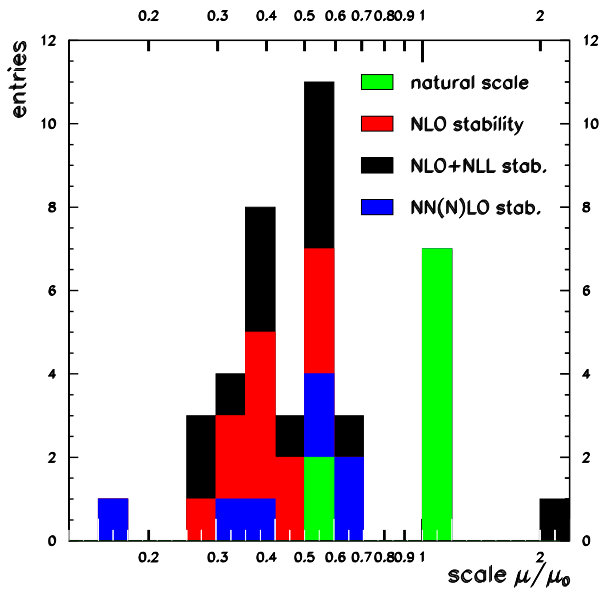}
\vspace{-0.5cm}
\caption{\small
Summary of optimised scales derived as described in the text.
\vspace{-0.5cm}
}
\label{fig:scale}
\end{figure}
In each case the natural scale as defined above was used as a reference. 
In addition, wherever possible, the optimal scales
from both the FAC and PMS principles, evaluated at NLO (NLO stability), NLO+NLL
(NLO+NLL stability), and/or NNLO/NNNLO (NNLO stability) were evaluated 
separately. Figure \ref{fig:scale} shows the result of this evaluation.
Each crossing point, maximum, or minimum in Fig. \ref{fig:scalevar} yields one 
entry into this figure, and similarly for all the other processes.
The conclusion is that the FAC and PMS principles tend to favour scales which
are around 25-60\% of the natural scale. Amazingly, this seems to be 
independent of whether these principles are applied at NLO, NLO+NLL, or NNLO 
level. For the jet \cite{TevaJet} or b-jet \cite{Tevabjet} cross sections at the 
Tevatron, it has in part already become customary to use half the natural scale
as the central scale.

Using the natural scale as the default and varying it by a factor two, which
is the choice adopted for most data/theory comparisons, covers only about half
the entries, while the other half lies entirely below this range.
Instead, using half the natural scale as the default and varying it 
by a factor two, thus still including the natural scale in the variation, 
covers about 95\% of all the entries.    

This yields the following conclusions.
\begin{itemize}

\item
Obviously, whenever an NNLO calculation is available, it should be used.

\item
Whenever possible, a dedicated scale study should be made for each 
process for the kinematic range in question.
Although there is no proof that the FAC and PMS principles should work, 
in practice they seem to give self-consistent and almost universal answers
for processes at fixed target energies, HERA, the Tevatron, and the LHC.

\item
In the absence of either of the above, the default scale should be chosen
      to be {\sl half} the natural scale, rather than the natural scale,
      in particular before claiming a discrepancy between data and theory.
      Empirically, this should enhance the chance that the NNLO calculation, 
      when it becomes available, will actually lie within the quoted error 
      band. To evaluate uncertainties, the customary variation of the central 
      scale by a factor 2 up and down remains unaffected by this choice. 
\end{itemize}

The latter principle has already been applied to a few of the results 
covered in this review. Of course, choosing the natural scale as the 
central value, which is still the default for most calculations (or making 
any other reasonable scale choice), is perfectly
legitimate and should also describe the data within the theoretical 
uncertainties. However, if it does not, it might be useful to consider 
alternative choices as discussed above before claiming evidence for the 
failure of QCD, and hence for new physics. 

Further complementary information, in particular on the related theory aspects,
is available elsewhere~\cite{Wu:2013ei}.

\subsection{Summary}

The theory of heavy flavour production in the framework of perturbative 
QCD, and in particular the occurrence of different
possibilities to treat the heavy quark masses in the PDF, matrix 
element and fragmentation parts of the calculation, introduces a significant
level of complexity into the corresponding QCD calculations, in addition to 
the usual scheme and scale choices. Confronting different choices with data
can be helpful to understand the effects of different ways to 
truncate the perturbative series.

The majority of the available MC calculations for the analysis of HERA data 
is based on leading order (plus parton shower) approaches, combining
a massive approach for the core boson-gluon fusion process, and the massless
approach for tree level higher order corrections. It will be demonstrated
in the later chapters that this is fully adequate for acceptance corrections.
For comparisons of the measured differential cross sections with 
QCD predictions, a next-to-leading order massive approach (fixed flavour 
scheme) is the state of the art. 

In some cases massless calculations are still in use, e.g. to facilitate
the perturbative treatment of fragmentation, or to implement resummation
of some of the logarithms arising when the mass competes with other 
hard scales occurring in a process. In particular for the prediction 
of the inclusive heavy flavour structure functions in DIS, a variety 
of so-called general-mass variable-flavour schemes are available, 
merging massive calculations at low scales with massless calculations 
at high scales. These are particularly useful for the extraction of 
PDFs over very large ranges in energy scale.
Partial NNLO calculations are also available for such inclusive quantities,
both in the fixed and variable flavour number schemes. 
Due to the absence of extra semi-arbitrary parameters, the fixed flavour 
number scheme is particularly well suited for the extraction of QCD 
parameters like the heavy quark masses. 

In general, QED corrections are nonnegligible, and available both at leading
and next-to-leading order. Since $\alpha$ is much smaller than $\alpha_s$,
the leading order precision is often sufficient. 
Several competing fragmentation models are in use, and the  
perturbative treatment of fragmentation in the massive approach is 
still in its infancy. 

Since higher order corrections are large, the uncertainties reflected
by the QCD scale variations are often dominant. Until full NNLO calculations
become available, a careful consideration of the choice of these scales 
can be helpful to avoid premature conclusions concerning potential 
discrepancies between the theory predictions and the data.  

\newpage

% HERA accelerator and experiments
\section{The HERA collider and experiments}
\label{sect:experiments}

In this section, the HERA collider, the H1 and ZEUS experiments, as well
as the reconstruction of the data from these experiments will be briefly 
described, with focus on aspects relevant for heavy flavour production.

\subsection{HERA}

\begin{figure}[hbtp]
	\centering
	\includegraphics[width=0.5\textwidth]{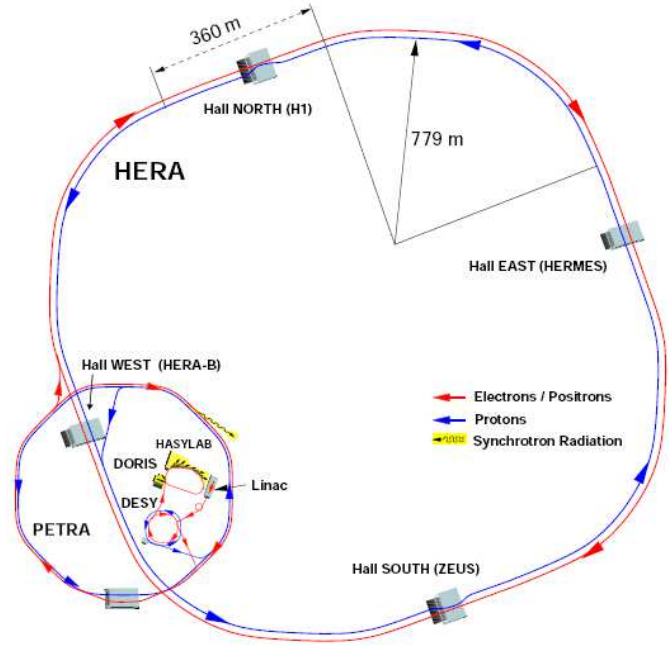}
	\caption{A schematic view of the HERA $ep$ storage rings 
	with the H1 and ZEUS experiments}
	\label{fig:heraview}
\end{figure}

HERA (German: \underline{H}adron-\underline{E}lektron-\underline{R}ing-\underline{A}nlage)
was the first and so far the only electron--proton collider. 
It was located at DESY in Hamburg, Germany.
The circumference of the HERA ring (see \fig{heraview}) was $6.4\km$. 
The accelerator was in operation from 1992 to 2007. 
Electrons or positrons
and protons were accelerated in two separate rings
to final energies of $27.5-27.7\gev$ and $920\gev$ ($820\gev$ before 1998),
respectively, leading to a centre-of-mass energy of $\sqrt{s}=318-319\gev$
($300\gev$ before 1998).
Both beams were stored in 180 bunches.
The bunch-crossing rate was $10\MHz$. 
Electrons and protons collided in two interaction regions,
where the H1 and ZEUS detectors were located.

In the years $2001\rnge2002$ the HERA collider was upgraded to increase the
instantaneous luminosity.
At the same time a number of upgrades of the H1 and ZEUS detectors were put 
in place, as described below.
Therefore, the data taking was subdivided into two phases:
``HERA I'' and ``HERA II'' corresponding to the data taking  
periods $1992\rnge2000$ and $2003\rnge2007$, respectively.
In 2007, a few months were dedicated to data taking at lower 
centre-of-mass energies.

%%%%%%%%%%%%%%%%%%%%%%%%%%%%%%%%%%%%%%%%%%%%%%%%%%%%%%%%%%%%%%%%%%%%%%%%%%%%%%%%%%%%%%%%%%%%%%%%%%%
\subsection{H1 and ZEUS Detectors}

\begin{figure}[p]
\begin{center}
{\includegraphics[trim = 50pt 200pt 45pt 230pt,clip,height=0.45\textheight]{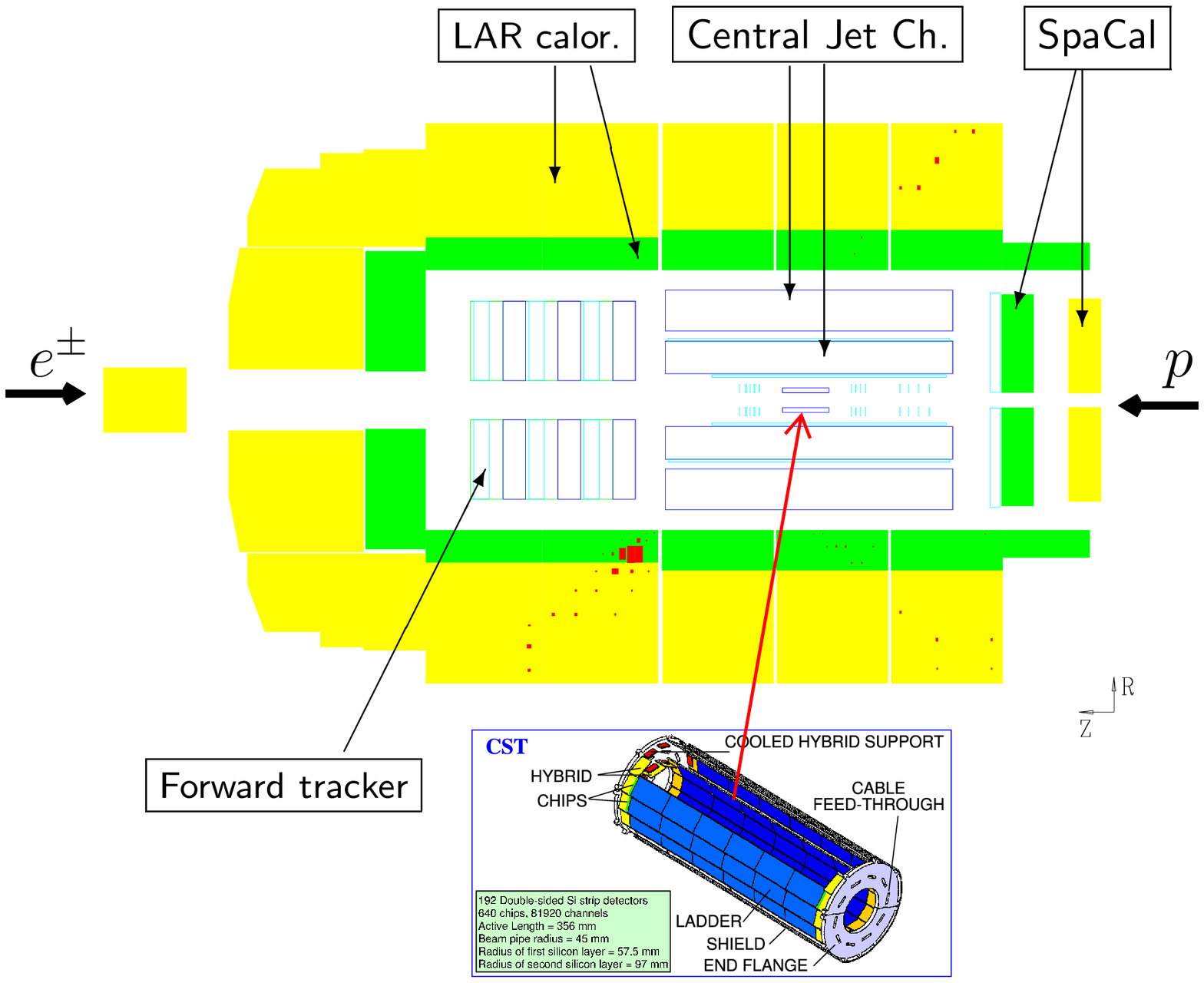}}\\
{\includegraphics[trim = 50pt 200pt 50pt 250pt,clip,height=0.45\textheight]{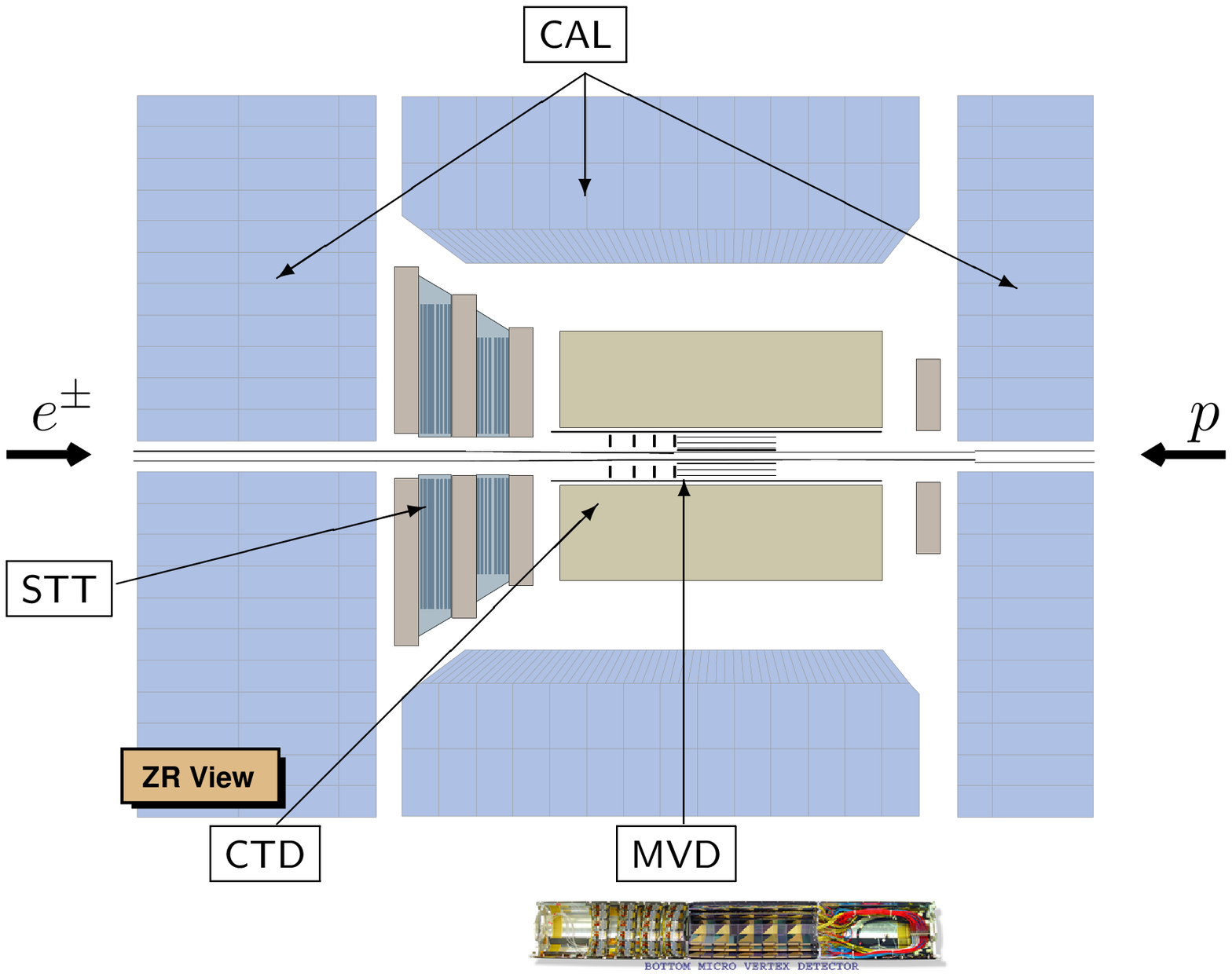}}
\caption{
The $rz$-view of the tracking system and calorimeters of the H1 (top) and ZEUS (bottom) detectors. 
The layout of the central silicon tracker (CST) and 
the microvertex detector (MVD) are shown  
separately below.
The electron beam enters from the left, while the proton beam enters 
from the right.
}
\label{fig:h1zeusdet}
\end{center}
\end{figure}

%
%////////////////////////////////////////////////////////////////
%
The H1 and ZEUS detectors were typical modern multi-purpose collider 
experiments and are described in detail 
in~\cite{h1p1,h1p2,H1SpaCal1,zeus}. 
\Fig{h1zeusdet} visualises 
the layout of the H1 and ZEUS detectors\footnote{The right-handed 
Cartesian coordinate system used at H1 and ZEUS has
the $Z$ axis pointing in the nominal proton beam direction,
referred to as the ``forward direction'', 
and the $X$ axis pointing towards the centre of HERA. 
Its coordinate origin is at the nominal center 
of the respective detector, which
coincided with the nominal interaction point in the \herai period. 
The pseudorapidity is defined as $\eta=-\ln\left(\tan(\theta/2)\right)$,
where the polar angle, $\theta$, is measured with respect to the proton beam
direction.
The $x$-$y$- or $r$-$\phi$-plane is also denoted as the transverse plane. }.
Due to significantly higher energy of the protons 
there was more detector hardware installed in the direction
of the outgoing proton beam.
The figures show the key parts of the main detectors that were
used for tagging and reconstruction of heavy-flavour events.
Some of the most important benchmarks 
of the H1 and ZEUS detectors, such as polar angle
coverage and momentum, the energy resolution and the resolution 
of the impact 
parameter $\delta$\footnote{Also referred to as the transverse distance of 
closest approach to the nominal vertex}, are listed
in \tab{det1}.
%
%%%%%%%%%%%%%%%%%%%%%%%%%%%%%%%%%%%%%%%%%%%%%%%%%
%---- H1 and ZEUS detectors: Benchmark 
%
\begin{table}[bt]
\begin{center}
\renewcommand{\arraystretch}{1.2}
\begin{tabular}{|r|c|c|}
\hline
{} & {\bf H1}  & {\bf ZEUS} \\
\hline
{\bf Silicon vertex detector:} & CST & MVD \\[1mm]
\# layers & 2 & $3$\\
$\theta$-coverage   & $[30^{\circ},150^{\circ}]$ & $[7^{\circ},150^{\circ}]$\\
$\eta$-coverage   & $[-1.3,1.3]$ & $[-1.3,2.8]$ \\
$\sigma(\delta)$ &
$43 \oplus 51/p_{T}\mum$ &
$46 \oplus 122/p_{T}\mum$\\
%first term- intrinsic resolution+misalignment, 
%second- multiple scattering in the beampipe and strip detectors 
\hline
{\bf Drift chambers:} & CTD & CTD \\[1mm]
%H1 "central" differs from 15..165 to 25..155 in different papers
$\theta$-coverage   & $[20^{\circ},160^{\circ}]$ &   $[15^{\circ},164^{\circ}]$ \\
$\eta$-coverage     & $[-1.74, 1.74]$ & $[-2.0, 2.0]$ \\
%Momentum resolution
$\sigma(p_{T})/p_{T}$ &
$0.002p_{T}\oplus 0.015$ &
$0.0029p_{T}\oplus0.0081\oplus0.0012/p_{T}$ \\
\hline
{\bf Calorimeters:} & LAr & CAL (\herai) \\[1mm]
$\theta$-coverage   & $[4^{\circ},154^{\circ}]$ &   $[2.6 ^{\circ},176.2^{\circ}]$ \\
$\eta$-coverage   & $[-1.46,3.35]$ &  $[-3.4,3.8]$ \\
El.-magn.
$\sigma(E)/E$ & $0.12/\sqrt{E} \oplus 0.01$ & $0.18/\sqrt{E}$ \\
Hadronic
$\sigma(E)/E$ & $0.50/\sqrt{E} \oplus 0.02$ & $0.35/\sqrt{E}$ \\
\cline{2-2}
&SpaCal&\\
$\theta$-coverage   & $[153^{\circ},178^{\circ}]$ &   \\
$\eta$-coverage   & $[-3.95, -1.43]$ & \\
El.-magn.
$\sigma(E)/E$ & $0.07/\sqrt{E} \oplus 0.01$ & \\
\hline
{\bf Muon systems:} & CMD &  R/B/FMUON+BAC \\[1mm]
$\theta$-coverage   & $[4^{\circ},171^{\circ}]$ & $[5^{\circ},171^{\circ}]$  \\
$\eta$-coverage   & $[-2.5, 3.4]$ & $[-2.5, 3.1]$\\
\hline
\end{tabular}
\caption{Parameters and performance of the
H1 and ZEUS subdetectors which are relevant for the
heavy-flavour physics analyses presented
in this review.
The benchmarking is shown for the \heraii run conditions, unless stated 
otherwise.
Transverse momenta $p_{T}$ and energies $E$
are in units of GeV.
$\delta$ is the transverse distance of closest approach of 
tracks to the nominal vertex. 
$\sigma(\delta)$ is the resolution of $\delta$,
averaged over the azimuthal distribution of tracks.
}
\label{tab:det1}
\end{center}
\end{table}
%//////////////////////////////////////////////////////////////////////////
%
%
%
In the following the main components of the H1 and ZEUS detectors
are discussed with emphasis on the advantages 
of the respective designs:
%%%%%%%%%%%%%%%%%%%%%%%%%%%%%%%%%%%%%%%%%%%%%%%%%%%%%%%%%%%%%%%%%%%%%%%%%%%%%%%%%%%%%%%%%%%%%%%%%%%
\paragraph{Tracking Chambers:}
Tracks from charged particles were reconstructed based on the 
position measurements
in the large {\bf Central Drift Chambers}.
The pulse height on the sense wires was used to measure the energy
loss in the detector medium, $dE/dx$. The $dE/dx$ 
measurements were used for particle identification,
distinguishing between electrons, pions, kaons and protons in a limited 
momentum range.
The important differences  between H1 and ZEUS are:
\begin{itemize}
\item
The H1 tracking detector comprised two chambers CJC1 and CJC2 \cite{h1p2} 
while the ZEUS CTD~\cite{ZEUSCTD} was a single chamber
that was divided into nine superlayers.
\item
For ZEUS a superconducting coil surrounded the tracking detectors
and provides a magnetic field of 1.43 Tesla.
This is considerably higher than the 1.15 Tesla delivered by the
H1 superconducting coil, situated outside the calorimeter.
In both experiments the magnetic field within the tracking 
system was parallel to the $Z$ axis.
\item
Four of the nine superlayers of the ZEUS chambers were equipped with stereo
wires, which were tilted $\sim 5^{\circ}$ with respect
to the beam axis.
This provided $z$-measurement 
points for tracks with
a resolution of $\sim 1.5\mm$.
At H1 the sense wires were strung parallel to the beam axis
and the track $Z$-position measurement was obtained by the division
of the charges recorded at both wire ends, yielding a moderate
resolution of a few centimetres.
Two additional $Z$-drift-chambers were installed
to provide for each track a few $Z$-measurement points
with typically $300\mum$ resolution.
\end{itemize}

In the forward region H1 and ZEUS 
have installed a {\bf Forward Tracking Detector}~(FTD)~\cite{H1forwardtr} 
and a
{\bf Straw Tube Tracker} (STT)~\cite{ZEUSSTT}, 
respectively, that are based on drift-chambers. 
Their main purpose is to extend the polar angular
coverage to angles smaller than $20^{\circ}$,
outside the acceptance of the central drift chambers. 
However, for both experiments these detectors have not been
used for momentum reconstruction due to a large 
amount of dead material in front of them.
Nevertheless, the forward detectors were partially used
in the pattern recognition.

%%%%%%%%%%%%%%%%%%%%%%%%%%%%%%%%%%%%%%%%%%%%%%%%%%%%%%%%%%%%%%%%%%%%%%%%%%%%%%%%%%%%%%%%%%%%%%%%%%%
\paragraph{Vertex-detector:}
\hspace*{-4mm}
The {\bf Central Silicon Tracker}~(CST)~\cite{H1cst}
and the {\bf Micro Vertex Detector} (MVD)~\cite{ZEUSMVD} were located
in the heart of the H1 and ZEUS experiments.
The MVD was installed only for the \heraii data taking.
The vertex detectors allowed the determination of trajectories of
charged particles in the vicinity of the primary vertex.
The achieved precision was sufficient to resolve vertices from 
secondary decays.
This is essential for the tagging of weakly-decaying
heavy-flavour hadrons with a typical lifetime $c\tau\simeq100\rnge300\mum$.
%From the H1 tracker, CAL and muon paper

The CST (MVD) consisted of two (three) $36\cm$ ($63\cm$) long
concentric cylindrical layers of double-sided silicon-strip detectors
The innermost layer of the CST and MVD was located at $57.6\mm$ 
and $\sim45\mm$, respectively.
The most important benchmarking parameters are given in \tab{det1}.
The following intrinsic hit resolutions were achieved: 
$\sim 11\mum$ for the CST and $\sim 24\mum$ for the MVD.
The CST had a somewhat better average 
transverse impact parameter resolution mainly due to
less material, hence less multiple scattering, but was essentially 
restricted to track reconstruction in the transverse plane.
The MVD contained four wheels of double-sided silicon-strip
detectors in the forward region 
that extended the polar-angle coverage from $20^{\circ}$ to $7^{\circ}$.
Furthermore, it allowed 3D standalone pattern recognition.
At H1 the CST was supplemented with the forward and backward strip detectors,
FST and BST, that extended the polar-angle coverage of CST to $[7^{\circ},173^{\circ}]$.
However, these were not used in H1 heavy-flavour analyses.
%
%%%%%%%%%%%%%%%%%%%%%%%%%%%%%%%%%%%%%%%%%%%%%%%%%%%%%%%%%%%%%%%%%%%%%%%%%%%%%%%%%%%%%%%%%%%%%%%%%%%
\paragraph{Calorimeters:}
The tracking detectors were surrounded by calorimeter systems,
which covered almost the full solid angle.
Their main tasks were to identify and measure the
scattered electron, to reconstruct the hadronic
final state (e.g. jets) and photons 
and to separate electrons from hadrons.
At H1 a fine-grain
liquid-argon (LAr) sandwich calorimeter \cite{h1p2,H1Argon} 
was installed in
the central and forward region. It was supplemented in
the backward region with the  
lead-scintillating fibre calorimeter
SpaCal~\cite{H1SpaCal1,H1SpaCal2}.
In the ZEUS detector the solenoid was surrounded 
by a high-resolution compensating uranium--scintillator calorimeter 
(CAL)~\cite{ZEUSCAL}.
The calorimeters had inner electromagnetic and outer
hadronic sections.
The electron and hadron energy scales of the calorimeters were known 
at the level of $1\%$ and $2\%$, respectively.
The calorimeters were calibrated from the data using kinematic constraints.
Overall, both calorimeter systems performed very well. 

Additionally, for detection of very-low-$Q^2$ events, where the electron
is scattered at a small angle, ZEUS installed the beampipe calorimeter 
(BPC)~\cite{ZEUSBPC}.
This calorimeter was in operation in \herai and was located just $4.4\cm$
from the beam line. 
It allowed the extension of the phase-space coverage to $0.05 < Q^2 < 0.7\gev^2$
\paragraph{Electron taggers:}
\label{sect:taggers}
Both H1 and ZEUS were equipped with special detectors, called electron 
taggers, which were able to detect electrons scattered at very small 
angles. Especially during the \herai
period, these could be used to explicitly identify photoproduction events
in specific ranges of $W$.

%%%%%%%%%%%%%%%%%%%%%%%%%%%%%%%%%%%%%%%%%%%%%%%%%%%%%%%%%%%%%%%%%%%%%%%%%%%%%%%%%%%%%%%%%%%%%%%%%%%
\paragraph{Muon systems:}
To identify muons 
both experiments installed large arrays of limited-streamer tubes~\cite{h1p2,ZEUSMU}
inside and outside
the magnetic return yoke 
(not shown), which covered a wide 
range in polar angle 
and measured muons efficiently for transverse momenta
above $\sim 2\gev$, with significant partial acceptance also at lower $p_T$.
The return yoke of the ZEUS detector was also equipped with drift tubes 
providing complementary muon identification and serving as a backing
calorimeter (BAC). 
A forward muon system completed the coverage 
of the tracking detectors. The H1 detector had a similar muon coverage, 
including the usage of the liquid argon calorimeter as a tracking 
calorimeter. 
%%%%%%%%%%%%%%%%%%%%%%%%%%%%%%%%%%%%%%%%%%%%%%%%%%%%%%%%%%%%%%%%%%%%%%%%%%%%%%%%%%%%%%%%%%%%%%%%%%%
\paragraph{Luminosity measurement:}
In both experiments the luminosity was measured using 
the photon bremsstrahlung process $ep\,\rightarrow\, e\gamma p$.
The photons were detected by dedicated detectors~%
\cite{ZEUSlumi,ZEUSlumi2}
%no reference for H1 lumi system found in H1 papers
about $100\m$
away from the interaction points in the $e$-beam direction.
In addition, H1 also used the SpaCal to measure 
the large-angle QED compton scattering~\cite{H1lumi2new}.
The ultimate precision of the luminosity measurement by H1 (ZEUS) is 
$2.3\%$ ($1.8\%$) for the \heraii period and 
$1.5\%$ ($2.2\%$) for the \herai period.
%H1 HERAII: H1lumi2new; H1 HERAI: from DESY-06-240, 99-00 data (same for vertexins analyses)
% ZEUS HERAII: all new papers, average over periods; ZEUS HERAI: DESY-03-115, 98-00 data (D* in DIS)
% 
%%%%%%%%%%%%%%%%%%%%%%%%%%%%%%%%%%%%%%%%%%%%%%%%%%%%%%%%%%%%%%%%%%%%%%%%%%%%%%%%%%%%%%%%%%%%%%%%%%%%
\paragraph{Trigger and readout system:}
Both H1~\cite{H1trig} and ZEUS~\cite{zeus,ZEUSgtt} have used a 
multi-level trigger system to select interesting $ep$ events online 
and to suppress background from beam--gas interactions.
The H1 trigger system consisted of two hardware layers
and one software filter.
It was supplemented for the \heraii period  
by an additional track trigger.
The ZEUS trigger was based on one hardware and two software
levels.
The first two levels mostly operated with the energy sums in the calorimeter, 
timing and limited tracking information.
On the third level a complete reconstruction of the event is performed,
using a simplified version of the offline 
reconstruction software, to select more sophisticated objects like 
jets, tracks and even D-meson candidates.
The triggers reduced the rate from the nominal HERA
bunch-crossing rate $10\MHz$ to the storage rate $\sim10\Hz$.
While the topology of DIS events allowed triggering on the scattered electron
inclusively already at the first level of the trigger chain,
triggering on heavy-flavour photoproduction was more challenging 
and required reconstruction of leptons, tracks, hadronic activity 
in the calorimeter or even explicitly charm hadrons.
Both experiments had capabilities to include limited tracking 
information already at the first trigger level, 
for instance on the number of tracks and the fraction which originates
from the $ep$ interaction vertex.

%%%%%%%%%%%%%%%%%%%%%%%%%%%%%%%%%%%%%%%%%%%%%%%%%%%%%%%%%%%%%%%%%%%%%%%%%%%%%%%%%%%%%%%%%%%%%%%%%%%
\subsection{Event reconstruction}

Various heavy-flavour tagging techniques (see Section \ref{sect:Tagging}) 
exploit different measured quantities,  
like tracks, vertices, energy-flow objects, jets and muons.
The reconstruction of these quantities is described in the following:

%%%%%%%%%%%%%%%%%%%%%%%%%%%%%%%%%%%%%%%%%%%%%%%%%%%%%%%%%%%%%%%%%%%%%%%%%%%%%%%%%%%%%%%%%%%%%%%%%%%%
\paragraph{Tracking:}

\begin{figure}[tb]
\begin{center}

\includegraphics[trim = 0pt 0pt 0pt 650pt,clip,%
height=0.15\textheight]{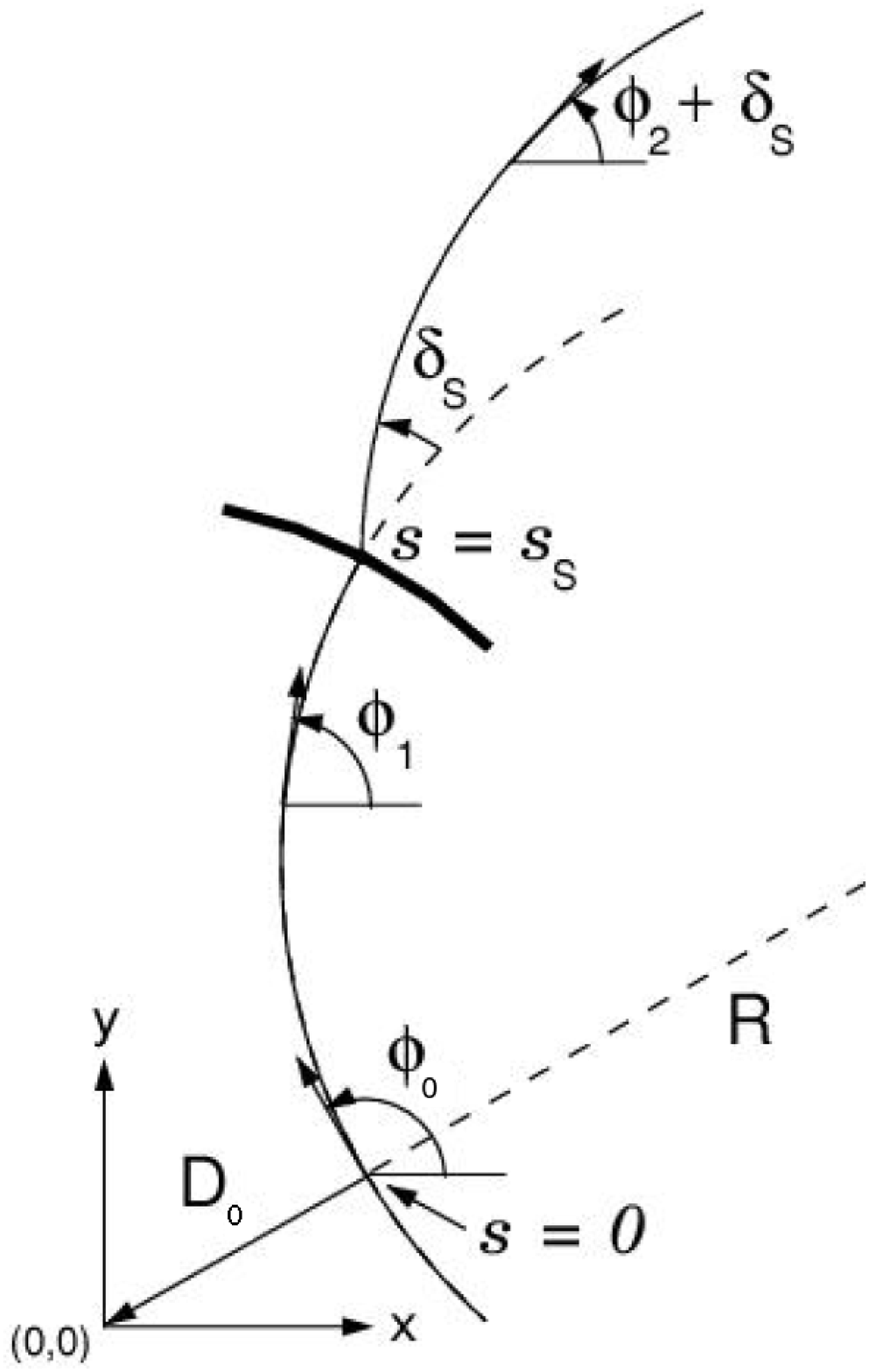}
\caption{
Helix parameterisation in the track-fit procedure. 
The initial parametrisation at the point of closest approach 
as well as effects of multiple scattering are shown. 
}
\label{fig:trkhelix}
\end{center}
\end{figure}

Tracks were reconstructed combining hits from the central tracking chambers,
silicon-strip detectors and forward/backward trackers for high $|\eta|$.
In both experiments tracks were parametrised with a helix 
defined by 5 parameters (\fig{trkhelix} shows the $r$-$\phi$-view): 
the curvature $\kappa=Q/R$, which is the signed inverse radius, 
$D_0$, the \emph{dca} distance of closest approach in the $XY$ plane, 
$\phi_0$, the azimuth angle, 
$z_0$, the distance of closest approach along the $Z$ axis, 
and the polar angle $\theta_0$.
To account for multiple scattering and energy loss in the
material along the track trajectory and for inhomogeneities of the magnetic 
field,
 the track parametrisation was refined in a track-refit process based on a 
broken-lines algorithm \cite{Blobel:2006yi} in H1 
and a Kalman filter \cite{zeusTrkfit} in ZEUS.
Additionally, the parameterisation of all tracks that were fitted to the
primary or to a secondary vertex (see below) was further improved by a 
track refit 
using the vertex position as a constraint.

Analyses that aimed at the best tracking precision used tracks with 
typically $|\eta|<1.7$.
%1.75  ZEUS: based on D* in DIS in HERAII and 98-00 
%H1: D* in DIS @ low Q2: extended eta<20, pt(pis)>0.08, nominal: eta<1.7 (h1th-504), also 1.7 for high-Q2 D*
%
This coverage corresponds to the region 
where efficiencies and resolutions are high and well known. 
Performance benchmarking for the two tracking systems is shown in 
\tab{det1}.

At low momenta, the $dE/dx$ measurement for each track allowed the separation
of pions, kaons and protons, while at high momenta electron/hadron 
separation was possible to some degree.

%%%%%%%%%%%%%%%%%%%%%%%%%%%%%%%%%%%%%%%%%%%%%%%%%%%%%%%%%%%%%%%%%%%%%%%%%%%%%%%%%%%%%%%%%%%%%%%%%%%
\paragraph{Vertexing:}
Reconstructed tracks
were used as an input for the primary vertex in each event.
In addition, if silicon-strip information was available, 
the time-averaged mean $XY$ position of the $ep$ interaction
region, the beam spot, was used to further confine the position of 
the primary vertex
in the event.
The beam spot was measured by the experiments as a function of time for
each $\sim1000$ events.
The beam spot size in the transverse plane was measured to be 
$\sigma_x=145\mum$ and $\sigma_y=25\mum$ in H1 \herai data,
% numbers taken from DESY-THESIS-2006-035 
$\sigma_x=110\mum$ and $\sigma_y=30\mum$ in H1 \heraii data,
% numbers taken from DESY-THESIS-2012-007
and $\sigma_x=85\mum$ and $\sigma_y=23\mum$ in ZEUS \heraii data.
% averaging over e+p and e-p data with rounding from 84 to 85
%
The beam-spot size along the $Z$ axis was much larger,
$\sigma_Z\sim10\cm$, and therefore was not used as a constraint.
In H1 the vertex fitting was performed in the $XY$ plane%
\footnote{An iterative procedure to determine the $Z$ position 
of the primary vertex was used.},
while ZEUS did a full vertex fit in $XYZ$.
Nevertheless, both experiments used only the $XY$ projections of
decays in heavy-flavour analyses due to superior resolution.

In addition, in the context of some ZEUS heavy-flavour analyses,
selected tracks were removed from the primary vertex fit 
and the fit was re-done.
Combinations of such tracks were fitted to a displaced secondary vertex 
that was associated with a decay of a heavy-flavour hadron.
Procedures similar to those used in the primary-vertex fit%
\footnote{For obvious reasons, no beamspot constraint is used.} are used
to fit the secondary vertices as well.
Alternatively, combinations of impact parameters of several tracks were 
used in H1.
Secondary vertices give an important handle:
\begin{itemize}
\item
to test the hypothesis that selected tracks originate from 
a decay of the same particle by evaluation of 
the $\chi^2$ of the secondary vertex;
\item
to evaluate the flight distance of that particle, which is related
to the particle $c\tau$.
\end{itemize}
This will play an essential role (together with the track impact parameter)
in the heavy-flavour lifetime tagging (cf. Section \ref{sect:Tagging}).

%%%%%%%%%%%%%%%%%%%%%%%%%%%%%%%%%%%%%%%%%%%%%%%%%%%%%%%%%%%%%%%%%%%%%%%%%%%%%%%%%%%%%%%%%%%%%%%%%%%
\paragraph{Electron reconstruction:}
Electron identification was needed to reconstruct DIS events
as well as to measure semi-leptonic decays of beauty and charm quarks.
Electrons were separated from hadrons using the shape of clusters
in the calorimeter and $dE/dx$ information from the 
central drift chambers.
A typical phase-space coverage for electrons in
a beauty-production measurement 
in the semi-leptonic electron decay channel was
$p_T>1\gev$ ($p_T>0.9\gev$) and
% 1 for H1 DESY-12-072 and 0.9 for ZEUS DESY-11-005, DESY-08-056
$-1 < \eta < 1.7$ ($|\eta|<1.5$)
% H1: DESY-12-072; ZEUS: DESY-11-005, DESY-08-056
for the H1 (ZEUS) measurements.

%%%%%%%%%%%%%%%%%%%%%%%%%%%%%%%%%%%%%%%%%%%%%%%%%%%%%%%%%%%%%%%%%%%%%%%%%%%%%%%%%%%%%%%%%%%%%%%%%%%
\paragraph{Muon reconstruction:}
Muons were identified by combining information from
the tracking systems, calorimeters and muon chambers.
For $p_T > 1.5-2\gev$ ($p>2$ GeV) the information from the muon chambers
was exploited and the fraction of hadrons which 
were misidentified as muons was typically less than $1\%$.
For momenta $1\rnge2\gev$ isolated muons can be efficiently 
identified in the calorimeters, using ``minimum ionising particle'' (mip)
signatures, which however leads to reduced purity.
The efficiency to identify high momentum isolated muons in 
the H1 and ZEUS muon systems was $90\%$ and $55\%$~\cite{h1zeusMUperf}, 
respectively.

The wide coverage of the muon chambers allowed the 
extension of the phase space of the measurements up to
$-1.6<\eta<2.3$.

%%%%%%%%%%%%%%%%%%%%%%%%%%%%%%%%%%%%%%%%%%%%%%%%%%%%%%%%%%%%%%%%%%%%%%%%%%%%%%%%%%%%%%%%%%%%%%%%%%%
\paragraph{Hadronic system:}
Energy flow objects (EFOs) were used in both experiments to reconstruct
the hadronic final state~\cite{Adloff:1997mi,briskin:thesis98}.
These objects were based on a combination of information 
from the calorimeter and the tracking system optimising energy resolution.
Track information is superior for low-energy EFOs, while the calorimeter 
measurement
is preferred at high energy as well as to measure neutral particles.

Jets at HERA have been reconstructed with 
the inclusive $k_T$ clustering algorithm~\cite{Ellis:1993tq,Catani:1993hr}%
\footnote{Mostly in the longitudinally-invariant mode with 
the massless $P_T$ and massive $E_T$ recombination 
schemes~\cite{Butterworth:2002xg} in the
H1 and ZEUS experiments, respectively}.
The $R$ parameter was set to $R=1$, which is larger than the values used 
typically at $pp$ and $p\bar p$ experiments ($0.4\rnge0.7$), since at HERA 
jets with relatively low transverse momenta were analysed.
The chosen jet algorithm is infrared and collinear safe
to all orders in perturbation theory.
It was checked~\cite{Abramowicz:2010ke} that at HERA the $k_T$, 
anti-$k_T$~\cite{Cacciari:2008gp} and SIScone~\cite{Salam:2007xv} algorithms
produce very similar measurement results and that the precision
of NLO QCD calculations for the anti-$k_T$ algorithm is very similar to 
that of the $k_T$ algorithm.

The final precision of the jet energy scale uncertainty of the H1
and ZEUS calorimeters was $1\rnge2\%$ (see \cite{SchornerSadenius:2012de}
for a recent review of jet results from HERA).

\subsection{Summary}

HERA was the first and so far only high energy $ep$ collider.
The results discussed in this review were obtained with the H1 and ZEUS
detector in two different running periods, denoted ``HERA I'' and 
``HERA II''. The main detector parts relevant for the detection and 
reconstruction of heavy flavour events were the 
electromagnetic part of the calorimeters for the 
reconstruction of the scattered electron (if detectable), the 
calorimeters and tracking
systems for the reconstruction of the decay products of heavy flavoured
particles, and the muon systems for the detection of semileptonic decay
final states.  

\newpage

% HQ tagging techniques
\clearpage
\section{Charm and Beauty detection at HERA}
\label{sect:Tagging}

%
%\subsection{Basic tagging elements}
The large charm- and beauty-quark masses result in kinematical 
suppression of their production compared to the light-flavour 
cross sections.
Therefore, special techniques have to be employed to
separate charm and beauty ``signal'' from the dominating light-flavour
``background''.
These techniques utilise distinct properties 
of the charm and beauty hadrons (see \fig{tag1} for illustration):
%
%%%%%%%%%%%%%%%%%%%%%%%%%%%%%%%%%%%%%%%%%%%%%%%%%%%%%%%%%%%%%%%%%%%%%%%%%
\begin{figure}[hb]
\centering
\includegraphics[width=0.9\linewidth]{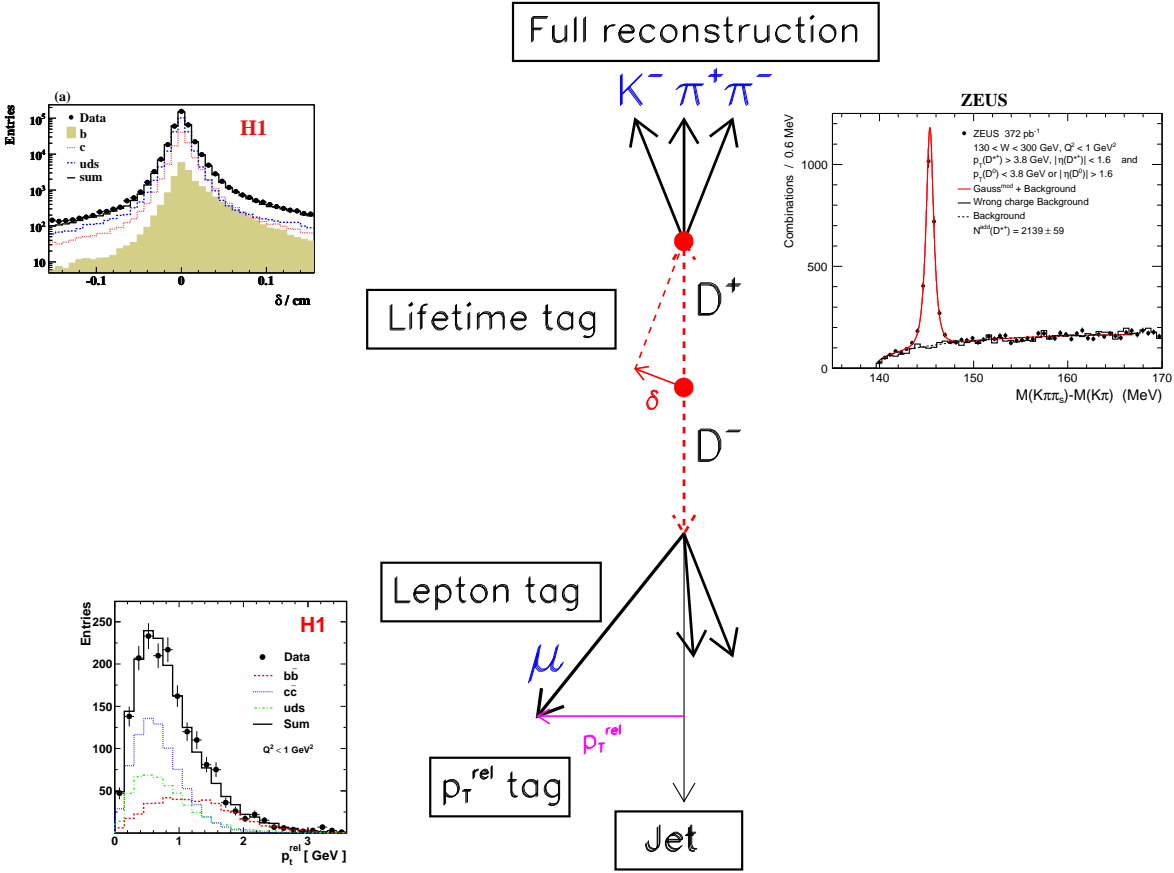}
\caption{Overview of tagging methods for heavy-flavour events.
Each method is accompanied by an illustrative distribution %
\pcite{H1cbPHP2006,H1bPHP2005,ZEUSfrag2013}.}
\label{fig:tag1}
\end{figure}
%%%%%%%%%%%%%%%%%%%%%%%%%%%%%%%%%%%%%%%%%%%%%%%%%%%%%%%%%%%%%%%%%%%%%%%%%
%
\begin{enumerate}
\item {\bf Flavour tagging.} 
The tagging of the quark flavour is done either by 
\emph{full reconstruction} of decays of heavy-flavoured hadrons 
or by \emph{lepton tagging} from the semi-leptonic 
decays of those hadrons. 
The former was used at HERA
only for charm tagging, since low production rates and small 
branching ratios for useful decays led to insufficient statistics for
fully-reconstructed beauty hadrons. 
The latter was mostly used for beauty tagging, since $b$-hadron decays 
produce leptons with sufficiently high momenta for 
efficient identification in the detectors.

\item {\bf Lifetime tagging.}
This method exploits the relatively long
lifetimes of weakly-decaying heavy-flavour hadrons through the 
reconstruction of tracks with large impact 
parameter $\delta$ or displaced secondary vertices.
In addition, the information about the flight direction, extracted from
either the track, hadron or jet momentum, can be used
to construct a \emph{signed impact parameter} or a
\emph{signed decay length} (see later).
This is a powerful tool to separate charm and beauty 
from light-flavour events, in which tracks mostly originate
from the primary vertex.
\item {\bf Mass tagging.}
The tagging using the mass of the heavy quark or meson is performed
either explicitly by a full reconstruction of the mass from all decay products
(also see flavour tagging), by a partial reconstruction via the mass of a jet 
or of all tracks at a secondary vertex,
or indirectly by measuring the relative transverse momentum of a particle 
with respect to 
the axis of the associated jet, \ptrel.
The latter was mostly used to separate beauty events from production
of other flavours, since the large quark mass produces 
large \ptrel values.
%
% For some measurements presented in this article, 
% a heavy quark tag is combined with requiring
% in the event a jet, which is not associated to the 
% tagged heavy quark, but represents another hard parton
% which may be the partner heavy quark or a light quark
% or a gluon.
% %
% This additional parton allows to study the production dynamics
% in heavy flavour events in more detail.
%
\end{enumerate}

The above methods are all based on measuring
the decay particles of one heavy quark (\emph{single tag}).
%and thus they are {\em single heavy quark tags}.
%
Several methods can be combined to increase the purity
at the cost of statistics.
Both heavy quarks (\emph{double tag}) in an event can be tagged,
by applying one method to tag one heavy quark 
and another (or the same) for the other heavy quark.
This allows a more detailed study of the heavy-quark production
mechanisms, 
but the double tagging efficiencies are low.

In the following the different tagging methods are discussed
in more detail with emphasis on the
advantages and disadvantages of each method.

%%%%%%%%%%%%%%%%%%%%%%%%%%%%%%%%%%%%%%%%%%%%%%%%%%%%%%%%%%%%%%%%%%%%%%%%%
%%%%%%%%%%%%%%%%%%%%% D mesons %%%%%%%%%%%%%%%%%%%%%%%%%%%%%%%%%%%%%%%%%%
%%%%%%%%%%%%%%%%%%%%%%%%%%%%%%%%%%%%%%%%%%%%%%%%%%%%%%%%%%%%%%%%%%%%%%%%%

\subsection{Charm tagging using full
reconstruction of charm hadrons.}
\label{sect:Dmestag}
\begin{figure}[tb!]
\centering
\bmp{c}{0.4\linewidth}
\includegraphics[width=0.99\linewidth]{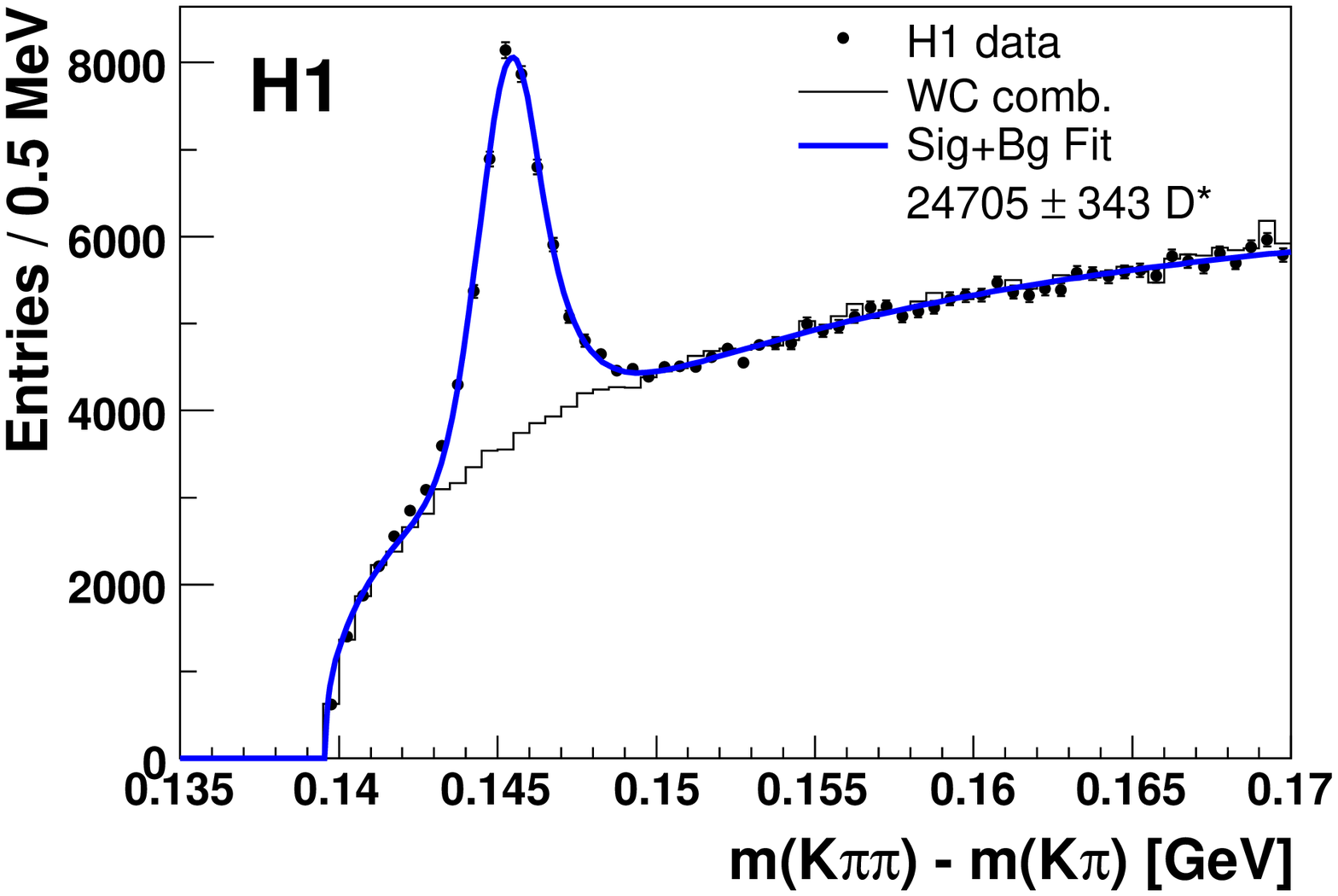}
\put(-45,60){\Large (a)}
\put(-80,40){\tiny $\mathsf{p_T(D^{*})>1.25\,Ge\eVdist{V}}$}
\emp%
\bmp{c}{0.4\linewidth}
\includegraphics[width=0.99\linewidth]{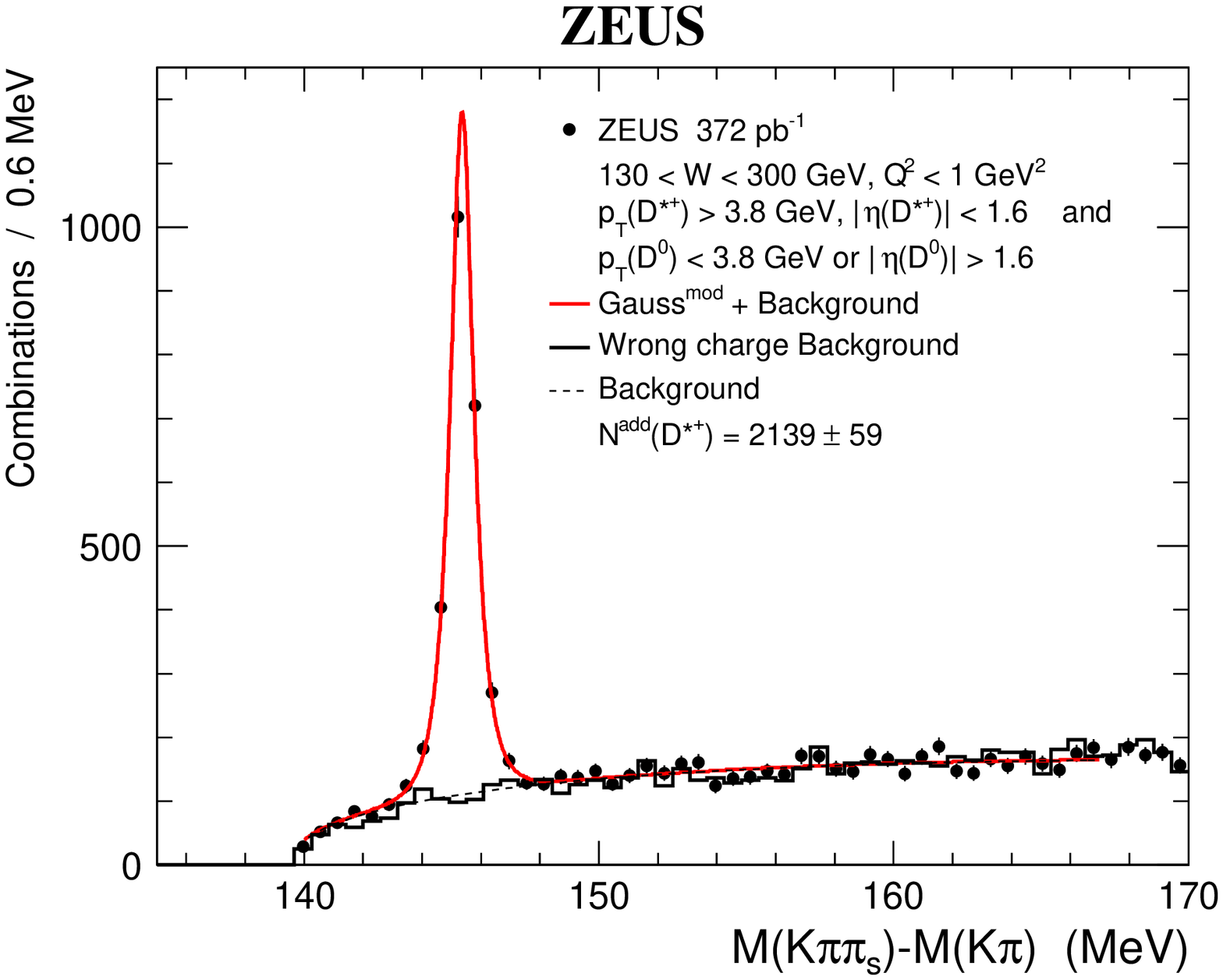}
\put(-45,70){\Large (b)}
\emp\\
\bmp{c}{0.4\linewidth}
\includegraphics[width=0.99\linewidth]{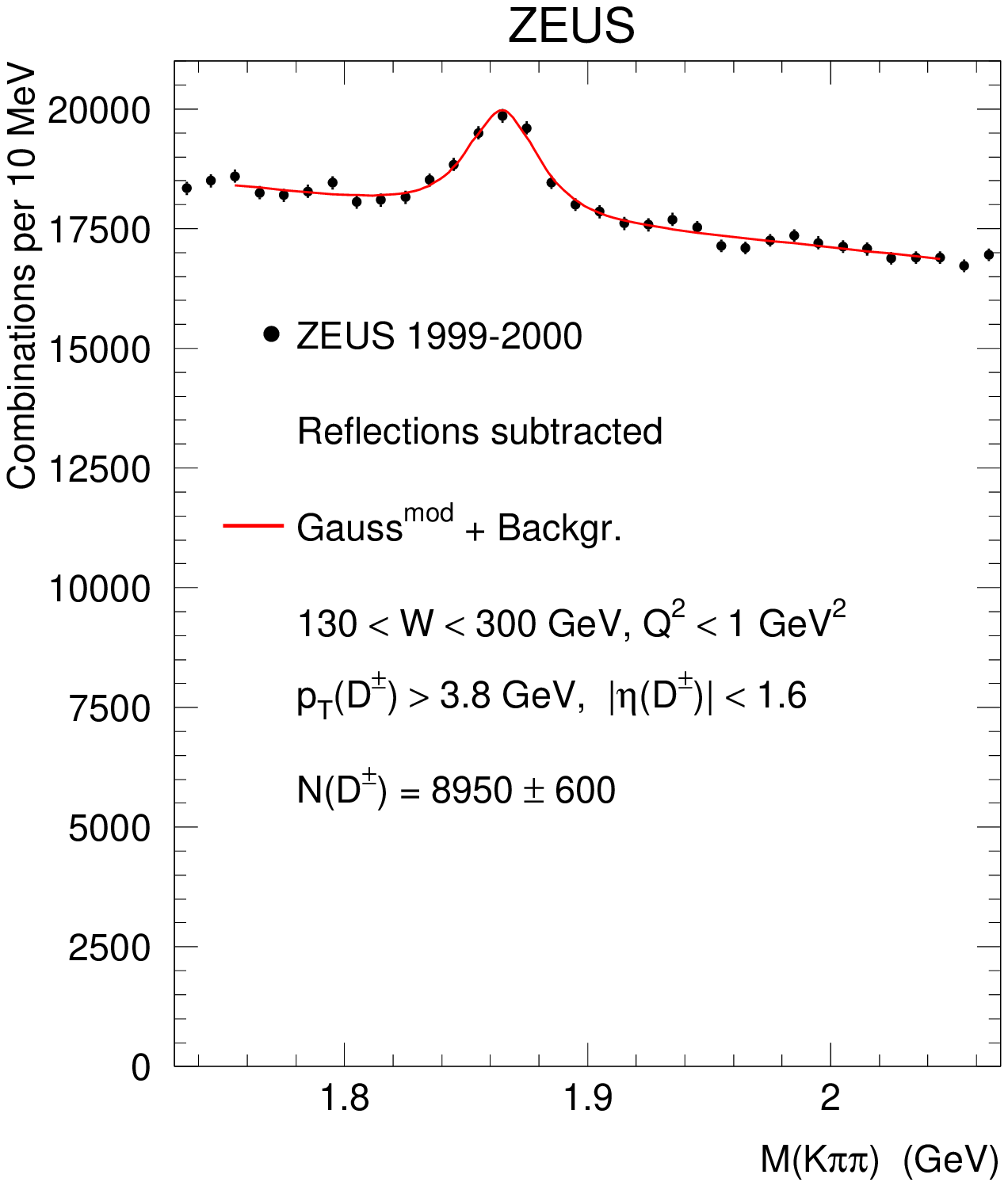}
\put(-60,120){\Large (c)}
\emp%
\bmp{c}{0.4\linewidth}
\includegraphics[width=0.99\linewidth]{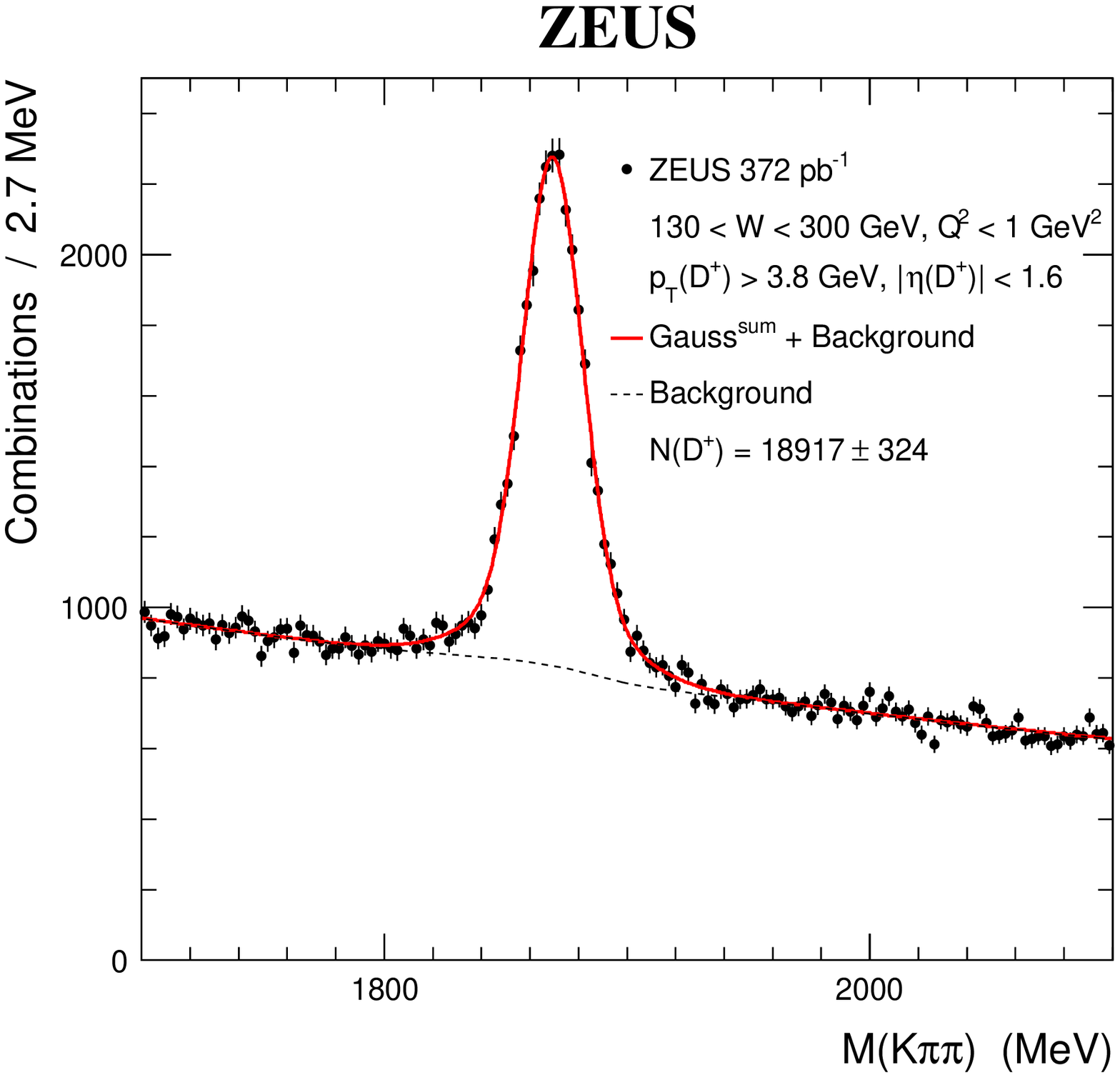}
\put(-35,100){\Large (d)}
\emp
\caption{Full reconstruction of mass spectra for charm mesons
in the data.
The reconstructed mass difference $\Delta M$ for \dst 
(a)~\pcite{h1dstar_hera2} and (b)~\pcite{ZEUSfrag2013} 
and the \dplusp mass 
(c)~\pcite{ZEUSfrag2005} and (d)~\pcite{ZEUSfrag2013} are shown.}
\label{fig:Dmass}
\end{figure}

Most of the HERA 
charm results have been made using the {\em golden decay channel}
$D^{*+} \rightarrow D^0 \pi^+_{s} \rightarrow (K^- \pi^+) \pi^+_{s}$
(see \fig{Dmass})\footnote{Throughout the paper the \dstarp is mostly
referred to as \dst. The full reconstruction of $D^{*0}$'s at HERA was 
impossible since the resulting decay photon or $\pi^0$ could not be reliably
measured.}.
%----------------------------------------------------------------------
%----------------------------------------------------------------------
Occasionally, also the $D^0 \to K^- \pi^+\pi^-\pi^+$ decay channel 
was used to increase statistics.
Due to the small energy release in the decay
$D^{*+} \rightarrow D^0 \pi^+_{s}$ 
($M(D^{*+})-M(D^0)-M(\pi^+) \approx 6\mev$)
the phase space for combinatorial background is 
suppressed, and the resolution for the $D^*-D$ mass difference is strongly 
enhanced, providing an excellent signal to background ratio.
This also leads to a small momentum of the produced pion, 
which is often called the ``slow'' pion, $\pi_s$.
The capability of a detector to measure very-low-momentum tracks
defines the accessible region of $p_T(\pi_s)$ and therefore the
phase space of the \dst:
a typical $p_T(\pi_s) > 0.1\gev$ restriction leads to 
$p_T(D^{*}) > 1.5\gev$.
For some analyses the $p_T(D^{*})$ cut was raised to suppress 
the combinatorial background, which rises
at low $p_T$ (cf. \fig{Dmass} (a) and (b)).
For the signal extraction, the observable 
$\Delta M = M(K^{-}\pi^{+}\pi^{+}_{s})-M(K^{-}\pi^{+})$
was chosen.
The number of signal events is determined 
either by counting the number of events in the peak region 
after subtracting the combinatorial background, which is estimated from the
$\Delta M$ distribution of ``wrong charge'' $K^{+}\pi^{+}\pi^{-}_s$ 
combinations,
by fitting the spectrum with a Gaussian-like shape for the signal and
a phenomenological function for the background, or by a combination of these
two methods.

Besides the \dst golden decay channel, 
the $D^0 \to K^- \pi^+$, $D^+ \to K^- \pi^+ \pi^+$, 
$D_s^+ \to K^+ K^- \pi^+$ 
and $\Lambda_c^+ \to p K^- \pi^+$ decays of charm hadrons were used to tag
charm in the events.
These charm hadrons feature much larger background 
(see \fig{Dmass}(c) for an example of a \dplusp measurement).
However, a lifetime tag can be added, exploiting the
relatively large values $c\tau \sim 100\rnge300\mum$ for the weakly-decaying charm
hadrons.
The decay length is reconstructed by fitting a displaced
secondary vertex to selected tracks of decay products
(see \Sect{LTtag} for more details).
This combined approach allowed to significantly reduce combinatorial
background, most noticeably for the \dplusp that has the largest lifetime
(cf. \fig{Dmass}(c) and (d)), to a level that is still 
somewhat worse than but comparable to the \dst.
Nevertheless, due to the low boost, lifetime tagging is inefficient in the 
phase space $p_T(D) < m(D)$ that is also not accessible with \dst's.
In this region it is beneficial to study particular decay channels 
with neutral strange hadrons $K^0_S$ or $\Lambda$, for example
$D^+ \to K^0_S \pi^+$.

The H1 and ZEUS experiments achieved a similarly good 
signal mass-peak resolution for charm hadrons.
Note that the distributions in \fig{Dmass} can not be 
compared directly due to different kinematic regions.

In summary, the advantages ($+$) and disadvantages ($-$) 
of full hadron reconstruction are:
\begin{itemize}  \itemsep5pt \parskip0pt
\item[$+$]
The full reconstruction allows an accurate determination of the momentum
of the charm hadron, which is correlated with the 
kinematics of the charm quark and can be used to study
the fragmentation process.
\item[$+$]
An excellent signal to background ratio of $\sim 1:1$ can be
achieved using the \dst golden decay channel or the reconstruction of 
other weakly-decaying mesons including a lifetime tag~(\fig{Dmass}).
\item[$+$]
The combinatorial background can be parametrised with an empirical
function and does not depend on Monte Carlo simulations of
the light-flavour background.
\item[$+$]
The signal mass peak is a clear signature which can be used 
in the online filtering of events to reduce the trigger rates.
This requires the usage of advanced tracking information in the trigger logic.
\item[$-$] 
The typical probabilities for a charm quark to hadronise 
into a specific $D$ meson are $\approx 0.15 \rnge 0.25$ 
and the branching ratios, $BR$, for the commonly used 
decay channels are $\approx 0.05 \rnge 0.10$; 
thus, only $\sim 1 \rnge 2 \%$ of all \emph{c} quarks can be tagged.
The kinematic and geometric  acceptances reduce the visible fraction
even further, in particular in the case of the additional use of 
lifetime tagging, which reduces the detection efficiency by a factor
$\sim 2 \rnge 5$.
Therefore, only a very small fraction of charm quarks can be tagged
with the full reconstruction method.
\end{itemize}

%%%%%%%%%%%%%%%%%%%%%%%%%%%%%%%%%%%%%%%%%%%%%%%%%%%%%%%%%%%%%%%%%%%%%%%%%
%%%%%%%%%%%%%%%%%%%%% semi-leptonic %%%%%%%%%%%%%%%%%%%%%%%%%%%%%%%%%%%%%
%%%%%%%%%%%%%%%%%%%%%%%%%%%%%%%%%%%%%%%%%%%%%%%%%%%%%%%%%%%%%%%%%%%%%%%%%

\subsection{Heavy-flavour tagging 
 with lepton + $\mathbf{p_T^{\mathrm rel}}$}
\label{sect:bmujjx}

\begin{figure}[tb]
\centering
\includegraphics[width=0.48\linewidth]{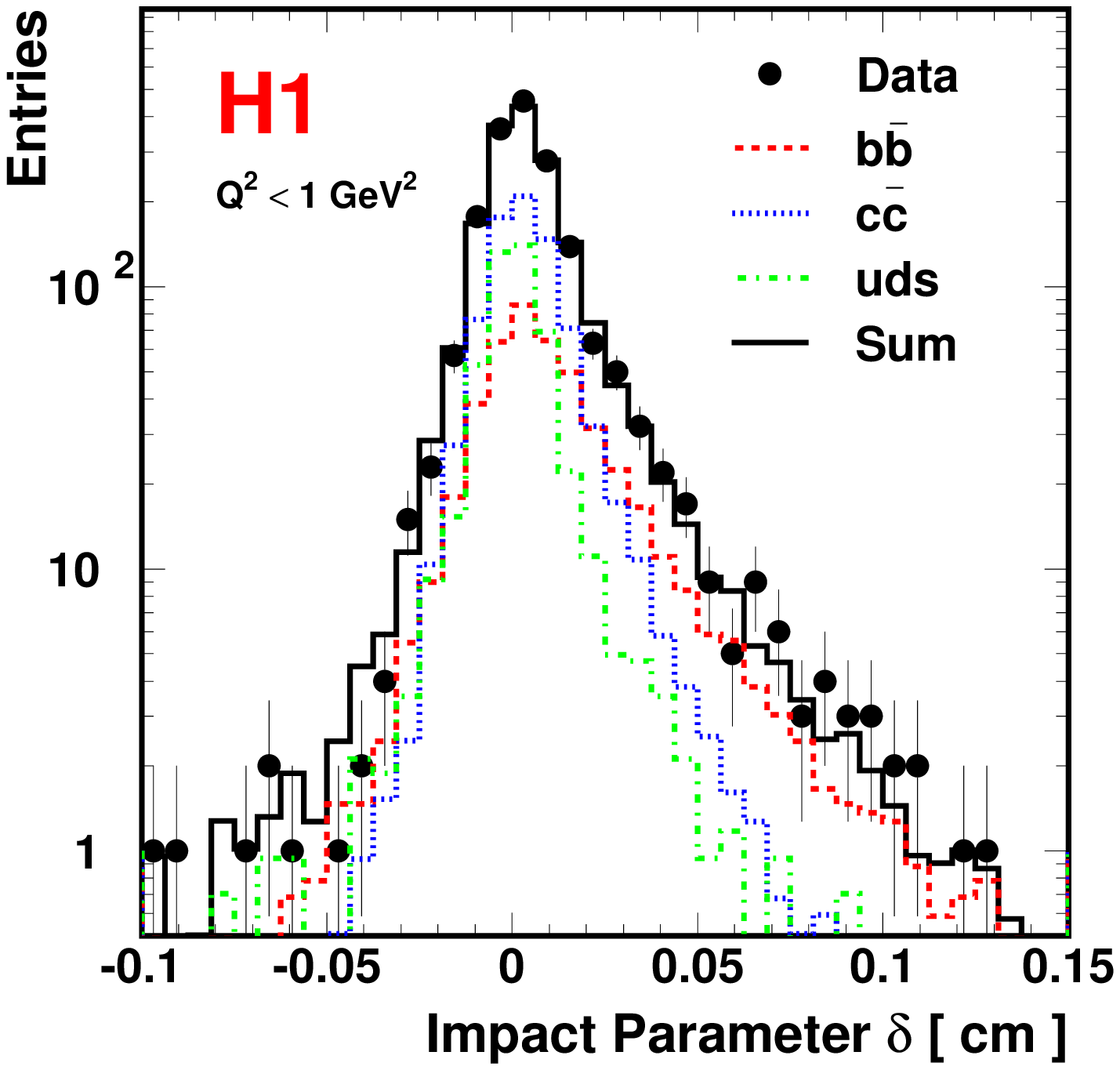}%
\includegraphics[width=0.48\linewidth]{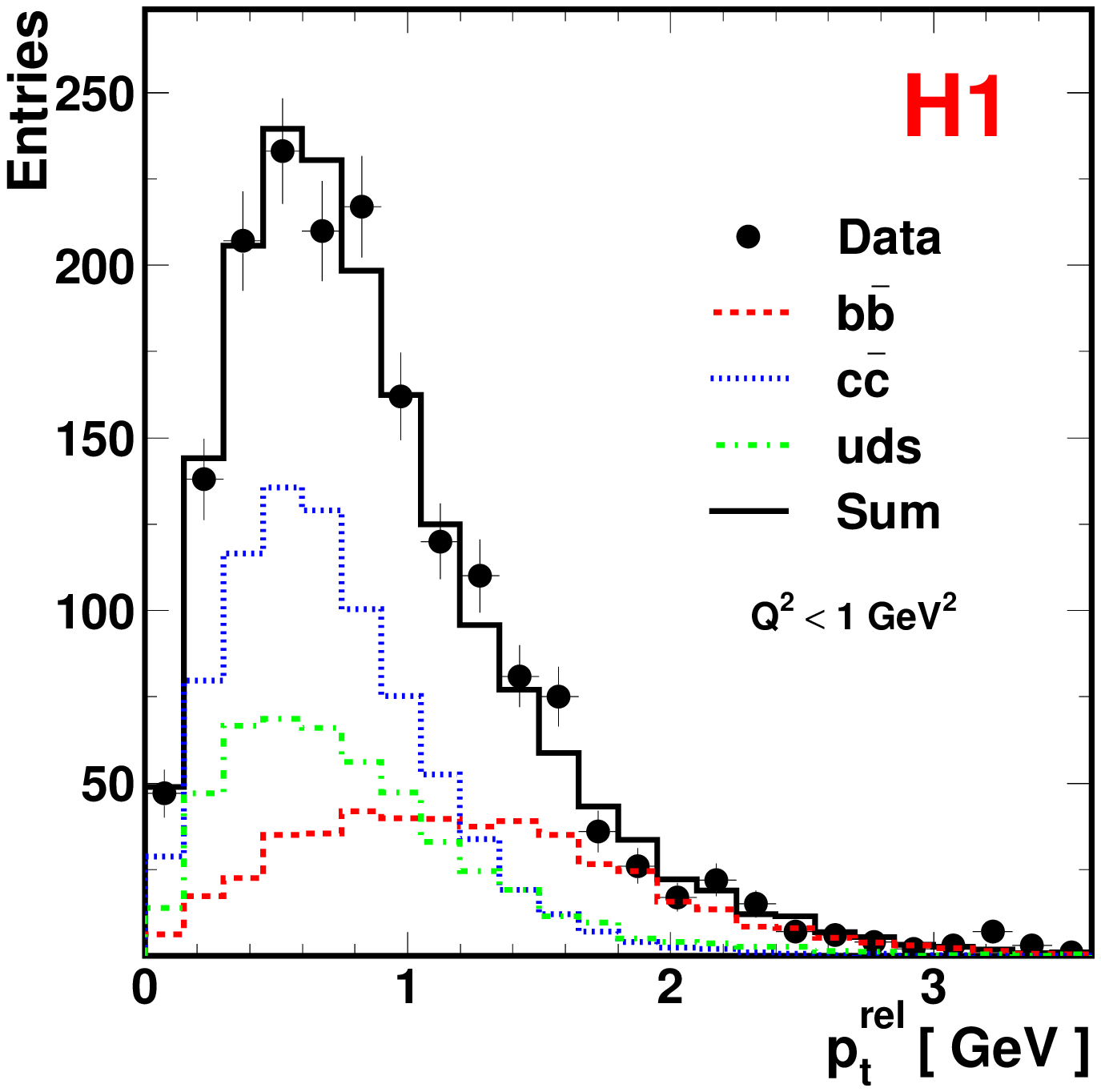}
\caption{
Distributions of the signed
impact parameter $\delta$ of the muon track (left) 
and the transverse muon momentum \ptrel relative
to the axis of the associated jet (right) 
for the photoproduction event sample of the H1 
beauty analysis~\cite{H1bPHP2005}.
%\cite{Aktas:2005zc}.
%
The data (dots) are compared to Monte Carlo predictions
(solid line).
Separate contributions of events arising from
$b$-quark (dashed line), \mbox{$c$-quarks} (dotted line)
and the light-quark (dash-dotted line) production are shown.
}
\label{fig:ptrel}
\end{figure}

A well established method to identify beauty%
\footnote{Also charm, but less efficiently.} 
quarks is to select a muon with
high transverse momentum of typically above $1.5-2\gev$ 
from semileptonic $b$-quark decay,
which is associated to a jet that represents the beauty quark
and consists of the muon
and further final-state particles.
The background to this signature is composed of
charm production with a genuine muon from semi-leptonic
decays and of light-flavour events with a hadron
misidentified as a muon (mainly due to in-flight 
$\pi^+$ and $K^+$ decays and hadronic energy leakage).
To separate beauty from the background contributions for a single tag,
the \ptrel observable is used which, due to the large 
beauty-quark mass, extends to much larger values than 
for the other sources.
Additionally, a lifetime tag can be added by reconstructing
the signed impact parameter, $\delta$, of the muon track
(see \Sect{LTtag} for more details).
This further improves beauty separation and also allows 
charm-event tagging, since long-lived heavy-flavour hadrons
lead to larger $\delta$ values than in light-flavour events.
\Fig{ptrel} shows the distribution of both variables.
%----------------------------------------------------------------------
%----------------------------------------------------------------------
The beauty component has a distinct shape in both that 
allows one to disentangle it from the others.
The fractions of light-flavour, charm and beauty events 
in the data are determined from template fits to the 
discriminating variables extracting template shapes 
from Monte Carlo simulations with
data-driven corrections.
In some analyses also the missing energy in the detector, 
which is associated with the undetected neutrino,
was considered to improve charm/light-flavour separation.

A similar technique can be followed using electrons.
This allows going down to $p_T(e)> 1\gev$, 
however the available $\eta$ region is narrower than for muons
(see \Sect{experiments}) and the lepton signature is
more complex.

In summary, the advantages ($+$) and disadvantages ($-$) 
of the use of semileptonic hadron decays are:
\begin{itemize}  \itemsep5pt \parskip0pt
\item[$+$]
The relatively large branching ratio
$BR(b \rightarrow \ell X) \sim 21\% $~\cite{PDG12}, 
which includes $b\rightarrow c X \rightarrow \ell X$ and other cascade decays, 
provides a reasonable tagging efficiency for $b$ quarks.
\item[$+$]
Muon tagging extends the phase space of heavy-quark measurements, due to
additional coverage outside of the polar acceptance of the central 
tracking systems.
Thus measurements of beauty production in the forward and backward
regions are possible with muons.
\item[$+$]
Semi-isolated leptons provide a clean experimental 
signature for the trigger system.
They allow one to efficiently select beauty events suppressing
charm and light-flavour production by $p_T(\ell)$ cuts.
\item[$-$]
The requirement of a jet associated with the lepton in order to use 
\ptrel or lifetime tagging 
%restricts measurements to relatively high $p_T^{\mathrm{jet}}$, which 
cuts into the low-$p_T$ phase space of the \emph{b}
quarks.
\item[$-$]
The usage of semi-leptonic tagging for charm studies is very complicated due to
weak separation power.
\end{itemize}

%%%%%%%%%%%%%%%%%%%%%%%%%%%%%%%%%%%%%%%%%%%%%%%%%%%%%%%%%%%%%%%%%%%%%%%%%
%%%%%%%%%%%%%%%%%%%%% inclusive lt %%%%%%%%%%%%%%%%%%%%%%%%%%%%%%%%%%%%%%
%%%%%%%%%%%%%%%%%%%%%%%%%%%%%%%%%%%%%%%%%%%%%%%%%%%%%%%%%%%%%%%%%%%%%%%%%

\subsection{Charm and Beauty with inclusive lifetime tagging}
\label{sect:LTtag}
The aforementioned tagging methods suffer from 
the fact that only a fraction of the charm or beauty
quark decays ends up in the selected final state.
This can be avoided by using an
{\em inclusive tagging method}, based
on the long lifetime of charm and 
beauty quarks: $c\tau_c \simeq 100\rnge300\mum$ and
$c\tau_b \simeq 500\mum$.
This approach relies on silicon-strip detectors to
accurately measure the track parametrisation in the vicinity of the 
interaction vertex~(see \Sect{experiments}).
Therefore, it was not yet available in ZEUS for the \herai
data set and was pioneered at HERA by the H1 collaboration.
Lifetime tagging can be based either on impact parameters of 
individual tracks or on decay lengths of displaced 
secondary vertices fitted to selected tracks.
Both techniques typically use tracks with transverse momenta
$p_T>0.5\gev$ to limit multiple scattering 
and hits in at least two layers of the vertex 
detectors.
The lifetime tagging is often applied based on information in 
the transverse $XY$ plane, since the profile of the interaction 
region and also the detector layouts do not allow for sufficiently 
high resolution of the tracks in the coordinate along the beam line.

The impact parameter distribution allows the separation of long-lived
heavy-flavour hadrons from short-lived light-flavour hadrons.
%
%----------------------------------------------------------------------
\begin{figure}[htb]
\centering
\bmp{c}{0.45\linewidth}
\centering
\includegraphics[width=0.6\linewidth]{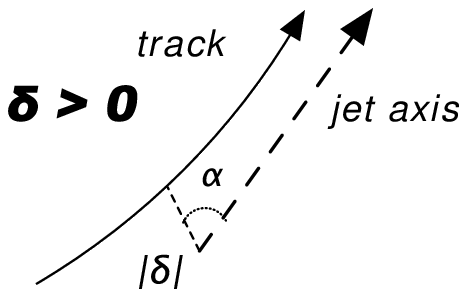}\\
\includegraphics[width=0.6\linewidth]{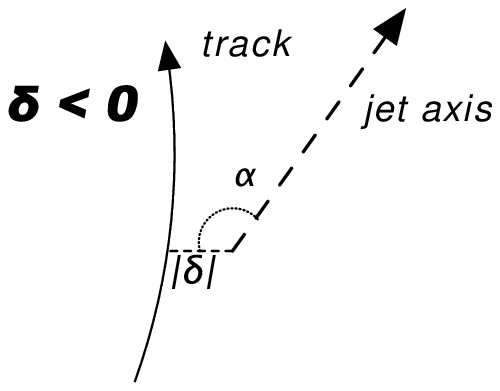}
\emp
\bmp{c}{0.45\linewidth}
\includegraphics[trim = 226pt 5pt 5pt 20pt,clip,%
width=0.9\linewidth]{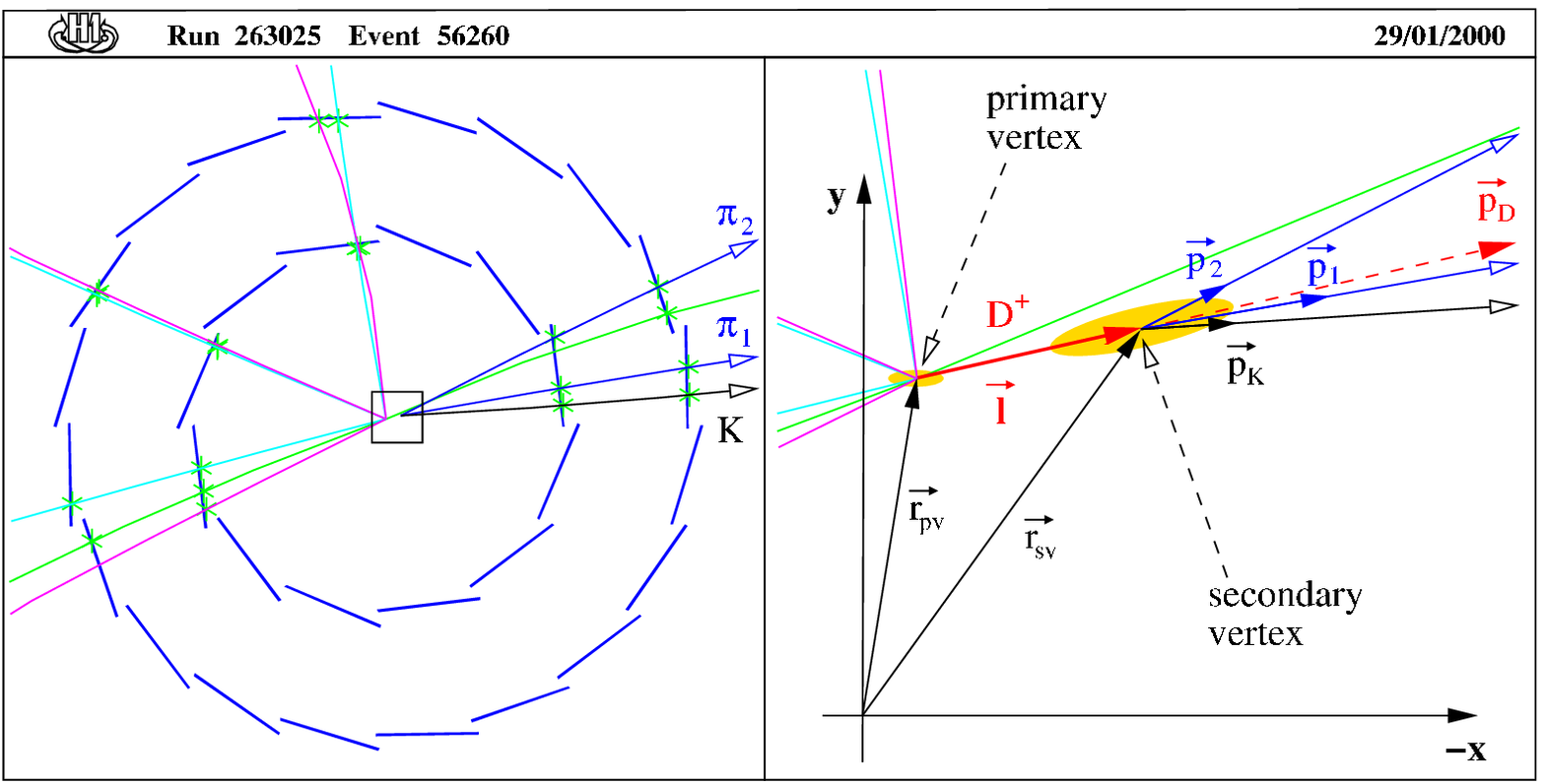}
\emp
\caption{
Illustration of the positive and negative signed 
impact parameter (left). 
A reconstructed vertex for a heavy-flavour decay
shown in the $XY$ plane~\cite{h1dmesons} (right).
The errors of the primary and secondary vertex positions 
(shaded ellipses) have been blown up by a factor 
of 10 for illustrative purposes. 
}
\label{fig:inclLT}
\end{figure}
%----------------------------------------------------------------------
\Fig{inclLT} illustrates how the signed impact parameter is
defined with respect to the jet to which the track is 
associated.
A positive sign is assigned to $\delta$ if 
the angle $\alpha$ between the jet axis and the 
line joining the primary vertex and the point of 
closest approach is less than $90^{\circ}$, 
and a negative sign otherwise.
\Fig{ptrel}(left) shows the distribution of the signed 
impact parameter of muon tracks in the data and the flavour decomposition 
in the Monte Carlo simulations.
The light-flavour component is characterised by 
a very small lifetime and the observed symmetric distribution 
is caused by the detector resolution.
%
%*** olaf: "small lifetime" sounds a bit silly, because
%    for strongly decaying light flavour hadrons it is 
%    effectively (compared to det. resolutions) zero , but ok, it is correct
%
In contrast, the charm and beauty contributions exhibit
a pronounced tail for large positive $\delta$ values.
To further improve the separation power, 
the \emph{impact parameter significance} 
$S=\delta/ \sigma(\delta)$ can be used.
This allows the rejection of candidates with an insignificant
measurement of the impact parameter.

\begin{figure}[htb]
\centering
%\fbox{\includegraphics[bb=0 200 260 400,clip=true,% 
%width=0.45\linewidth]{DESY-14-083_1_1}}
\includegraphics[width=0.62\linewidth]{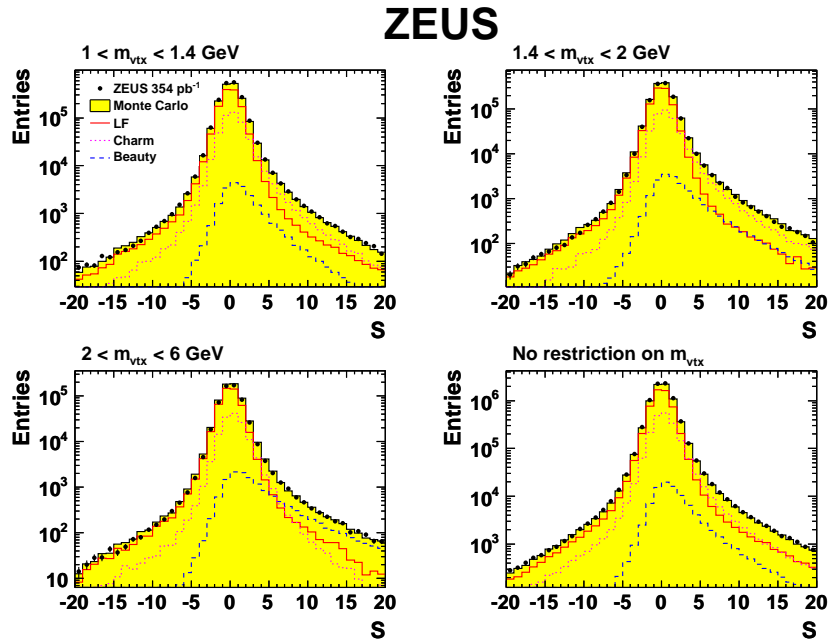}
\caption{
%*** Needs to be trimmed, but doesn't work for some reason.
%To be fixed or to be included as it is***
Distributions of the decay-length significance
%for $1 < \mvtx < 1.4\gev$ (left) and $2 < \mvtx < 6\gev$ (right)%
for different ranges of the vertex mass $\mvtx$.
~\pcite{zeusltt_hera2}.
The data (points) are compared to Monte Carlo simulations 
(filled area).
The individual contributions of beauty (dashed line), 
charm (dotted line) and light flavours (solid line) are shown.
}
\label{fig:secvtx}
\end{figure}

In the vertexing approach, a displaced secondary vertex 
is fitted to all tracks that are associated to a
selected jet\footnote{Here, a jet can be either a
real jet in the detector or a set of tracks corresponding 
to a $D$-meson candidate.}.
The distance between the primary and the secondary 
vertex is sensitive to the lifetime of the hadron that
initiated the jet.
Similar to the impact parameter, the flight direction 
can be introduced to form the \emph{signed decay length} 
as used by H1 or the \emph{projected decay length} 
in ZEUS.
The former is defined similarly to the signed 
impact parameter, whereas the latter is defined
as $$l = \frac{(\vec{r}_{\mathrm{SV}}-\vec{r}_{\mathrm{PV}}) \cdot \vec{p}} {|\vec{p}\,|}$$ (see \fig{inclLT}), 
i.e. the projection of the vector from the primary 
to the secondary vertex on the jet momentum.
Finally, the ratio of such a quantity over its uncertainty, $S$,
provides the optimal separation power.
A kinematic reconstruction of the mass of the vertex that 
corresponds to the jet, $\mvtx$, provides an additional handle
on flavour separation, since contributions of 
light-flavours, charm and beauty are expected to
populate predominantly the small ($\mvtx \ll m_D$), 
medium ($\mvtx~\lapprox~m_D$) and large ($m_D < \mvtx~\lapprox~m_B$)
mass domains, respectively.
\Fig{secvtx} illustrates the lifetime tagging with 
secondary vetrices.
%----------------------------------------------------------------------
%----------------------------------------------------------------------
%
The light-flavour contribution is symmetric around zero,
while the charm and beauty components exhibit a pronounced
asymmetry in the region of large \emph{decay-length significance}.
The beauty contribution dominates at large vertex mass and
large significance values.

Lifetime tagging can be used either as an add-on to other 
tagging techniques (as described before) or as a separate
tagging tool.
The dominant background is light-flavour production, 
which is symmetric in the signed impact parameter significance 
or the projected decay length significance.
The contents of the negative bins of the significance 
distribution can be subtracted from the contents of the 
corresponding positive bins, yielding a subtracted
significance distribution.
This way, the contribution from light-flavour quarks
is minimised.

\begin{figure}[htb]
\centering
\includegraphics[width=0.33\linewidth]{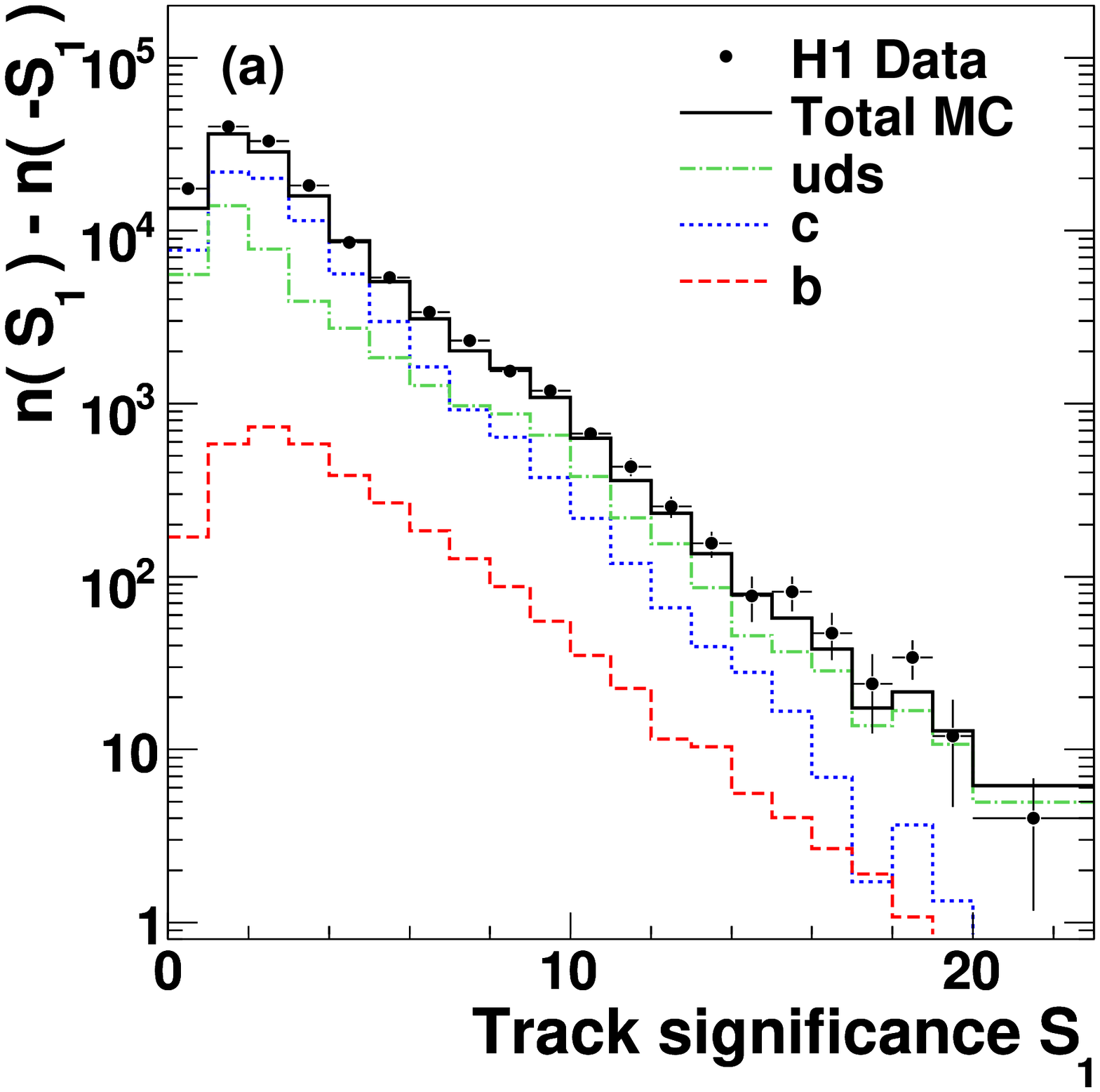}%
\includegraphics[width=0.33\linewidth]{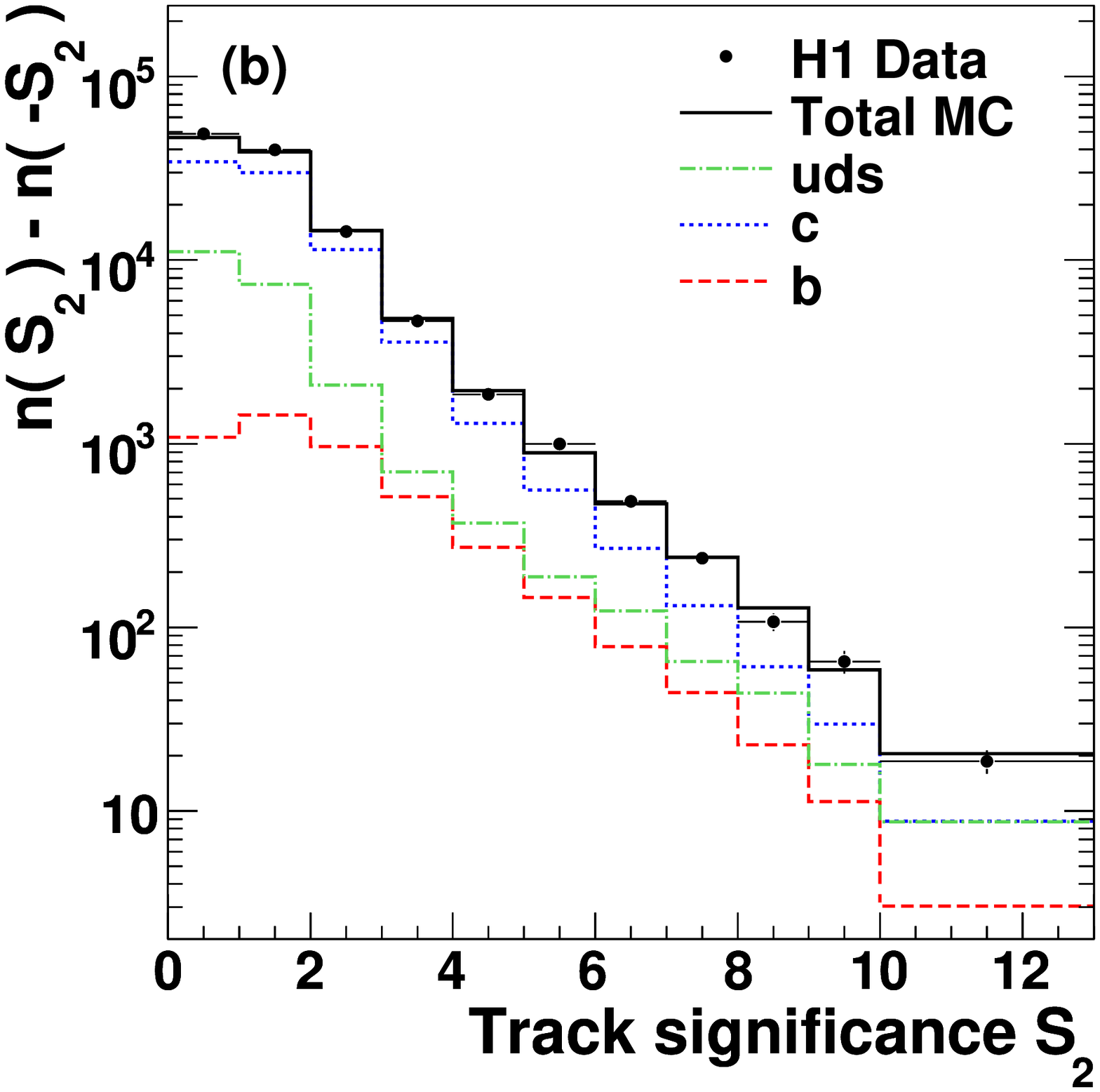}%
\includegraphics[width=0.33\linewidth]{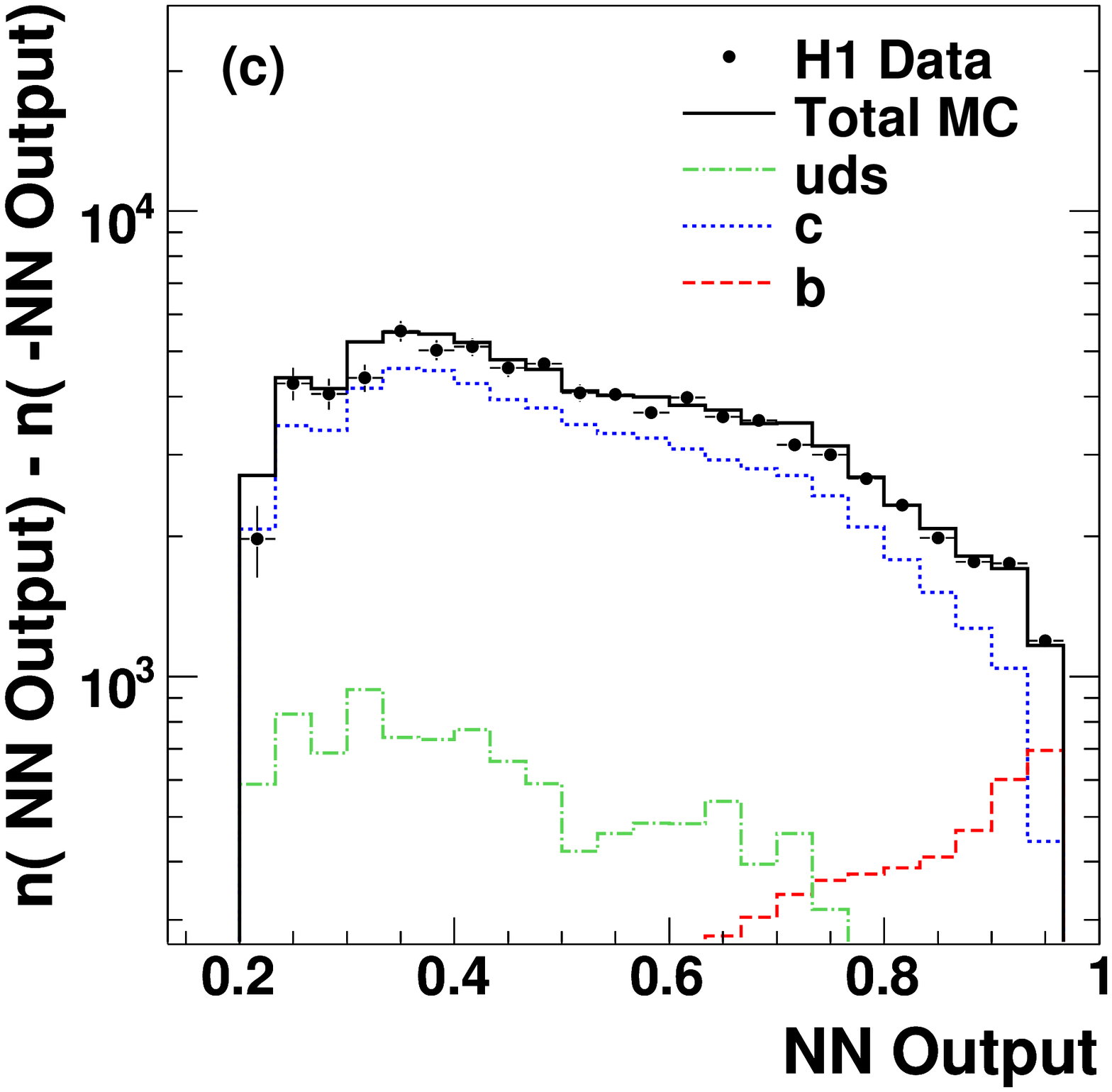}
\caption{
The subtracted distributions of (a) $S_1$, (b) $S_2$
and (c) the neural-network output~\pcite{h1ltt_hera2}.
The beauty- (dashed line), charm- (dotted line) 
and light-flavour (dashed-dotted line) contributions are shown.
}
\label{fig:h1vtx}
\end{figure}

The H1 collaboration has chosen to use a combination of 
the signed impact parameter significance of individual 
tracks and the signed decay-length significance to tag 
heavy-flavour production.
Events are exclusively categorised according to the 
number of tracks in the event.
The significances $S_1$, $S_2$ and $S_3$ are defined 
as the significance of the track with the highest, 
second highest and the third highest absolute 
significance, respectively.
The $S_1$ and $S_2$ significance distributions (Fig. \ref{fig:h1vtx}) are 
used for events with one and two selected tracks, respectively.
%----------------------------------------------------------------------
%----------------------------------------------------------------------
%
For events with three or a higher number of tracks various 
sensitive variables including 
$S_1$, $S_2$, $S_3$ and the signed decay-length significance
of the reconstructed secondary vertex are combined using
an artificial neural network.
In general, $S_2$ has a better discrimination between 
light- and heavy-flavour contributions than $S_1$, since the chance 
of reconstructing two high 
significance tracks is further reduced for light-flavour.
The neural network (Fig. \ref{fig:h1vtx}(c)) provides separation 
between \emph{c} and \emph{b} events.
For all distributions the negative part was subtracted 
from the positive one to minimise the light-flavour component
and a least-squares fit was performed simultaneously to all
three distribution.
The charm and residual light-flavour components were found to be 
very strongly anti-correlated in such fits (typical 
correlation coefficients are $C_{lc}<-0.95$), while the 
correlation with beauty is weaker due to the more 
distinct shape of the beauty distributions
($C_{cb}\approx-0.65$ and $C_{lb}\approx0.55$).
Also alternative  approaches to lifetime tagging have been 
studied, but were found to be more sensitive to systematics 
from track resolution and efficiency.

\begin{figure}[htb]
\centering
\includegraphics[width=0.55\linewidth]{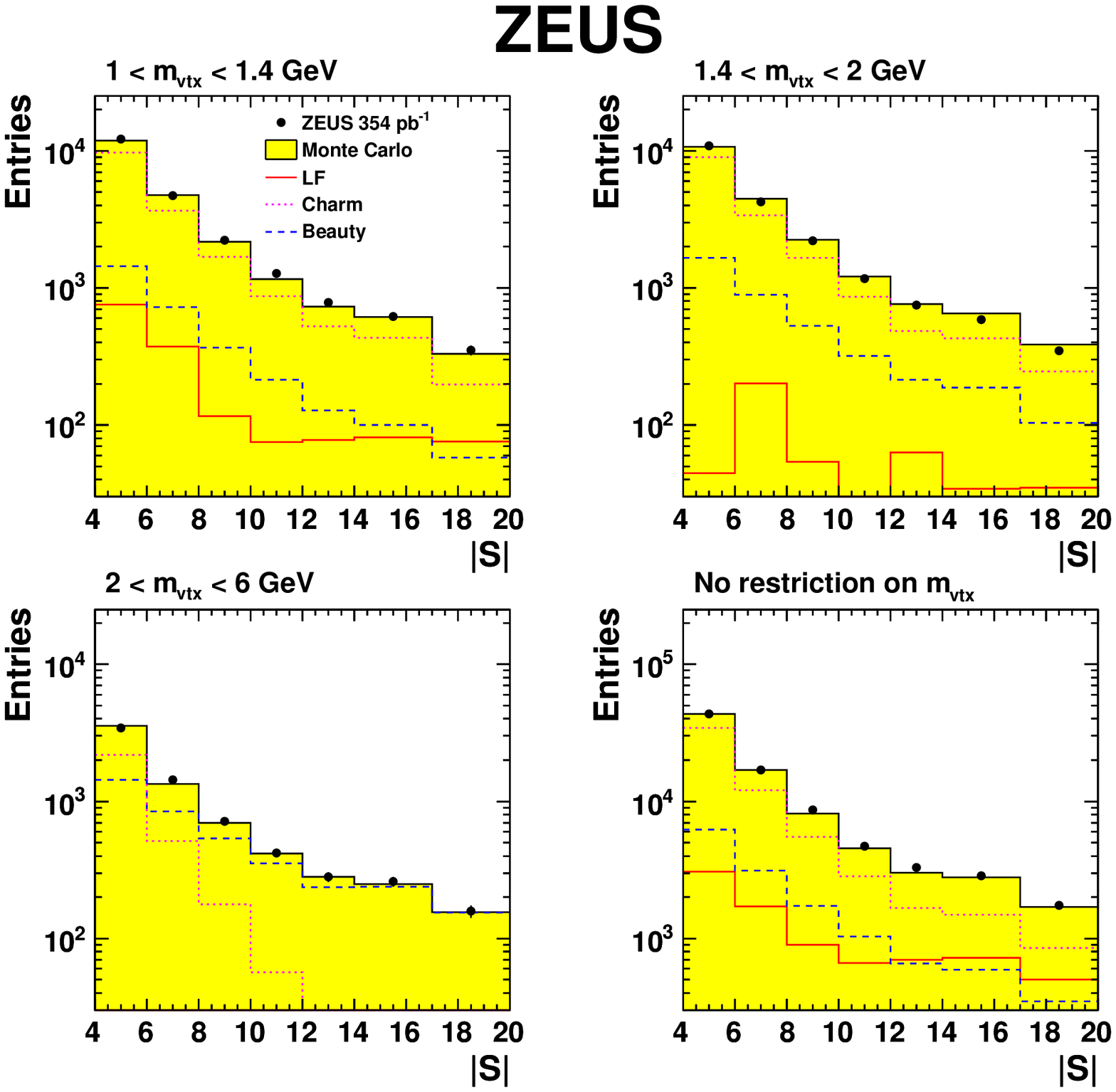}
\caption{
%*** Need to be trimmed, but doesn't work for some reason.
%To be fixed or to be included as it is***
The subtracted distributions of the  decay-length significance
%for $1 < \mvtx < 1.4\gev$ (a), $1.4 < \mvtx < 2\gev$ (b) 
%and $2 < \mvtx < 6\gev$ (c)%
for different ranges of the vertex mass $\mvtx$
~\pcite{zeusltt_hera2}.
The data (points) are compared to Monte Carlo simulations 
(filled area).
Individual contributions of beauty (dashed line), 
charm (dotted line) and light flavours (solid line) are shown.
}
\label{fig:zeusvtx}
\end{figure}

The ZEUS collaboration, on the other hand, made inclusive 
charm and beauty measurements exploiting vertexing for tagging.
The projected decay-length significance and 
the reconstructed mass of the fitted secondary vertices (Fig. \ref{fig:secvtx})
were used as discriminating variables.
A $\chi^2$ fit of the subtracted significance distribution
was performed in the three vertex-mass bins simultaneously.
The correlation pattern between components was found to be 
very similar to the one in the H1 analyses.
\Fig{zeusvtx} shows the subtracted significance distributions.
%----------------------------------------------------------------------
%----------------------------------------------------------------------
With optimised cuts after subtraction one can get samples
with very high charm and beauty enrichment of roughly $80 \%$ and
$90 \%$, respectively.
Such selection resulted in $\sim 26000$ charm and $\sim1500$
beauty events after negative subtraction in the 
recent ZEUS measurement~\cite{zeusltt_hera2}.

In summary, the advantages ($+$) and disadvantages ($-$)
of this inclusive lifetime tagging method are:
\begin{itemize}  \itemsep5pt \parskip0pt
\item[$+$]
This tagging method gives access to the largest statistics
due to the inclusive selection of the final state.
\item[$+$]
The technique provides strong discrimination power and 
is often combined with other tagging methods.
\item[$+$]
With the applied track minimal transverse momentum cut of
$0.5\gev$ one obtains a good acceptance for 
low heavy-quark momenta, which is of high importance
for measuring the charm and beauty contributions
to inclusive $ep$ scattering.
The additional typical jet cut $E_T^{\mathrm{jet}} > 5\gev \sim m_b$
retains a high acceptance for beauty production 
near threshold.
\item[$-$]
The total achieved effective
signal to background ratio is typically 
not better than 1:10 for both charm
and beauty.
%checked with the recent ZEUS DIS paper using B/S=S*rel.err.**2-1
%
This can be estimated from the 
numbers of charm and beauty events in the
positive subtracted significance spectra,
which effectively represent the numbers of tagged events,
and from the errors achieved for the charm and beauty 
components in the fit.
\item[$-$]
The method requires the track resolutions and efficiencies 
to be thoroughly scrutinised.
\item[$-$]
With the typical cuts on the jet transverse momentum
one actually cuts strongly into the kinematic phase space for charm.
A requirement of a jet with $E_T^{\mathrm{jet}} > 4\gev$ 
corresponds to a cut $p_T^{D} > 2.4\gev$ 
assuming that $\sim 60 \%$ of the quark transverse momentum
is transfered to the $D$ meson.

\end{itemize}

\subsection{Charm and Beauty with double tagging}

The various heavy-flavour tagging methods outlined above 
can be combined, aiming towards tagging both heavy quarks in the event.
At HERA, 
%for double tagging the flavour tags were used: \dst $\mu$ and 
%pairs of muons or electrons.
%
\dst $\mu$ combinations were used to tag both charm and beauty
events, while a di-lepton tag was used for beauty only.
The usage of two flavour tags significantly reduces the 
light-flavour background, which allows omitting any additional mass or 
lifetime tags.
Furthermore, it gives access to correlations between the quarks of 
heavy flavour pair.

\begin{figure}[th]
\centering
\includegraphics[width=0.51\linewidth]{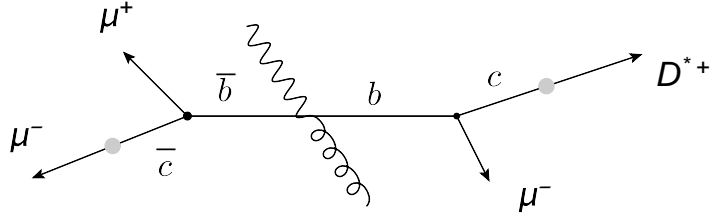}
\caption{Various possible ways to produce \dst $\mu$ or
a $\mu \mu$ pair 
from the decays of the $b$- and the $\bar{b}$-quark produced in the
photon gluon fusion process at HERA.
} 
\label{fig:Dstarmu}
\end{figure}

In the example of the photon-gluon fusion process, 
%*** A.G. double tagging is not 
%restricted to this configuration ***
$\gamma g \rightarrow c\bar{c}$ or $b\bar{b}$, the
two heavy quarks are produced back-to-back in the 
$\gamma g$ frame as illustrated in \fig{Dstarmu}
for the beauty case.
%
%%%%%%%%%%%%%%%%%%%%%%%%%%%%%%%%%%%%%%%%%%%%%%%%%%%%%%%%%%%%%%%%%%%%%
%////////////////////////////////////////////////////////////////////
%
Unlike-sign combinations such as $\mu^{+}\mu^{-}$ or $D^{*+}\mu^{-}$ can be produced from either the same or different
$b$ quarks, while like-sign combinations originate always from 
different $b$ quarks (combination of $b$ and $\bar c$ decays + charged 
conjugate, or $B^0-\bar B^0$ mixing).
In charm events only oppositely-charged combinations are produced.
In analyses using \dst $\mu$ tags the charm and beauty components
were separated\footnote{
Light-flavour production is suppressed by \dst 
reconstruction} 
based on the sign combination 
and the angular distance in azimuth.
In charm events mostly back-to-back
unlike-sign \dst and $\mu$ are produced,
while for beauty both like- and unlike-sign combinations
are possible and more complicated angular distributions arise.
The di-lepton analyses have also used the information about 
charge combination
and angular separation between leptons, and additionally
the mass of the lepton pair.

In summary, the advantages ($+$) and disadvantages ($-$) 
of double tagging are:
\begin{itemize} \itemsep5pt \parskip0pt 
\item[$+$]
For a large fraction of the events double tagging gives
access to the kinematics of \emph{both} heavy quarks.
This information can be used to investigate 
the $c\bar{c}$\/ and $b\bar{b}$\/ production processes in detail.
\item[$+$]
Since light flavours are efficiently suppressed by the
requirement of two flavour tags, the leptons
can be selected in transverse momentum
down to $\sim 1\gev$ with a reasonable 
purity.
For beauty this gives access to much lower
quark momenta than the lepton + \ptrel tag that 
was discussed in \Sect{bmujjx}.
\item[$-$]
The total tagging efficiency is very low.
\item[$-$]
The lepton tagging is well suited for the
measurement of beauty production but has relatively
small acceptance for charm, where, due to softer fragmentation,
the leptons take a smaller fraction of the quark transverse momentum 
than in the beauty case.
\item[$-$]
Due to their low $p_T$,
the correlation of the \dst and the $\mu$ momenta with those of the
parent quarks is not as good as for jets.
\end{itemize}

\subsection{Summary}  

Various heavy-flavour tagging methods have been used at HERA.
Each of them has advantages and disadvantages, 
which results in different tags being optimal 
for different purposes.
The most commonly used tags have been 
\dst reconstruction and inclusive lifetime tags for charm 
and
lepton + \ptrel and inclusive lifetime tags for beauty.
Whenever possible, a comparison between (and potentially 
a combination of) measurements performed with different
techniques allows improved constraints on the measurements, 
due to cross-calibration of systematics of different 
nature for independent tags.
Often a combination of tags yields an increased
purity of the heavy-flavour sample, at the cost 
of reduced efficiency and additional systematics.
In general, the choice of the tagging method(s) is a trade-off between
statistical and systematic uncertainties.

\newpage

% Single top searches
\section{Search for single top-quark production}
\label{sect:top}

Already before the start of HERA data taking, it became clear from the 
lower limits of order $70\rnge80\gev$ on the top-quark mass obtained from $\bar p p$ 
collisions by the 
UA1/UA2 \cite{UA12top} and CDF \cite{CDFtop} collaborations, that top-quark 
pair production would probably be outside the kinematic reach of HERA. 
This was confirmed
by indirect constraints from LEP \cite{topLEP} and by the direct observation 
of the top quark at the Tevatron at a mass of $174\gev$ \cite{topTevatron}.
Single top quark production in the charged current 
reaction \cite{singletopHERA} 
$$ e^+ + b \to \bar \nu_e + t$$
(and its charged conjugate) 
remained kinematically possible, but the expected Standard Model cross section 
of less than $1\fb$ \cite{HERAttheory} 
is too small to be experimentally accessible. This is due to the fact 
that the occurrence of initial state $b$ quarks is strongly suppressed at 
high $x$, since it would need to originate from the splitting of ultra-high-$x$
gluons in the proton (in analogy to Fig. \ref{fig:intro6}), 
which are known to be very rare \cite{DISreview}.
Charged current reactions on light initial state quarks are strongly 
suppressed by the very small corresponding CKM matrix elements \cite{CKM}.
      
\begin{figure}[htbp]
     \hspace{0.6 cm}
\centering
\includegraphics[width=0.23\linewidth]{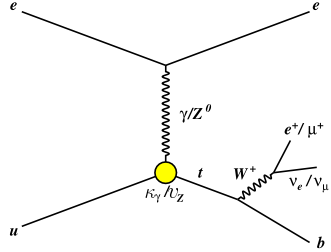}
\caption{
Feynman graph for anomalous single top production \cite{ZEUSsingletop}.}
\label{fig:singletopfeyn}
\end{figure}

If at all, single top quarks could thus be produced at HERA only via a 
process beyond the Standard Model \cite{FCNC}. 
One such process is the transition of a $u$ quark into a $t$ quark via a 
flavour-changing neutral current \cite{FCNCgZ} (Fig. \ref{fig:singletopfeyn}) 
%%%%%%%%%%%%%%%%%%%%%%%%%%%%%%%%%%%%%%%%%%%%%%%%%%%%%%%%%%%%%%%%%%%%%%
%
% single top limits
%
%
%
%/////////////////////////////////////////////////////////////////
%
%
caused by non-Standard Model couplings of the photon or $Z$ boson.
This possibility was investigated in particular due to an excess observed 
by the H1 collaboration in the single isolated lepton + jets final state 
\cite{H1isolept}, which was however not confirmed by a corresponding ZEUS 
analysis \cite{ZEUSisolept}, and greatly reduced in significance by a common 
analysis of the ZEUS and H1 data \cite{H1ZEUSisolept}.

%%%%%%%%%%%%%%%%%%%%%%%%%%%%%%%%%%%%%%%%%%%%%%%%%%%%%%%%%%%%%%%%%%%%%%
%
% single top limits
%
%
\begin{figure}[h]
     \hspace{0.6 cm}
\centering
\includegraphics[width=0.5\linewidth]{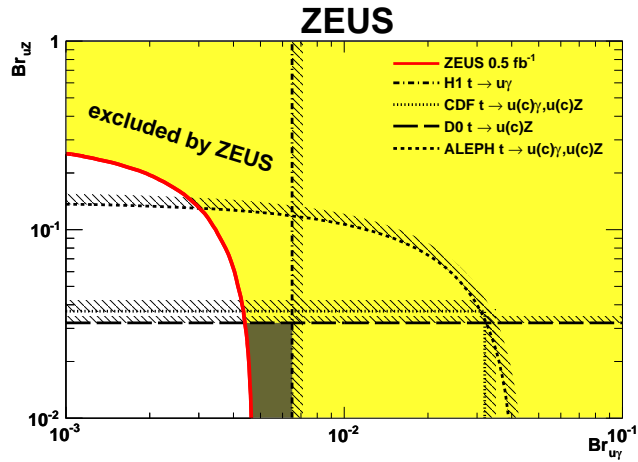}
\caption{
Limits on anomalous couplings for single top production, translated to 
branching fractions (Br) for top decay into $uZ$ or $u\gamma$ 
\cite{ZEUSsingletop}.}
\label{fig:singletop}
\end{figure}
%
%/////////////////////////////////////////////////////////////////

No significant signal was observed. 
Fig. \ref{fig:singletop} shows the exclusion contours obtained by the 
H1 \cite{H1singletop} and ZEUS \cite{ZEUSsingletop} collaborations for 
the anomalous couplings to photons or $Z$ bosons for this process, translated
to branching ratios for anomalous top quark decays.
The H1 limit is a bit looser than the 
the one from ZEUS due to the small excess mentioned above.
These limits are competitive with
limits obtained from other colliders, also shown in Fig. \ref{fig:singletop},
and currently represent the best limit for the anomalous photon coupling.
They can be improved further in future analyses using data from 
the LHC \cite{LHCsingletop}.
Single top quarks were also searched for in the full-hadronic top
decay channel, 
with a similar sensitivity as for the lepton channel \cite{tophad}. 
However, the corresponding analysis was performed on 
the \herai data set only, so the resulting limit is no longer competititive.

% Open charm and beauty, DIS and PHP
\section{Charm photoproduction}
\label{sect:charmphoto}

The study of open charm- and beauty-production cross sections provides 
stringent tests of perturbative QCD from several perspectives.
On one hand, the size of the charm- and beauty-quark masses ensures 
that for all 
such final states the production cross sections 
are in the perturbatively calculable regime, 
since $m_c, m_b \gg \Lambda_{QCD}$.
On the other hand, the QCD scales obtained from these masses compete with other
potential scales like the quark tranverse momentum or, in the DIS case, 
the virtuality of the exchanged photon (Fig. \ref{fig:intro5}). The 
treatment of such a multi-scale problem is theoretically challenging and a 
comparison of different theoretical schemes to data can shed light on the 
strengths and weaknesses of the respective perturbative 
approximations. Furthermore, from the theoretical point of view, the presence 
of final-state heavy-flavour hadrons ensures that these processes will not 
interfere with corresponding final states involving only 
gluons and light quarks, and that they can thus be treated independently 
for each flavour. Thus, in the following, charm and beauty production will
be treated separately.

In general, photoproduction processes have many similarities to corresponding
processes in hadroproduction, ``simply'' interchanging the incoming quasi-real 
photon with a quasi-real gluon. 
Since the S$p\bar p$S and 
Tevatron $p\bar p$ colliders went into operation almost a decade before the 
HERA collider, perturbative NLO QCD calculations 
have often first been obtained for hadroproduction, 
although the incoming photon diagrams are 
somewhat easier to calculate. 
Furthermore, since the virtuality of the incoming photon 
($Q^2 < 1$~GeV$^2$) is of the order of typical hadron masses or lower, 
the photon can have a hadron-like structure (``hadron-like resolved photon'',
Fig. \ref{fig:fey1}b). 
Thus, the cross sections get a contribution
from the convolution of perturbative hadroproduction diagrams with this 
photon structure.
This in turn complicates the photoproduction cross-section calculations.    
However, this hadron-like photon contribution is small 
(of order 10\% or less) in the case of the treatment of heavy-flavour 
production in the massive scheme.

An important variable for photoproduction analyses at HERA is the event 
kinematic observable $y$, which represents the fraction of the electron 
energy entering the hard interaction (Section \ref{sect:kinvar}). This 
variable can be  
reconstructed from the hadronic final state in the main detector
(Section \ref{sect:experiments}).
An overview of all charm photoproduction measurements in H1 and ZEUS
is given in Table~\ref{tab:r1}.
%
%%%%%%%%%%%%%%%%%%%%%%%%%%%%%%%%%%%%%%%%%%%%%%%%%%%%%%%%%%%%%%%%%%%%%%%%%%%%%%%%%%%%%%%%%%%%%%%%%%%%%%%%%%%%%%%%%%%%%%%%%%%%%%%%%%%%%%%%%%%%%

%
%
% Charm photoproduction measurement table:
% ----------------------------------------
%
\begin{sidewaystable}[htbp]
\setlength{\tabcolsep}{1.47mm}
\renewcommand{\arraystretch}{1.1}
\begin{center}
\footnotesize
\it
\begin{tabular}{||c|l|l|c|c|c|c|l|c||l|l|c||l|l|l||}
\hline
\hline
%   1         2          3       4       5        6         7                       8             9          10            11
%  12          13      14     15
   No. & Analysis &   c-Tag  &  Ref. &  Exp. &  Data &  $\cal{L}$ $[pb^{-1}]$   & $Q^2\;[\mbox{GeV}^2]$ &        $y$     & Particle  & $p_T\;[\mbox{GeV}]$   &
 $\eta$     & Events &
\begin{tabular}{c}
effect.\\
s:b \\
\end{tabular}
&
\begin{tabular}{l}
bgfree \\
events\\
\end{tabular}
\\
%
%--- First D* analysis from ZEUS:
%
\hline
1 & 
$D^*$ incl. & $K\pi\pi_s$  & \cite{ZEUS95} & ZEUS &  93     & $0.5$                      & $<4$  & $[0.15,0.84]$ & $D^*$   & $>1.7$ &    
$[-1.5,1.5]$  & $48 \pm 11$ & $1:1.5$ & $19$   \\ 
\cline{1-15}
%
%--- First D* analysis from H1:
%
2 & 
$D^*$ tagged & $K\pi\pi_s$  & \cite{H194} & H1 &  94     & $2.8$                      & $<0.01$  & $[0.28,0.65]$ & $D^*$   & $>2.5$ & 
$[-1.5,1.0]$  & $119 \pm 16$ & $1:1.2$ & $55$   \\
\cline{7-9}
\cline{13-15} 
  & \qquad incl. & &  &  &  & $1.3$                      & $<4$     & $[0.10,0.80]$   &  &  & 
   &  $97 \pm 15$ & $1:1.3$ & $42$   \\ 
%
%--- First differential D* analysis from ZEUS:
%
\cline{1-15}
3 & 
$D^*$ incl. & $K\pi\pi_s$  & \cite{ZEUS97a} & ZEUS &  94     & $3.0$                      & $<4$  & $[0.15,0.87]$ & $D^*$   & $>3$ &    
$[-1.5,1.0]$  & $152 \pm 16$ & $1:0.7$ & $90$   \\ 
\cline{3-3}
\cline{13-15}
  & & $K3\pi\pi_s$  &   &   &  &  &  &  &  &  &  & $199 \pm 29$ & $1:3.2$ & $17$   \\ 
%
%--- tagged PhP analysis from H1:
%
\cline{1-15}
4 & 
$D^*$ tagged & $K\pi\pi_s$  & \cite{H1gluon} & H1 &  95-96  & $10.2$                      & $<0.009$  & $[0.02,0.32]$ & $D^*$   & $>2$ &    
$[-1.5,1.5]$  & $299 \pm 75$ & n.a. & $16$   \\ 
\cline{6-9}
\cline{11-11}
\cline{13-15}
  &  &  &   &   & 94-96 & $10.7$ & $<0.01$ & $[0.29,0.62]$ & & $>2.5$ & $(\hat y(D^*))$  & $489 \pm 92$ & n.a. & $28$   \\ 
%
%--- ZEUS D* +dijets 96/97
%
\hline
5 & 
$D^*$ incl. & $K\pi\pi_s$  & \cite{ZEUScPHP1999} & ZEUS &  96-97     & $37$                     & $<1$  & $[0.19,0.87]$ & $D^*$   & $>2$ &    
$[-1.5,1.5]$  & $3702 \pm 136$ & $1:4.0$ & $741$   \\ 
\cline{3-3}
\cline{13-15}
  &  & $K3\pi\pi_s$  &   &   &  &  &  &  &  & $>4$ &  & $1397 \pm 108$ & $1:7.3$ & $167$   \\ 
\cline{2-3}
\cline{10-15}
  & 
$D^*$ + dijet   & $K\pi\pi_s$  &  &   &   &  &  &  & 
\renewcommand{\tabcolsep}{-1.mm}
\begin{tabular}{l}  
$D^*$ \\
Jet1(2) \\
\end{tabular}
 & 
\renewcommand{\tabcolsep}{0mm}
\begin{tabular}{l}  
$>3$ \\
$>7(6)$ \\
\end{tabular}
&
\renewcommand{\tabcolsep}{0mm} 
\begin{tabular}{l}  
$[-1.5,1.5]$ \\
$[-2.4,2.4]$ \\
\end{tabular}
& $587 \pm 41$ & $1:1.9$ & $205$  \\ 
%
%--- ZEUS prel. 2002
%
\cline{1-15}
6 & D* incl.
         & $K\pi\pi_s$  & \cite{zeus-ichep02-786}   & ZEUS &  98-00  & $79$                       & $<1$  & $[0.17,0.77]$ & $D^*$    & $[1.9,20]$ & $[-1.6,1.6]$  & $10350 \pm 190$ & $1:2.5$ & $2970$   \\ 
\hline
%
% H1 D* + jet + dijet combining several H1 preliminaries
%
7 & 
$D^*$ tagged &  $K\pi\pi_s$ & \cite{H1cPHP2007} & H1  & 99-00  & $51$   & $<0.01$  & $[0.29,0.65]$ & 
$D^*$ & $>2$ & $[-1.5,1.5]$ & $1166 \pm 82$ & $1:4.8$ & $202$ \\ 
\cline{10-15}
  &  +jet    & & & & & & & & Jet & $>3$ & $[-1.5,1.5]$ & $592 \pm 57$ & $1:4.5$ & $108$ \\ 
\cline{10-15}  
  &  +dijet    & & & & & & & & Jet 1(2) & $>4 (3)$ &  $[-1.5,1.5]$ & $496 \pm 53$ & $1:4.7$ & $88$ \\ 
%
%--- ZEUS D* +dijets costheta* 
%
\hline
8 & 
$D^*$ + dijet   & $K\pi\pi_s$  & \cite{ZEUScPHP2003} & ZEUS   & 96-00  & $120$ & $<1$  & $[0.17,0.77]$ & 
\renewcommand{\tabcolsep}{-1mm}
\begin{tabular}{l}  
$D^*$ \\
Jet1(2) \\
\end{tabular}
 & 
\renewcommand{\tabcolsep}{0mm}
\begin{tabular}{l}  
$>3$ \\
$>7(6)$ \\
\end{tabular}
& 
\renewcommand{\tabcolsep}{0mm}
\begin{tabular}{l}  
$[-1.5,1.5]$ \\
$[-1.9,1.9]$ \\
\end{tabular}
& $1092 \pm 43$ & $1:0.7$ & $650$  \\ 
%
%--- ZEUS D* charm jet + dijet cross-sections 2005
%
\hline
9 & 
$D^*$ + jet    & $K\pi\pi_s$  & \cite{ZEUScPHP2005}    & ZEUS & 98-00  & $79$                       & $<1$  & $[0.17,0.77]$ &   
$D^*$ & $>3$ &  $[-1.5,1.5]$ & $4891 \pm 113$ & $1:1.6$ & $1870$  \\ 
\cline{13-15}
  & \quad + dijet    &   &  & & & & & & 
Jet1(2) & $>6(7)$ & $[-1.5,2.4]$ & $1692 \pm 70$ & $1:1.6$ & $584$  \\ 
\hline
%
%---- H1 dijets + vertex
%
10 & 
lifet.+dijet & imp.par. & \cite{H1cbPHP2006} & H1   & 99-00  & $57$                       & $<1$  & $[0.15,0.80]$ & 
\renewcommand{\tabcolsep}{-1.4mm}
\begin{tabular}{l}  
Track \\
Jet1(2) \\
\end{tabular}
 & 
\renewcommand{\tabcolsep}{0mm}
\begin{tabular}{l}  
$>0.5$ \\
$>11(8)$ \\
\end{tabular}
& 
\renewcommand{\tabcolsep}{0mm}
\begin{tabular}{l}  
$[-1.3,1.3]$ \\
$[-0.9,1.3]$ \\
\end{tabular}
& $4600 \pm 460$ & $1:45$ & $100$ \\ 
%
%--- H1 D* + muon
%
\hline
11 &
$D^*$ + $\mu$    & 
\begin{tabular}{l}
$K\pi\pi_s$ \\
 + $\mu$ \\
\end{tabular} 
&  \cite{H1cbPHP2005}  & H1 & 98-00  & $89$                       & $<1$  & $[0.05,0.75]$ & 
\begin{tabular}{l}  
$D^*$ \\
$\mu$ \\
\end{tabular}
 &
\begin{tabular}{l}  
$>1.5$ \\
$p>2$ \\
\end{tabular}
 & 
\begin{tabular}{l}  
$\ \ [-1.5,1.5]$ \\
$[-1.74,1.74]$ \\
\end{tabular}
& $53 \pm 13$ & $1:2.2$ & $17$  \\ 
\hline
%
% ZEUS e + dijets
%
12 & 
$e$ + dijet   & $e + {\not E}_T$  & \cite{ZEUScbPHP2008} & ZEUS   & 96-00  & $120$ & $<1$  & $[0.2,0.8]$ & 
\renewcommand{\tabcolsep}{-1mm}
\begin{tabular}{l}  
$\ e$ \\
Jet1(2) \\
\end{tabular}
 & 
\renewcommand{\tabcolsep}{0mm}
\begin{tabular}{l}  
$>0.9$ \\
$>7(6)$ \\
\end{tabular}
& 
\renewcommand{\tabcolsep}{0mm}
\begin{tabular}{l}  
$[-1.5,1.5]$ \\
$[-2.5,2.5]$ \\
\end{tabular}
& $\sim 8000$ & n.a. & $70$  \\ 
% unsafe to quote any number of signal events: pTrel gives ~ 10k, while lnT ~ 6k.
\hline
%
% ZEUS dijets + vtx
%
13 & 
lifet.+dijet  & sec. vtx.  & \cite{ZEUScbPHP2011} & ZEUS   & 05  & $133$ & $<1$  & $[0.2,0.8]$ & 
\renewcommand{\tabcolsep}{-1mm}
\begin{tabular}{l}  
tracks \\
Jet1(2) \\
\end{tabular}
 & 
\renewcommand{\tabcolsep}{0mm}
\begin{tabular}{l}  
$>0.5$ \\
$>7(6)$ \\
\end{tabular}
& 
\renewcommand{\tabcolsep}{0mm}
\begin{tabular}{l}  
$[-1.6,1.4]$ \\
$[-2.5,2.5]$ \\
\end{tabular}
& $\sim 20000$ & n.a. & $2320$  \\ 
\hline
%
% H1 dijets + mu
%
14 & 
$\mu$ + dijet  & 
\renewcommand{\tabcolsep}{-1mm}
\begin{tabular}{l}
$\mu$ + \\
imp.par. \\
\end{tabular}  
& 
\cite{H1cbPHP2012} & H1   & 06-07  & $179$ & $<2.5$  & $[0.2,0.8]$ & 
\renewcommand{\tabcolsep}{-1mm}
\begin{tabular}{l}  
$\ \mu$ \\
Jet1(2) \\
\end{tabular}
 & 
\renewcommand{\tabcolsep}{0mm}
\begin{tabular}{l}  
$>2.5$ \\
$>7(6)$ \\
\end{tabular}
& 
\renewcommand{\tabcolsep}{0mm}
\begin{tabular}{l}  
$[-1.3,1.5]$ \\
$[-1.5,2.5]$ \\
\end{tabular}
& $3315 \pm 170$ & $1:7.7$ & $380$  \\ 
\hline
%
% H1 D* + dijet
%
15 & 
$D^*$ incl &  $K\pi\pi_s$ & \cite{H1cPHP2012} & H1  & 06-07  & $31\rnge93$  & $<2$  & $[0.1,0.8]$ & 
$D^*$ & $>1.8$ & $[-1.5,1.5]$ & $8232 \pm 164$ & $1:2.3$ & $2520$ \\ 
\cline{10-15}
  &  +dijet    & & & & & & & & Jet 1(2) & $>3.5$ &  $[-1.5,2.9]$ & $3937 \pm 114$ & $1:2.3$ & $1200$ \\ 
\hline
%
% ZEUS D* + HER/MER/LER
%
16 & 
$D^*$ incl &  $K\pi\pi_s$ & \cite{ZEUSHMLER} & ZEUS  & 06-07  & $144$   & $<1$  & $[0.167,0.802]$ & 
$D^*$ & $[1.9,20]$ & $[-1.6,1.6]$ & $12256 \pm 191$ & $1:2.0$ & $4120$ \\ 
\cline{6-7}
\cline{13-15}
  &  MER    & & & & 07 & $6.3$ & & & & & & $417 \pm 37$ & $1:2.3$ & $127$ \\ 
\cline{6-7}
\cline{13-15}  
  &  LER    & & & & 07 & $13.4$ & & & & & & $859 \pm 49$ & $1:1.8$ & $307$ \\ 
\hline
\end{tabular}
\end{center}
\vspace{-4mm}
\caption{
\footnotesize
{\bfseries Charm photoproduction cross-section measurements} at HERA.
Information is given for each analysis on the
charm tagging method, the experiment,
the data taking period, integrated luminosity, $Q^2$ 
and $y$ ranges
and the cuts on transverse momenta and pseudorapidities of selected
final state particles.
The last three columns provide information on the number of
tagged charm events, the effective signal-to-background ratio
%, as calculated from
%$\,s:b = events: (\sigma(events)^2-events)$ 
and the
equivalent number of background-free events.
%, as calculated from
%$\,[events/\sigma(events)]^2.$
%
The centre-of-mass energy of all data taken up to 1997 ($6^{th}$ column)
was $300\gev$, while it was $318\rnge319\gev$ for all subsequent runs, with the 
exception
of the analyses marked ``MER'' and ``LER'' (entry 16), for which the data were 
taken at 251 and 225 GeV.
}
\label{tab:r1}
\end{sidewaystable}

%%%%%%%%%%%%%%%%%%%%%%%%%%%%%%%%%%%%%%%%%%%%%%%%%%%%%%%%%%%%%%%%%%%%%%%%%%%%%%%%%%%%%%%%%%%%%%%%%%%%%%%%%%%%%%%%%%%%%%%%%%%%%%%%%%%%%%%%%%%%

\normalsize 

\subsection{$D^*$ inclusive measurements}

Despite the large loss in statistics through fragmentation fractions and 
branching ratios, it is clear
from Table \ref{tab:r1} that the very clean explicit reconstruction of 
$D^*$-meson final states offers the 
best effective signal sensitivity for charm photoproduction.

\subsubsection{Charm total cross sections}

\begin{figure}[bhtp]
\centering
\includegraphics[width=0.5\linewidth]{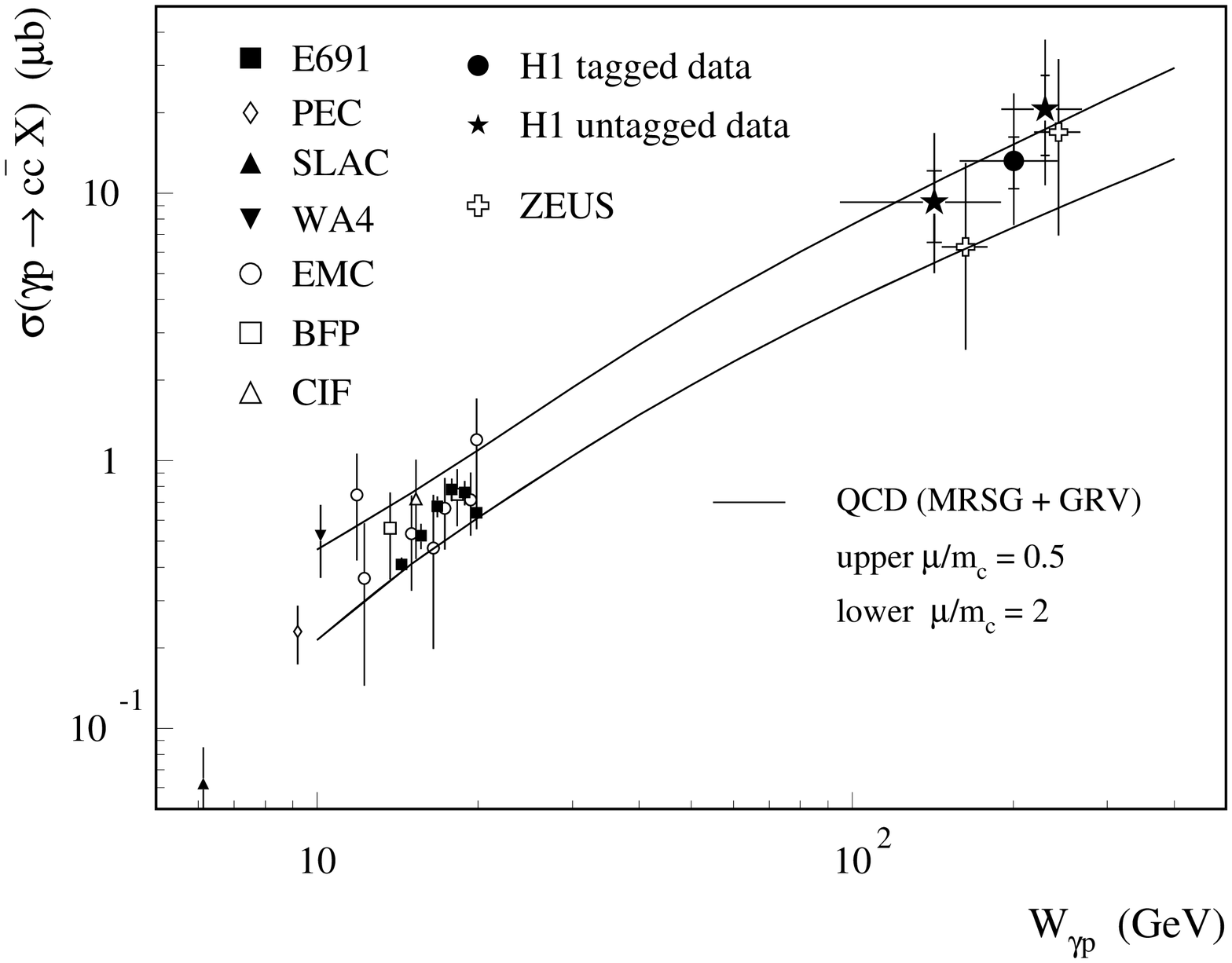}
\includegraphics[width=0.45\linewidth,bb=0 -40 567 439,clip]{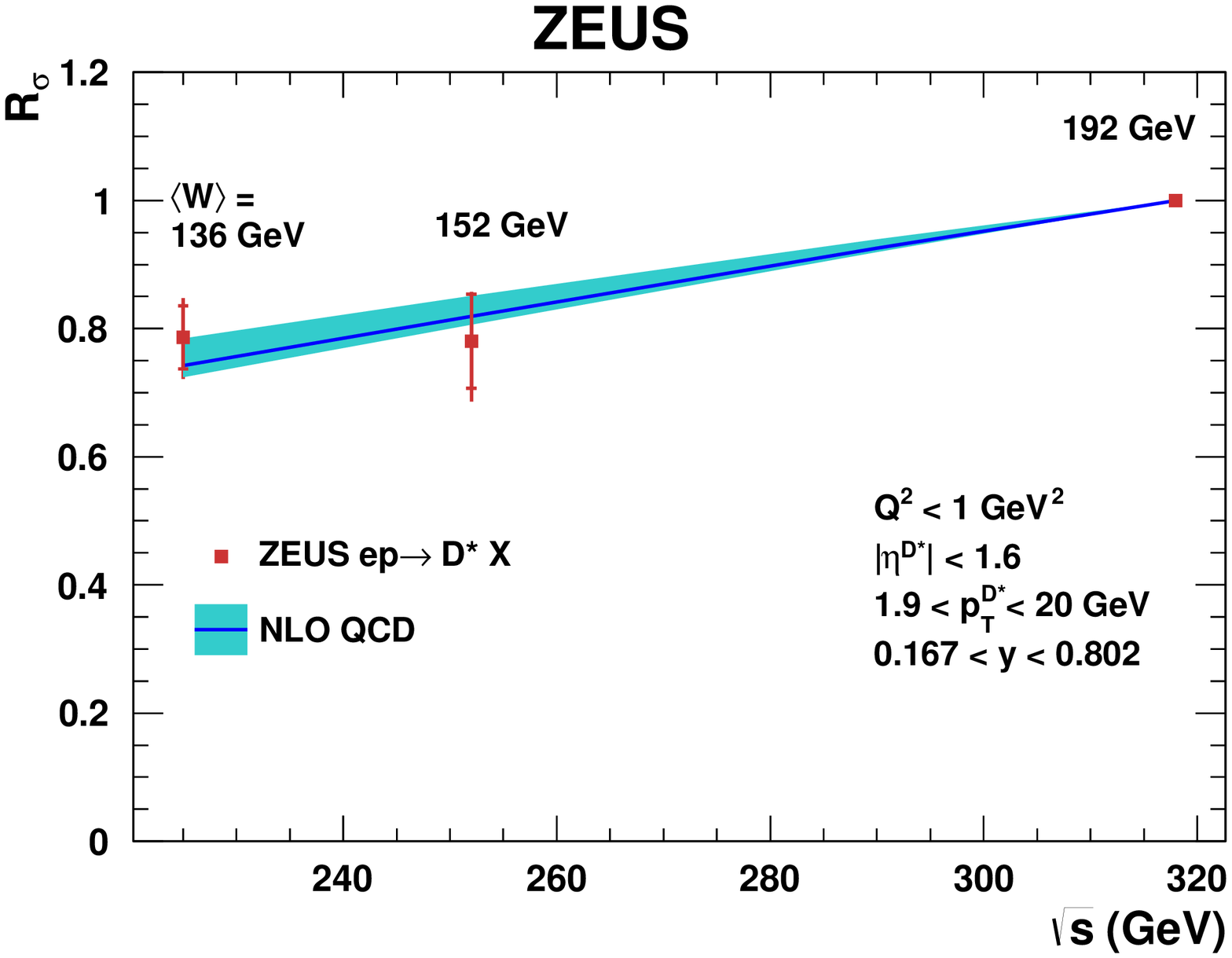}
\vspace{-0.3cm}
\caption{Left: Total charm-photoproduction cross section as a function
of centre-of-mass energy $W_{\gamma p}$ \cite{H194}.
The data shown are from the first 
H1 and ZEUS publications on open charm production
and from previous fixed-target
experiments.
Right: Inclusive charm-photoproduction cross section as a function
of $ep$ centre-of-mass energy \cite{ZEUSHMLER}, normalised to the cross 
section at 318 GeV.}
\label{fig:r1}
\end{figure}

In the very first ZEUS~\cite{ZEUS95} and H1~\cite{H194} measurements 
on open-charm production (entries 1 and 2 in Table \ref{tab:r1})
the inclusive $D^*$ results
were extrapolated to obtain total charm-photoproduction 
cross sections.
The results are shown in Fig.~\ref{fig:r1}(left)
as a function of the photon-proton centre-of-mass energy $W_{\gamma p}$.
%
%%%%%%%%%%%%%%%%%%%%%%%%%%%%%%%%%%%%%%%%%%%%%%%%%%%%%%%%%%%%%%%%%%%%%%%%%%%%%%%%%%%%%%%%%%%%%%%%%%%%%%%%%%%%%%%%%%%%%%%%%
%
% Charm vs W_gp
%
% Charm vs center-of-mass
%
%
%
%//////////////////////////////////////////////////////////////////////////////////////////////////////////////////////
%
Measurements from fixed-target experiments from the pre-HERA era are also
shown.
%
%%%%%%%%%%%%%%%%%%%%%%%%%%%%%%%%%%%%%%%%%%%%%%%%%%%%%%%%%%%%%%%%%%%
%
At HERA, both $W_{\gamma p}$ values
and the observed cross sections
are roughly one order 
of magnitude larger. 
The steep cross-section rise 
reflects the fact that with
increasing $W_{\gamma p}$ 
gluons with smaller and smaller 
proton momentum fractions are accessible 
for charm production via
the photon-gluon-fusion process 
(Fig.~\ref{fig:t1}).
The data in Fig.~\ref{fig:r1} are compared to 
a massive scheme NLO prediction~\cite{Frixione:1995qc},
which is able to describe both the 
fixed-target data at lower  $W_{\gamma p}$
and the HERA data at higher $W_{\gamma p}$.
Despite the large uncertainties, this demonstrated early on that the basic 
charm-production mechanism
in photoproduction is at least reasonably well understood.

Figure \ref{fig:r1}(right) shows the latest HERA measurement in 
photoproduction \cite{ZEUSHMLER} (entry 16 in Table \ref{tab:r1}), 
focusing on the dependence of the inclusive visible cross section on the 
centre-of-mass energy. This makes use of the very 
last HERA running period, in which the proton beam energy was lowered. 
This result was obtained and published based on a ZEUS master 
thesis \cite{masterHMLER}, which 
was made possible by the strong simplification of the data format and 
calibration procedure implemented as part of a long-term high energy 
physics data preservation project \cite{DPHEP}.
The result is presented as a ratio to the highest centre-of-mass energy 
cross section, such that both experimental and theoretical correlated
uncertainties cancel. While the data uncertainties remain dominated by 
statistical uncertainties (inner error bars), 
the theoretical uncertainties are dramatically
reduced with respect to the absolute predictions in Fig. \ref{fig:r1}.  
The massive NLO prediction \cite{Frixione:1997ma} agrees 
well with the data, indicating that the extrapolation of the 
energy dependence to even higher 
centre-of-mass energies such as those at a future LHeC 
collider \cite{LHeC} can be reliably predicted. 
In addition, since different centre-of-mass energies 
correspond to different $x$ ranges, such a ratio 
potentially provides constraints  on the gluon PDF
in the proton.

\subsubsection{{\bf $D^{*}$} single-differential cross sections} 
%
%%%%%%%%%%%%%%%%%%%%%%%%%%%%%%%%%%%%%%%%%%%%%%%%%%%%%%%%%%%%%%%%%%%%%%%%%%%%%%%%%%%%%%%%%%%%%%%%%%%%%%%%%%%%%%%%%%%%%%%%
%
%---- Ds ICHEP2002 ZEUS vs NLO and NLL
%     H1 2012 paper
%
\begin{figure}[htbp]
\centering
\includegraphics[width=0.7\linewidth,bb=0 268 510 550,clip]{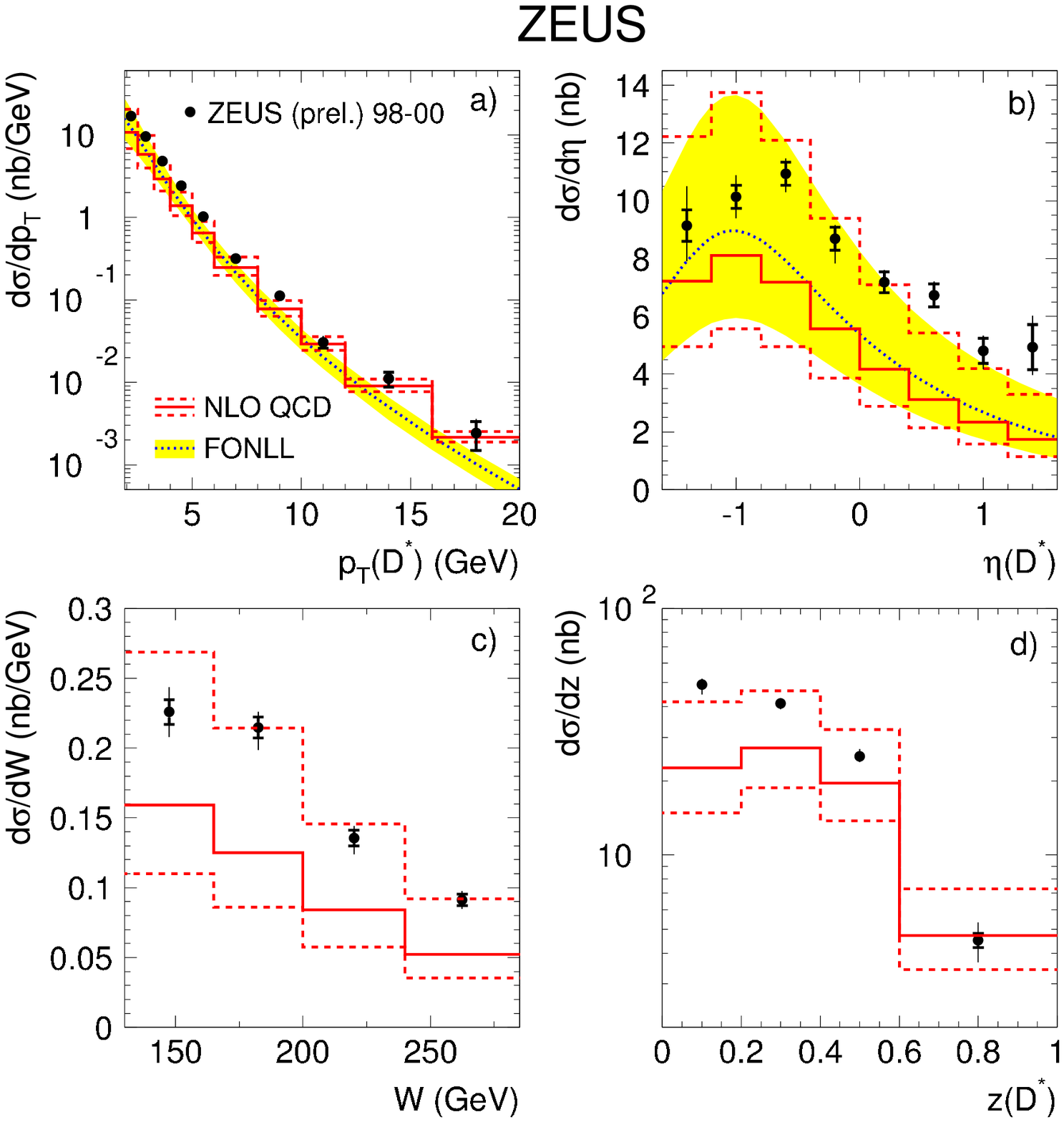}
\includegraphics[width=0.7\linewidth,bb=0 268 512
520,clip]{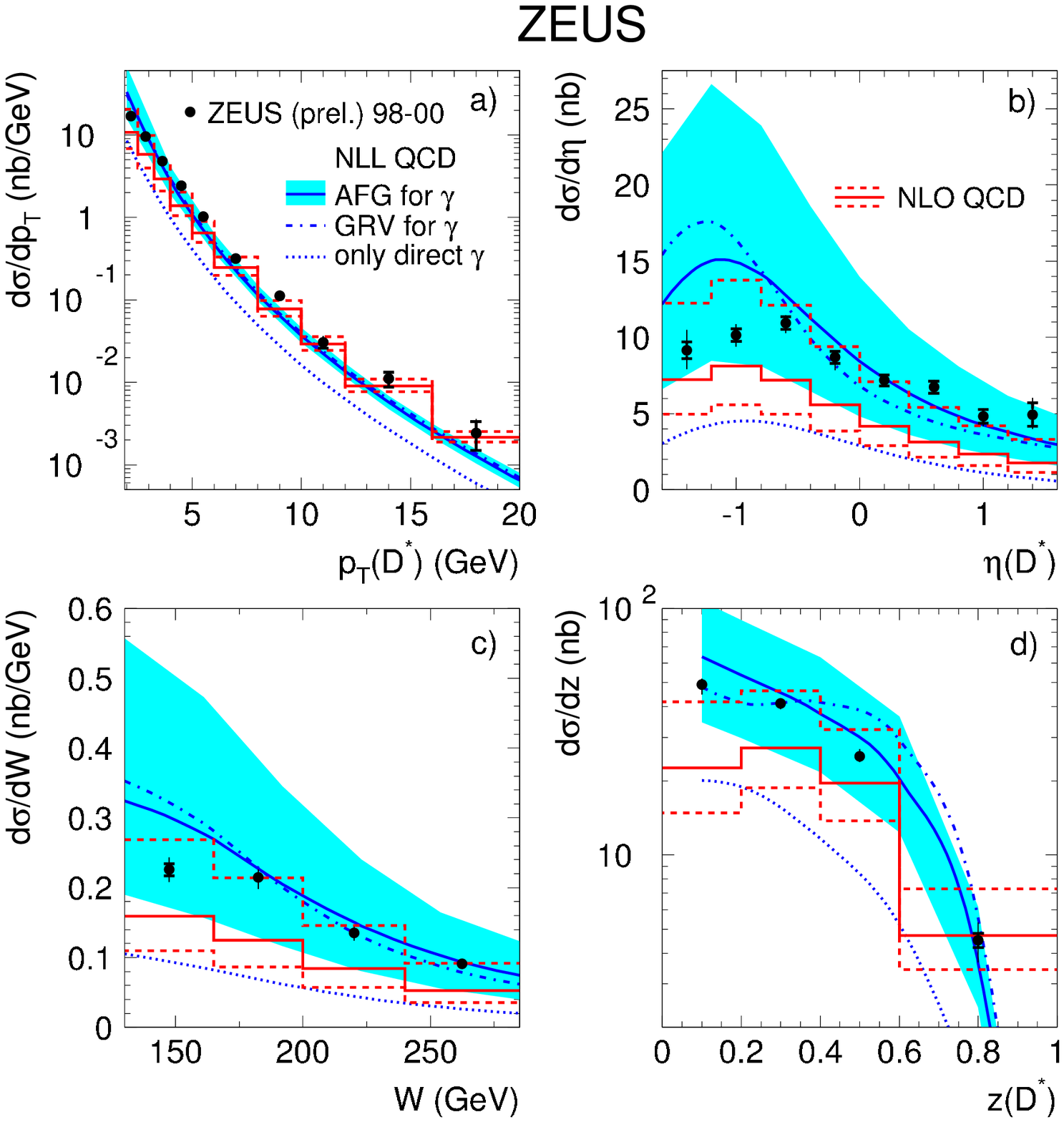} \\
\hspace{.9 cm}
\includegraphics[width=0.337\linewidth]{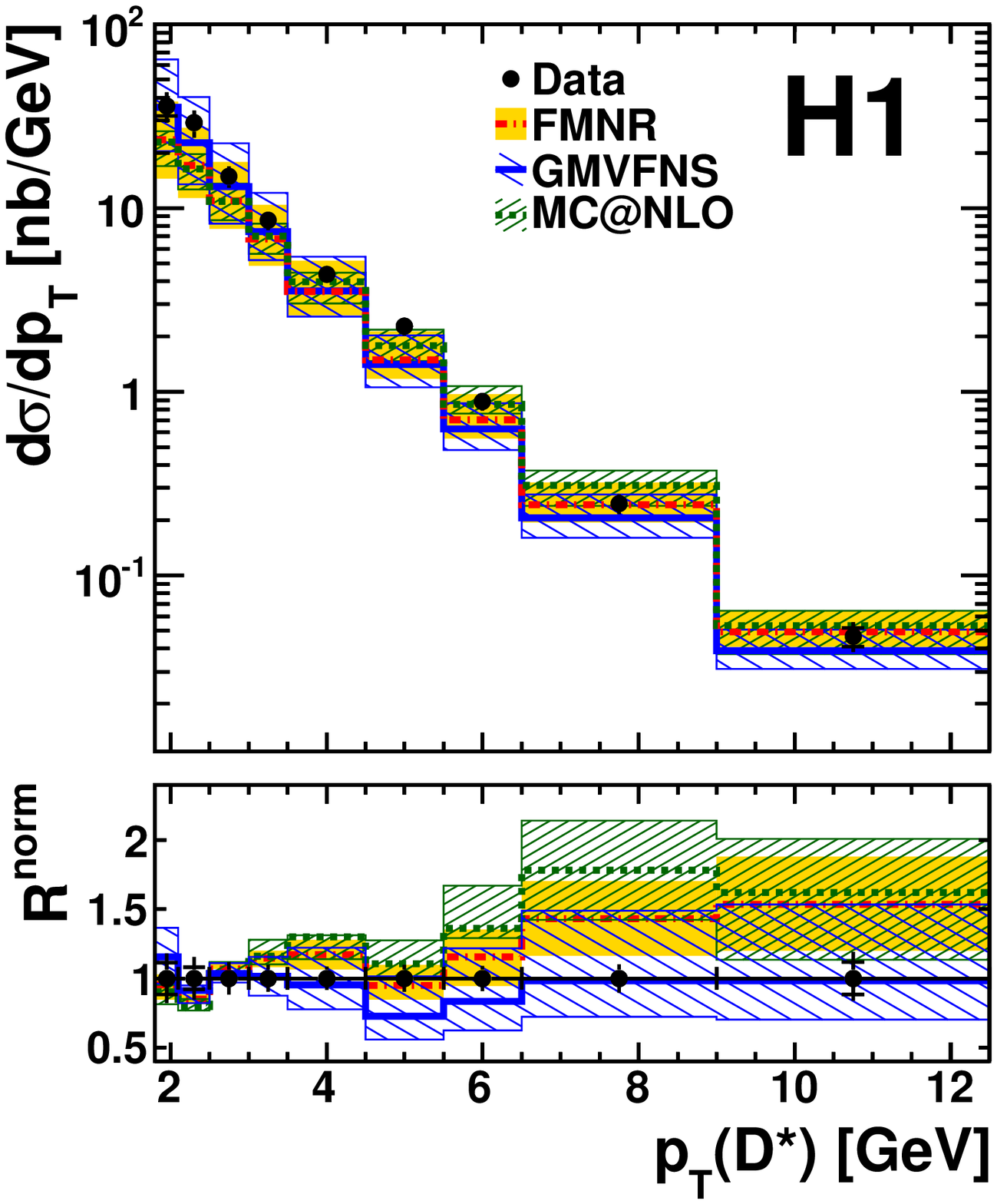}
\includegraphics[width=0.337\linewidth]{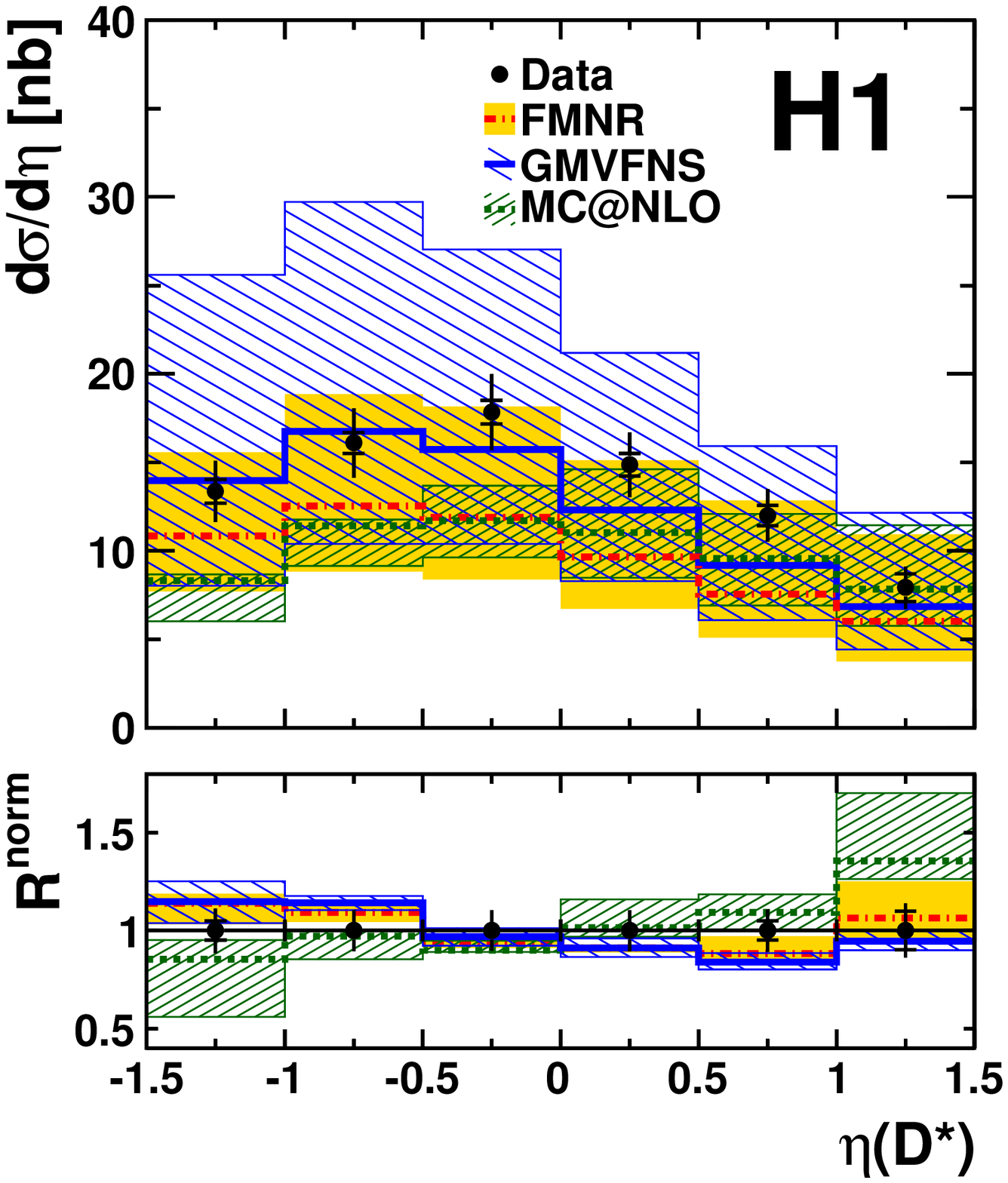} 
\vspace{-2cm}
\caption{
$D^*$ single differential cross sections in photoproduction 
as function of the $D^*$ transverse momentum (left) and pseudorapidity (right),
from ZEUS~\cite{zeus-ichep02-786} (top and center) and H1~\cite{H1cPHP2012}
(bottom).
The measurements are compared to five  
NLO predictions: the massive scheme calculations 
from Frixione {\it et al.}~\cite{Frixione:1995qc} without (NLO, FMNR) and 
with (MC@NLO) \cite{MCNLO} interface to LL parton showering,
the massless scheme predictions from Kniehl {\it et al.}~\cite{Heinrich:2004kj}
(NLL) and the general mass variable flavour scheme calculations from 
Cacciari {\it et al.}~\cite{Cacciari:2001td} (FONLL) and 
Kniehl {\it et al.}~\cite{massfrag} (GMVFNS).
The NLL and GMVFNS predictions include a perturbative treatment of $D^*$ 
fragmentation.
}
\label{fig:r2}
\end{figure}
%
%////////////////////////////////////////////////////////////////////////////////////////////////////////
%
Figure~\ref{fig:r2} shows 
the results for the ZEUS HERA I \cite{zeus-ichep02-786} and H1 HERA II 
\cite{H1cPHP2012} $D^*$ analyses (entries 6 and 15 of Table \ref{tab:r1}) as 
a function of the $D^*$ transverse momentum 
and pseudorapidity. 
These results have been selected since the data samples used in
these analyses are among those with the highest
statistical significance of all heavy flavour
measurements at HERA, as can be seen from the last column of 
Table~\ref{tab:r1} (entry 6). 
The data span a large kinematic range from 
$p_T(D^*) =1.8\;\mbox{GeV} \sim m_c$ to $p_t = 20\;\mbox{GeV} \gg m_c$.
Over this range the cross section falls off by about four
orders of magnitude.
The measurements are compared to five  
NLO predictions: massive fixed-flavour scheme (NLO,FMNR) calculations 
from Frixione {\it et al.}~\cite{Frixione:1995qc}, a variant of these 
calculations matched to parton showers 
(MC@NLO) \cite{MCNLO}, 
massless scheme (NLL) predictions from Kniehl {\it et al.}~\cite{Heinrich:2004kj}, 
general mass variable flavour scheme (FONLL) calculations from 
Cacciari {\it et al.}~\cite{Cacciari:2001td}, and a different GMVFNS variant
(GMVFNS) from Kniehl {\it et al.}~\cite{massfrag}. Both calculations from 
Kniehl {\it et al.} include a perturbative treatment of 
the charm fragmentation function. 
At first glance, 
all five predictions 
are able to describe 
the spectrum over the complete $p_T(D^*)$ range within
a factor of two.
However, looking more in detail, one observes:
\begin{enumerate}
\item
The uncertainty of the measurements is generally much smaller than those of the theory,
dominated by QCD scale variations, the variation of the charm mass, and the variation of 
the charm fragmentation parameters. 
Especially for low transverse momenta $p_T(D^*)<3\;\mbox{GeV}$
the scale uncertainties reach a factor of two.
This indicates
that in this kinematic region 
the hard scales provided by the charm mass and
the transverse momentum of the charm quarks are not yet
large enough to ensure a fast convergence of the 
QCD perturbation series at next-to-leading order. 
\item 
As to be expected, the FONLL prediction is very close to the massive 
NLO/FMNR prediction at low $p_T$, and  for the $\eta$ distribution, which is 
dominated by the low $p_T$ $D^*$ contribution.
The measured cross sections are higher than the central prediction, but the 
predictions are consistent with the data within the large uncertainties.
At high transverse momenta, contrary to many people's expectations originally 
based on leading-order studies \cite{Catani}, 
the FONLL prediction is actually lower than the NLO prediction. Thus the 
final state resummation corrections originating from higher-order log terms 
in the massless part of the calculation 
reduce the prediction, rather than enhancing it. The data are closer
to the pure
NLO prediction. At least within the HERA kinematic regime there is thus 
no evidence for the claim \cite{Catani} that the massive fixed-order 
calculation should fail at large values of charm transverse momentum
unless final state resummation corrections are applied. 
Both predictions give a reasonable but not perfect description of the shape 
of the $\eta$ distribution. 
\item
The massless NLL prediction, which, in contrast to the massive predictions 
discussed in the previous item, incorporates a proper perturbative treatment 
of charm fragmentation \cite{massless}, fits the data well at low $p_T$,  
while it is a bit too low for high $p_T$,
where it is expected to work best. As expected, it is similar to the FONLL 
prediction in this region.
The theoretically superior treatment of fragmentation does not lead to a 
smaller uncertainty, as can be seen from the $\eta$ distribution.
Also, the shape of the $\eta$ distribution is a bit less well described than 
with the massive prediction.  
In this approximation, a large fraction of the cross section arises from the 
(massless) charm contribution to the photon parton density function 
(using the AFG \cite{AFG} or GRV \cite{GRV} parametrisations), in contrast to 
the ``direct'' contribution, which is also shown separately.  
\item
The partially massive GMVFNS prediction, which incorporates a perturbative 
treatment of the charm fragmentation function, has a larger uncertainty than 
the traditional massive predictions, similar to the NLL prediction.
The shape of this prediction describes the data better than the NLL prediction.
\item
The MC@NLO prediction has the same core parton-level cross section as the 
NLO/FMNR predictions by definition. The differences seen w.r.t. the latter 
must thus arise from the addition of the HERWIG-type parton showers and the 
different fragmentation treatment. It exhibits slighty smaller uncertainties,
but, surprisingly, fits the data less well than the original NLO/FMNR 
predictions. This offers room for potential retuning of some of the MC 
parameters entering this calculation.  
\end{enumerate}

A similar inclusive $D^*$ photoproduction 
measurement as the above is available from
H1~\cite{H1cPHP2007}, 
performed in a more restricted $W_{\gamma p}$
region (entry 7 in table~\ref{tab:r1})
with a roughly ten times smaller data sample. The narrower kinemetic range 
and smaller statistics are due 
to explicit detection of the electron scattered at very low angles in 
dedicated forward electron taggers
(section \ref{sect:taggers}), which was part of the trigger requirement.  
The conclusions are very similar.

ZEUS has also recorded such tagged photoproduction samples, 
but they were found 
not to be statistically competitive with results from data sets triggered on 
inclusive $D^*$ production. 

\subsubsection{$D^*$ double-differential cross sections} 
\begin{figure}[htbp]
\centering
\includegraphics[width=0.65\linewidth]{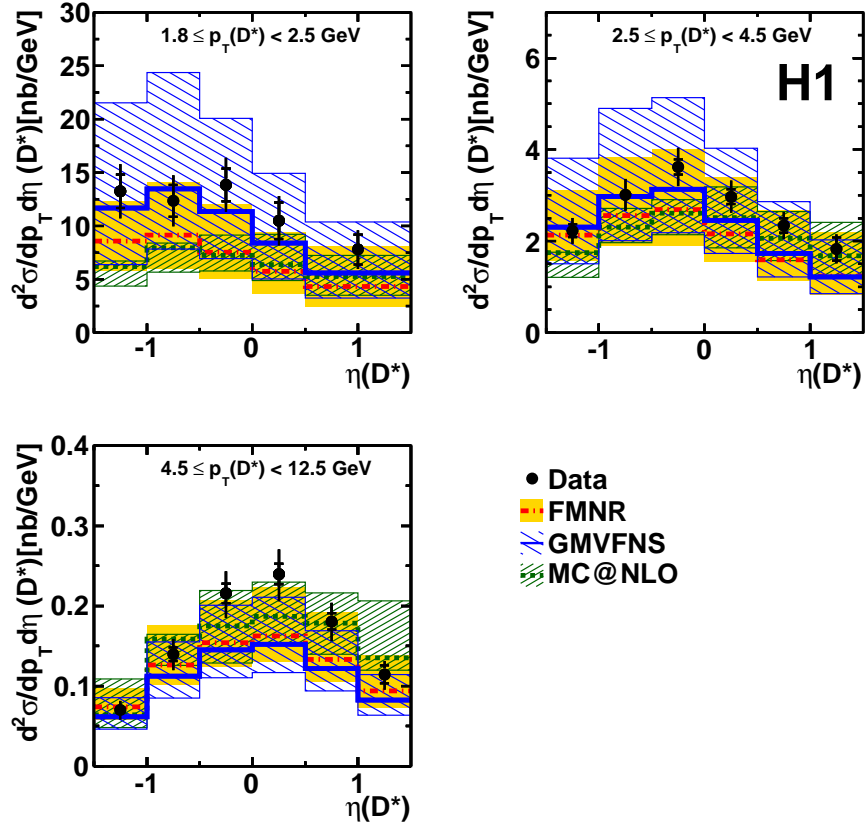}
\caption{
$D^*$ double differential cross sections in photoproduction 
as function of the $D^*$ transverse momentum and pseudorapidity
from H1~\cite{H1cPHP2012}.
The measurements are compared to three out of the five  
NLO predictions also shown in Fig. \ref{fig:r2}.
}
\label{fig:doubdiff}
\end{figure}

Double-differential cross-section measurements as a function of
the $D^*$ transverse momentum and pseudorapidity have been performed by ZEUS 
in ~\cite{zeus-ichep02-786} and also in a previous  
charm milestone paper~\cite{ZEUS97a} (entries 6 and 5 in Table \ref{tab:r1}),
and by H1 \cite{H1cPHP2012} (entry 15 in Table \ref{tab:r1}). The results of
the latter are shown in Fig. \ref{fig:doubdiff}, together with some of the 
predictions already discussed for the single-differential case.
%
%%%%%%%%%%%%%%%%%%%%%%%%%%%%%%%%%%%%%%%%%%%%%%%%%%%%%%%%%%%%%%%%%%%%%%%%%%%%%%%%%%%%%%%%%%%%%%%%%%%%%%%%%%%%%%%%%%%%%%%%
%
%     H1 2012 paper 
%
%
%////////////////////////////////////////////////////////////////////////////////////////////////////////
% 
In general, the conclusions are similar to those from the single-differential 
cross sections. At high $p_T$ the uncertainty of the theory predictions 
reduces as expected, such that the comparisons become more meaningful.
Reasonable agreement with the data is observed for all predictions in this 
high-$p_T$ region, while MC@NLO undershoots the data at low $p_T$ and $\eta$,
similar to what was observed in the single-differential case 
(Fig. \ref{fig:r2}).

\subsection{Inclusive measurements using other tagging methods} 

Although the $D^*$ channel generally yields the best signal-to-background 
ratio and therefore the best effective overall statistics 
(last column of Table \ref{tab:r1}), the small 
branching ratio limits the statistics in regions in which the cross section
is small. In such regions, more inclusive tagging techniques can be an 
advantage. Furthermore, the consistency of results obtained with different
tagging methods enhances confidence in the results.

\begin{figure}[htbp]
\centering
\includegraphics[width=0.44\linewidth,bb=34 -170 563 402,clip]{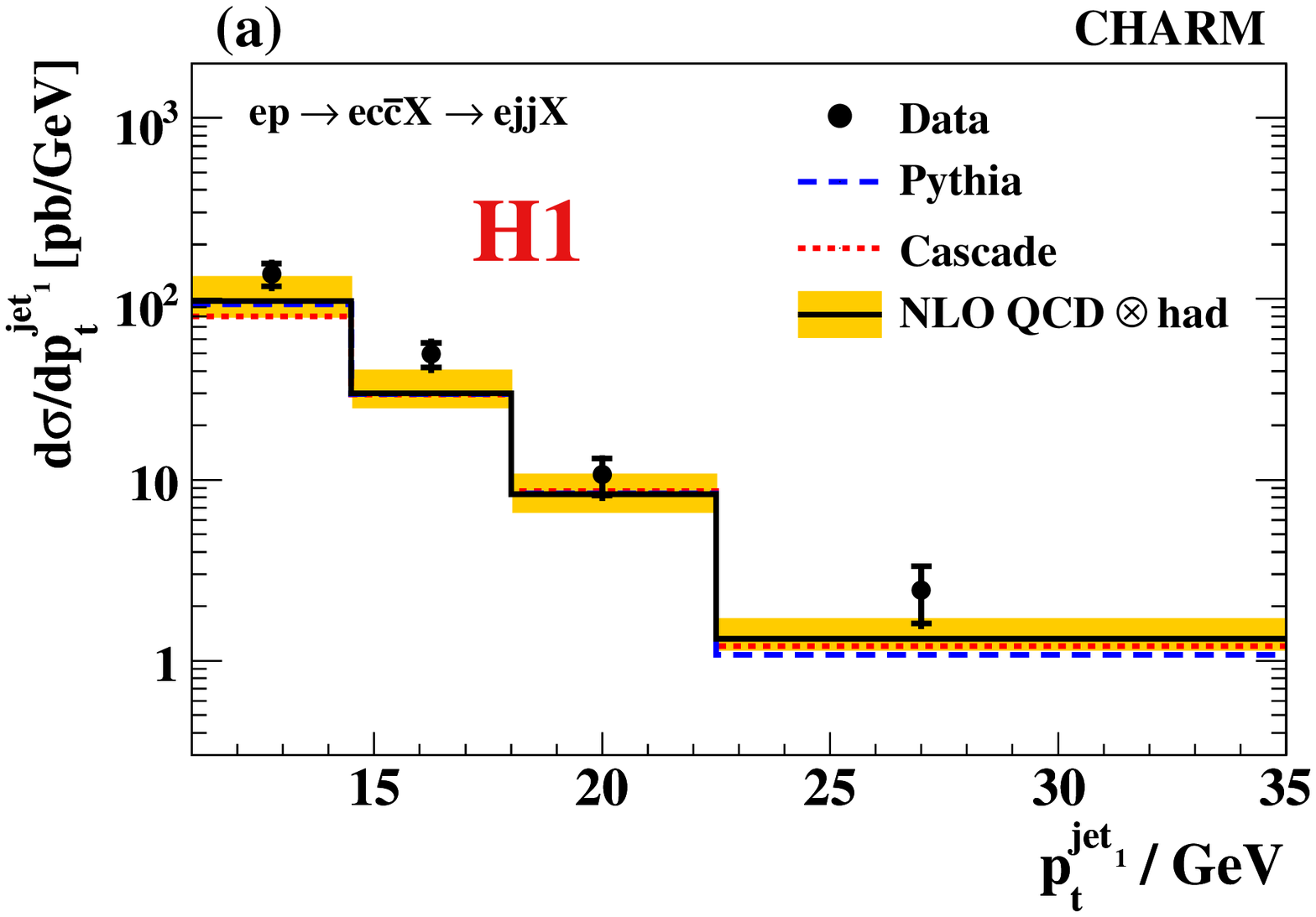}
\includegraphics[width=0.55\linewidth]{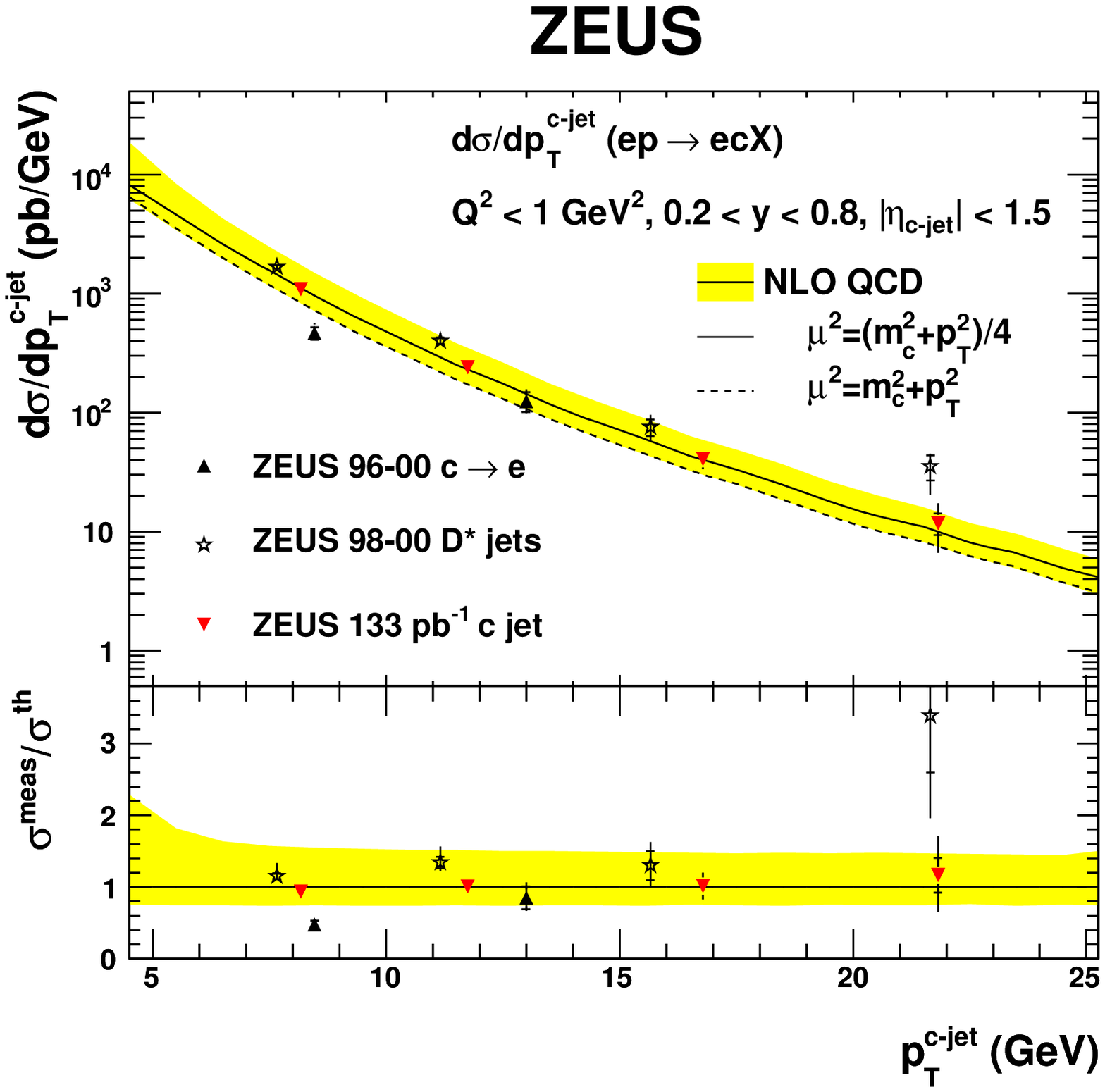}
\vspace{-1.3cm}
\caption{
Left: Differential cross sections for the process
$ep\rightarrow ec\bar{c}X \rightarrow ejjX$ 
as function of the transverse momentum $p_t^{jet1}$ of the leading jet,
from the H1 analysis~\cite{H1cbPHP2006}.
The data are compared to an NLO calculation~\cite{Frixione:1995qc} 
in the massive scheme, and to LO+PS MC predictions from PYTHIA \cite{PYTHIA}
and CASCADE \cite{CASCADE}.
Right: Summary of differential $c$-quark jet cross sections as a function 
of the jet transverse momentum, as measured by the ZEUS 
collaboration \cite{ZEUScbPHP2011}.
The data are compared to an NLO calculation~\cite{Frixione:1995qc} 
in the massive scheme, for two different QCD scale choices.
}
\label{fig:r3}
\end{figure}

H1 has performed a measurement based on inclusive lifetime 
tagging~\cite{H1cbPHP2006} (entry 10 in table \ref{tab:r1}),  
which extends to the highest charm transverse momenta $p_T^c=35\;$GeV
reached so far.
Here events with two jets in the central
rapidity region are used (cuts are listed in table~\ref{tab:r1}).
Due to the high jet transverse momenta the events 
are efficiently triggered using the deposits of the jet
particles in the calorimeter.
An inclusive
lifetime tagging is applied, based on the displaced impact parameters of
jet-associated charged
tracks from charm and beauty decays.
Details of the tagging method are discussed in \Sect{Tagging}.
Figure~\ref{fig:r3}(left) shows the 
measured charm-production cross sections as 
function of the transverse momentum of the 
leading jet.
%
%%%%%%%%%%%%%%%%%%%%%%%%%%%%%%%%%%%%%%%%%%%%%%%%%%%%%%%%%%%%%%%%%%%%%%%%%%%%%%%%%%%%%%%%%%%%%%%%%%%%%%%%%%%%%%%%%%%%%
% 
% c jet summary plot
%
%
%%%%%%%%%%%%%%%%%%%%%%%%%%%%%%%%%%%%%%%%%%%%%%%%%%%%%%%%%%%%%%%%%%%%%%%%%%%%%%%%%%%%%%%%%%%%%%%%%%%%%%%%%%%%%%%%%%%%%
%
The data are compared to a massive scheme NLO 
prediction~\cite{Frixione:1995qc},
which describes the data reasonably and equally well 
up to the highest jet transverse momenta.
To compare this result with the above $D^*$ measurement
(Fig.~\ref{fig:r2})
one has to take into account that the jet gives 
a direct approximation of the charm quark kinematics,
while on average the $D^*$ takes only about $70\%$ 
of the charm quark momentum
in the fragmentation (after cuts).
Thus, the kinematic range tested with the 
leading jet $p_T$ from $11$ to $35\gev$ roughly corresponds
to a $D^*$ transverse-momentum region from 8 to $25\gev$.
For $D^*$ \/ transverse momenta from 8 GeV up to the highest covered
value of 20 GeV the $D^*$\/ data are similarly
well described by the NLO calculation as the dijet data
at their correspondingly higher momenta.
So the two independent measurements using different
tagging techniques give consistent results.

A similar and more direct comparison is shown in Fig. \ref{fig:r3}(right) for 
several measurements from ZEUS (entries 8,12,13 in table \ref{tab:r1}).
Here, the measurements have already been translated to cross sections for 
inclusive $c$-jet
production. The results obtained from $D^*$ and inclusive-vertex tagging agree
well with each other and with theory. 
Since in the core of a jet 
electrons are not easily separated from $\pi/\pi^0$ overlaps, 
charm tagging using semileptonic decays into electrons is experimentally 
difficult\footnote{Several other such 
charm analyses were eventually not published due to insufficient control of 
systematics.}
and the corresponding $c\to e$ result, which was a byproduct of an analysis 
focusing on beauty production, might not include all relevant systematic 
uncertainties. 
To compare with Fig. \ref{fig:r3}(left), the prediction with scale choice
$m^2 + p_T^2$ (dashed line in Fig. \ref{fig:r3}(right)) should be considered.
Good agreement is observed between the results of the two experiments.

\subsection{Studies with a $D^*$ and one other hard parton}
\begin{figure}[htbp]
\centering
\includegraphics[width=0.35\linewidth]{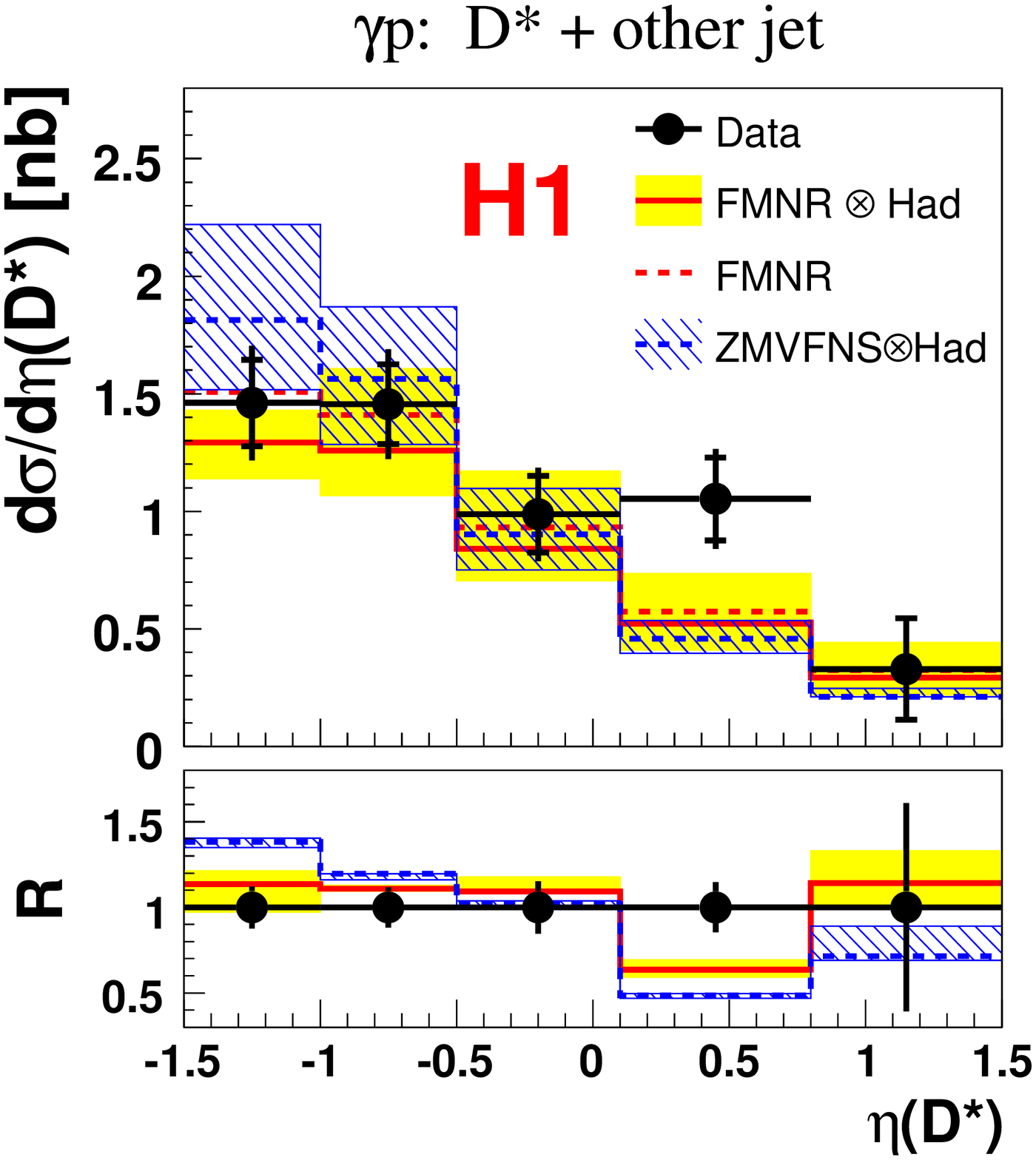}
\includegraphics[width=0.35\linewidth]{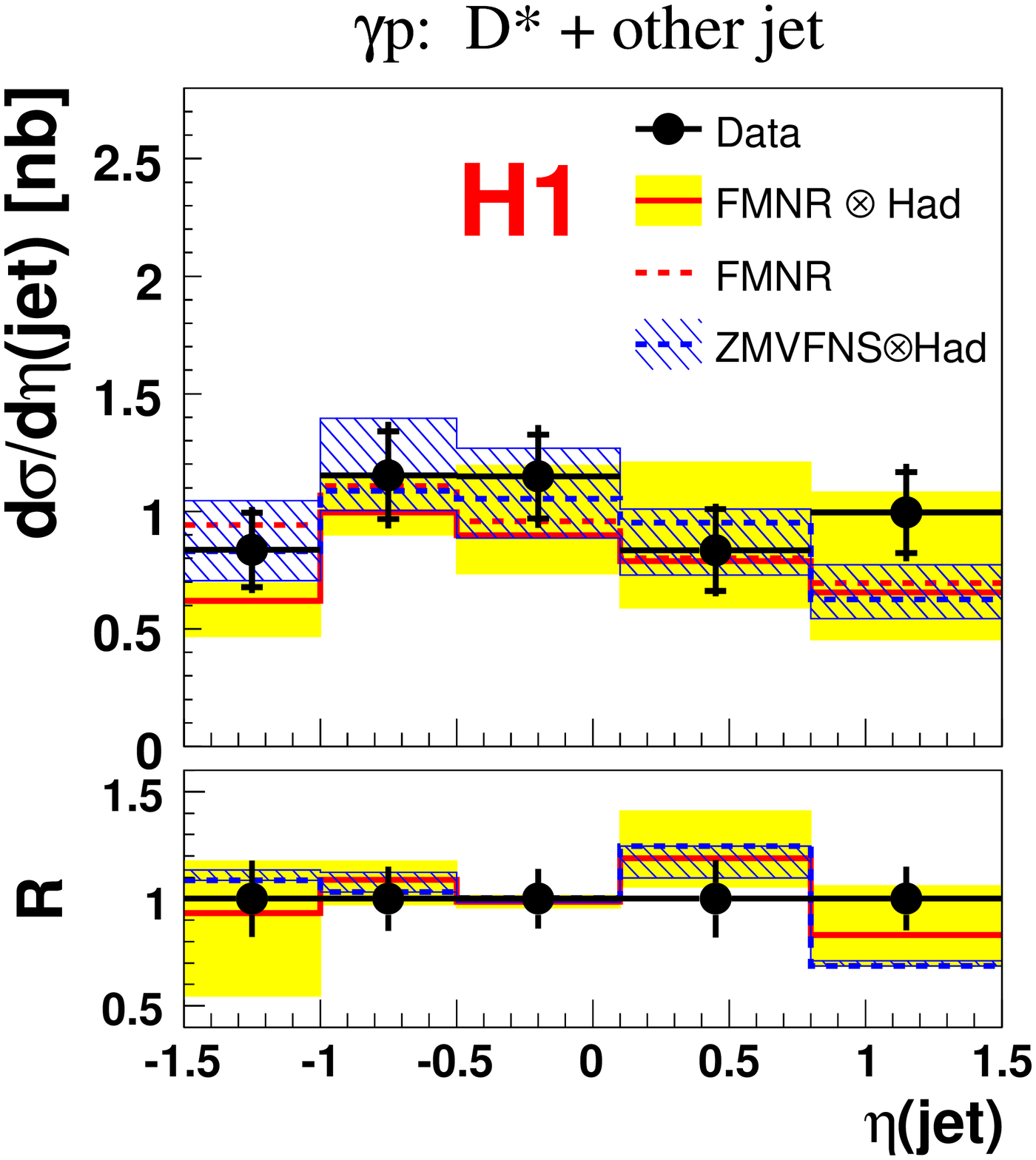}
\hspace{5mm}
\caption{$D^*$ + jet\/
 cross sections as function of the pseudorapidities of the 
$D^*$ (left) and the jet (right), from the H1 analysis~\cite{H1cPHP2007}.
The measurements are compared to two  
NLO predictions, the massive scheme (FMNR) calculations 
from Frixione {\it et al.}~\cite{Frixione:1995qc}
and the massless scheme (ZMVFNS) predictions from Heinrich and Kniehl 
\cite{Heinrich:2004kj}.} 
\label{fig:r4}
\end{figure}

To obtain more information on the
charm-photoproduction process, one possibility is to 
require the presence of a jet in the final state
in addition to the $D^*$, 
which is not associated to the $D^*$.
This means that the jet and the $D^*$ are well
separated in their directions and that the
jet tags another hard parton in the process.
This parton can be the other charm quark or a gluon 
or light quark.
In one analysis \cite{H1cPHP2007} a very soft jet momentum cut 
$p_T>3\;\mbox{GeV}$ was applied.
The jets were restricted to the central 
pseudorapidity region $|\eta|<1.5$, thus covering the same kinematic range 
as the $D^*$s.
In Fig.~\ref{fig:r4}
the differential cross sections
are shown as function of the
pseudorapidities of the $D^*$ and the jet.
%
%%%%%%%%%%%%%%%%%%%%%%%%%%%%%%%%%%%%%%%%%%%%%%%%%%%%%%%%%%%%%%%%%%%%%%%%%%%%%%%%%%%%%%
%
% dstar and jet pseudorapidity spectra
%
%
%%%%%%%%%%%%%%%%%%%%%%%%%%%%%%%%%%%%%%%%%%%%%%%%%%%%%%%%%%%%%%%%%%%%%%%%%%%%%%%%%%%%%%%
%
For the leading-order boson-gluon-fusion process
it is expected that 
the $D^*$ tags one charm quark and the jet the other.
Since similar momentum cuts are applied 
for the $D^*$ and the jet, one would expect
very similar pseudorapidity distributions for the
$D^*$ and the jet. 
However, the observed pseudorapidity spectrum 
for the jet (Fig.~\ref{fig:r4})  
is significantly shifted towards the more forward direction
compared to that of the $D^*$.
This indicates that, as expected from higher-order contributions, 
the jet often tags another parton, 
i.e. a gluon or a light quark.  
This effect is predicted by 
the massive and massless scheme NLO calculations
to which the data are compared in Fig.~\ref{fig:r4},
and these calculations describe the data reasonably well.
Also, the additional jet requirement significantly reduces
the theoretical uncertainties w.r.t. Fig.~\ref{fig:r2}. 

\begin{figure}[htb]
\centering
\includegraphics[width=0.8\linewidth]{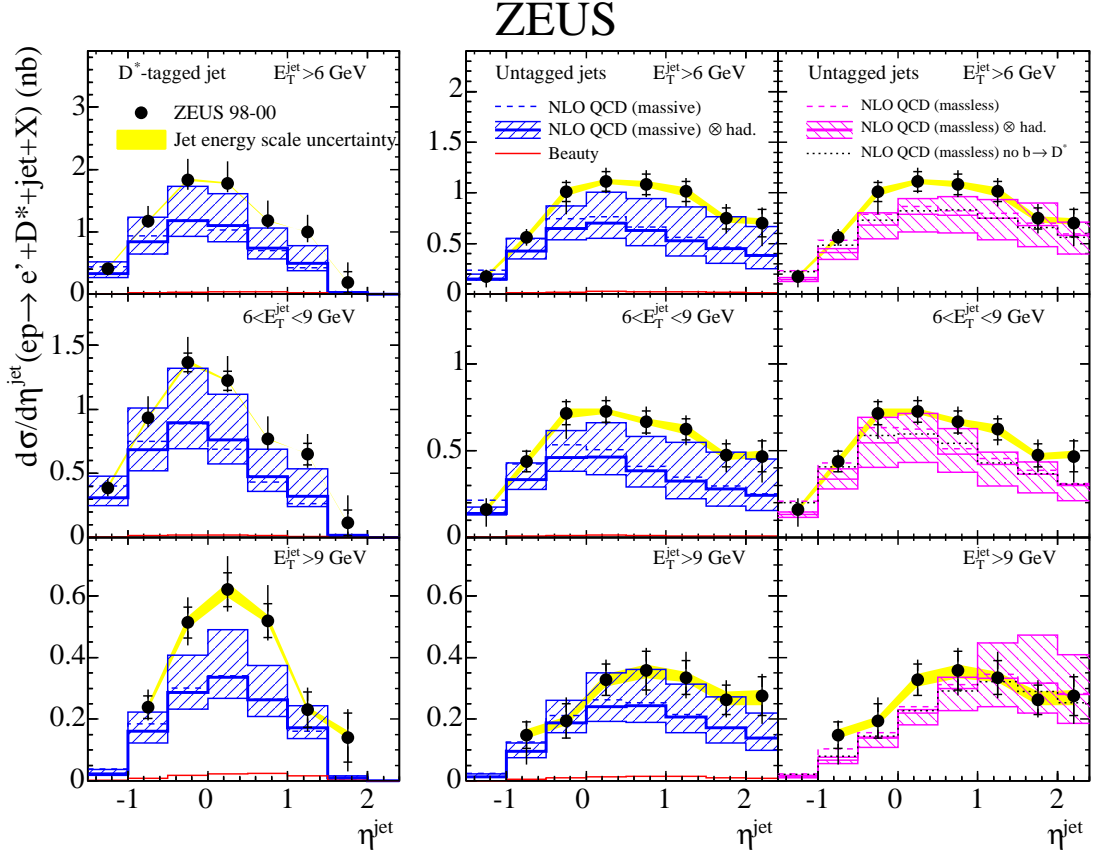}
\hspace{5mm}
\caption{$D^*$ + jet\/
 cross sections as a function of the pseudorapidities of 
$D^*$-tagged (left) and untagged (center and right) jets, 
from the ZEUS analysis~\cite{ZEUScPHP2005}.
The measurements are compared to two  
NLO predictions, the massive scheme calculations 
from Frixione {\it et al.}~\cite{Frixione:1995qc}
and the massless scheme predictions from Heinrich and 
Kniehl~\cite{Heinrich:2004kj}.} 
\label{fig:ZEUSpseudo}
\end{figure}

In addition to jets not associated to the $D^*$,
the corresponding ZEUS measurement~\cite{ZEUScPHP2005}
also selected events in which the 
$D^*$ is associated to the jet. 
In the latter case,
one does obtain information only about one hard
parton in the event, which is a charm quark.
Furthermore, the jet tranverse momentum cut $p_T^{jet}>6\gev$ is
much harder and a much wider pseudorapidity range $-1.5<\eta^{jet}<2.4\,$
is covered.
Good agreement with NLO predictions is observed for all 
single-jet distributions (not shown). In particular, the $E_T$ spectra for 
\dst-tagged jets (from charm quarks) and untagged jets (from charm, gluons, 
or light quarks) are similar. The pseudorapity distributions for \dst-tagged 
and untagged jets (Fig. \ref{fig:ZEUSpseudo}) show differences consistent 
with those 
of the H1 analysis. As expected, the average jet pseudorapity increases with 
increasing jet $E_T$. Again, the theoretical uncertainties are 
reduced with respect to those of Fig. \ref{fig:doubdiff}. 
At high jet $E_T$, the shape of the massive calculation describes the data 
somewhat better than the massless one.

%%%%%%%%%%%%%%%%%%%%%%%%%%%%%%%%%%%%%%%%%%%%%%%%%%%%%%%%%%%%%%%%%%%%%%%%%%%%%%%%%%%%%%
%
% ZEUS D*+jet: dstar and jet pseudorapidity spectra
%
%
%%%%%%%%%%%%%%%%%%%%%%%%%%%%%%%%%%%%%%%%%%%%%%%%%%%%%%%%%%%%%%%%%%%%%%%%%%%%%%%%%%%%%%%

The selection of events with a $D^*$ and a muon from a semileptonic charm decay,
not associated with the $D^*$ \cite{H1cbPHP2005}, allows explicit tagging of 
both charm quarks. The small statistics (entry 11 in Table \ref{tab:r1}, 
where also the visible phase space cuts are given) do not allow differential
distributions, but the observed
total visible cross section for $D^*\mu$ production from double-tagged 
$c\bar c$ final states
of $250 \pm 57 \pm 40$ pb is consistent with the prediction from the massive 
NLO calculation \cite{FMNR} of $256^{+159}_{-59}$ pb.

\subsection{Parton-parton-correlation studies in charm-tagged events}
Analyses using tagged charm events with
two identified hard partons in the final 
state and studying the correlations of the two 
partons~\cite{ZEUScPHP1999,H1cPHP2007,ZEUScPHP2003,ZEUScPHP2005, 
H1cbPHP2006,H1cbPHP2012,H1cPHP2012}
provide the most detailed
information on the charm-production mechanism.
Similar to the previous subsections, 
there are two different experimental approaches:
\begin{itemize}
\item
The $D^*$ tag is used for charm tagging.
For the two hard partons either 
the reconstructed $D^*$ plus an additional non-associated jet are 
used \cite{H1cPHP2007,ZEUScPHP2005}, or alternatively
two jets are identified, \
one of which is tagged by the $D^*$~\cite{ZEUScPHP1999,ZEUScPHP2003,
ZEUScPHP2005,H1cPHP2012}.
\item
Alternatively, dijet events are selected and one jet is tagged as a charm jet 
using the displaced impact parameters of jet-associated charged 
tracks \cite{H1cbPHP2006} or by a muon from a charm semileptonic 
decay \cite{H1cbPHP2012}.
\end{itemize}

With the two identified partons 
three correlation observables are constructed,
which will be discussed in the following:
\begin{enumerate}
\item 
The observable $\xgobsm$, which allows the separation, 
in the leading order picture, 
of direct- and resolved-photon interactions. In the NLO picture, 
it separates 3-parton from 2-parton final states.
\item
The azimuthal correlation $\Delta \phi$ of the
two partons, which is sensitive to higher-order effects. 
Combined with $\xgobsm$, it can distinguish between 2-parton, 
3-parton and 4-parton final states.
\item 
The hard-scattering angle $cos\theta^{*}$ of the two
partons, which allows the distinction of contributions
with quark or gluon propagators in the hard scattering.
\end{enumerate}

\subsubsection{\xgobs studies}
\label{sect:xgammaPHP}

The case of two jets is assumed in the following 
for the two hardest partons.
The observable $\xgobsm$ is defined as
\begin{equation}
\xgobsm = \frac{ \sum_{Jet_1}(E-p_Z) +
                 \sum_{Jet_2}(E-p_Z)}{\sum_{h}(E-p_Z)}.
\label{eq:xgobs}
\end{equation}
The sums in the numerator run over
the particles associated with the two jets
and those in the denominator over all
detected hadronic final state particles.
$E$ and $p_Z$ denote the particle energy, and the
momentum parallel to the proton beam, respectively.

In the leading-order pQCD picture (Section \ref{sect:leading}, 
2 partons + potential photon remnant + 
proton remnant)
this variable is an estimator of the
fraction of the photon energy entering the hard interaction.
For the direct boson-gluon-fusion process (Fig.~\ref{fig:fey1}(a))
$\xgobsm$ approaches unity, as the hadronic final state consists of
only the two hard jets and the proton remnant in the
forward region, which contributes little to $\sum_{h}(E-p_Z)$.
In resolved processes (Figs. \ref{fig:fey1}(b-d)) 
the photon remnant significantly contributes to the 
denominator but not to the numerator, so $\xgobsm$ can be small.
The addition of parton showering can somewhat dilute this simple picture.

$\xgobsm$ is also smaller than unity for next-to-leading-order  
processes with a third hard outgoing parton 
(Fig.~\ref{fig:intro4}). In the massive NLO case for charm production
this often coincides with the other quark originating from initial-state 
photon splitting into a $c\bar c$ pair, which would be classified as 
a photon remnant in the 
leading-order picture. Since in the fixed-flavour NLO case there are at most 
three partons, $\xgobs$ separates 2-parton from 3-parton final states.
In the variable-flavour NLO case the two pictures described above get mixed, 
since in the case of an initial-state $c$ quark from the photon the other
$c$ quark can be a fourth hard parton.  
Thus, in general, the observable $\xgobsm$ is sensitive
to the resolved-photon structure (if any) and to tree-level higher-order 
processes (if any).

\begin{figure}[htb]
\centering
\includegraphics[width=0.5\linewidth]{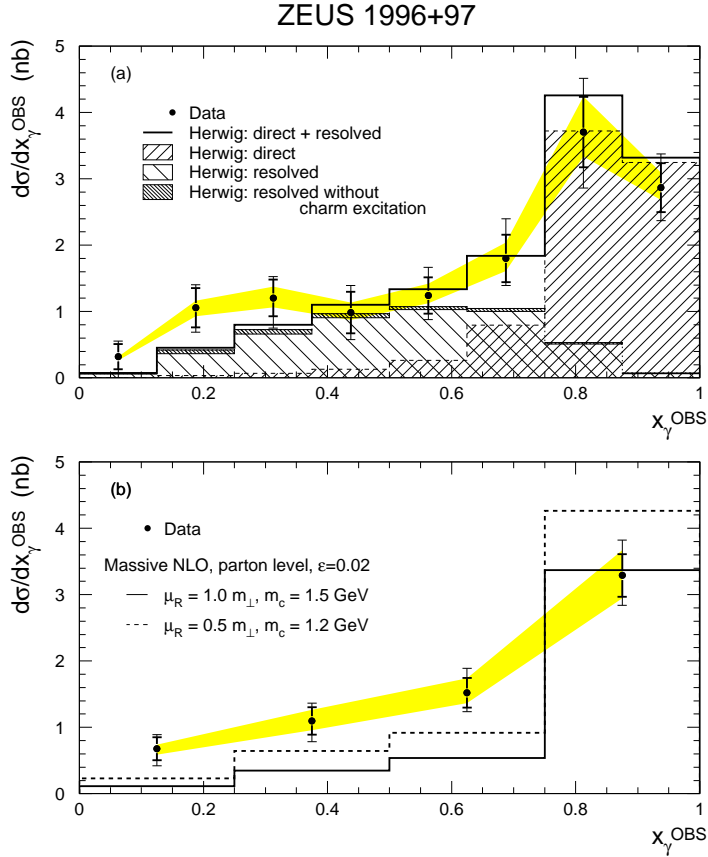}
\caption{
Differential cross section as a function of \xgobs
for dijet events with an associated $D^*$ meson, from a
ZEUS analysis~\cite{ZEUScPHP1999}. The shaded band indicates the energy 
scale uncertainty.
The same data are compared in the lower plot to an NLO calculation~\cite{Frixione:1995qc}, and 
in the upper plot to Monte Carlo predictions from 
HERWIG~\cite{HERWIG} with direct and resolved photon
contributions shown separately. The latter is dominated by the 
charm excitation component.}
\label{fig:r5}
\end{figure}

One of the milestone papers on charm
photoproduction at HERA was the ZEUS analysis
\cite{ZEUScPHP1999},
where 
$\xgobs$ studies are performed
using events with a $D^*$ and two jets.
The jets are required to have transverse momenta 
$p_T^{jet1(2)} > 7(6)\;$GeV and are selected in a
wide rapidity range $|\eta^{jet}|<2.4$.
In most events the $D^*$ is associated to one
of the two jets.
Figure~\ref{fig:r5}
shows the measured single-differential
cross section as a function of $\xgobs$.
A peak at large 
$\xgobs>0.75$ is observed,  which reflects the 
direct-photon/2-hard-parton component.
Roughly 50\% of the data are observed at $\xgobs<0.75$,
indicating large contributions from
resolved-photon/3-hard-parton or other higher-order contributions.

%
%%%%%%%%%%%%%%%%%%%%%%%%%%%%%%%%%%%%%%%%%%%%%%%%%%%%%%%%%%%%%%%%%%%%%%%%%%%%%%%%%%%%%%%%%%%%%%%%%%%%
%
% ZEUS 96/97 Ds + dijet xgamma:
%
%
%
%
%
%
%///////////////////////////////////////////////////////////////////////////////////////////////////
%

In the lower plot in Fig.~\ref{fig:r5}\/
the data are compared to predictions from a
massive scheme NLO calculation~\cite{Frixione:1995qc}.
Not all theoretical uncertainties are shown here. The calculation has a 
tendency to underestimate the data cross sections at $\xgobs<0.75$, where it
is effectively a leading order calculation.
This might indicate the need
for even higher-order corrections.

A much better shape description is obtained with the
LO+PS HERWIG~\cite{HERWIG} Monte Carlo program as shown in the upper
plot of Fig.~\ref{fig:r5}.
In this calculation a large part of the NLO photon splitting diagram 
in Fig. \ref{fig:intro7}(c) is included in the form of a charm excitation component, 
where the charm quark is treated as a massless constituent of the
resolved photon, as shown in Figs.~\ref{fig:fey1}(c) 
and~\ref{fig:fey1}(d).
This gives the dominant contributions for $\xgobs<0.75$. Combined with 
parton showering, which also pulls the ``direct'' contribution towards lower
$\xgobs$ values, this provides a reasonable data description. 
This LO+PS MC approach thus provides an effective 
way to describe the small $\xgobs$ region, 
although the charm quark is treated as massless in a kinematic region where
this is probably not a good approximation.
Note that the total cross section with $D^*$ + dijets 
is only about $18\%$ of the $D^*$ cross section without
the dijets, also measured in~\cite{ZEUScPHP1999}, 
for the same $D^*$ cuts applied
($p_T(D^*)>3\;$GeV and $|\eta(D^*)|<1.5$).
Thus the problematic (for NLO) $\xgobsm<0.75$ region in the $D^*$ + dijet 
sample 
contributes only a relatively small part
to the inclusive-$D^*$ cross section.

\begin{figure}[htbp]
\centering
\includegraphics[width=0.7\linewidth,bb=0 0 560 510,clip]{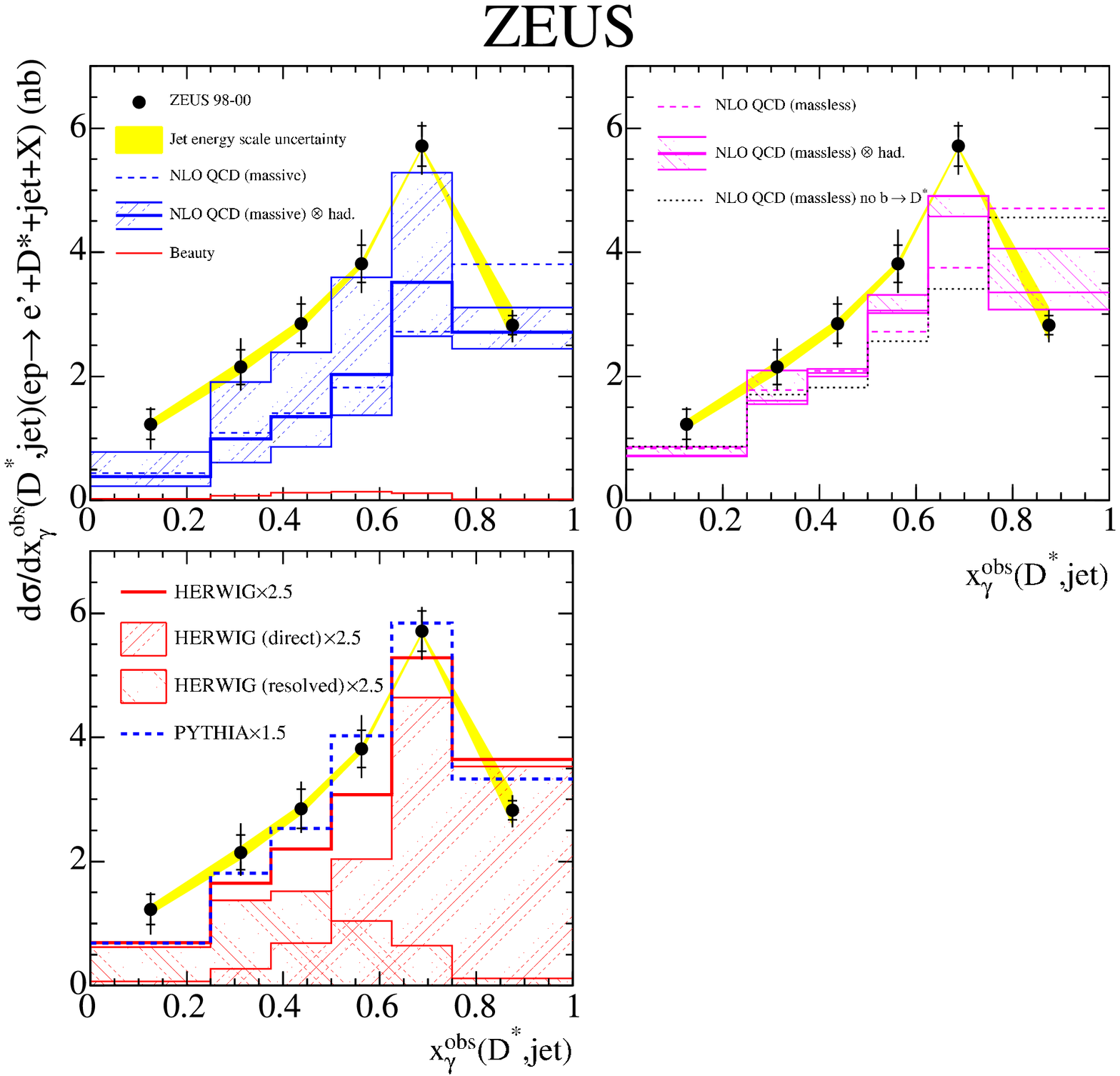}
\caption{
Differential cross sections as a function
of $\xgobs$ for events with a jet and a
$D^*$ meson, which is not associated to the jet, from
the ZEUS analysis~\cite{ZEUScPHP2005}.
In the left (right) upper plots the data are compared to 
a NLO calculation
in the massive~\cite{Frixione:1995qc} (massless~\cite{Heinrich:2004kj}) scheme.
The bottom plot shows a comparison to the PYTHIA and HERWIG MC, which were used
to calculate the hadronisation corrections in the upper left plot. 
}
\label{fig:r6}
\end{figure}

Another ZEUS analysis~\cite{ZEUScPHP2005}
using events with a $D^*$ and at least one jet
compares the measured  
\xgobs cross sections to both massive and 
massless scheme NLO calculations.
Here the $D^*$ and a jet, to which the $D^*$ is not
associated, are taken 
as estimators for the two 
leading partons and used for the \xgobs reconstruction\footnote{
Note that the available massless scheme calculations~\pcite{Heinrich:2004kj} 
provide only cross sections for 
a $D^*$ + jet final state but not for two jets.}
in Eq.~(\ref{eq:xgobs}).
The jet is required 
to have transverse momentum $p_T>6\;$GeV 
in a pseudorapidity range  $-1.5<\eta<2.4$.
Figure~\ref{fig:r6} shows the differential cross sections
as a function of \xgobs.
In the left (right) plot the data are compared to 
the massive (massless) scheme NLO predictions.
Both predictions
are a bit too low for the 3-or-more-parton final state region 
$\xgobs<0.75$, but are still compatible with the data within their 
uncertainties. Note that in the massless calculation, which absorbs the 
initial state photon splitting to $c\bar c$ into the photon PDF, this 
contribution is effectively calculated to NLO (one-loop virtual corrections),
while it is only calculated to LO (0 loop) in the massive case. This partially
explains why the uncertainty of the massless calculation is much smaller
in this region.
At high $\xgobs$ both calculations are effectively NLO (1-loop) calculations, 
and the uncertainties are similar. 
%
%/////////////////////////////////////////////////////////////////////////////////
%

\subsubsection{Azimuthal correlations $\Delta \phi$}
\begin{figure}[htbp]
\centering
\hspace*{-1.4cm}
\includegraphics[width=0.7\linewidth]{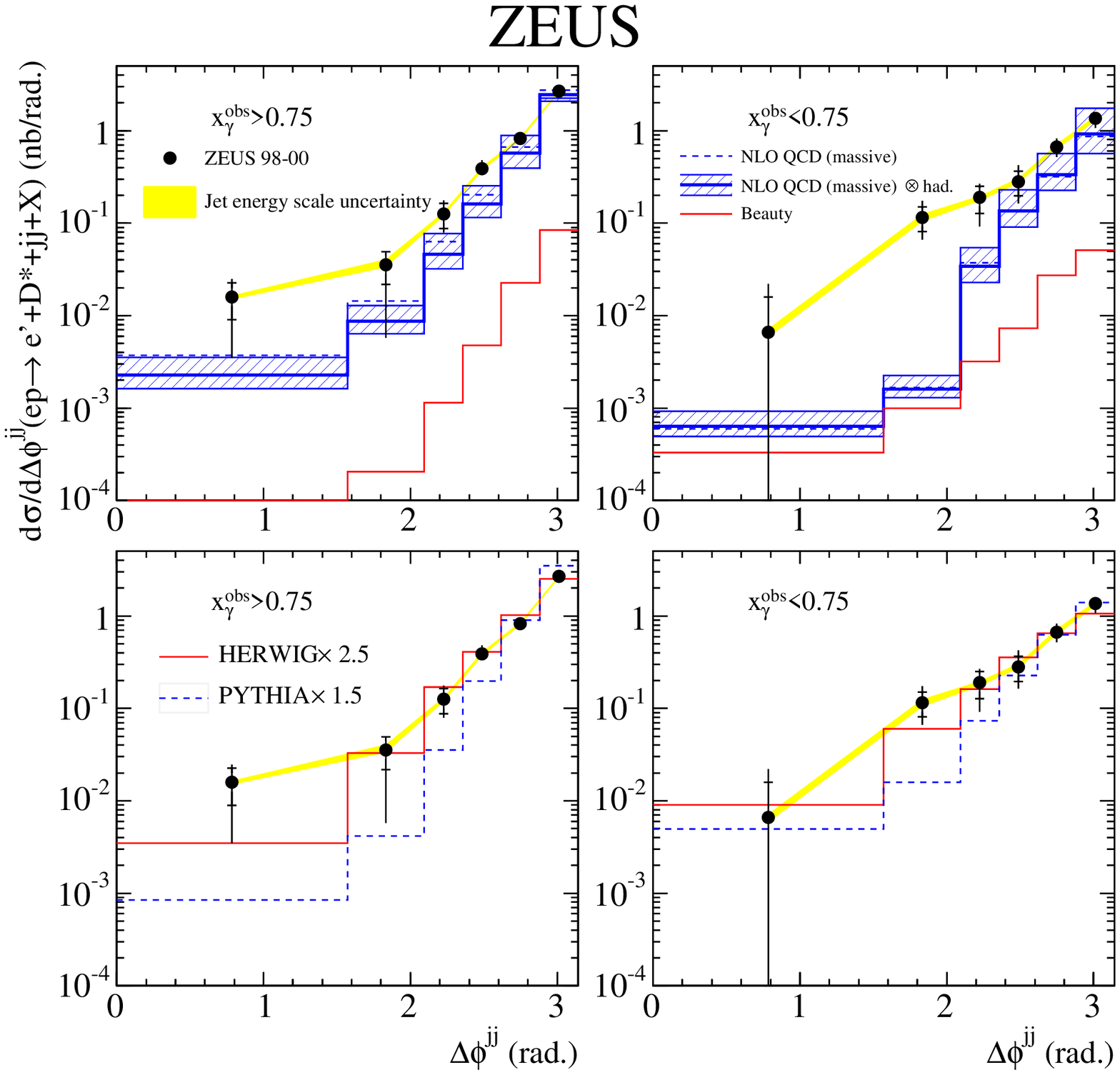}
\includegraphics[width=0.3\linewidth]{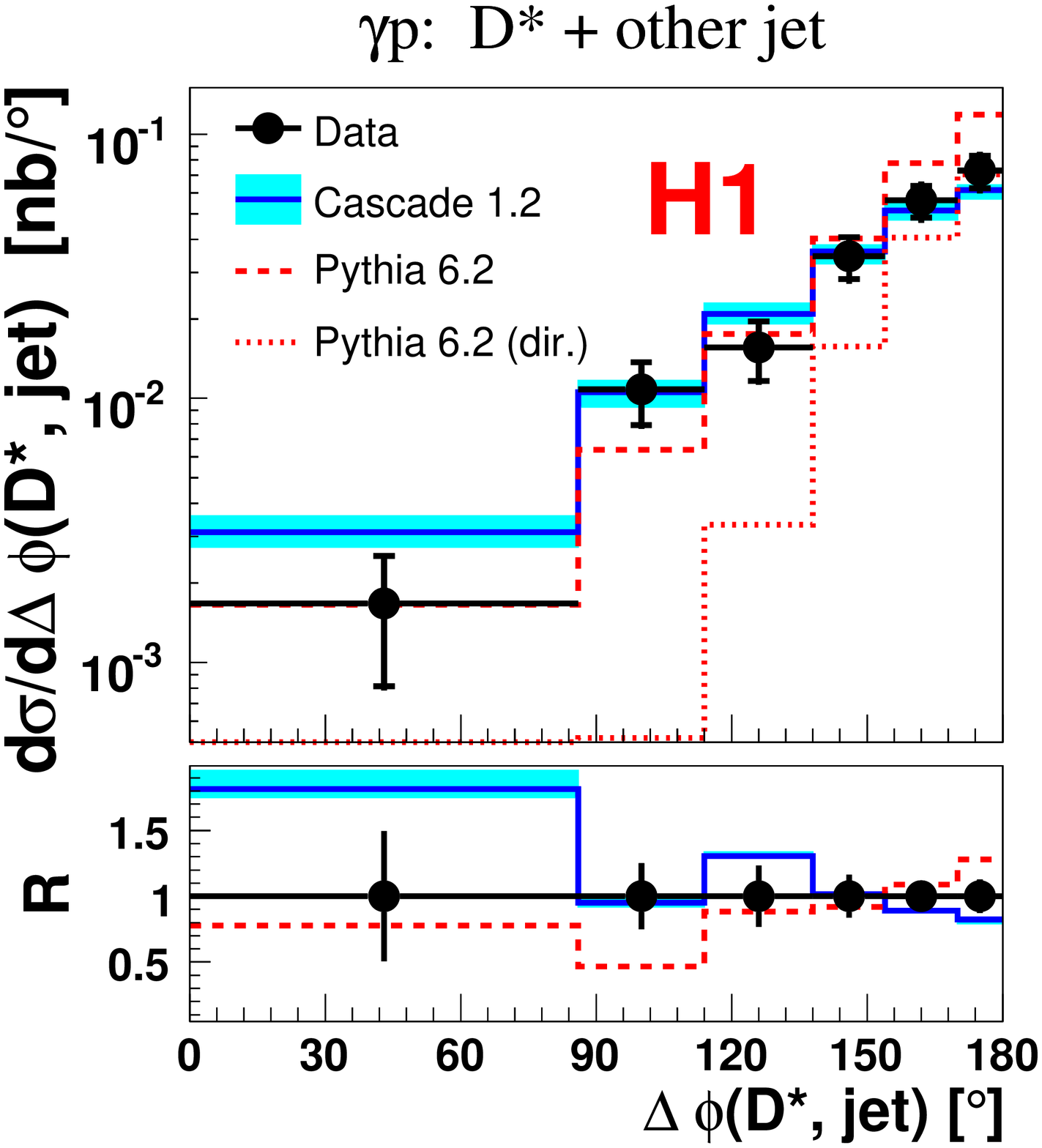}
\hspace*{-0.8cm}
\includegraphics[width=0.34\linewidth]{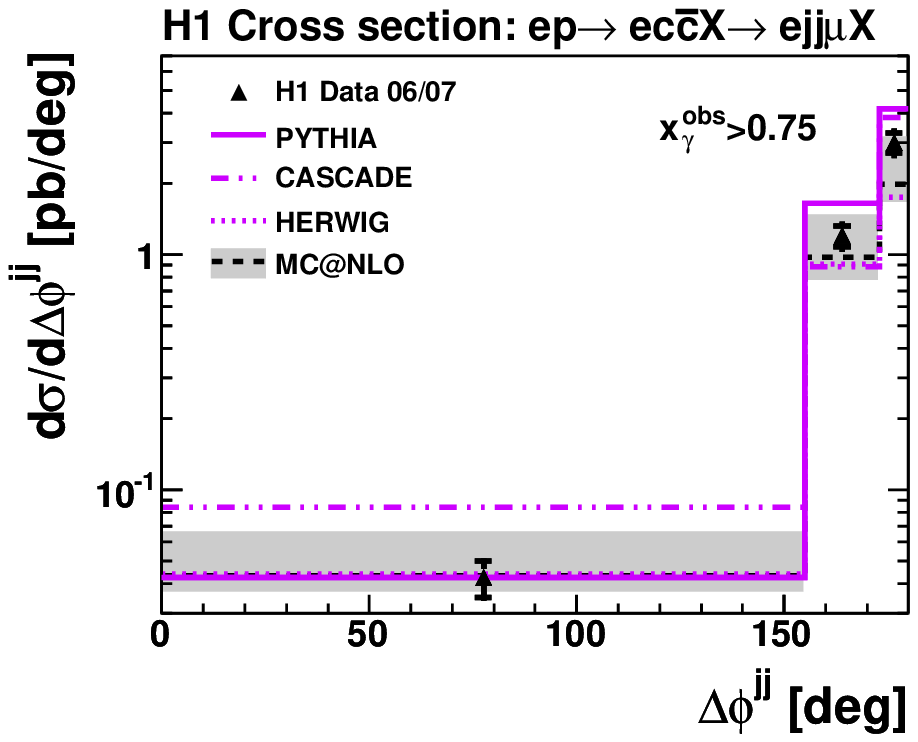}
\includegraphics[width=0.34\linewidth]{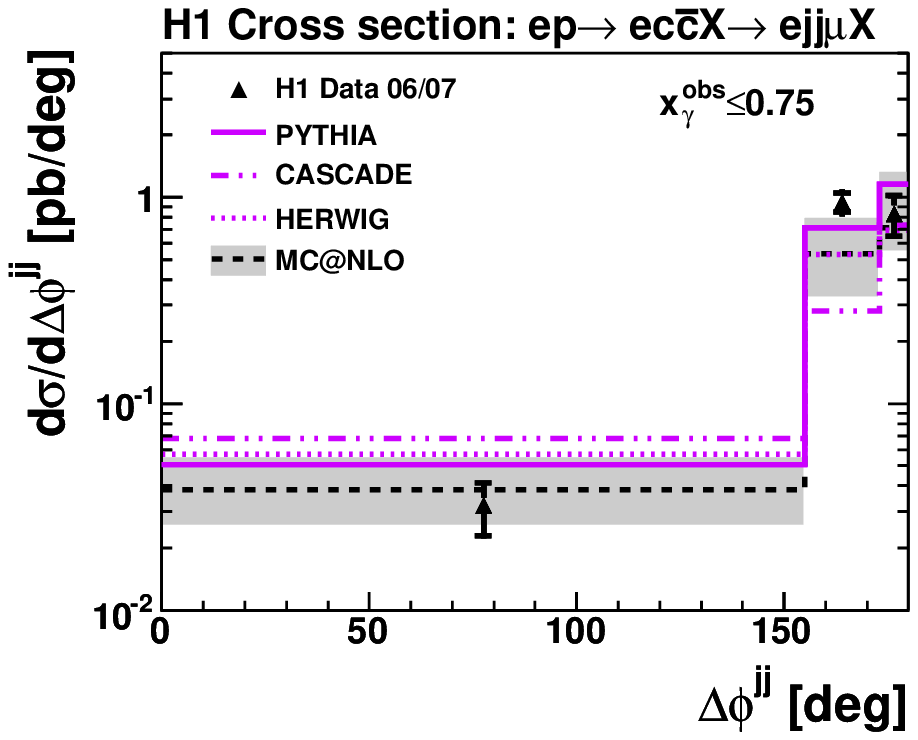}
\hspace*{6cm} \\
% place last plot in upper right corner
\vspace*{-17.1cm}\hspace*{11.7cm}
\includegraphics[width=0.3\linewidth]{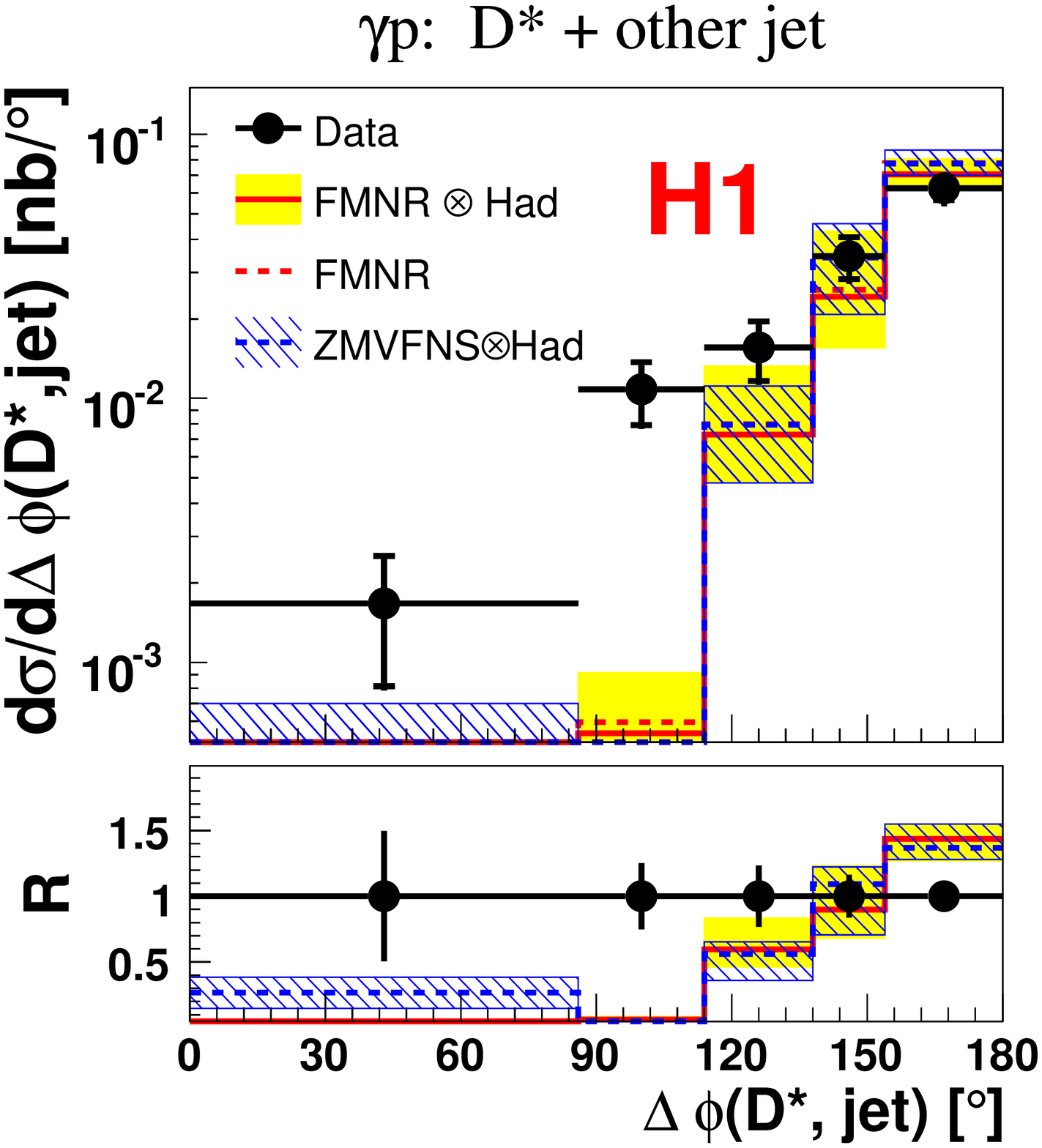}
\vspace*{11cm}
\caption{
Azimuthal differences of two outgoing hard partons in
charm events.
The upper row shows the $D^*$-tagged jet-jet azimuthal difference from 
ZEUS \cite{ZEUScPHP2005} separately for the high (left) and low (middle) 
$\xgobs$ region and the $D^*$-other~jet azimuthal difference from 
H1 \cite{H1cPHP2007} without cut on $\xgobs$ (right). 
Also shown are massive (NLO/FMNR)\cite{FMNR} and massless 
(ZMVFNS)\cite{Heinrich:2004kj} NLO predictions. 
The middle row shows the same data compared to HERWIG\cite{HERWIG}, 
PYTHIA\cite{PYTHIA} and CASCADE\cite{CASCADE}
LO+PS MC predictions.
The bottom row shows the muon-tagged jet-jet azimuthal difference from 
H1 \cite{H1cbPHP2012}
compared to these same MCs and to the NLO+PS MC@NLO \cite{MCNLO} prediction.
}
\label{fig:r10}
\end{figure}

In a ZEUS analysis~\cite{ZEUScPHP2005}
using events with a $D^*$ and two jets (entry 9 in Table \ref{tab:r1}),
and in the H1 measurement~\cite{H1cPHP2007} 
with a $D^*$ and a non-associated jet (entry 7),
the azimuthal correlation of the two hard partons is 
investigated.
In the leading-order picture of direct-photon interactions 
(Fig.~\ref{fig:fey1}a), the two charm quarks are produced 
back-to-back in the azimuthal plane of the 
lab frame, i.e. with $\Delta \phi = 180^{\circ}$.
Smaller $\Delta \phi$ can be due to 
higher-order processes, such as gluon radiation,
or due to a non-zero transverse momentum
of the partons that enter the hard interaction,
e.g. from a flavour-excitation process in which the $c$ quark
gets a finite transverse momentum in the backwards parton showering step.
In the ZEUS analysis jets were selected with
harder transverse-momentum requirements but in a wider $\eta$
range than in the H1 analysis (cf. entries 9 and 7 in \Tab{r1}).
Figure~\ref{fig:r10} shows the differential cross sections
as a function of the azimuthal difference $\Delta \phi$ 
between the $D^*$ and the jet for the H1 analysis and between the two
jets for the ZEUS measurement.
%
%%%%%%%%%%%%%%%%%%%%%%%%%%%%%%%%%%%%%%%%%%%%%%%%%%%%%%%%%%%%%%%%%%%%%%%
%
% H1 and ZEUS Delta phi
%
%
%
%////////////////////////////////////////////////////////////
%
The H1 result is shown in the 
two rightmost plots. 
The cross sections are highest for $
\Delta \phi = 180^{\circ}$,
i.e. for the back-to-back configuration, and
drop off towards smaller angles.
NLO calculations in the massive scheme~\cite{Frixione:1995qc} 
and in the massless scheme~\cite{Heinrich:2004kj}
are compared to the data.
Both calculations drop off more 
steeply than the data towards smaller 
opening angles.
Below $\Delta \phi \approx 100^{\circ}$
the two calculations predict 
very small contributions, while 
there are still sizeable ones in the data.
A better description is obtained with
the LO+PS programs PYTHIA~\cite{PYTHIA} 
and CASCADE~\cite{CASCADE}.  
PYTHIA includes charm excitation processes in resolved-photon events,
which give the dominant contribution
for $\Delta \phi <140^{\circ}$ and provide
a reasonable data description in this region. 

The results of the ZEUS analysis~\cite{ZEUScPHP2005}
are shown in the left and central plots of Fig.~\ref{fig:r10}.
Here the azimuthal correlation is measured
separately in the 2-parton region 
$\xgobs>0.75$ and in the 3-or-more-parton
region $\xgobs<0.75$.
The data are compared 
to an NLO calculation~\cite{Frixione:1995qc}
in the massive scheme.
For the high-$\xgobs$ region the description 
is satisfactory.
However, in the low-$\xgobs$ region
the NLO calculation is clearly falling below the
data for $\Delta \phi <120^{\circ}$.
This is straightforward to understand since a 3-parton final state can not 
produce an angle between the two leading $p_T$ partons of less than 
$120^\circ$~\footnote{However, 
in the 3-parton topology one of the leading jets can 
escape outside of the kinematic region of the measurement. 
Thus, the softest jet is used instead, which leads to strongly
suppressed but non-zero charm contribution
for $\Delta \phi <120^{\circ}$}. 
Correspondingly, at least four partons are needed to populate 
this region. 
A massive NLO calculation produces at most three, so an NNLO 
calculation is needed to fill the gap.
Again, a better shape description is obtained by
PYTHIA and HERWIG, which can provide several extra partons through parton 
showering (of which flavour excitation is a part).
Thus, the conclusion is again very similar to 
the above studies with the \xgobs observable:
the NLO calculation is missing a component 
in the data, which can be effectively described by
a LO+PS calculation.
As to be expected from this explanation, in the two
lower plots the MC@NLO calculation, which 
complements the 2- and 3-parton NLO matrix elements by parton showering, 
is able to describe these data well. 

An H1 analysis using $\mu$+dijet final states \cite{H1cbPHP2012} 
(entry 14 in table \ref{tab:r1}) further supports these 
conclusions. 

\subsubsection{Study of hard-scattering angle $cos\,\theta^{*}$}

\begin{figure}[htb]
\centering
\includegraphics[width=0.5\linewidth]{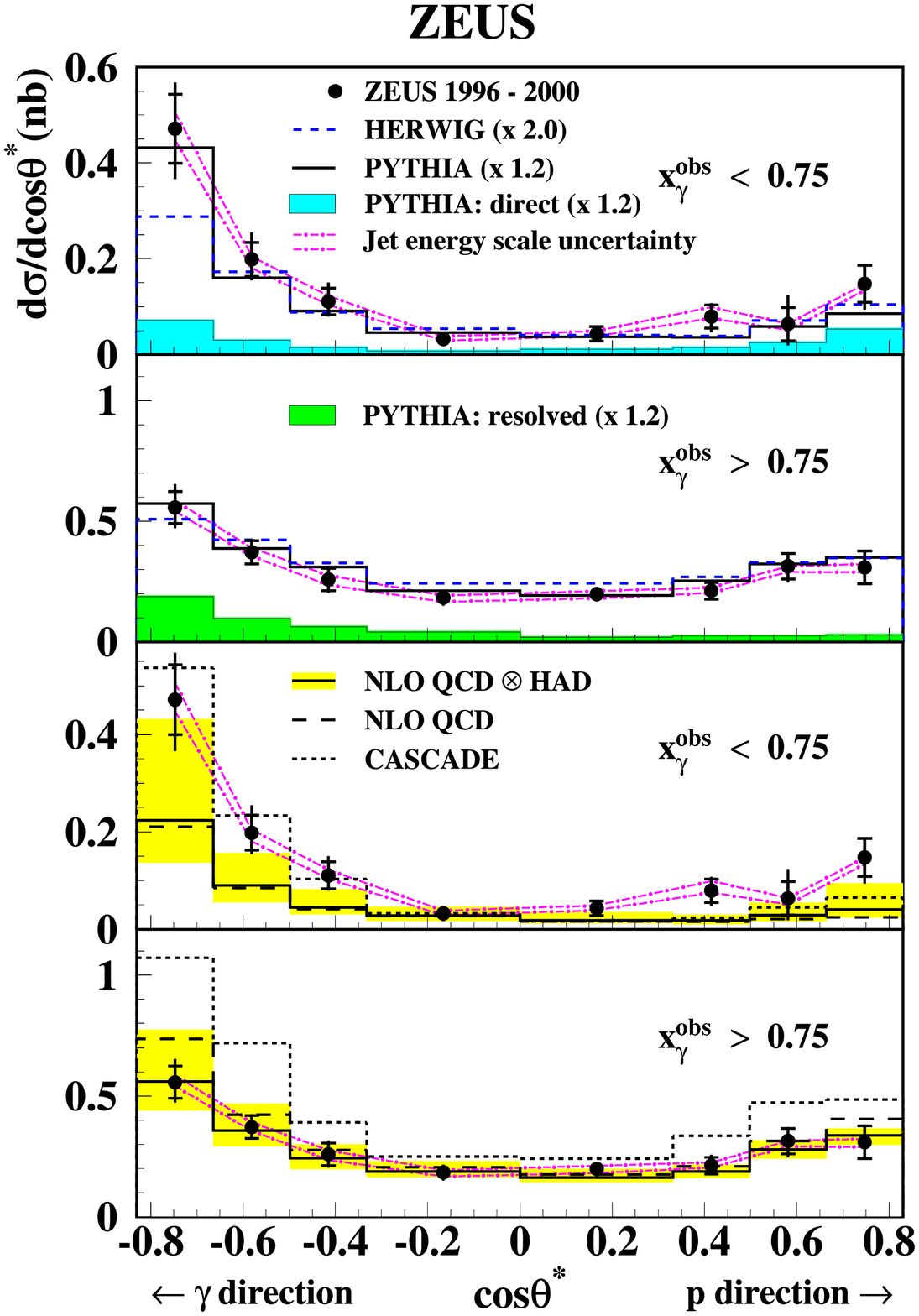}
\caption{
Differential cross sections as function of $cos\,\theta^{*}\,$ for 
dijet events with an associated $D^*$ meson~\cite{ZEUScPHP2003}.
Results are given separately for samples enriched in 
direct ($\xgobs>0.75$) and 
resolved photon events ($\xgobs<0.75$).
The data are compared in the lower two plots to NLO 
predictions~\cite{Frixione:1995qc} 
in the massive scheme
using the parameter settings and variations 
listed in table~\ref{tab:r1}.
The predictions of the CASCADE~\cite{CASCADE} 
MC are also shown.
In the upper two plots the data are compared to 
predictions using the HERWIG~\cite{HERWIG} 
and PYTHIA~\cite{PYTHIA} MCs.}
\label{fig:r8}
\end{figure}

In a dedicated 
analysis~\cite{ZEUScPHP2003}, using events with
a $D^*$ and two jets (entry 8 in Table \ref{tab:r1}), 
ZEUS has investigated 
the scattering angle $\theta^{*}$ 
of the charm quark with respect to the proton
direction in the dijet rest frame.
The charm quark is identified by the jet to which
the reconstructed $D^*$ is associated.
The $cos \,\theta^{*}\,$ distribution 
strongly reflects the type of the propagator particle exchanged
in the $2 \rightarrow 2$ hard interaction: 
\begin{itemize}
\item
For a charm quark propagator  
$cos \, \theta^{*}$  
should follow a $(1 - |cos(\theta^{*})|)^{-1}$ distribution.
The direct photon (Fig.~\ref{fig:fey1}(a)) 
and the resolved process with a gluon from 
the photon structure (Fig.~\ref{fig:fey1}(b))
belong to this class of processes and also one of the 
charm excitation diagrams (Fig.~\ref{fig:fey1}(c)).
\item
For a gluon propagator
$cos \, \theta^{*}$  
should follow a $(1 - |cos(\theta^{*})|)^{-2}$ distribution,
i.e. a much steeper rise for $|cos\, \theta^{*}| \rightarrow 1$.
For leading-order processes only 
the charm-excitation mechanism provides such a contribution
(Fig.~\ref{fig:fey1}(d)).
\end{itemize}
The main idea of the analysis is to 
look for such effects directly in the data.
Special cuts are applied in order 
to ensure a flat acceptance for the 
$cos\,\theta^{*}$ distribution over a wide range,
extending to as large values of $|cos\,\theta^{*}|$
as possible. 
The invariant mass of the two jets is required to be
above 18 GeV.
The average pseudorapidity of the two jets, defined
as $ \frac{\eta^{jet1} + \eta^{jet2}}{2}$ is required
to be smaller than 0.7.
Note that these cuts yield a much smaller (but still sizeable)
contribution at $\xgobs<0.75$ than the 
one shown in Fig.~\ref{fig:r5}, mainly because they implicitly
restrict the two jets to the pseudorapidity region
$\eta^{jet}<1.9$. As intended, this analysis mainly probes 2-parton
final states, plus a potential 3rd parton from the photon remnant.
In Fig.~\ref{fig:r8} the 
differential cross sections are shown as 
a function of $cos \theta^{*}$, separately
for $\xgobs<0.75$ and $\xgobs>0.75$.
%
%%%%%%%%%%%%%%%%%%%%%%%%%%%%%%%%%%%%%%%%%%%%%%%%%%%%%%%%%%%%%%%%%%%%%%%%%%%%%%%%%%%%%%%%%%%%%%
%
%---- Costheta*
%
%
%
%//////////////////////////////////////////////////////////////////////////////////
%
In the lower half of Fig.~\ref{fig:r8} the
data are compared to massive scheme 
NLO predictions~\cite{Frixione:1995qc}.
For the $\xgobs>0.75$ region the NLO calculation provides
a good description of the data over the whole 
range of $cos\, \theta^{*}$, with reasonably small uncertainty.
The relatively shallow $cos\, \theta^{*}$ dependence is consistent with the 
expectation that this region is dominated by the boson-gluon-fusion graph, 
where the propagator particle is a charm quark, and for which the prediction 
is stabilised at NLO by 1-loop virtual corrections.
In the $\xgobs<0.75$ region 
a much stronger rise is visible 
towards more negative $cos\, \theta^{*}$ 
values, and the central region is more strongly depleted.
This can be interpreted as a direct proof
for sizeable contributions from gluon propagator
exchanges such as the charm excitation process
(Fig.~\ref{fig:fey1}(d)), which at NLO is a tree-level process
(Fig.~\ref{fig:intro4}(d,e) ).
Correspondingly, the NLO uncertainty is much larger. The upper edge of the 
uncertainty band describes the data reasonably.
The strong asymmetry can be attributed to the fact that the charm jet will
preferentially be correlated with the incoming photon direction. 

The plots in the upper half of Fig.~\ref{fig:r8}
show that the PYTHIA and HERWIG LO+PS MCs with
their large excitation contributions are able to describe
the data well everywhere. This is particularly true for PYTHIA.
For the NLO calculation this means that
contributions beyond NLO would probably further improve the
description.
On the other hand, the CASCADE MC, which attempts to describe hard 
higher-order topologies by allowing initial state partons 
to have sizeable transverse momentum, reasonably describes 
the shapes, but fails to describe the relative 
normalisation of the low- and high-$\xgobs$ regions.

\subsection{Summary}

The charm mass provides a semi-hard QCD scale which already allows the 
application of perturbative calculations to all of phase space, but which 
also competes with other, often even harder perturbative scales.
Total cross sections for charm photoproduction are reasonably described 
by such calculations.
Single-differential cross sections already provide a good handle 
to test the applicability of different QCD approximations, although the 
theoretical uncertainties are mostly much larger than the experimental ones. 
The theory predictions agree with the data up to the highest accessible
transverse momenta, showing no indications that final state resummation
corrections are needed for massive calculations.

Double-differential cross sections, in particular those including jets, 
reveal a partial failure
of the massive scheme NLO predictions for the three independent 
parton-parton kinematic 
observables $\xgobs$, $cos\, \theta^{*}$ and $\Delta \phi$,
which were studied in charm events with a $D^*$ and one
or two jets. For certain kinematic regions this can be traced back to the 
absence of 4-or-more-parton final states in the calculation. 
The partially large theoretical
uncertainties can be explained by the absence of stabilizing virtual
corrections for 3-parton final states at this order.
The NLO calculations in the massless scheme, 
where available, do mostly not provide a
better description for the observables.
The LO+PS MCs PYTHIA and HERWIG, which are often used for acceptance 
corrections, are able to describe all topologies reasonably, often 
even very well. The CASCADE MC performs somewhat less well on average.

\newpage
\section{Beauty photoproduction}
\label{sect:beautyphoto}

From the theoretical point of view, the only differences between charm 
photoproduction as discussed in the previous section and beauty photoproduction
are the beauty-quark mass and electric charge. The first suppresses the 
cross section w.r.t. charm at low values of transverse momentum, while 
the second suppresses it by about a factor 4 everywhere (Eq. \ref{eq:cfrac}).
Experimentally, the signal-to-background ratio is thus more challenging, 
and the available statistics is smaller. Together with the small branching
ratio to specific final states, this precludes any attempt to use 
fully-reconstructed beauty-hadron final states at HERA. On the other hand, the 
higher mass and longer lifetime compared to charm hadrons increases the 
tagging efficiency for inclusive tagging methods. 

Table \ref{tab:bPHP} summarises all H1 and ZEUS beauty photoproduction 
measurements. For the reasons explained above, the first such measurements
(entries 1 and 2) came several years after the first measurements of charm,
focused on beauty jet production, and were severely limited by statistics.

%%%%%%%%%%%%%%%%%%%%%%%%%%%%%%%%%%%%%%%%%%%%%%%%%%%%%%%%%%%%%%%%%%%%%%%%%%%%%%%%%%%%%%%%%%%%%%%%%%%%%%%%%%%%%%%%%%%%%%%%%%%%%%%%%%%%%%%%%%%%%

%
%
% Beauty photoproduction measurement table:
% ----------------------------------------
%
\begin{sidewaystable}[htbp]
\setlength{\tabcolsep}{1.47mm}
\renewcommand{\arraystretch}{1.1}
\begin{center}
\footnotesize
\it
\begin{tabular}{||c|l|l|c|c|c|c|l|c||l|l|c||l|l||}
\hline
\hline
%   1         2          3       4       5        6         7                       8             9          10            11
%  12          13      14     15
   No. & Analysis &   $b\,$ Tag  &  Ref. &  Exp. &  Data &  $\cal{L}$ $[pb^{-1}]$   & $Q^2\;[\mbox{GeV}^2]$ &        $y$     & Particle  & $p_T\;[\mbox{GeV}]$   &
 $\eta$     & Events &
\begin{tabular}{l}
bgfree \\
events\\
\end{tabular}
\\
%
%--- H1 first measurement of beauty at HERA 
%
\hline
1 & $\mu$ + dijets
& $\mu$ + $p_T^{rel}$   & \cite{H1bPHP1999} & H1   & 96  & $6.6$                       & $<1$  & $[0.1,0.8]$ & 
\setlength{\tabcolsep}{0mm}
\begin{tabular}{l}  
$\mu$ \\
jet1(2) \\
\end{tabular}
 & 
\setlength{\tabcolsep}{0mm}
\begin{tabular}{l}  
$>2$ \\
$>6(6)$ \\
\end{tabular}
& 
\setlength{\tabcolsep}{0mm}
\begin{tabular}{l}  
$[-0.9,1.1]$ \\
             \\
\end{tabular}
& $470 \pm 43$ & $120$ \\ 
%
%--- ZEUS electron measurement
%
\hline
2 & 
$e$ + dijets & $e$ + $p_T^{rel}$   & \cite{ZEUSbPHP2001}  & ZEUS   & 96-97  & $38.5$                       & $<1$  & $[0.2,0.8]$ & 
\setlength{\tabcolsep}{0mm}
\begin{tabular}{l}  
$e$ \\
jet1(2) \\
\end{tabular}
 & 
\setlength{\tabcolsep}{0mm}
\begin{tabular}{l}  
$>1.6$ \\
$>7(6)$ \\
\end{tabular}
& 
\setlength{\tabcolsep}{0mm}
\begin{tabular}{l}  
$[-1.1,1.1]$ \\
$[-2.4,2.4]$  \\
\end{tabular}
& $140 \pm 35$ & $16$ \\ 
\hline
%
%--- ZEUS bmujj in gammap 2003
%
3 
& 
$\mu$ + dijets   & $\mu$ + $p_T^{rel}$  & \cite{ZEUSbPHP2004} & ZEUS   & 96-00  & $110$                       & $<1$  & $[0.2,0.8]$ & 
\setlength{\tabcolsep}{0mm}
\begin{tabular}{l}  
$\mu$ \\
jet1(2) \\
\end{tabular}
 & 
\setlength{\tabcolsep}{0mm}
\begin{tabular}{l}  
$>2.5$ \\
$>7(6)$ \\
\end{tabular}
%}
& 
\setlength{\tabcolsep}{0mm}
\begin{tabular}{l}  
$[-1.6,2.3]$ \\
$[-2.5,2.5]$ \\
\end{tabular}
& $834 \pm 65$ & $165$ \\ 
%
%--- H1 bphan
%
\hline
4 & $\mu$ + dijets 
& $\mu$ + $p_T^{rel}$ + $\delta$  &  \cite{H1bPHP2005} & H1   & 99-00  & $50$                       & $<1$  & $[0.2,0.8]$ & 
\setlength{\tabcolsep}{0mm}
\begin{tabular}{l}  
$\mu$ \\
jet1(2) \\
\end{tabular}
 & 
\setlength{\tabcolsep}{0mm}
\begin{tabular}{l}  
$>2.5$ \\
$>7(6)$ \\
\end{tabular}
& 
\setlength{\tabcolsep}{0mm}
\begin{tabular}{l}  
$[-0.55,1.1]$ \\
$[-2.5,2.5]$ \\
\end{tabular}
& $1745$ & $128$ \\ 
\hline
%
%---- H1 dijets + vertex
%
5 & 
lifet.+dijets & imp. par. & \cite{H1cbPHP2006} & H1   & 99-00  & $57$                       & $<1$  & $[0.15,0.8]$ & 
\renewcommand{\tabcolsep}{-1.4mm}
\begin{tabular}{l}  
Track \\
Jet1(2) \\
\end{tabular}
 & 
\renewcommand{\tabcolsep}{0mm}
\begin{tabular}{l}  
$>0.5$ \\
$>11(8)$ \\
\end{tabular}
& 
\renewcommand{\tabcolsep}{0mm}
\begin{tabular}{l}  
$[-1.3,1.3]$ \\
$[-0.9,1.3]$ \\
\end{tabular}
& $\sim 80000$ & $78$ \\ 
\hline
%
% ZEUS e + dijets
%
6 & 
$e$ + dijets   & $e + p_T^{rel}+{\not E}_T$  & \cite{ZEUScbPHP2008} & ZEUS   & 96-00  & $120$ & $<1$  & $[0.2,0.8]$ & 
\renewcommand{\tabcolsep}{-1mm}
\begin{tabular}{l}  
$\ e$ \\
Jet1(2) \\
\end{tabular}
 & 
\renewcommand{\tabcolsep}{0mm}
\begin{tabular}{l}  
$>0.9$ \\
$>7(6)$ \\
\end{tabular}
& 
\renewcommand{\tabcolsep}{0mm}
\begin{tabular}{l}  
$[-1.5,1.5]$ \\
$[-2.5,2.5]$ \\
\end{tabular}
& $\sim 6000$ & $129$  \\ 
\hline
%
% ZEUS mu + dijets + imp. par.
%
7 & 
$\mu$+dijets  & $\mu + p_T^{rel}+\delta$  & \cite{ZEUSbPHP2009b} & ZEUS   & 05  & $126$ & $<1$  & $[0.2,0.8]$ & 
\renewcommand{\tabcolsep}{-1mm}
\begin{tabular}{l}  
$\mu$ \\
Jet1(2) \\
\end{tabular}
 & 
\renewcommand{\tabcolsep}{0mm}
\begin{tabular}{l}  
$>2.5$ \\
$>7(6)$ \\
\end{tabular}
& 
\renewcommand{\tabcolsep}{0mm}
\begin{tabular}{l}  
$[-1.6,1.3]$ \\
$[-2.5,2.5]$ \\
\end{tabular}
& $7351$ & $122$  \\ 
\hline
%
% ZEUS dijets + vtx
%
8 & 
lifet.+dijets  & sec. vtx.  & \cite{ZEUScbPHP2011} & ZEUS   & 05  & $133$ & $<1$  & $[0.2,0.8]$ & 
\renewcommand{\tabcolsep}{-1mm}
\begin{tabular}{l}  
tracks \\
Jet1(2) \\
\end{tabular}
 & 
\renewcommand{\tabcolsep}{0mm}
\begin{tabular}{l}  
$>0.5$ \\
$>7(6)$ \\
\end{tabular}
& 
\renewcommand{\tabcolsep}{0mm}
\begin{tabular}{l}  
$[-1.6,1.4]$ \\
$[-2.5,2.5]$ \\
\end{tabular}
& $\sim 70000$ & $1050$  \\ 
\hline
%
% H1 dijets + mu
%
9 & 
$\mu$ + dijets  & $\mu$+imp.par.  & \cite{H1cbPHP2012} & H1   & 06-07  & $179$ & $<2.5$  & $[0.2,0.8]$ & 
\renewcommand{\tabcolsep}{-1mm}
\begin{tabular}{l}  
$\ \mu$ \\
Jet1(2) \\
\end{tabular}
 & 
\renewcommand{\tabcolsep}{0mm}
\begin{tabular}{l}  
$>2.5$ \\
$>7(6)$ \\
\end{tabular}
& 
\renewcommand{\tabcolsep}{0mm}
\begin{tabular}{l}  
$[-1.3,1.5]$ \\
$[-1.5,2.5]$ \\
\end{tabular}
& $6807$ & $425$  \\ 
\hline
%
%--- H1 D* + muon
%
\hline
10 &
$D^*$ + $\mu$    &  
$K\pi\pi_s$  + $\mu$ 
&  \cite{H1cbPHP2005}  & H1 & 98-00  & $89$                       & $<1$  & $[0.05,0.75]$ & 
\begin{tabular}{l}  
$D^*$ \\
$\mu$ \\
\end{tabular}
 &
\begin{tabular}{l}  
$>1.5$ \\
$p>2$ \\
\end{tabular}
 & 
\begin{tabular}{l}  
$\ \ [-1.5,1.5]$ \\
$[-1.74,1.74]$ \\
\end{tabular}
& $56 \pm 17$ & $15$  \\ 
%
%--- ZEUS D* + muon
%
\hline
11 & 
$D^*$ + $\mu$ 
& 
$K\pi\pi_s$  + $\mu$ 
&  \cite{ZEUSbPHP2007} & ZEUS & 96-00  & $114$                       & $<1$   & $[0.05,0.85]$ & 
\setlength{\tabcolsep}{0mm}
\begin{tabular}{l}  
$D^*$ \\
$\mu$ \\
\end{tabular}
 & 
\setlength{\tabcolsep}{0mm}
\begin{tabular}{l}  
$>1.9$ \\
$>1.4$ \\
\end{tabular}
& 
\setlength{\tabcolsep}{0mm}
\begin{tabular}{l}  
$[-1.5,1.5]$ \\
%$[-1.75,1.3]$ \\
$[-1.8,1.3]$ \\
\end{tabular}
& $232$ & $16$  \\ 
%
%--- ZEUS Dimuon
%
\hline
12 & 
dimuon & $\mu$ + $\mu$      & \cite{ZEUSbPHP2009a} & ZEUS   & 96-00  & $114$ & all   & all  & 
$\mu$1(2) & $>1.5(0.75)$ & $[-2.2,2.5]$ 
& $4146$ & $86$ \\ 
%
%--- H1 Dielectron
%
\hline
13 & 
dielectron & $e$ + $e$      & \cite{H1bPHP2012} & H1   & 07  & $48$ & $<1$   & $[0.05,0.65]$  & 
$e$ & $>1$ & $[-1.0,1.74]$ 
& $\sim 1500$ & $51$ \\ 
\hline
\end{tabular}
%\label{tab:r1}
\end{center}
\vspace{-4mm}
\caption{
\footnotesize
{\bfseries Beauty photoproduction cross-section measurements} at HERA.
Information is given for each analysis on the
beauty tagging method, the experiment,
the data taking period, integrated luminosity, $Q^2$ 
and $y$ ranges
and the cuts on transverse momenta and pseudorapidities of selected
final state particles.
The last two columns provide information on the number of
events in the analysis (number of signal events if an uncertainty is given) 
%, as calculated from
%$\,s:b = events: (\sigma(events)^2-events)$ 
and the
equivalent number of background-free events.
%, as calculated from
%$\,[events/\sigma(events)]^2.$
%
The centre-of-mass energy of all data taken up to 1997 ($6^{th}$ column)
was $300\gev$, while it was $318\rnge319\gev$ for all subsequent runs.
}
\label{tab:bPHP}
\end{sidewaystable}

\subsection{Total cross section for beauty production}

Due to their high mass, even beauty quarks at rest in the centre-of-mass 
system of the partonic interaction still produce reasonably
high-momentum muons or electrons, which can be detected, when decaying 
semileptonically. 
The forward and backward muon systems allow the detection 
of such beauty quarks even when they are strongly boosted along the beam 
direction.
Furthermore, the requirement of two such muons, i.e. a double tag, strongly 
reduces both the light flavour and charm backgrounds.

In a ZEUS analysis \cite{ZEUSbPHP2009a} 
these properties were used to measure the total cross section 
for beauty production in $ep$ collisions without any 
cuts, i.e. including both photoproduction and DIS, 
by pushing the measureable muon phase space to the limit 
(entry 12 of Table \ref{tab:bPHP}). After correcting for muon acceptance 
and semileptonic branching ratios the resulting
total cross section for $b\bar b$ pair production in $ep$ collisions at HERA 
for $\sqrt{s}=318$ GeV was determined to be
\begin{equation}
\label{eq:sigq}
 \sigma_{\rm tot}(ep \to b\bar b X) =
   13.9 \pm 1.5({\rm stat.}) ^{+4.0}_{-4.3}({\rm syst.}) {\rm\ nb} ,
\end{equation}
where the first uncertainty is statistical and the second systematic.
The total cross section predicted by next-to-leading-order QCD
calculations
was obtained in the massive approach by adding the predictions from
FMNR~\cite{FMNR} and HVQDIS~\cite{hvqdis}
for $Q^2$ less than or larger than 1 GeV$^2$, respectively. 
The resulting cross section for $\sqrt{s} = 318$ GeV, using the scale 
choice $\mu = \sqrt{m_b^2+p_{Tb}^2}$
\begin{equation}
\label{eq:sigqNLO}
\sigma_{\rm tot}^{\rm NLO}(ep \to b\bar b X)
   = 7.5 ^{+4.5}_{-2.1} {\rm\ nb} 
\end{equation}
is a factor 1.8 lower than the measured value, although compatible
within the large uncertainties.

Compareable measurements were obtained in reduced regions of phase space 
from $D^*$ + muon and 
dielectron final states (entries 10, 11 and 13 in table \ref{tab:bPHP}), 
and similar results were obtained for the ratio of measured to predicted
cross sections (see also corresponding entries in Fig. \ref{fig:ptb}).
Since $D^*$ mesons and semi-isolated electrons could only be measured in the 
more central rapidity range, total $b\bar b$ cross sections were not 
extracted.  

\subsection{Single-differential cross sections}

\begin{figure}[tbp]
\centering
\includegraphics[width=0.7\linewidth]{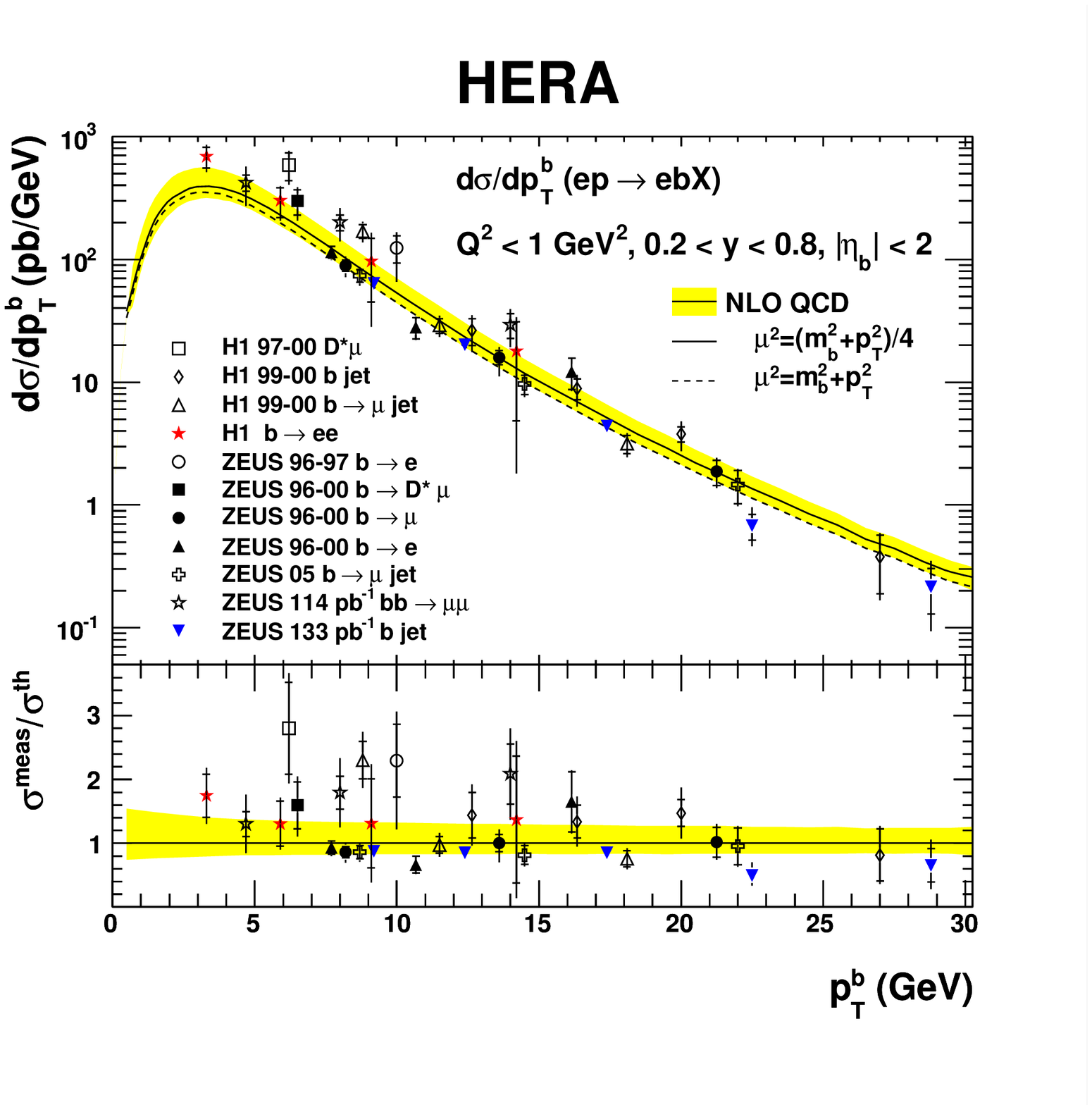}
\vspace{-1cm}
\caption{
Differential cross section as a function of the transverse momentum of 
$b$ quarks for the kinematic range indicated in the figure.
The bottom inset shows the ratio of the measured cross sections to the 
central NLO QCD prediction.
For more details see the text.}
\label{fig:ptb}
\end{figure}

In order to make them compareable with each other, in Fig. \ref{fig:ptb}
almost all available beauty-pho\-to\-pro\-duc\-tion 
cross sections\footnote{entries 
10,5,4,13,2,11,3,6,7,12,8 in table \ref{tab:bPHP}, following the 
order in the figure legend} have been translated, using NLO massive QCD 
calculations, into
cross sections for inclusive $b$-quark production as a function of 
$p_{Tb}$ in the kinematic range $Q^2 <1$ GeV$^2$, $0.2<y<0.8$ and 
beauty pseudorapidity $|\eta_b| < 2$. 
Each data point is displayed at the 
centroid of the $p_T$ distribution of the $b$ quarks entering the measurement
bin of the respective analysis, which is mostly a bin in $b$-jet $E_T$, or, 
where not available, a bin in muon or electron $p_T$. 
%
%%%%%%%%%%%%%%%%%%%%%%%%%%%%%%%%%%%%%%%%%%%%%%%%%%%%%%%%%%%%%%%%%%%%%%%%%%%%%%%%%%%%%%%%%%%%%%
%
%---- ptb plot
%
%
%//////////////////////////////////////////////////////////////////////////////////
The $b$-quark $p_T$ rather than
the $b$-jet $p_T$ has been chosen here because the measurements extend down 
to very low $p_T$ at which jets can not be usefully defined any more.  
Two massive NLO \cite{FMNR} theory predictions are given: one with scale
choice $\mu = \sqrt{m_b^2 + p_{Tb}^2}$ (dashed), and one with scale choice
$\mu = \sqrt{m_b^2 + p_{Tb}^2}/2$. The full theory uncertainty band 
is shown for the latter 
(for the scale choice see also Section \ref{sect:scale}).  
It is dominated by the scale variations (independent variation of 
renormalisation and factorisation scales by factor 2) and by the variation
of the pole mass ($m_b = 4.75 \pm 0.25$ GeV).
Where not provided directly in the original publications, the data points 
were obtained using the data/theory ratio of the respective original 
measurements and rescaling them to the theory prediction in Fig. \ref{fig:ptb},
properly accounting for differences in the respective theory calculation 
settings.

Within the large uncertainties, reasonable agreement between theory and data 
is observed over the complete $p_T$ range covering 3 orders of magnitude in 
the cross section. In particular, as in the charm case, there is no 
indication for a failure of the predictions at large $p_T$.
There might be a trend that on average, the measurements of the double 
tagging analyses ($D^*\mu$, $ee$ and $\mu\mu$), which were already briefly 
discussed in the total cross-section subscetion, tend to lie a bit above 
the other measurements which typically require dijet final states. 
The effect is not very significant, but if taken serious, might indicate 
that the contribution of $b$ quarks not associated to jets might be
underestimated by the theory. Unfortunately, currently no measurement is 
available which directly tests this hypothesis by considering both 
topologies in a single analysis framework. 

All available beauty photoproduction results are represented in this plot, 
except the results of the 
very first H1 analysis \cite{H1bPHP1999}
(entry 1 in \Tab{bPHP}), which has been declared superseded by 
a more recent analysis \cite{H1bPHP2005}, and the results of
one of the latest H1 analyses \cite{H1cbPHP2012} (entry 9), for which 
no comparison to pure NLO predictions was provided. 

Double-differential cross sections have not been measured so far. In the future
they could best be extracted using the inclusive vertexing approach
\cite{ZEUScbPHP2011}, which offers the best effective statistics
(entry 8 in \Tab{bPHP}).  

%%%%%%%%%%%%%%%%%%%%%%%%%%%%%%%%%%%%%%%%%%%%%%%%%%%%%%%%%%%%%%%%%%%%%%%%%%%%%%%%%%%%%%%%%%%%%%%%%%%%%%%%%%%%%%%%%%%%%%%%
%
%---- ZEUS dimuon paper
%
%
\begin{figure}[htbp]
\centering
\vspace{.6cm}
\hspace{.6cm}
\includegraphics[width=0.27\linewidth,bb=68 418 511 712]{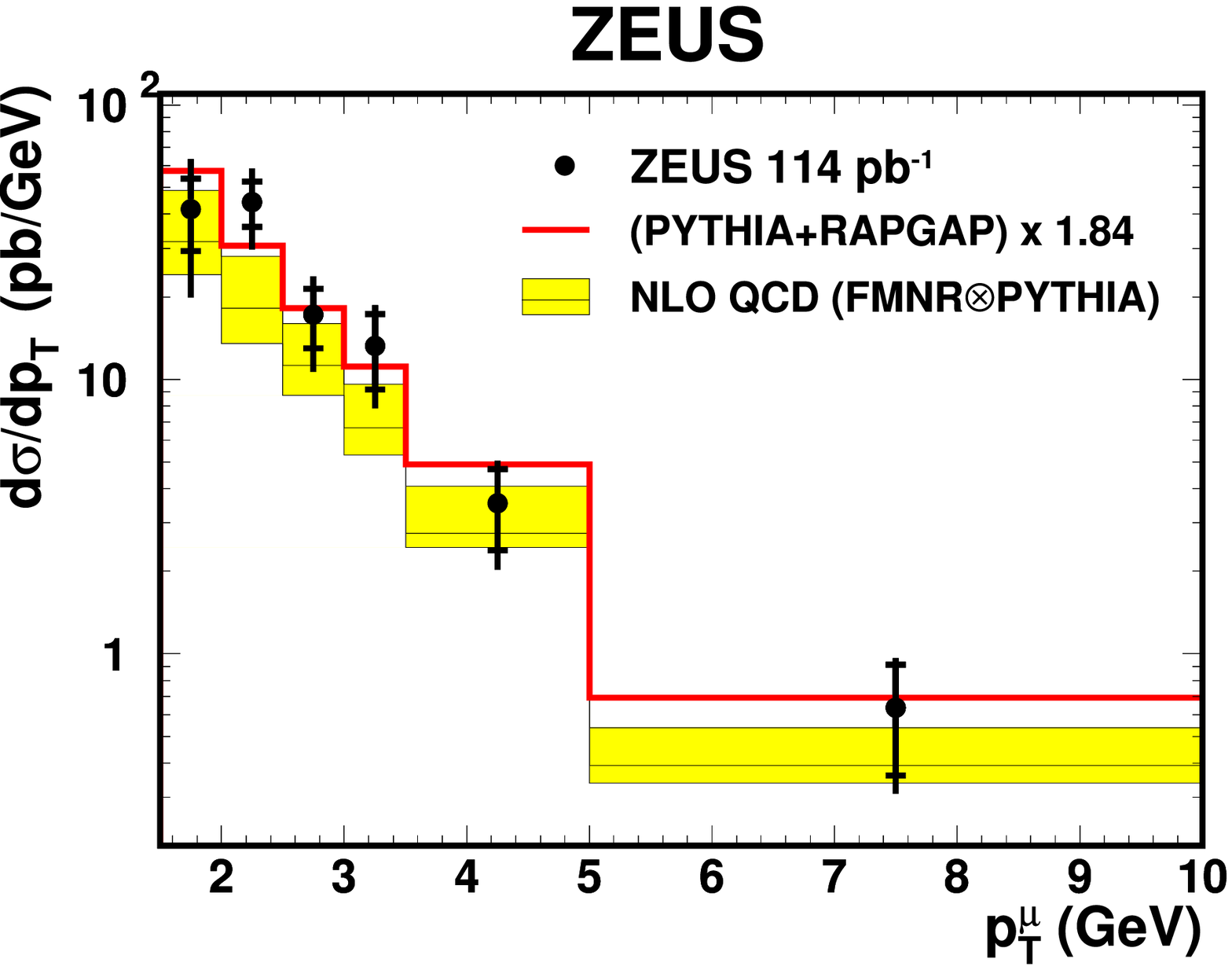}
\hspace{1cm}
\includegraphics[width=0.27\linewidth,bb=68 418 511 712]{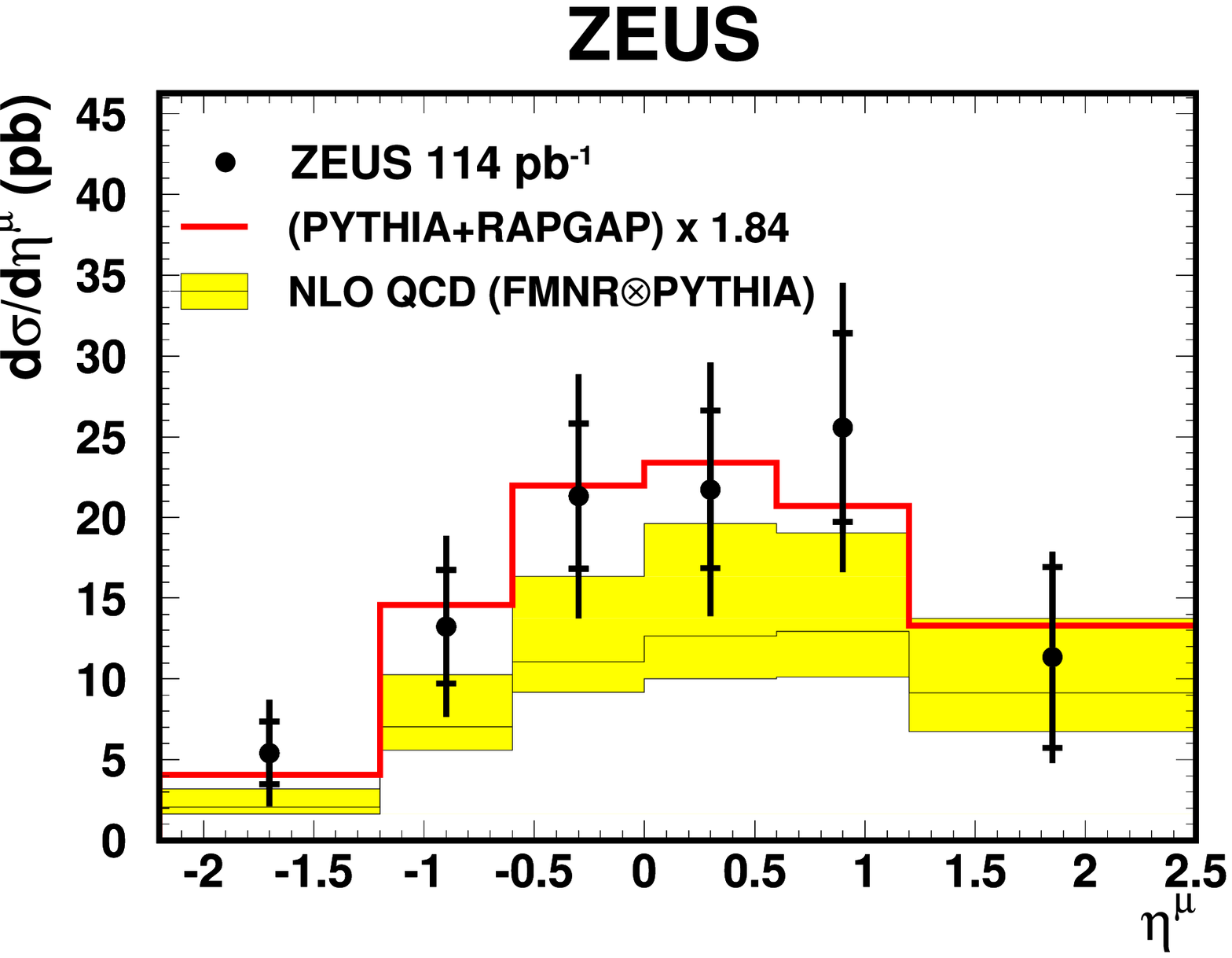} 
\hspace{1cm}
\includegraphics[width=0.27\linewidth,bb=68 418 511 712]{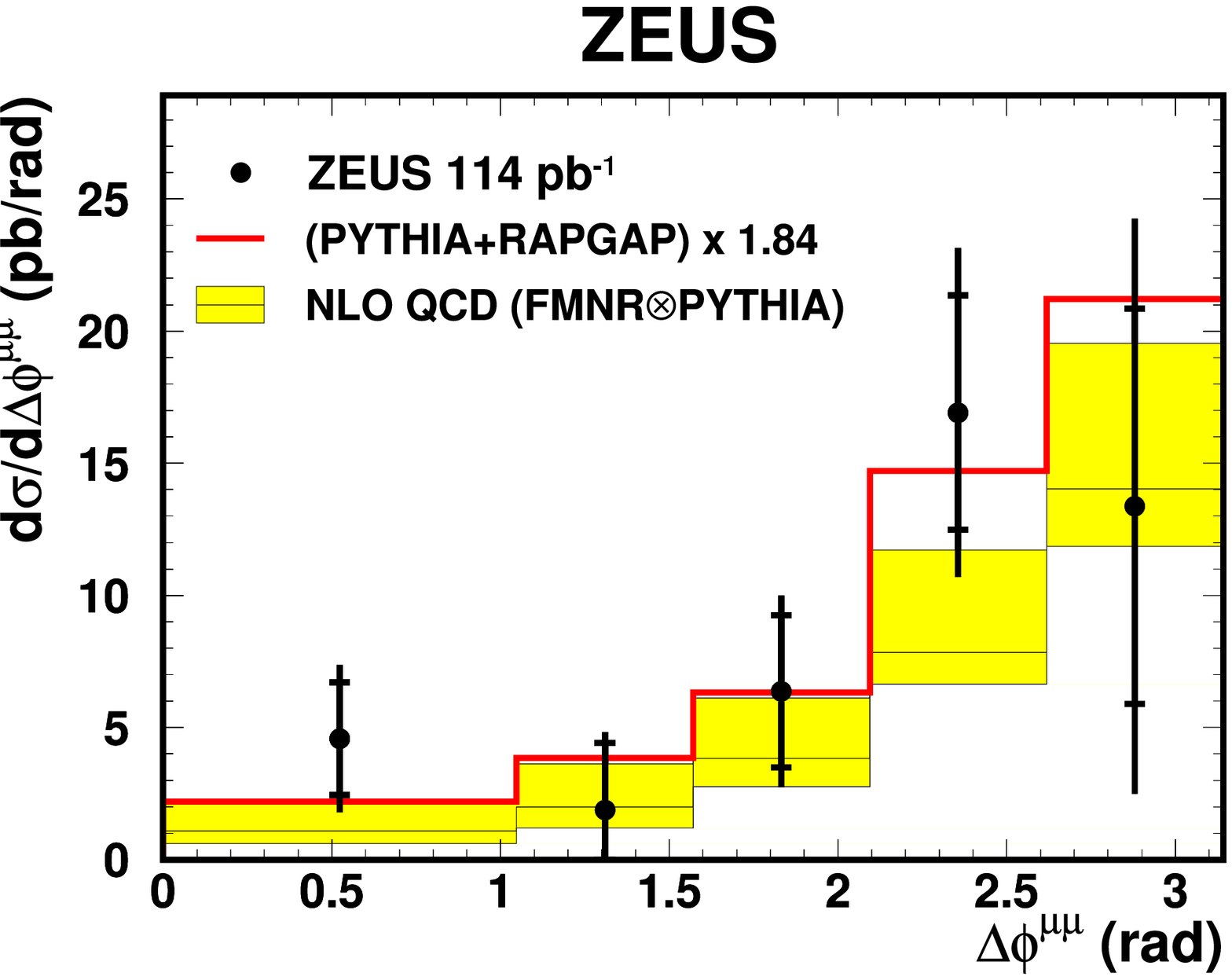} 
\vspace{.6cm}
\caption{
Single-differential cross sections for muons from $b\bar b$ decays to dimuons, 
as a function of the muon transverse momentum (left), 
pseudorapidity (center),
and dimuon azimuthal angle difference (right), from ZEUS~\cite{ZEUSbPHP2009a}.
The measurements are compared to massive NLO predictions with the same 
settings as the band in Fig. \ref{fig:ptb}, and to the PYTHIA \cite{PYTHIA}
prediction scaled to the data.
}
\label{fig:dimu}
\end{figure}
%
%////////////////////////////////////////////////////////////////////////////////////////////////////////
%

\subsection{Measurements of $b\bar b$ and jet-jet correlations}

Several results give insight into correlations between two final state partons
in $b\bar b$ events. The ZEUS analysis of dimuon final states 
(Fig. \ref{fig:dimu}, entry 11 in \Tab{bPHP}) studies the azimuthal 
angle difference between muons 
originating from different $b$ quarks, in addition to single-differential 
distributions. Both the massive NLO predictions and the PYTHIA MC predictions 
used for acceptance correction show reasonable agreement with 
the data, in particular in shape.

\begin{figure}[htbp]
\centering
\includegraphics[width=0.32\linewidth]{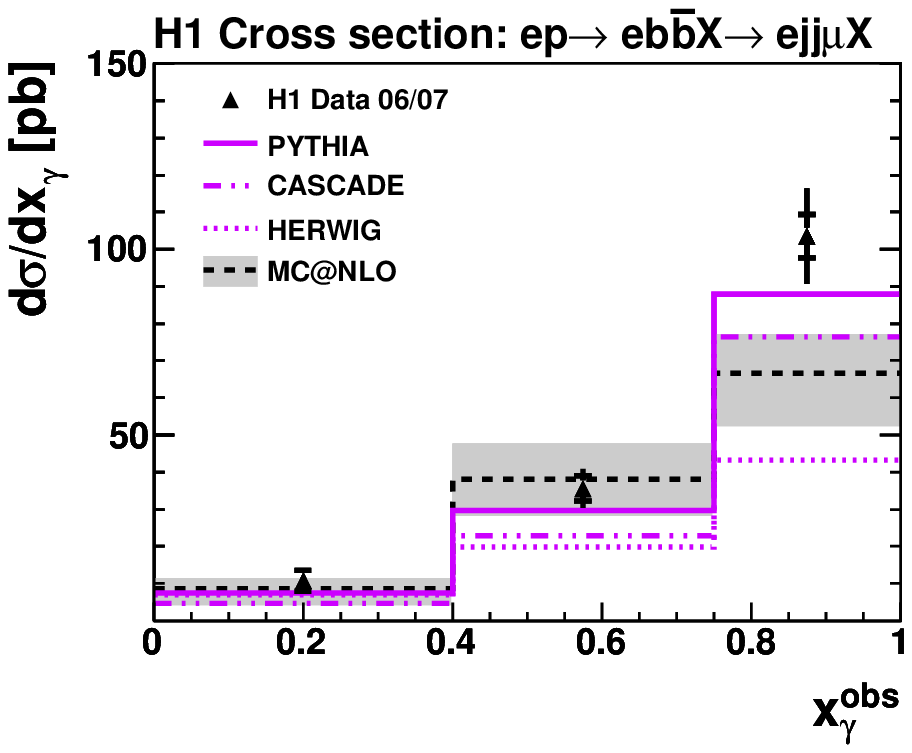}
\includegraphics[width=0.32\linewidth]{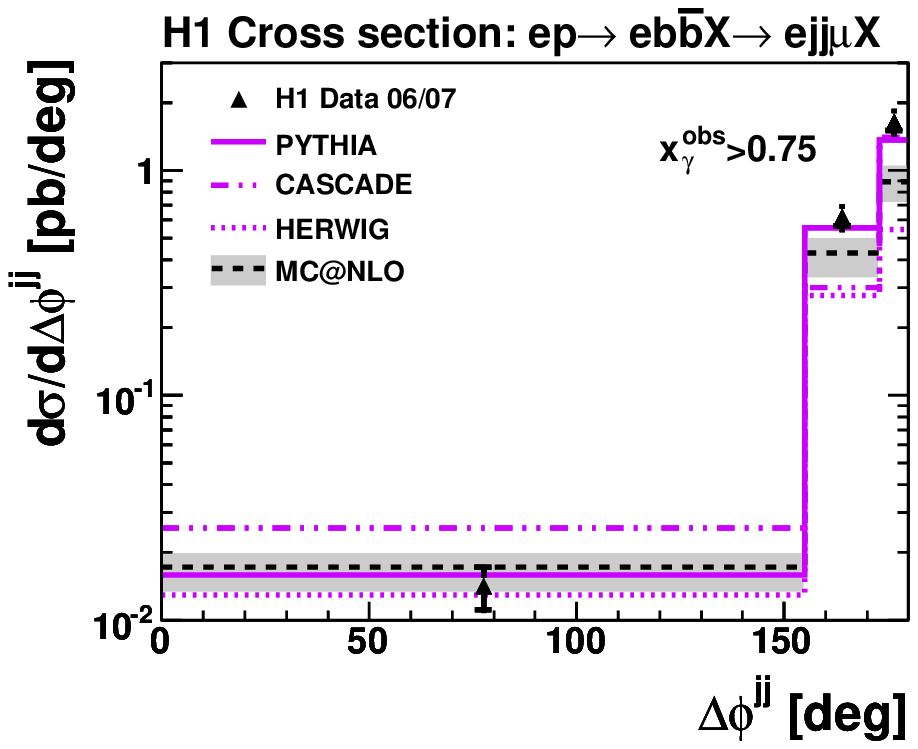} 
\includegraphics[width=0.32\linewidth]{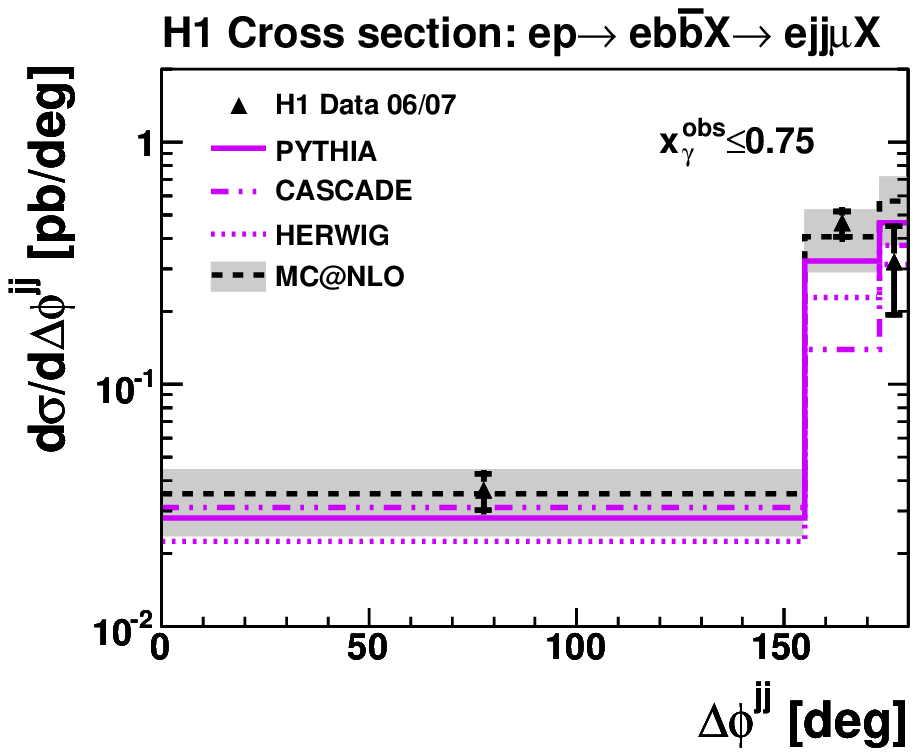} 
\caption{
Beauty cross sections as a function of $\xgobs$ (left) and as a function of
the jet-jet azimuthal angle difference $\Delta\phi$ for high (center) and 
low (right) values of $\xgobs$, from a recent H1 analysis~\cite{H1cbPHP2012}.
The measurements are compared to MC@NLO \cite{MCNLO} predictions, as well as 
to predictions from the PYTHIA~\cite{PYTHIA}, HERWIG~\cite{HERWIG} and 
CASCADE~\cite{CASCADE} LO+PS Monte Carlos.
}
\label{fig:mudijet}
\end{figure}

An H1 analysis of dijet final states in which one of the jets is tagged by 
a muon from a semileptonic $b$ decay (Fig. \ref{fig:mudijet}, entry 9
in \Tab{bPHP}) studies the 
$\xgobs$ and $\Delta \phi$ variables described earlier in the charm section.
%%%%%%%%%%%%%%%%%%%%%%%%%%%%%%%%%%%%%%%%%%%%%%%%%%%%%%%%%%%%%%%%%%%%%%%%%%%%%%%%%%%%%%%%%%%%%%%%%%%%%%%%%%%%%%%%%%%%%%%%
%
%---- H1 mu+dijets paper
%
%
%////////////////////////////////////////////////////////////////////////////////////////////////////////
%
The MC@NLO prediction describes the data, except in the high $\xgobs$, high 
$\Delta \phi$ bin. The agreement is thus slightly worse than in the charm case
(Fig. \ref{fig:r10}). PYTHIA agrees everywhere, HERWIG describes the shape but 
not the normalisation, and CASCADE fails for both. The latter finding is 
again in qualitative agreement with the charm result. Thus the PYTHIA or HERWIG
MCs should preferentially be used for acceptance corrections.

A similar analysis by ZEUS \cite{ZEUSbPHP2009b} (not shown) compares the data 
directly to the massive NLO predictions. Not surprisingly, these predictions 
fail in the same kinematic regions as for charm (Fig. \ref{fig:r10}), for the 
same reasons as discussed there.  

Several other analyses \cite{ZEUSbPHP2001,ZEUSbPHP2004,H1bPHP2005} also 
studied $\xgobs$, with less statistics
than but similar conclusions as for charm.

\subsection{Summary}

Due do the suppression of the cross section by mass and charge, and small 
branching ratios to exclusive final states, only inclusive or semi-inclusive 
tagging methods can be used at HERA to measure beauty production.
The reasonable acceptance for the detection of $b$ hadron decays down to
0 transverse momentum and the coverage of almost the full physically relevant 
rapidity range allowed the measurement of the total beauty production 
cross section at HERA. This cross section is higher than, but still compatible
with, NLO QCD predictions.  
Several single differential beauty photoproduction cross sections have also
been measured. The measurements from H1 and ZEUS and from different final 
states agree reasonably well with each other and with QCD predictions 
from threshold up to the 
highest accessible transverse momenta. Double differential cross sections
have not yet been measured.

\newpage

%%%%%%%%%%%%%%%%%%%%%%%%%%%%%%%%%%%%%%%%%%%%%%%%%%%%%%%%%%%%%%%%%%%%%%%%%%%%%%%%%%%%%%
%%%%%%%%%%%%%%%%% c&b DIS %%%%%%%%%%%%%%%%%%%%%%%%%%%%%%%%%%%%%%%%%%%%%%%%%%%%%%%%%%%%
%%%%%%%%%%%%%%%%%%%%%%%%%%%%%%%%%%%%%%%%%%%%%%%%%%%%%%%%%%%%%%%%%%%%%%%%%%%%%%%%%%%%%%

\section{Charm and beauty production in DIS}
\label{sect:cbDIS}

In the previous \Sectand{charmphoto}{beautyphoto}
heavy-flavour production
in $ep$ collisions with the exchange of
quasi-real photons was discussed.
The production of charm and beauty quarks was also studied in 
the deeply inelastic scattering regime, which corresponds to
photon virtualities $Q^2 \gsim 1\gev^2$.
Large photon virtuality provides an additional 
hard scale in the calculations and allows probing the
parton dynamics inside the proton with high resolution.
An overview of all measurements is given in \Tab{cDIS} and 
\Tab{bDIS} for charm and beauty production, respectively.
\begin{sidewaystable}[p]
\setlength{\tabcolsep}{1.5mm}
\renewcommand{\arraystretch}{1.1}
\begin{center}
%\scriptsize
\footnotesize
\it
\begin{tabular}{|c|l|l|l|c|c|c|l|c||l|c|c||l|c|c|}
\hline
%   1         2          3       4       5        6         7                       8             9          10            11
%  12          13      14     15
   No. & Analysis &   c-Tag  &  Ref. &  Exp. &  Data &  $\mathcal{L}\;[\pbi]$   & $Q^2\;[\Gev^2]$ &        $y$     & Particle  & $p_T\;
[\Gev]$   &
 $\eta$     & Events &
\begin{tabular}{c}
effect.\\
s:b \\
\end{tabular}
&
%\scriptsize
\begin{tabular}{l}
%$\cong\,
%$
bgfree \\
events\\
\end{tabular}
\\
%
%
%--- First H1 measurement
%
\hline
1 &
\dst incl. & $K\pi\pi_s$  & \cite{h196}   & H1 &  94     & $3$                      & $[10,100]$  & $<0.53$ & \dst   &
%$\approx>1.5$
$>1.5$
& $[-1.5,1.5]$    & $103 \pm 13$ & $1:0.7$ &
$64$   \\
\cline{2-3} \cline{10-11} \cline{13-15}
 &
\dzero incl. & $K\pi$ & & & & & & & \dzero   & $>2.0$ & & $144 \pm 19$ & $1:1.5$ &
$57$   \\
%
%--- First ZEUS measurement
%
\cline{1-15}
2 &
\dst incl. & $K\pi\pi_s$  & \cite{zeusdstar97}   & ZEUS &  94     & $3$                      & $[5,100]$  & $<0.7$ & \dst   &
$[1.3,9.0]$
& $[-1.5,1.5]$    & $122 \pm 17$ & $1:1.4$ & $52$   \\
%
%--- H1 gluon density
%
\hline
3 &
\dst incl. & $K\pi\pi_s$  & \cite{H1gluon}  & H1 &  95-96  & $10$                     & $[2,100]$  & $[0.05,0.7]$ & \dst   & $[1.5,15]$ & $[-1.5,1.5]$    & $583 \pm 35$ & $1:1.1$ & $278$   \\
%
%---  ZEUS with 96-97 data
%
\hline
4 &
\dst incl. & $K\pi\pi_s$  & \cite{zd97}  & ZEUS &  96-97     & $37$                     & $[1,600]$  & $[0.02,0.7]$ & \dst   & $[1.5,15]$ & $[-1.5,1.5]$    & $2064 \pm 72$ & $1:1.5$ & $822$   \\
\cline{3-3} \cline{11-11} \cline{13-15}
 &
  & $K\pi\pi\pi\pi_s$  & & & & & & & & $[2.5,15]$ & & $1277 \pm 124$ & $1:11$ & $106$ \\
%
%---  H1 F2c with 97 data
%
\hline
5 &
\dst incl. & $K\pi\pi_s$  & \cite{h1f2c}  & H1 &  97     & $18$                     & $[1,100]$  & $[0.05,0.7]$ & \dst   & $>1.5$ & $[-1.5,1.5]$    & $973 \pm 40$ & $1:0.6$ & $590$   \\
%
%--- ZEUS F2c 98-00 Flagship
%
\hline
6 &
\dst incl. & $K\pi\pi_s$  & \cite{zd00}  & ZEUS &  98-00  & $82$                     & $[1.5,1000]$  & $[0.02,0.7]$ & \dst   & $[1.5,15]$ & $[-1.5,1.5]$  & $5545 \pm 129$ & $1:2$ & $1850$   \\
%
%---  H1 D mesons
%
\hline
7 &
$D$ incl. & $D$ mes. + $S$  & \cite{h1dmesons}  & H1 &  99-00     & $48$                     & $[2,100]$  & $[0.05,0.7]$ & $D$ mesons  & $>2.5$ & $[-1.5,1.5]$    & n.a. & n.a. & $263$   \\
%
%--- H1 D* + dijets
%
\hline
8 &
\dst incl. & $K\pi\pi_s$  & \cite{h1dstar_hera1}  & H1 &  99-00  & $47$                     & $[2,100]$  & $[0.04,0.7]$ & \dst   & $[1.5,15]$ & $[-1.5,1.5]$  & $2604 \pm 77$ & $1:1.3$ & $1140$   \\
\cline{2-2} \cline{10-12} \cline{13-15}
&+ dijet &  & & &  & & & & Jet1(2)   & $>4(3)$ & $[-1,2.5]$  & $668 \pm 49$ & $1:2.5$ & $186$   \\
%
%--- ZEUS D* vs Q2 incl. BPC
%
%
\hline
9 &
\dst incl. & $K\pi\pi_s$  & \cite{zdbpc}  & ZEUS &  98-00  & $82$                     & $[0.05,0.7]$  & $[0.02,0.85]$ & \dst   & $[1.5,9]$ & $[-1.5,1.5]$    & $253 \pm 25$ & $1:1.5$ & $100$   \\
%
%---  ZEUS D mesons
%
\hline
10 &
$D$ incl. & $D$ mes. & \cite{zeusdmesons}  & ZEUS &  98-00     & $82$                     & $[1.5,1000]$  & $[0.02,0.7]$ & $D$ mesons  & $>3$ & $[-1.6,1.6]$    & n.a. & n.a. & $1100$   \\
%
%--- ZEUS D+, Lambda_c at pT~0 in HERA I1
%
\hline
11 &
\dplusp incl. & $K\pi\pi$  & \cite{zeusdpluslambda}   & ZEUS &  96-00 & $120$ & $[1.5,1000]$  & $[0.02,0.7]$ & \dplusp  & $[0,10]$ & $[-1.6,1.6]$    & $691 \pm 107$ & $1:16$ & $42$   \\
\cline{2-3} \cline{10-10} \cline{13-15}
 & $\Lambda_c$ incl. & $pK^0_S$ & & & & & & & $\Lambda_c$   & & & $79 \pm 25$ & $1:7$ & $10$   \\
\cline{3-3} \cline{13-15}
 & & $\Lambda\pi^+$ & & & & & & & & & & $84 \pm 34$ & $1:13$ & $6$   \\
%
%---  H1 lifetime tag HERAI
%
\hline
12 &
incl. lifet. &  imp. par.    & \cite{h1vertex05} & H1   & 99-00  & $57$                       & $>150$     & $[0.1,0.7]$  & Track  & $>0.5$ & %$[-1.3,1.3]$ & $2000  \pm 164?$ & $1:20$ &  $150$ \\
$[-1.3,1.3]$ & $\sim 2300$ & $1:22$ &  $100$ \\
%recalculated N from histos
\hline
13 &
incl. lifet. &  imp. par.    & \cite{h1ltt_hera1} & H1   & 99-00  & $57$                       & $[6,120]$  & $[0.07,0.7]$ & Track  & $>0.5$ & %$[-1.3,1.3]$ & $18000 \pm 562$ & $1:18$ & $1024$ \\
$[-1.3,1.3]$ & $\sim 50000$ & $1:48$ & $1024$ \\
%recalculated N from histos
%
%---  ZEUS D+, D0 mesons
%
\hline
14 &
\dzero incl. & $K\pi$ + $S$ & \cite{zd0dp}  & ZEUS &  05     & $134$                     & $[5,1000]$  & $[0.02,0.7]$ & \dzero  & $[1.5,15]$ & $[-1.6,1.6]$    & $8274 \pm 352$ & $1:14$ & $550$   \\
%
%---  ZEUS mu+jet
%
\hline
15 &
$\mu$ + jet & \begin{tabular}{l} 
$\mu$ + \ptrel + \\ 
$\delta + {\not E}_T$ \\ 
\end{tabular} & 
\cite{zmu}  & ZEUS &  05     & $126$                     & $>20$  & $[0.01,0.7]$ & 
\begin{tabular}{l} 
$\mu$ \\ 
%Jet \\ 
\end{tabular} &
\begin{tabular}{l} 
$>1.5$ \\ 
%$>2.5$ \\ 
\end{tabular} &
\begin{tabular}{l} 
$[-1.6,2.3]$ \\ 
%$>2.5$ \\ 
\end{tabular} &
$\sim 5100$ & $1:20$ & $250$   \\
%total number of selected events given, fc=0.456+-0.029, fb=0.122+-0.013
%
%--- H1 D* high Q2 HERA II
%
\hline
16 &
\dst incl. & $K\pi\pi_s$  & \cite{h1dstarhighQ2}  & H1 &  04-07  & $351$ & $[100,1000]$  & $[0.02,0.7]$ & \dst   & $[1.5,15]$ & $[-1.5,1.5]$  & $\sim 600$ & $1:7$ & $260$  \\
%stat=6.222%
%
%--- H1 D* low Q2 HERA II
%
\hline
17 &
\dst incl. & $K\pi\pi_s$  & \cite{h1dstar_hera2}  & H1 &  04-07  & $348$ & $[5,100]$  & $[0.02,0.7]$ & \dst   & $>1.25$ & $[-1.8,1.8]$  & $24705 \pm 343$ & $1:3.8$ & $5200$  \\
%
%--- ZEUS D* HERA II
%
\hline
18 &
\dst incl. & $K\pi\pi_s$ & \cite{zeusdstar_hera2}  & ZEUS &  04-07  & $363$ & $[5,1000]$  & $[0.02,0.7]$ & \dst   & $[1.5,20]$ & $[-1.5,1.5]$  & $12893 \pm 185$ & $1:2.7$ & $4860$  \\
%
%--- ZEUS D+ HERA II
%
\hline
19 &
\dplusp incl. & $K\pi\pi$ + $S$ & \cite{zeusdplus_hera2}  & ZEUS &  04-07  & $354$ & $[5,1000]$  & $[0.02,0.7]$ & \dplusp   & $[1.5,15]$ & $[-1.6,1.6]$  & $8356 \pm 198$ & $1:3.7$ & $1800$  \\
%
%---  H1 lifetime tag HERAII
%
\hline
20 &
incl. lifet. &  $\delta$ + $S$   & \cite{h1ltt_hera2} & H1   & 06-07  & $189$                       & $[5,2000]$  & n.a. & Track  & $>0.3$ & $[-1.3,1.3]$ & $\sim 210000$ & n.a. & n.a.\\
%
%---  H1 jets with lifetime tag HERAII
%
\hline
21 &
incl. lifet. &  jet + $\delta$ + $S$   & \cite{h1cbjets} & H1   & 06-07  & $189$  & $>6$  & $[0.07,0.625]$ & Jet  & $>6$ & $[-1.0,1.5]$ & $\sim 85000$ & $1:17$ & $4800$\\
%stat_b=+-5.26%; stat_c=+-1.44%
%
%--- ZEUS sec. vtx. HERA II
%
\hline
22 &
incl. lifet. & jet + $S$ & \cite{zeusltt_hera2}  & ZEUS &  04-07  & $354$ & $[5,1000]$  & $[0.02,0.7]$ & Jet   & $>4.2$ & $[-1.6,2.2]$  &  $\sim 55000$ & $1:11$ & $4400$ \\
%k_c=0.940+-0.014; k_b=1.32+-0.05
\hline
%
%--- ZEUS D* +dijets    xgamma vs Q2
%
% \hline
% 6 &
% \dst + dijets & $K\pi\pi_s$  & \cite{zeus-eps01-495}  & ZEUS &  96-00  & 104                 & $<5\cdot 10^{3}$  & $[0.25,0.65]$ &
% \setlength{\tabcolsep}{0mm}
% \begin{tabular}{l}
% \dst \\
% Jets1(2) \\
% \end{tabular}
%  &
% \setlength{\tabcolsep}{0mm}
% \begin{tabular}{c}
% $>3.$ \\
% $>7.5(6.5)$ \\
% \end{tabular}
% &
% \setlength{\tabcolsep}{0mm}
% \begin{tabular}{c}
% $[-1.5,1.5]$ \\
% $[-2.4,2.4]$ \\
% \end{tabular}
% & $2600 \pm 70$ & $1:0.9$ & 1380   \\
% \hline
\end{tabular}
\end{center}
\caption{
{\bfseries Charm DIS measurements} at HERA.
Information is given for each analysis on the
charm tagging method, the experiment,
% (H stands for H1, Z for ZEUS),
the data taking period, integrated luminosity, $Q^2$ 
and $y$ ranges
and the cuts on transverse momenta and pseudorapidities of selected
final state particles.
The last three columns provide information on the number of
tagged charm events, the effective signal-to-background ratio
%, as calculated from
%$\,s:b = events: (\sigma(events)^2-events)$ 
and the
equivalent number of background-free events.
%, as calculated from
%$\,[events/\sigma(events)]^2.$
%
The centre-of-mass energy of all data taken up to 1997 ($6^{th}$ column)
was $300\gev$, while it was $318 \rnge 319\gev$ for all subsequent runs.
}
\label{tab:cDIS}
\end{sidewaystable}

\begin{sidewaystable}[p]
\setlength{\tabcolsep}{1.5mm}
\renewcommand{\arraystretch}{1.1}
\begin{center}
%\scriptsize
\footnotesize
\it
\begin{tabular}{|c|l|l|l|c|c|c|l|c||l|c|c||l|c|c|}
\hline
%   1         2          3       4       5        6         7                       8             9          10            11
%  12          13      14     15
   No. & Analysis &   c-Tag  &  Ref. &  Exp. &  Data &  $\mathcal{L}\;[\pbi]$   & $Q^2\;[\Gev^2]$ &        $y$     & Particle  & $p_T\;
[\Gev]$   &
 $\eta$     & Events &
\begin{tabular}{c}
effect.\\
s:b \\
\end{tabular}
&
%\scriptsize
\begin{tabular}{l}
%$\cong\,
%$
bgfree \\
events\\
\end{tabular}
\\
%
%---  ZEUS mu+jet 99-00
%
\hline
1 &
$\mu$ + jet & $\mu$ + \ptrel & 
\cite{zbmujet9900}  & ZEUS &  99-00     & $72$                     & $>2$  & $[0.05,0.7]$ & 
\begin{tabular}{l} 
$\mu$ \\ 
Jet \\ 
\end{tabular} &
\begin{tabular}{l} 
$>2$ \\ 
$E_T^{Br}>6$ \\ 
\end{tabular} &
\begin{tabular}{l} 
$[-1.6,1.3]$ \\ 
$[-2,2.5]$ \\ 
\end{tabular} &
$\sim 290$ & $1:4.5$ & $70$   \\
% fb= 30.2+-4.1 %, sigma= 40.9+-5.7, Ntot = 941
%
%---  H1 mu+jet 99-00
%
\hline
2 &
$\mu$ + jet & $\mu$ + \ptrel & 
\cite{H1bPHP2005}  & H1 &  99-00     & $50$                     & $[2,100]$  & $[0.1,0.7]$ & 
\begin{tabular}{l} 
$\mu$ \\ 
Jet \\ 
\end{tabular} &
\begin{tabular}{l} 
$>2.5$ \\ 
$p_T^{Br}>6$ \\ 
\end{tabular} &
\begin{tabular}{l} 
$[-0.75,1.15]$ \\ 
$[-2.5,2.5]$ \\ 
\end{tabular} &
$\sim 230$ & $1:2.5$ & $64$   \\
%sigma = 16.3 +- 2.0, Ntot = 776, 
%
%---  ZEUS mu+jet 96-00
%
\hline
3 &
$\mu$ + jet & $\mu$ + \ptrel & 
\cite{zbmujet9600}  & ZEUS &  96-00     & $114$                     & $>2$  & $[0.05,0.7]$ & 
\begin{tabular}{l} 
$\mu$ \\ 
Jet \\ 
\end{tabular} &
\begin{tabular}{l} 
$>1.5$ \\ 
$>5$ \\ 
\end{tabular} &
\begin{tabular}{l} 
$>-1.6$ \\ 
$[-2,2.5]$ \\ 
\end{tabular} &
$\sim 3000$ & $1:18$ & $160$   \\
%Ntot= 19700, fb = 0.16 +- 0.01, sigma = 70.4 +- 5.6
%
%---  ZEUS mu+jet
%
\hline
4 &
$\mu$ + jet & \begin{tabular}{l} 
$\mu$ + \ptrel + \\ 
$\delta + {\not E}_T$ \\ 
\end{tabular} & 
\cite{zmu}  & ZEUS &  05     & $126$                     & $>20$  & $[0.01,0.7]$ & 
\begin{tabular}{l} 
$\mu$ \\ 
%Jet \\ 
\end{tabular} &
\begin{tabular}{l} 
$>1.5$ \\ 
%$>2.5$ \\ 
\end{tabular} &
\begin{tabular}{l} 
$[-1.6,2.3]$ \\ 
%$>2.5$ \\ 
\end{tabular} &
$\sim 1300$ & $1:14$ & $90$   \\
%Ntot = 11126, fb = 0.122 +- 0.013
%
%---  ZEUS e+jet HERA II
%
\hline
5 &
$e$ + jet & \begin{tabular}{l} 
$e$ + \ptrel + \\ 
$\delta + {\not E}_T$ \\ 
\end{tabular} & 
\cite{zbe_hera2}  & ZEUS &  04-07     & $363$                     & $>10$  & $[0.05,0.7]$ & 
$e$ & $[0.9,8]$ & $[-1.5,1.5]$ & 
$\sim 2700$ & $1:15$ & $170$   \\
%k_b = 1.32 +- 0.11, sigma = 71.8 +- 5.5
%
%---  ZEUS D* mu
%
\hline
6 &
\dst + $\mu$ &  $K\pi\pi_s + \mu$    & \cite{ZEUSbPHP2007} & ZEUS   & 96-00  & $114$                       & $>2$     & $[0.05,0.7]$ &
\begin{tabular}{l} 
\dst \\
$\mu$ \\ 
\end{tabular} &
\begin{tabular}{l} 
$>1.5$ \\ 
$>1.4$ \\ 
\end{tabular} &
\begin{tabular}{l} 
$[-1.5,1.5]$ \\ 
$[-1.75,1.3]$ \\ 
\end{tabular} & 
$\sim 11$ & $1:1$ &  $4$ \\
% sigma = 58 +-29
%
%---  H1 lifetime HERA I
%
\hline
7 &
incl. lifet. &  imp. par.    & \cite{h1vertex05} & H1   & 99-00  & $57$                       & $>150$     & $[0.1,0.7]$  & Track  & $>0.5$ & %$[-1.3,1.3]$ & $2000  \pm 164?$ & $1:20$ &  $150$ \\
$[-1.3,1.3]$ & $\sim 760$ & $1:16$ &  $45$ \\
%recalculated N from histos, Pb= 1.62 +- 0.24
\hline
8 &
incl. lifet. &  imp. par.    & \cite{h1ltt_hera1} & H1   & 99-00  & $57$                       & $[6,120]$  & $[0.07,0.7]$ & Track  & $>0.5$ & %$[-1.3,1.3]$ & $18000 \pm 562$ & $1:18$ & $1024$ \\
$[-1.3,1.3]$ & $\sim 5800$ & $1:60$ & $100$ \\
%recalculated N from histos, Pb = 1.55 +- 0.16
%
%---  H1 lifetime tag HERAII
%
\hline
9 &
incl. lifet. &  $\delta$ + $S$   & \cite{h1ltt_hera2} & H1   & 06-07  & $189$                       & $[5,2000]$  & n.a. & Track  & $>0.3$ & $[-1.3,1.3]$ & $\sim 12000$ & n.a. & n.a.\\
%
%---  H1 jets with lifetime tag HERAII
%
\hline
10 &
incl. lifet. &  jet + $\delta$ + $S$   & \cite{h1cbjets} & H1   & 06-07  & $189$  & $>6$  & $[0.07,0.625]$ & Jet  & $>6$ & $[-1.0,1.5]$ & $\sim 10000$ & $1:12$ & $400$\\
%stat_b=+-4.8%; stat_c=+-1.44%
%
%--- ZEUS sec. vtx. HERA II
%
\hline
11 &
incl. lifet. & jet + $S$ & \cite{zeusltt_hera2}  & ZEUS &  04-07  & $354$ & $[5,1000]$  & $[0.02,0.7]$ & Jet   & $>5$ & $[-1.6,2.2]$  &  $\sim 13000$ & $1:14$ & $800$ \\
%k_c=0.940+-0.014; k_b=1.32+-0.05
\hline
\end{tabular}
\end{center}
\caption{
{\bfseries Beauty DIS measurements} at HERA.
Information is given for each analysis on the
beauty tagging method, the experiment,
% (H stands for H1, Z for ZEUS),
the data taking period, integrated luminosity, $Q^2$ 
and $y$ ranges
and the cuts on transverse momenta and pseudorapidities of selected
final state particles.
The last three columns provide information on the 
estimated number of
tagged beauty events, the effective signal-to-background ratio
%, as calculated from
%$\,s:b = events: (\sigma(events)^2-events)$ 
and the equivalent number of background-free events.
%, as calculated from
%$\,[events/\sigma(events)]^2.$
%
The centre-of-mass energy of all data taken up to 1997 ($6^{th}$ column)
was $300\gev$, while it was $318 \rnge 319\gev$ for all subsequent runs.
The "Br" label in the superscript refers to measurements in 
the Breit frame (see text).
}
\label{tab:bDIS}
\end{sidewaystable}

\subsection{Production mechanism}

\begin{figure}[hb]
\centering
\includegraphics[trim = 0pt 0pt 0pt 220pt,clip,%
 width=0.6\linewidth]{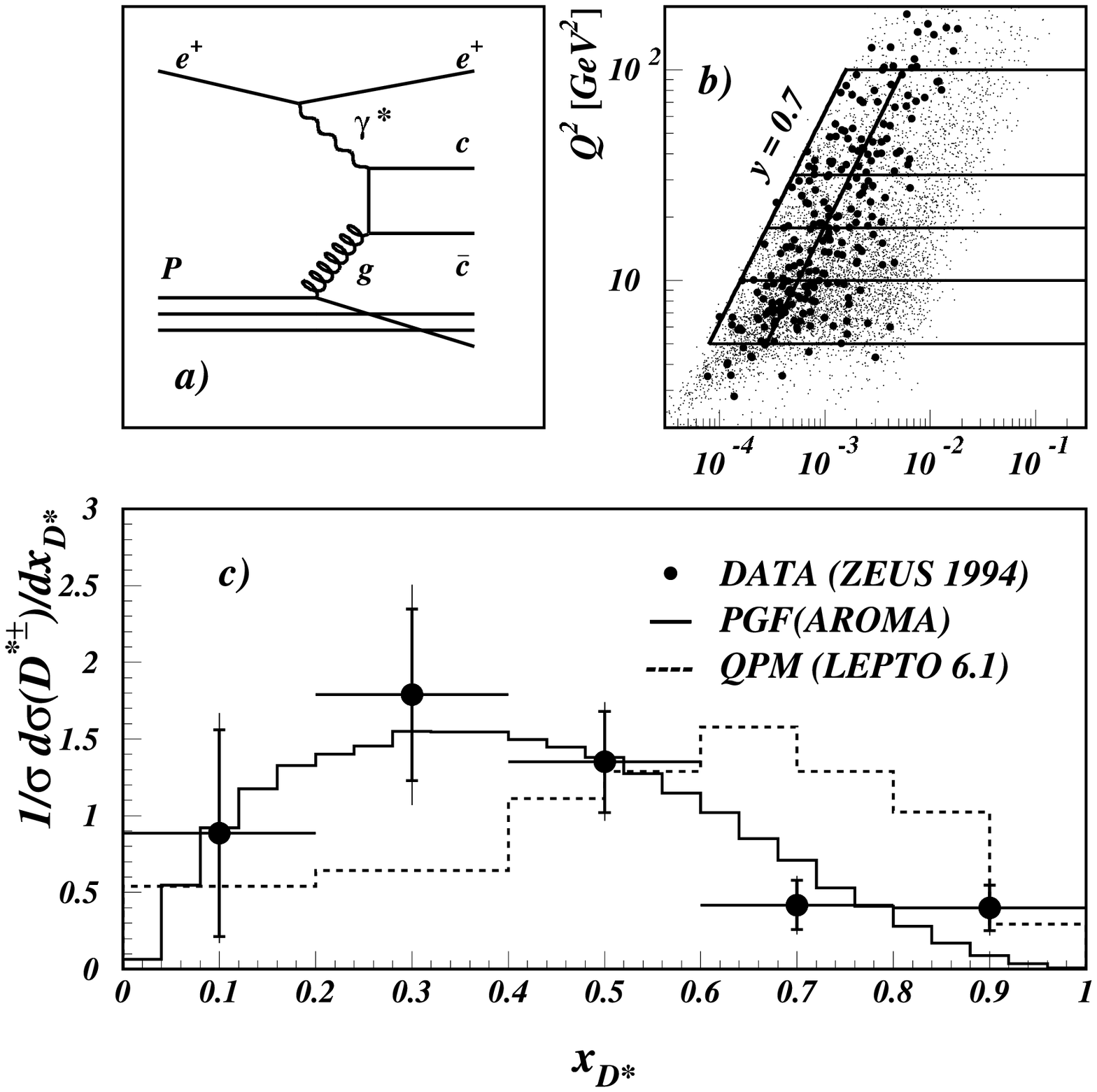}
\caption{
Normalised differential \dst-production cross section as a function of 
$x_D^*$~\pcite{zeusdstar97}.
The measurement was performed for $5< Q^2 < 100\gev^2$.
The points show the data, while solid and dashed lines show the 
BGF (PGF) and QPM predictions.
}
\label{fig:QPMfails}
\end{figure}

Already in the first H1~\cite{h196} and ZEUS~\cite{zeusdstar97} 
measurements in DIS 
(entries 1 and 2 in \Tab{cDIS}) boson-gluon
fusion was clearly identified to be the dominant
production mechanism for charm quarks.
This was investigated using the distribution of the fractional
momentum of \dst mesons in the $\gp$ system, 
$x_D = \frac{2p^*_D}{W_{\gp}}$, where $p^*_D$
denotes the \dst momentum measured in the
$\gamma^* p$ frame.
The data were used to disentangle between BGF and QPM-like models
(cf. \fig{intro1}(a) and \fig{intro6}).
The BGF process produces a $c \bar{c}$ pair that recoils 
against the proton remnant in the $\gamma^* p$ frame, while the (massless) QPM 
produces a single charm quark recoiling against the proton remnant
(which contains the other charm quark).
Since the $D$ meson carries a large fraction $x_D$ of the
charm quark momentum, the former model should lead to significantly 
softer distribution in $x_D$.
\Fig{QPMfails} shows a comparison of the two models against 
the ZEUS data. 
%%%%%%%%%%%%%%%%%%%%%%%%%%%%%%%%%%%%%%%%%%%%%%%%%%%%%%%%%%%%%%%%%%%%%%%%%%%%%%%
%%%%%%%%%%%%%%%%%%%%%%%%%%%%%%%%%%%%%%%%%%%%%%%%%%%%%%%%%%%%%%%%%%%%%%%%%%%%%%%
The observed shape of the cross section in the data 
proves that BGF is the dominant charm-production
process in DIS at HERA.
This was quantified in~\cite{h196} in the leading order QCD picture by 
setting an upper limit
for the fraction of the QPM-like contribution $f(QPM)$ 
to charm DIS production to be below $0.05$ at $95\%$ C.L.

\subsection{Single-differential cross sections}
\label{sect:ZMVFNSfail}

\begin{figure}[tb!]
\centering
\bmp{c}{0.45\linewidth}
\includegraphics[trim = 70pt 60pt 100pt 30pt,clip,%
 width=0.99\linewidth]{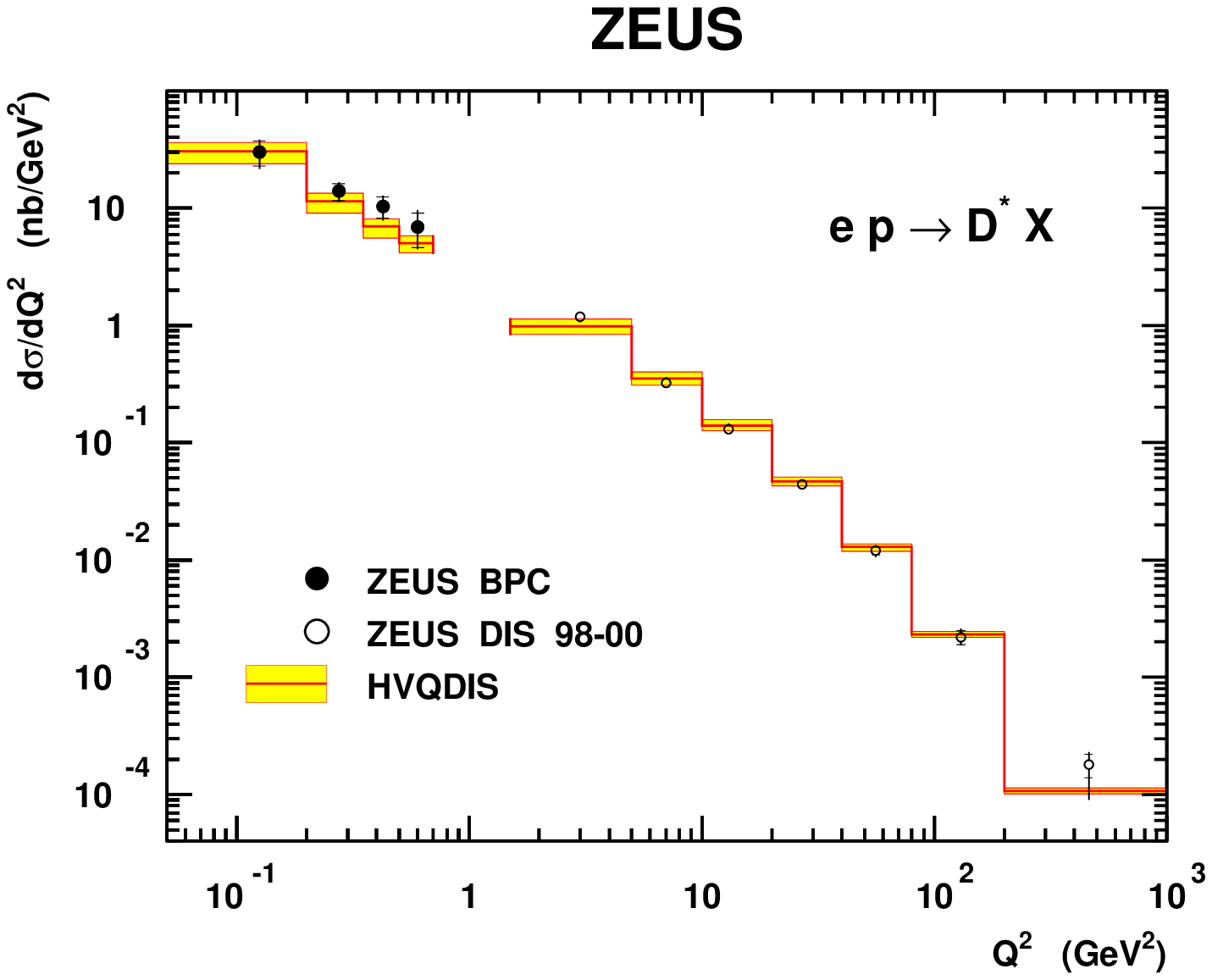}
\put(-50,130){\Large (a)}
\emp
\hspace{2mm}
\bmp{c}{0.45\linewidth}
\includegraphics[trim = 70pt 70pt 80pt 20pt,clip,%
 width=0.99\linewidth]{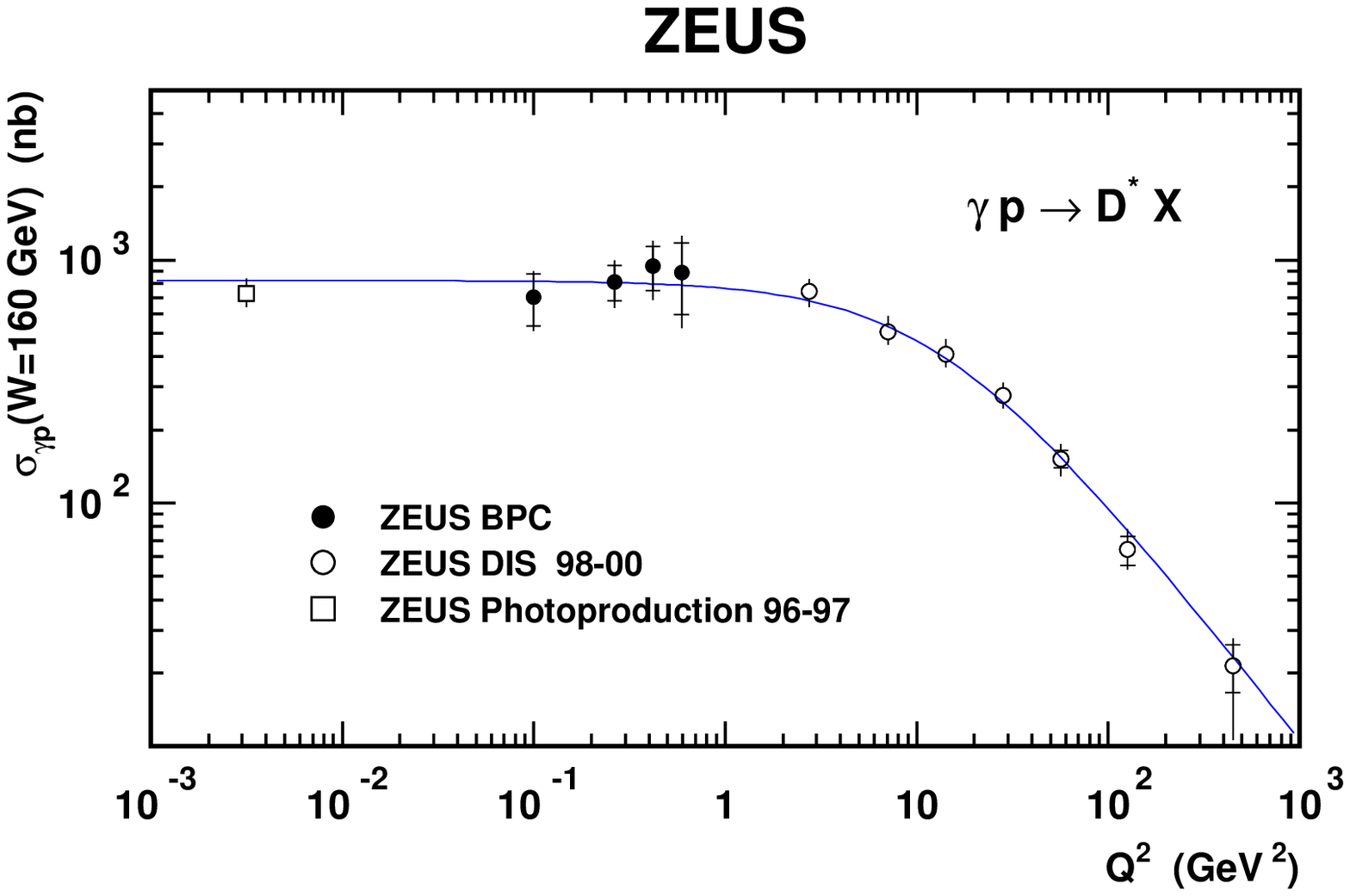}
\put(-50,90){\Large (b)}
\emp
\caption{
(a) Differential $ep$ cross section for \dst production 
as a function of $Q^2$~\pcite{zdbpc} in the kinematic region 
indicated in Table \ref{tab:cDIS}. 
The measurements~\pcite{zdbpc,zd00} are compared 
to massive-scheme NLO QCD predictions (HVQDIS)~\pcite{hvqdis}.
(b) Differential $\gp$ cross section for \dst production 
as a function of $Q^2$~\pcite{zdbpc}.
The \dst data are shown in the 
photoproduction~\pcite{ZEUScPHP1999}, 
transition~\pcite{zdbpc} and DIS~\pcite{zd00} regions.
The curve shows a fit to the data (see text).
}
\label{fig:DstarBPC}
\end{figure}

\paragraph{Transition from photoproduction to DIS:}
The ZEUS collaboration has studied~\cite{zdbpc} (entry 9 in \Tab{cDIS}) 
charm production in the intermediate $Q^2$ region between 
photoproduction and DIS:
$0.05 < Q^2 < 0.7\gev^2$.
The scattered electron was detected with the beampipe 
calorimeter (BPC) at very small scattering angles.
\Fig{DstarBPC}(a) shows a comparison of the massive-scheme NLO QCD
predictions~\cite{hvqdis} to these very-low-$Q^2$ as well as $Q^2>1.5\gev^2$ 
data~\cite{zd00} (entry 6 in \Tab{cDIS}).
%%%%%%%%%%%%%%%%%%%%%%%%%%%%%%%%%%%%%%%%%%%%%%%%%%%%%%%%%%%%%%%%%%%%%%%%%%%%%%%%%%%%%%%%
%%%%%%%%%%%%%%%%%%%%%%%%%%%%%%%%%%%%%%%%%%%%%%%%%%%%%%%%%%%%%%%%%%%%%%%%%%%%%%%%%%%%%%%%
The calculations provide a remarkable description of the drop of the
measured cross sections over 5 orders of magnitude from 
$Q^2 = 0.05\gev^2 \ll 4m_c^2$ to $Q^2 = 100\gev^2 \gg 4m_c^2$.
The slope of $d \sigma/d Q^2$ changes with $Q^2$: it is
steeper at high $Q^2 > 4 m_c^2$, where it is mainly dictated 
by the photon-propagator dependence $1/Q^4$,
than at low $Q^2 < 4 m_c^2$, where an asymptotic $1/Q^2$ dependence 
is expected.
To study this further, the measured \dst electroproduction cross sections were
converted into $\gamma^* p$ cross sections using the photon flux in the
improved Weizs\"acker-Williams approximation (see \Sect{QED}).
\Fig{DstarBPC}(b) shows the converted DIS as well as the photoproduction
cross sections~\cite{ZEUScPHP1999} (entry 5 in \Tab{r1}).
The very-low-$Q^2$ measurements are consistent with the photoproduction
cross section.
The data were fitted with a function 
$\sigma_{\gp} (Q^2) \propto M^2 / (Q^2+M^2)$.
The extracted value was $M^2 = 13 \pm 2\gev^2$, which is 
close\footnote{The actual kinematic threshold for a $D^*-D$ meson pair 
with only the $D^*$ detected is 
$(\sqrt{m(D^*)^2+p_T(D^*)^2}+m_D)^2-p_T(D^*)^2=17$ GeV$^2$.}  to 
$4 m_c^2$ and is significantly larger than the value obtained
from inclusive data, 
$M^2_0 = 0.52 \pm 0.04\gev^2 \simeq m_\rho^2$~\cite{ZEUSBPC}.

\begin{figure}[htbp]
\centering
\bmp{c}{0.45\linewidth}
\includegraphics[trim=0 50pt 0pt 0pt,clip,%
width=0.99\linewidth]{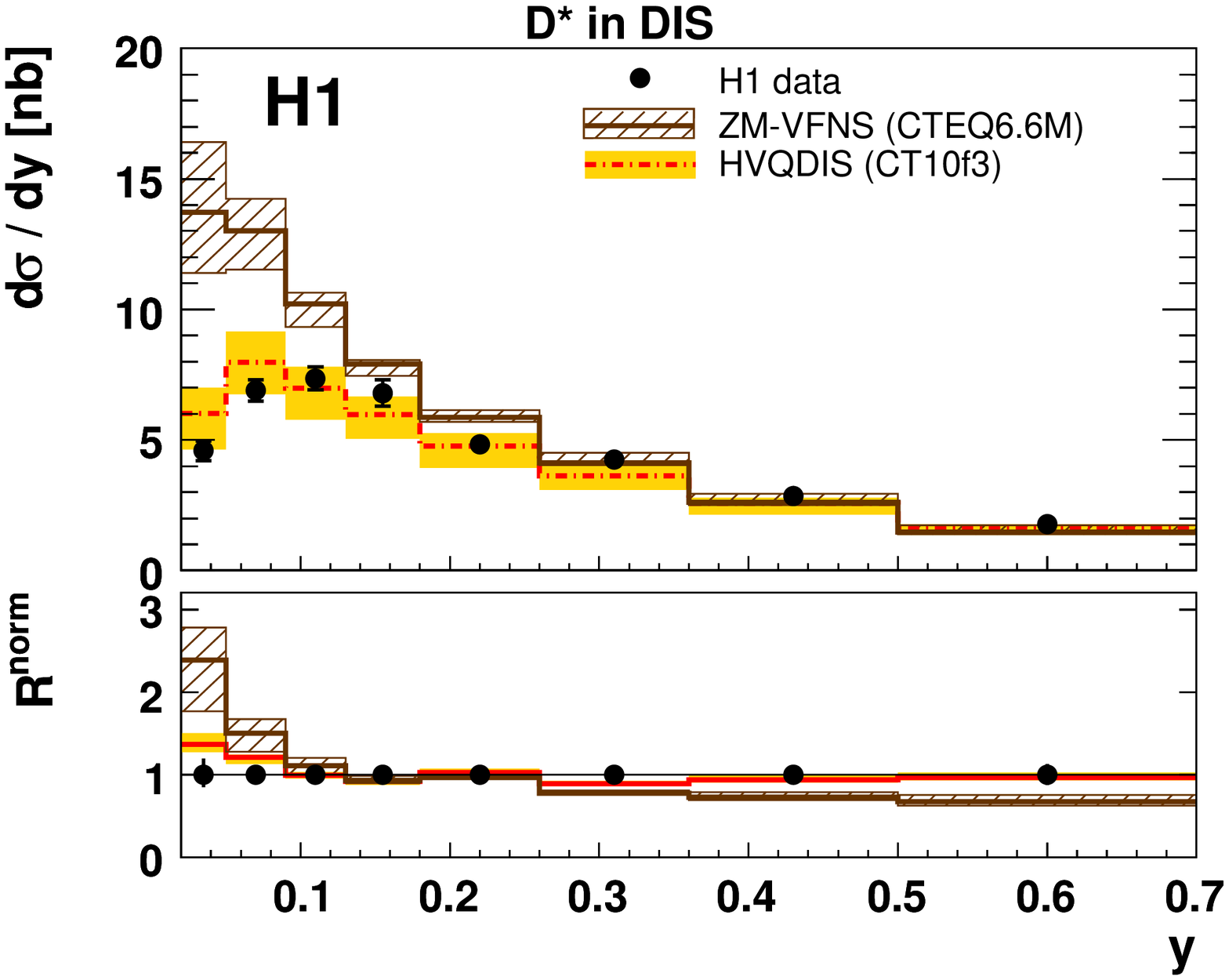}
\put(-60,140){\Large (a)}
\emp
\bmp{c}{0.4\linewidth}
\includegraphics[width=0.99\linewidth]{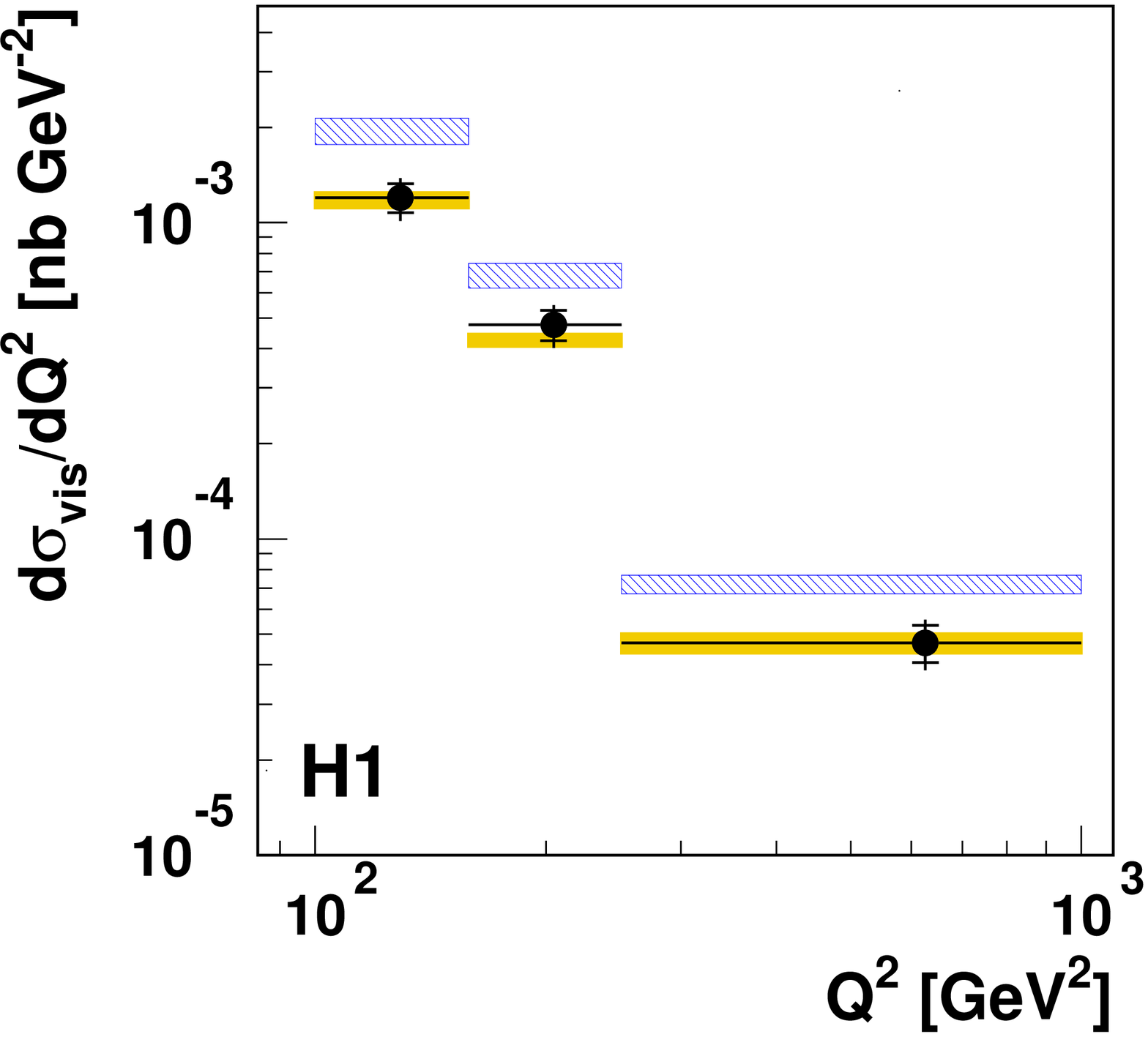}
\put(-100,100){\includegraphics[width=0.5\linewidth]{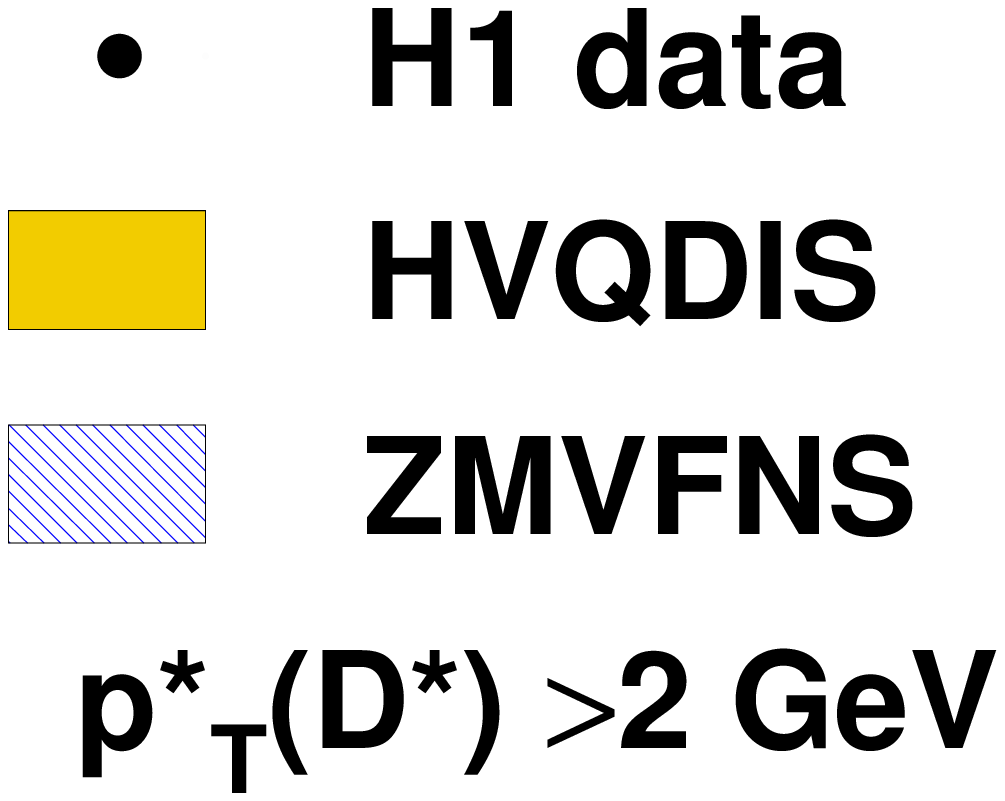}}
\put(-50,110){\Large (b)}
\emp
\caption{
Differential $D^*$-production cross section as a function of 
$y$ for $Q^2 < 100\gev^2$~\pcite{h1dstar_hera2} (a) 
and as a function of $Q^2$  for $Q^2 > 100\gev^2$~\pcite{h1dstarhighQ2} (b).
Also shown are the massive NLO prediction (HVQDIS)~\pcite{hvqdis} and
the massless NLO prediction (ZMVFNS)~\pcite{Heinrich:2004kj,SandovalZMinDIS}.
The ratio shown on the left, $R^{\mathrm{norm}}$, represents the ratio
of individually-normalised distributions to the data, thus allowing a
comparison of shapes only.
}
\label{fig:ZMfails}
\end{figure}

\paragraph{Performance of the ZMVFNS:}
\dst-production single-differential cross sections in DIS 
have also been used to test available
calculations in the massive and massless schemes. 
\Fig{ZMfails} shows a comparison of the most precise 
measurements from H1~\cite{h1dstarhighQ2,h1dstar_hera2} 
(entries 16 and 17 in \Tab{cDIS}) to 
NLO QCD calculations.
%
%%%%%%%%%%%%%%%%%%%%%%%%%%%%%%%%%%%%%%%%%%%%%%%%%%%%%%%%%%%%%%%%%%%%%%%%%%%%%%%%%%%%%%%%%%%
%---- H1 D* low-/high-Q2 data against ZMVFNS
%
%///////////////////////////////////////////////////////////////////////////////////////////
%
Since the ZMVFNS calculation is only valid in the regime 
where the charm-quark mass can be neglected, 
an additional restriction is needed on the \dst transverse
momentum in the $\gamma^* p$ frame, $p_T^*(D^*) > 2\gev$, 
on top of the selection outlined in \Tab{cDIS}.
This is compareable to the cuts used in the 
photoproduction analysis, for which the laboratory
frame approximately coincides with the $\gp$ system (see \fig{r2}).
The inelasticity is correlated with the centre-of-mass energy
in the $\gamma^* p$ frame, $W_{\gp}$ (see \eq{WysQ}), thus the low-$y$
region corresponds to the low-$W_{\gp}$ region.
Therefore, as expected, the ZMVFNS predictions deviate significantly 
from the massive-scheme calculations at low $y$, where 
$W_{\gp}$ is not $\gg 4m_c^2$, and come close to the FFNS calculations
at high $y$.
At low $y$ the massless-scheme calculations clearly fail 
to describe the data, while massive predictions are in agreement
with the measurement in the whole $y$ range.
Also for $Q^2 > 100\gev^2$ the massive predictions describe the data 
well within uncertainties, whereas the massless approach significantly 
overestimates the charm cross section.
The data also clearly establish that the ZMVFNS fails to describe 
heavy-flavour
production in DIS at HERA.
Similar conclusions were drawn in~\cite{h1dstar_hera1} 
(entry 8 in \Tab{cDIS}), but with a lower precision of the data.

\begin{figure}[tb!]
\centering
\bmp{c}{0.45\linewidth}
\includegraphics[width=0.99\linewidth]{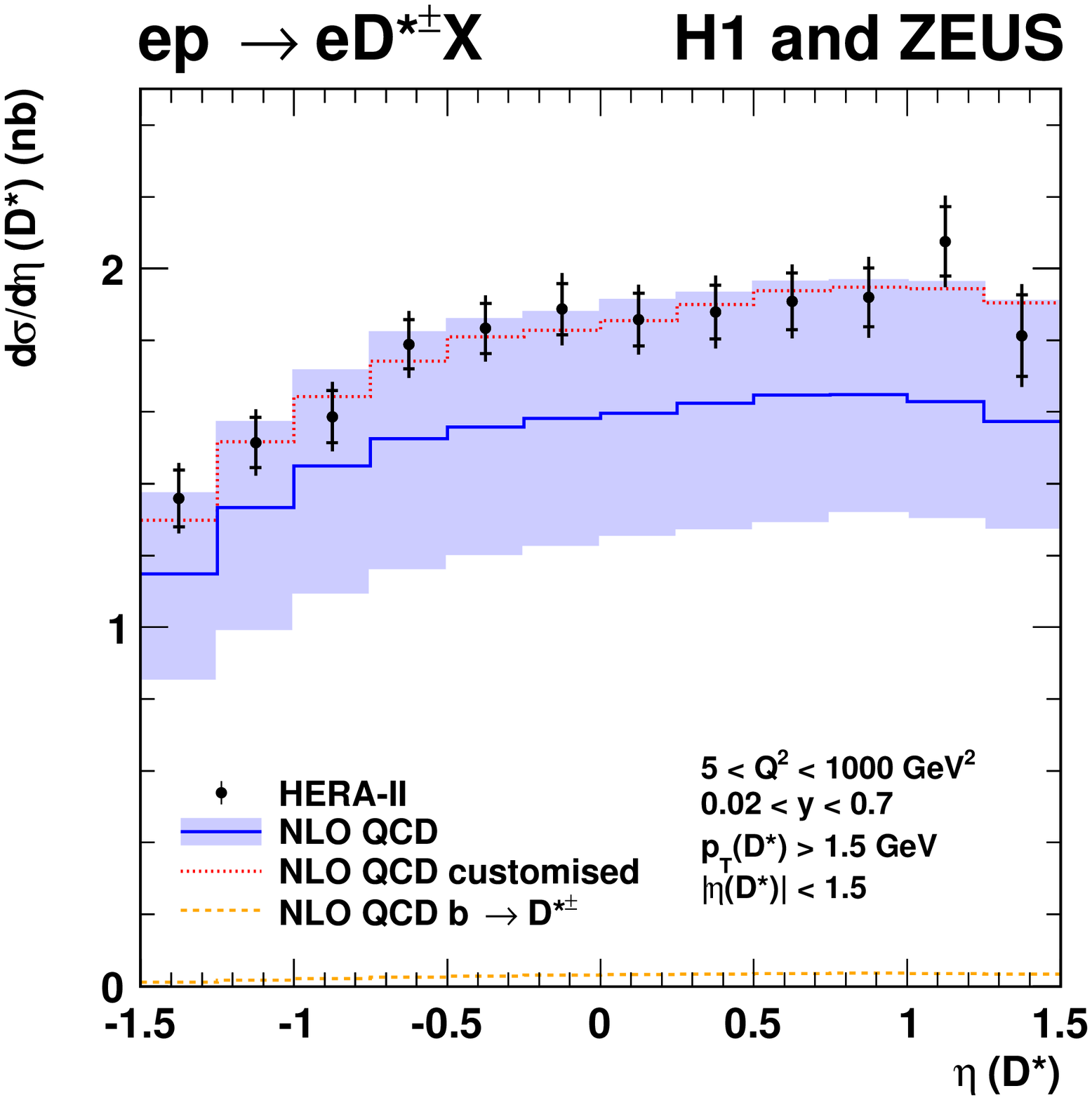}
\put(-50,90){\Large (a)}
\emp
\bmp{c}{0.45\linewidth}
\includegraphics[width=0.99\linewidth]{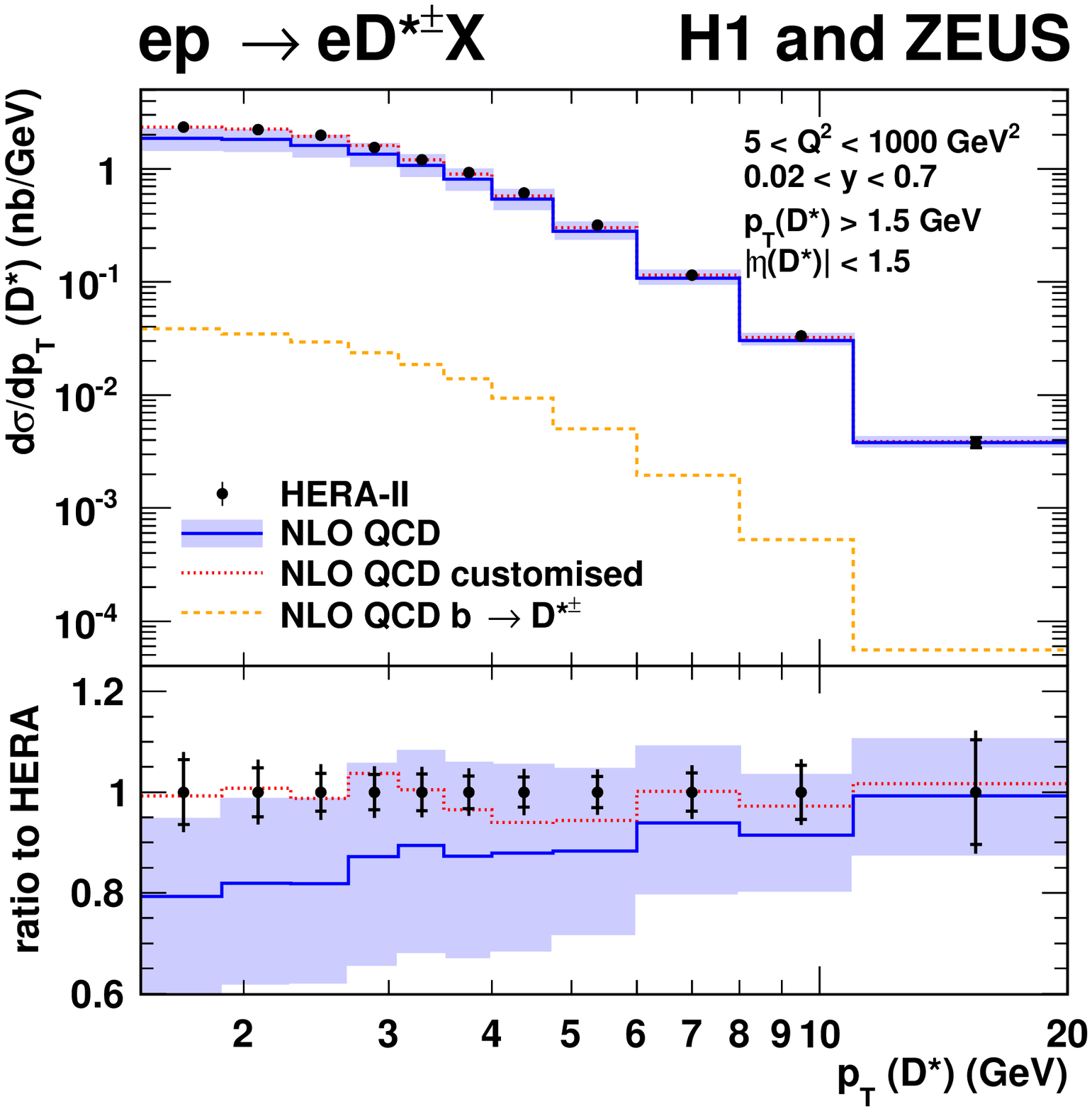}
\put(-50,150){\Large (b)}
\emp\\
\bmp{c}{0.45\linewidth}
\includegraphics[width=0.99\linewidth]{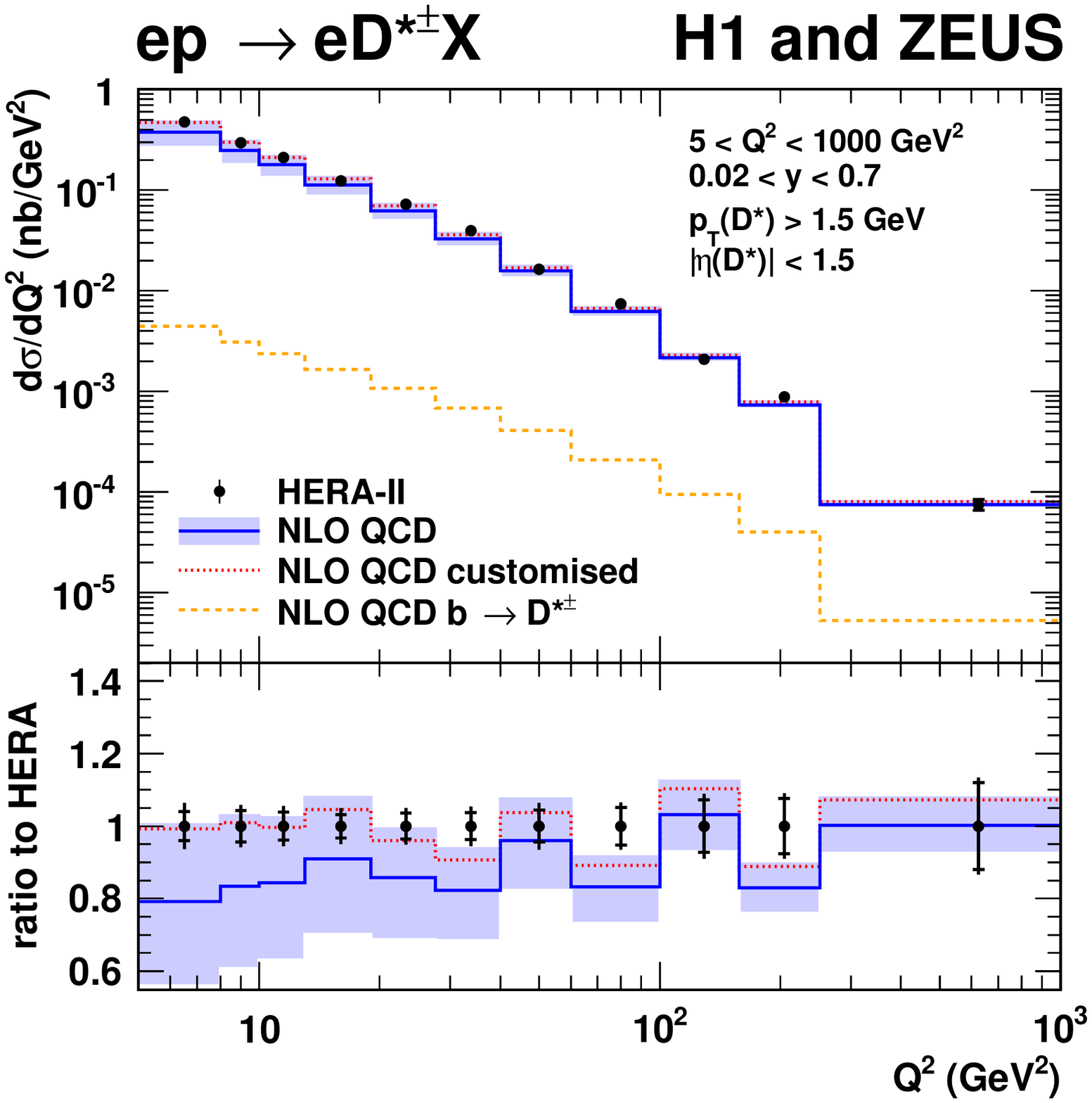}
\put(-50,150){\Large (c)}
\emp
\bmp{c}{0.45\linewidth}
\includegraphics[width=0.99\linewidth]{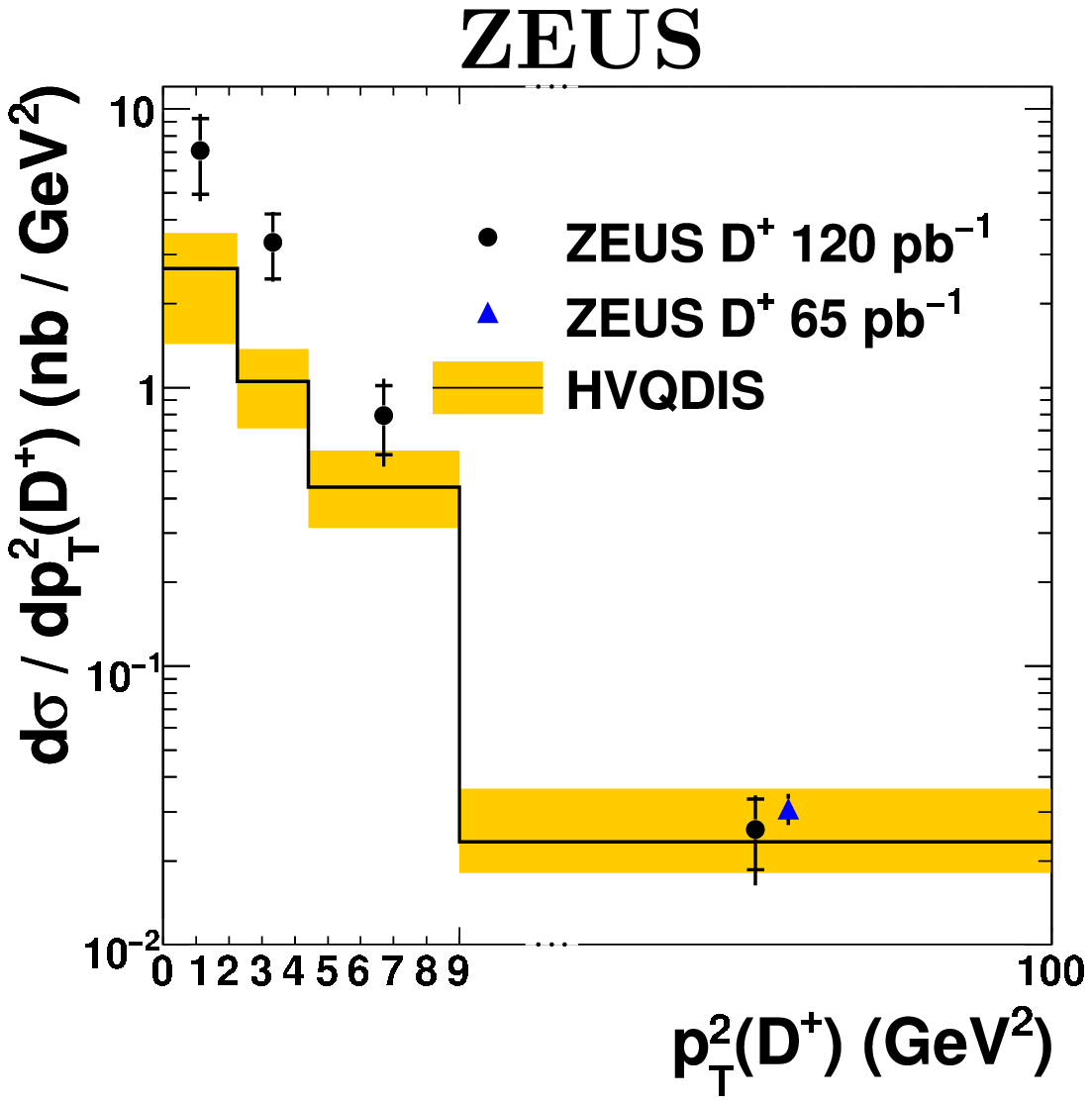}
\put(-60,110){\Large (d)}
\emp
\caption{
Differential \dst-production cross section~\pcite{HERAdstcomb} 
as a function of (a) $\eta(D^*)$, (b) $p_T(D^*)$ and (c) $Q^2$.
The data points are shown with uncorrelated (inner error bars)
and total (outer error bars) uncertainties.
Also shown are the NLO QCD predictions (HVQDIS)~\pcite{hvqdis}
with theory uncertainties indicated by the band.
The beauty-production contribution is included in the 
cross section definition and is plotted separately.
A customised NLO calculation (see text) is also shown.
(d) Differential \dplusp cross section as a function of 
$p_T^2(D^+)$ down to $p_T(D^+) = 0\gev$~\pcite{zeusdpluslambda}.
}
\label{fig:Dstcomb}
\end{figure}

\paragraph{Event and heavy-flavour kinematics:}
Most of the analyses summarised in \Taband{cDIS}{bDIS} 
studied event, charm and beauty kinematics differentially in
the respective fiducial phase spaces.
The most precise \dst measurements~\cite{h1dstarhighQ2,
h1dstar_hera2, zeusdstar_hera2} (entries 16--18 in
\Tab{cDIS}) were combined~\cite{HERAdstcomb} to 
obtain the most precise charm differential cross
sections with essentially no theory uncertainty due to
extrapolation to a common phase space.
The combination was done with a careful treatment of 
correlations.
As expected, the individual measurements were found to be consistent.
The uncorrelated uncertainties were reduced due to 
effective doubling of statistics, while the
correlated systematic uncertainties were reduced through 
cross-calibration effects between the two 
experiments.
\Fig{Dstcomb}(a)--(c) shows a comparison of massive-scheme
NLO QCD predictions~\cite{hvqdis} to the \dst combined 
single-differential cross sections.
%
%%%%%%%%%%%%%%%%%%%%%%%%%%%%%%%%%%%%%%%%%%%%%%%%%%%%%%%%%%%%%%%%%%%%%%%%%%%%%%%%%%%%%%%%%%%
%---- D meson 1D
%
%///////////////////////////////////////////////////////////////////////////////////////////
The predictions describe the data very well
within uncertainties.
However, the data reach $5 \%$ precision over a large
fraction of the measured phase-space, whereas the typical 
theory uncertainty ranges from $30 \%$ at low $Q^2$ to
$10 \%$ at high $Q^2$.
The theory uncertainty is dominated by the 
independent variation of the $\mu_R$ and $\mu_F$ scales, 
the uncertainty on the charm-quark pole mass and 
variations of the fragmentation model.
Therefore, higher-order massive-scheme NNLO
calculations and an improved fragmentation model for these
predictions are needed to fully exploit the 
potential of these data.
In addition, theory  uncertainties were studied in detail
and a ``customised'' prediction was obtained by a variation of the 
theory parameters within their uncertainties, to show that the calculations 
can simultanously provide a good description of the shape and normalisation
of all measured distributions with a single set of parameters.
This led to a renormalisation scale reduced by a factor 2
(see also \Sect{scale}), 
the charm-quark pole mass reduced to $m_c=1.4\gev$ and 
to a change of fragmentation parameters, all within the nominal uncertainties.

Moreover, differential cross sections of other $D$ mesons 
as well as of leptons from heavy-flavour decays and of heavy flavour jets
were measured.
In particular, \fig{Dstcomb}(d) shows the \dplusp 
differential cross section~\pcite{zeusdpluslambda} 
(entry 11 in \Tab{cDIS})
measured down to $p_T(D^+) = 0\gev$.
The measurement was done in the $D^+ \to K^0_S \pi^+$ decay
channel.
The presence of a neutral strange hadron in the decay resulted in 
a reasonable signal-to-background ratio even 
at very low transverse momentum of the \dplusp.
The data were found to be described by the
massive NLO QCD calculations within about two standard deviations.

Furthermore, parton-parton correlations have been studied in
\dst-tagged events~\cite{h1dstar_hera1} (entry 8 in 
\Tab{cDIS}). 
The conclusions are similar to those obtained from the respective 
photoproduction measurements. 
In general the massive QCD calculations provide 
a good description apart from the region of small $\Delta \phi$ 
and very large $|\Delta \eta|$ between the two
leading jets in the event (not shown).

Inclusive lifetime tagging (entries 21, 22 in \Tab{cDIS}) 
allowed the extension of the kinematic
range of charm measurements up to $\etjet = 35\gev$ (not shown), 
which roughly corresponds to $p_T(D) \approx 20\gev$, 
where the statistics of fully reconstructed charm mesons
becomes poor. Good agreement is again observed.

\begin{figure}[tb]
\centering
\bmp{c}{0.4\linewidth}
\includegraphics[width=0.99\linewidth]{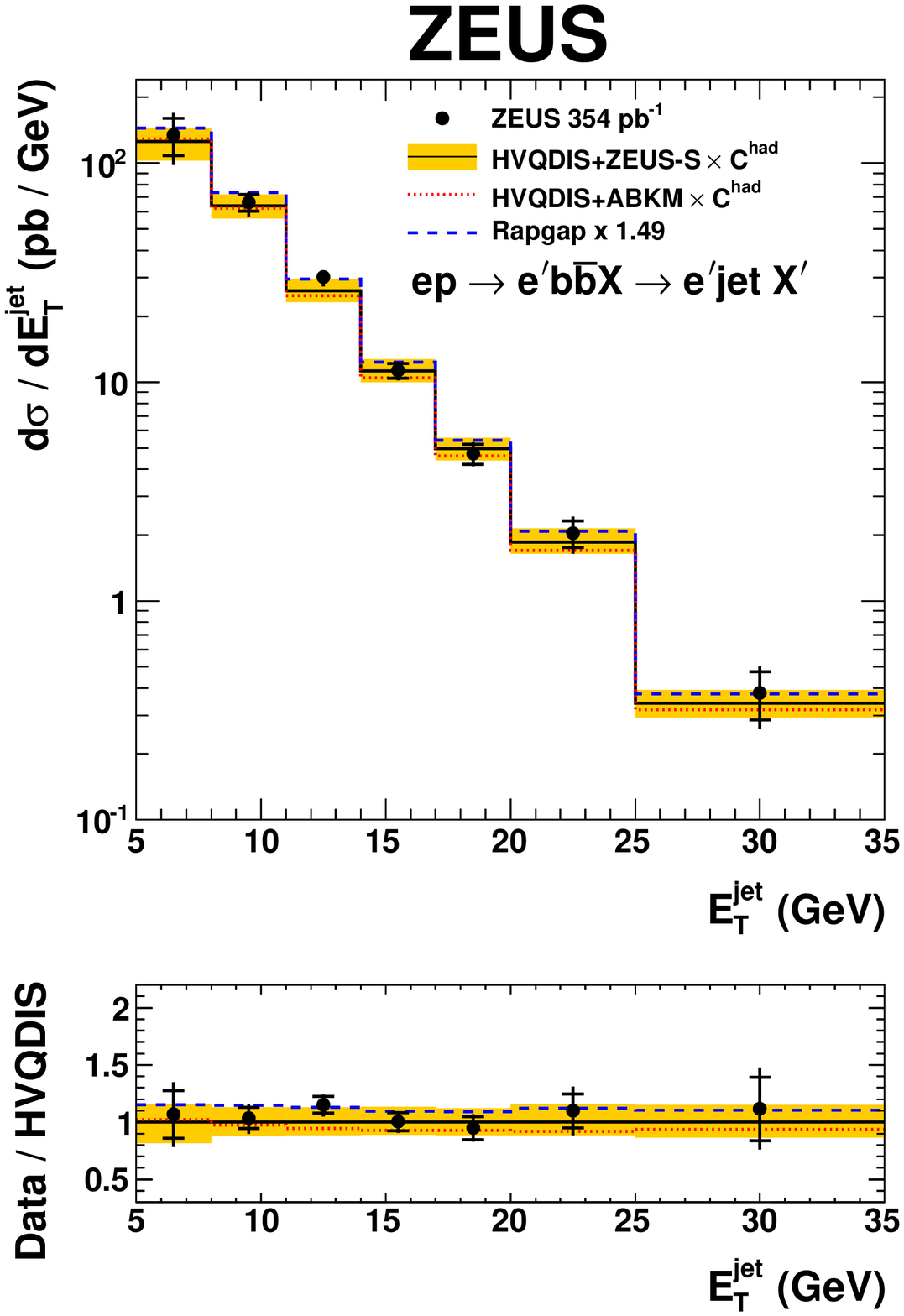}
\put(-40,180){\Large (a)}
\emp
\hspace{5mm}
\bmp{c}{0.4\linewidth}
\includegraphics[width=0.99\linewidth]{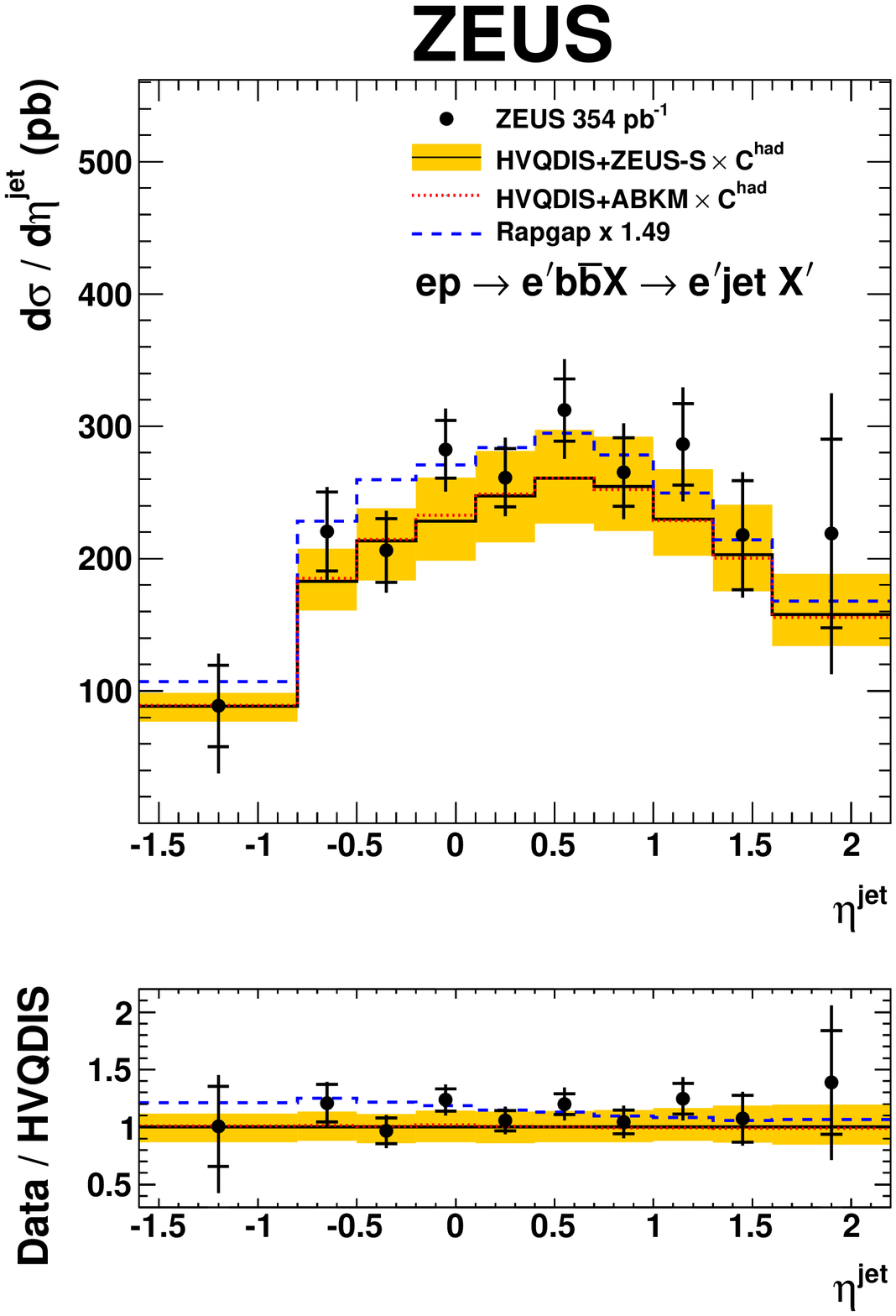}
\put(-40,130){\Large (b)}
\emp
\caption{
Differential cross section for inclusive-jet production in
beauty DIS events as a function of $\etjet$ (a) and
$\etajet$ (b)~\pcite{zeusltt_hera2}.
The data points are shown with statistical (inner error bars)
and total (outer error bars) uncertainties.
Also shown are the NLO QCD predictions (HVQDIS)~\pcite{hvqdis},
corrected for hadronisation effects,
with theory uncertainties indicated by the band.
The dashed line shows the prediction from the RAPGAP MC 
generator~\cite{RAPGAP} scaled to 
the measured integrated cross section.
}
\label{fig:bjet1d}
\end{figure}

\Fig{bjet1d} shows the corresponding single-differential jet
cross sections for beauty production in DIS~\cite{zeusltt_hera2}
(entries 11 and 22 in \Taband{bDIS}{cDIS}, respectively).
%%%%%%%%%%%%%%%%%%%%%%%%%%%%%%%%%%%%%%%%%%%%%%%%%%%%%%%%%%%%%%%%%%%%%%%%%%%%%%%%%%%%%%%%%%%
%---- ZEUS most precise b in DIS
%////////////////////////////////////////////////////////////////////////////////
%
The lifetime-tagging technique together with the 
reconstruction of the vertex mass were used to extract 
charm- and beauty-jet cross sections simultaneously.
This measurement was selected since it has the highest statistical 
significance for beauty-quark production, as can be seen
from the last column of \Tab{bDIS}.
The typical precision reached in the data is $10\rnge20 \%$
and is comparable to the theory uncertainties.
The massive-scheme NLO QCD calculations provide 
a good description of the shape and normalisation of the measured 
cross sections.

\subsection{Double-differential cross sections}

The large collected data samples allowed measurements of 
double-differential heavy-flavour cross sections, 
to study the correlations between various kinematic 
variables.

The H1 collaboration has studied~\cite{h1dstar_hera1} 
(entry~8 in \Tab{cDIS}) the cross section as a function of
$\xgobsm$ in different $Q^2$ ranges, complementing the measurements 
in the photoproduction regime discussed in \Sect{xgammaPHP}. 
It was shown that the amount of higher order contributions
included in the massive NLO calculations, including topologies which would
be called ``flavour excitation'' in the leading order picture, is enough to 
describe 
the data for different $Q^2$, while the BGF-only component of 
the RAPGAP Monte Carlo can describe the measurement for $Q^2 > 5\gev^2$
after rescaling, but fails to describe the shape observed 
in the data at lower $Q^2$ (not shown here). This is to be expected, since
in this ``photoproduction-like'' region (see \fig{DstarBPC}) 
the ``flavour excitation'' component will then be missing.

\begin{figure}[tb]
\centering
\includegraphics[height=0.55\textheight]{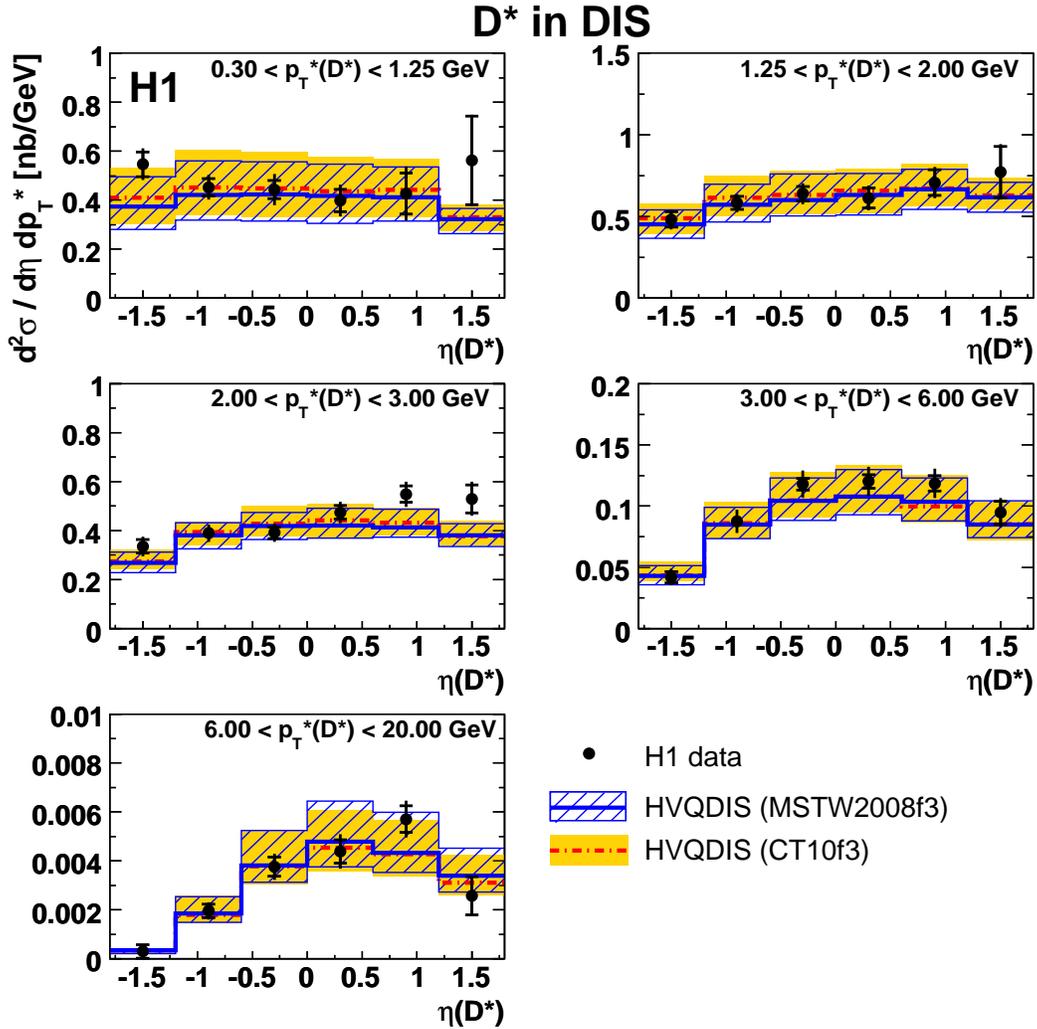}
\caption{
Double-differential \dst cross section as a function of 
$p_T^*(D^*)$ and $\eta(D^*)$~\pcite{h1dstar_hera2}.
The data points are shown with statistical (inner error bars)
and total (outer error bars) uncertainties.
NLO QCD calculation (HVQDIS)~\pcite{hvqdis}
with two different proton parton densities are compared to the data.
Theoretical uncertainties are indicated by the bands.
}
\label{fig:Dstar2dH1}
\end{figure}

Using the full \heraii data sample, cross sections have been measured by 
H1~\cite{h1dstar_hera2}~(entry~17 in \Tab{cDIS}) as a function
of the \dst pseudorapidity in the laboratory frame, $\eta(D^*)$, in bins of the \dst 
transverse momentum in the $\gp$ centre-of-mass frame, $p_T^*(D^*)$.
\Fig{Dstar2dH1} shows a comparison of the massive NLO QCD predictions 
to the data.
%%%%%%%%%%%%%%%%%%%%%%%%%%%%%%%%%%%%%%%%%%%%%%%%%%%%%%%%%%%%%%%%%%%%%%%%%%%%%%%%%%%%%%%%%%%
%---- H1 pT*-eta 2d measurement from low Q2 HERA II paper
%////////////////////////////////////////////////////////////////////////////////
%
At large $p_T^*(D^*)$, \dst production in the backward region is 
very suppressed, while at low $p_T^*(D^*)$ the $\eta(D^*)$ distribution
is rather flat in the phase space of the measurement.
The massive-scheme NLO predictions provide a good description of the data.
The predictions depend only very little on the proton PDFs used for the 
calculation. 

\begin{figure}[tb]
\centering
\includegraphics[height=0.55\textheight]{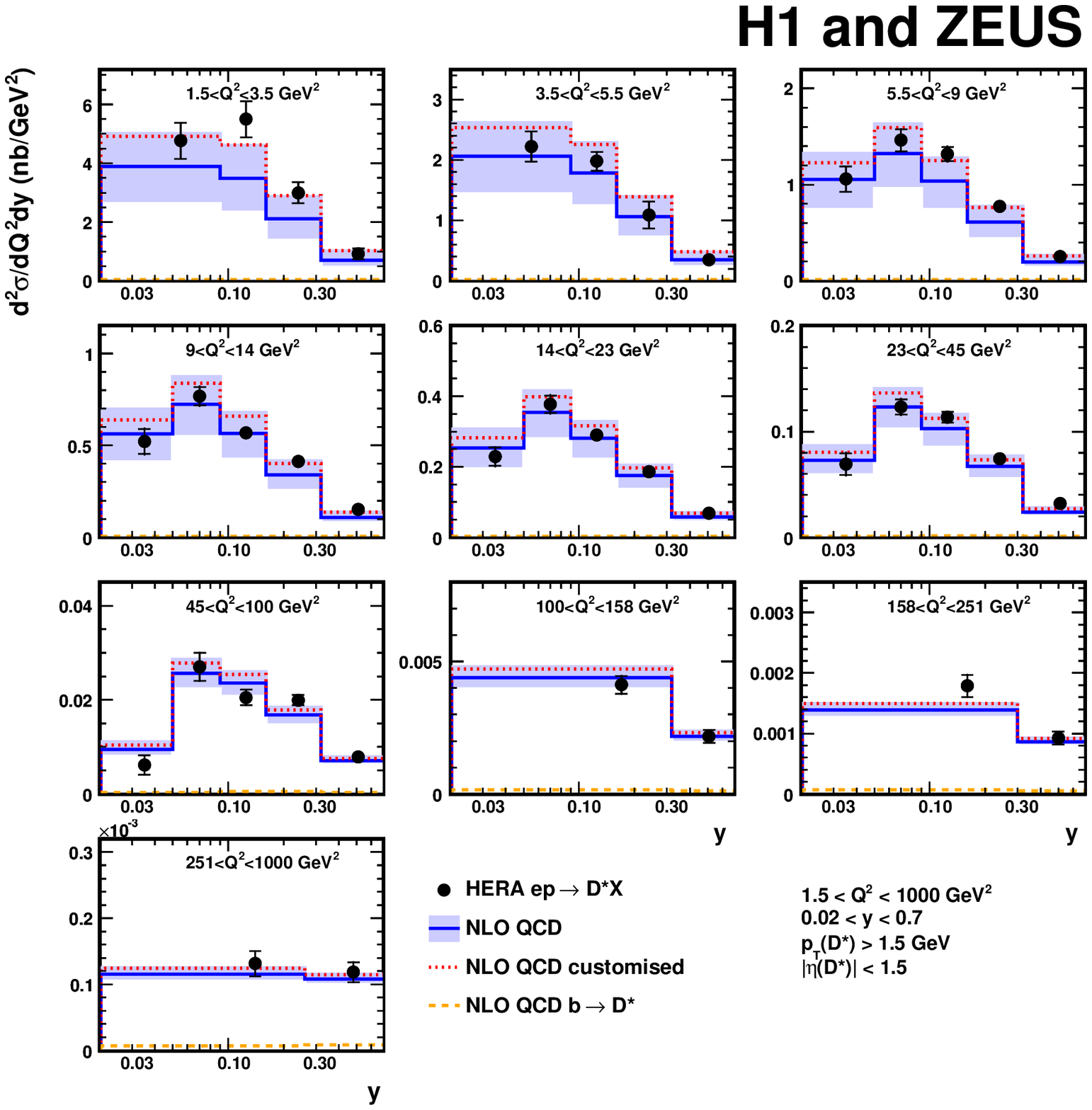}
\caption{
Double-differential \dst cross section as a function of 
$Q^2$ and $y$~\pcite{HERAdstcomb}.
The data points are shown with uncorrelated (inner error bars)
and total (outer error bars) uncertainties.
Also shown are the NLO QCD predictions (HVQDIS)~\pcite{hvqdis}
with theory uncertainties indicated by the band.
The beauty-production contribution is included in the 
cross-section definition and is plotted separately.
}
\label{fig:Dstcomb2d}
\end{figure}

In most of the analyses summarised in \Taband{cDIS}{bDIS} 
the double-differential cross sections in $Q^2$ and $y$ or $Q^2$ and $x$
were also measured.
These measurements allowed dedicated studies of the inclusive heavy-flavour 
event kinematics which can be expressed in terms of the charm 
reduced cross sections,
or of the charm contribution to the structure function $F_2$ 
(see \Sect{F2cc}).
\Fig{Dstcomb2d} shows the combined %of the most precise 
double-differential \dst cross sections as a function of $Q^2$ and $y$~\cite{HERAdstcomb}.
%%%%%%%%%%%%%%%%%%%%%%%%%%%%%%%%%%%%%%%%%%%%%%%%%%%%%%%%%%%%%%%%%%%%%%%%%%%%%%%%%%%%%%%%%%%
%---- ZEUS most precise D* in DIS
%////////////////////////////////////////////////////////////////////////////////
%
Massive-scheme NLO QCD predictions provide a good description of these 
cross sections in the full range in $Q^2$ between $1.5\gev^2$
and $1000\gev^2$.
The theoretical uncertainties decrease with increasing $Q^2$.
For $Q^2 \lsim 50\gev^2$ the theoretical uncertainties are larger 
than those of the measured cross sections.
Similar to the single-differential distributions shown in \fig{Dstcomb}, 
the theoretical uncertainties are dominated by the scale variations,
the uncertainty on the charm-quark pole mass and the variation of 
the fragmentation model.
A higher-order calculation with improved fragmentation model is needed
to achieve a theoretical precision similar to the data.

\subsection{Proton structure functions and reduced cross sections}
\label{sect:F2cc}

The measured double-differential DIS cross sections of heavy-flavour production 
as a function of $Q^2$ and $y$ or $Q^2$ and $x$ were used to extract 
the heavy-flavour reduced cross sections, \redq, or the heavy-flavour contribution
to the proton structure function $F_2$, \ftq, where $Q$ is either $c$ or $b$.
As discussed in \Sect{pdf}, the inclusive double-differential cross sections 
of heavy-flavour production can be expressed in terms of \redq or \ftq and 
\flq.
In measurements of \ftq the small contribution arising from \flq
was subtracted relying on theory, corresponding to a correction of up to $4 \%$.
The extraction from the measured cross sections requires an extrapolation
from the experimentally accessible kinematic region in $p_T$ and $\eta$
and a particular final state to the full phase space of heavy quarks.
The extrapolation was done either using the massive-scheme NLO
QCD calculations or LO+PS Monte Carlo simulations.
Since this procedure relies on the description of kinematic distributions
by predictions, a non-negligible theoretical uncertainty was introduced. 
This additional uncertainty was estimated by varying the parameters in the 
calculations which affect the shapes of the kinematic distributions.

The \redq and \ftq values extracted from measurements performed with 
different experimental techniques and different detectors can be directly compared.
Such measurements are complementary to each other due to different 
dominant sources of systematics, mostly independent statistics and 
different kinematic coverage, resulting in somewhat different 
theoretical uncertainties due to extrapolation.
For instance, for the \redc measurements the dominant systematics 
in the H1 inclusive vertexing analysis (entry 20 in \Tab{cDIS}) 
is due to the treatment of the light-flavour component, 
while in the H1 \dst \heraii measurement (entry 17 in \Tab{cDIS})
the dominant systematics is due to the modelling of the tracking efficiency.
The ZEUS analysis of charm semileptonic decays (entry 15 in \Tab{cDIS})
has yet completely different systematics.  
Therefore, a combination of measurements with such different techniques allows 
a significant reduction not only of statistical and uncorrelated but also 
of correlated systematic and extrapolation uncertainties.

\begin{figure}[htb]
\center
\includegraphics[width=0.7\textwidth]{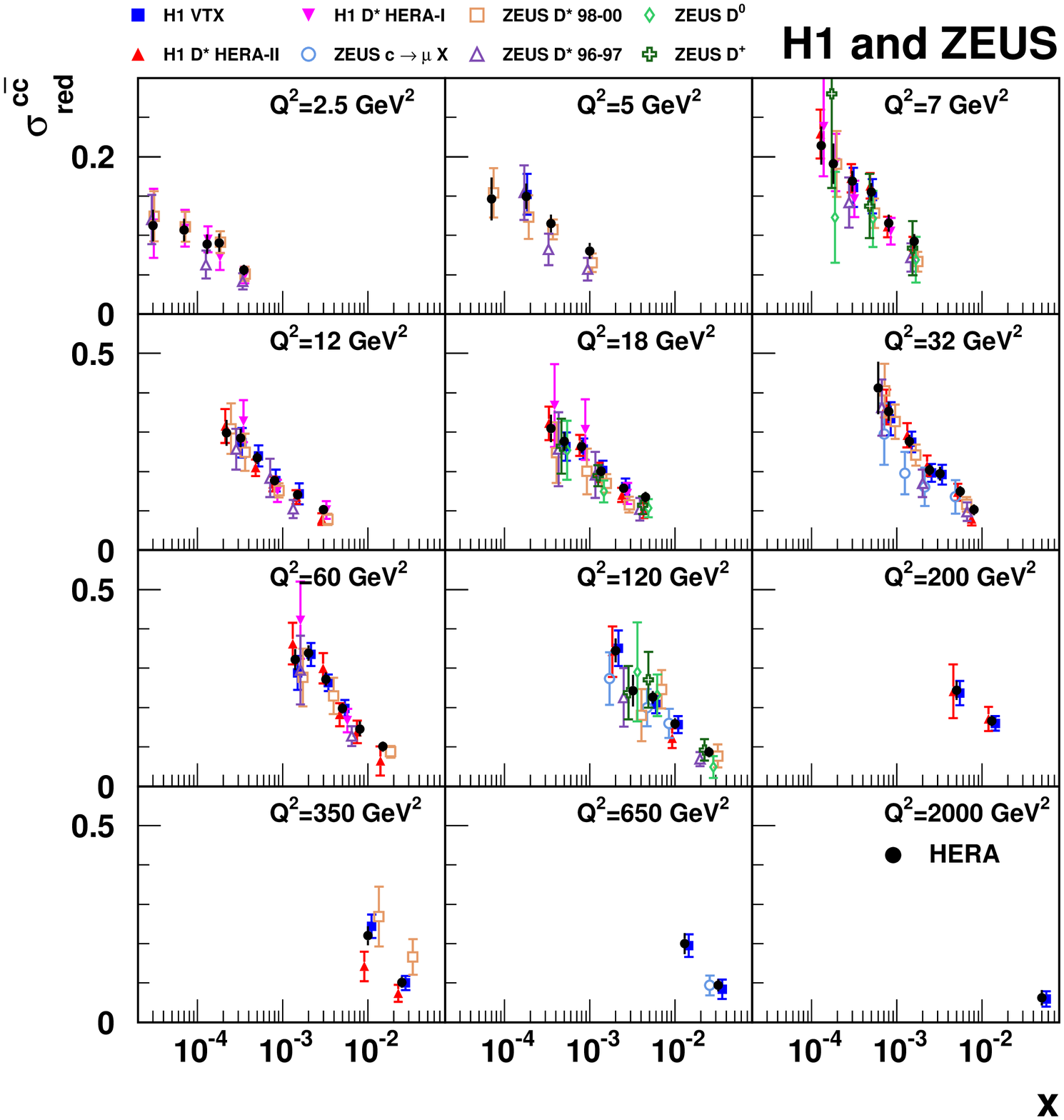}
\caption{Combined reduced cross sections~\pcite{HERAcharmcomb} \redc (filled circles) as a function of $x$ for fixed values of $Q^2$.
The input data are shown with various other symbols as explained in the legend.
The error bars represent the total uncertainty including uncorrelated, correlated and procedural uncertainties added in quadrature.
For presentation purposes each individual measurement was shifted in $x$.}
\label{fig:combined_vs_input} 
\end{figure}

\Fig{combined_vs_input} shows a comparison of H1 and ZEUS measurements 
of the charm reduced cross sections\footnote{entries 4,6,8,12,13,14,15,16,17,20 in \Tab{cDIS}}
as well as the milestone result of their combination~\cite{HERAcharmcomb}.
The combination accounts for correlations of the systematic uncertainties 
among the different input data sets.
The individual \redc measurements show good consistency, with a $\chi^2$ value of $62$
for $103$ degrees of freedom.
The combined data are significantly more precise than any of the 
input data sets.
\Fig{combined_vs_input} also highlights the advantages of different tagging techniques:
while \dst has superior precision at low $Q^2$ due to better signal-to-background ratio, 
the inclusive vertexing analysis with lifetime tagging dominates at 
high $Q^2$ due to the larger accessible statistics.
The final total precision of the combined charm reduced cross sections 
is $10 \%$ on average and reaches $6 \%$ at low $x$ and medium $Q^2$.
This corresponds to a factor 2 improvement over the most precise data 
set in the combination.
\begin{figure}[htbp]
\center
\hspace*{-1.4cm}
\includegraphics[height=0.49\textheight]{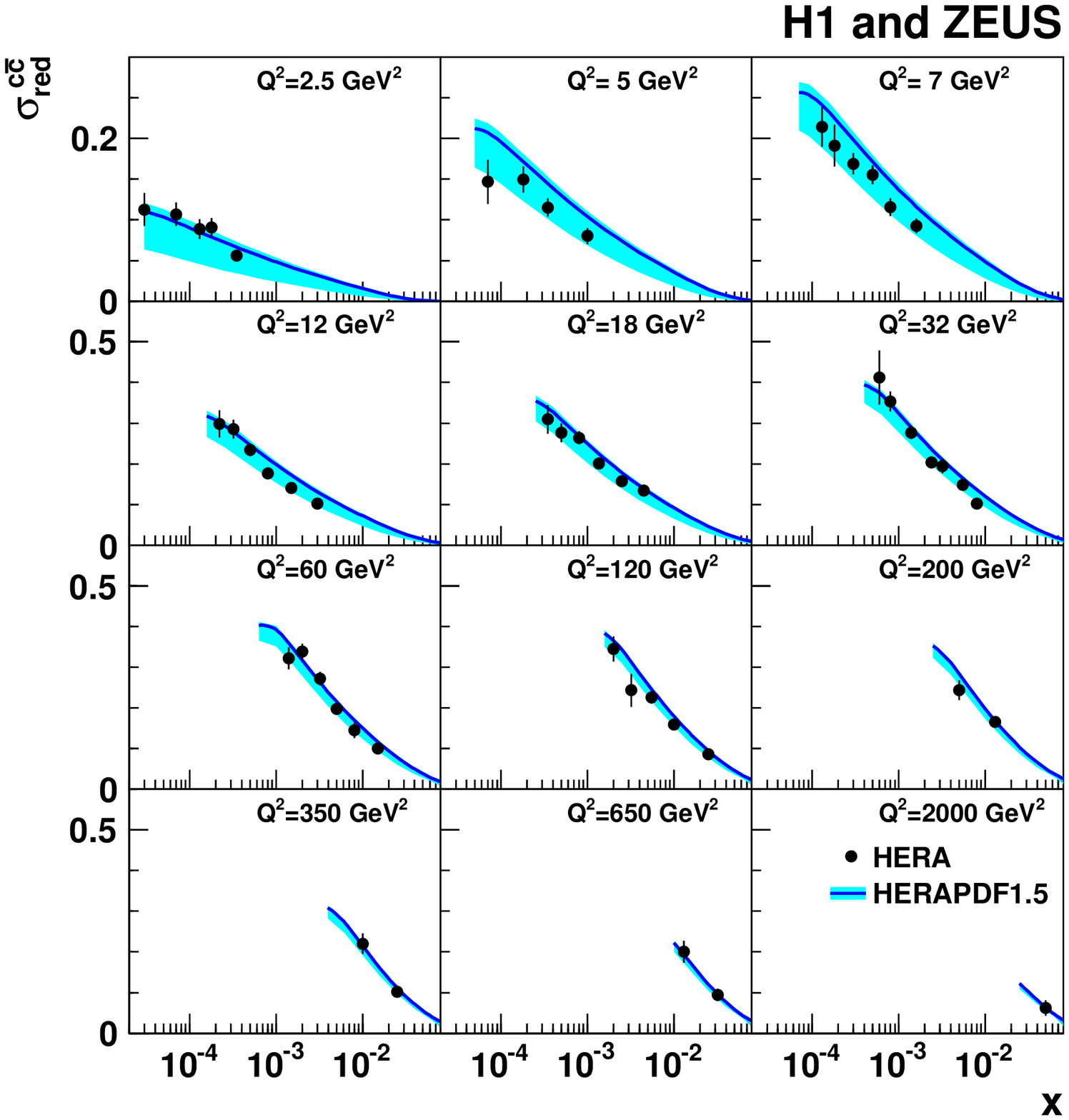}%
\hspace*{-0.7cm}
\includegraphics[height=0.49\textheight]{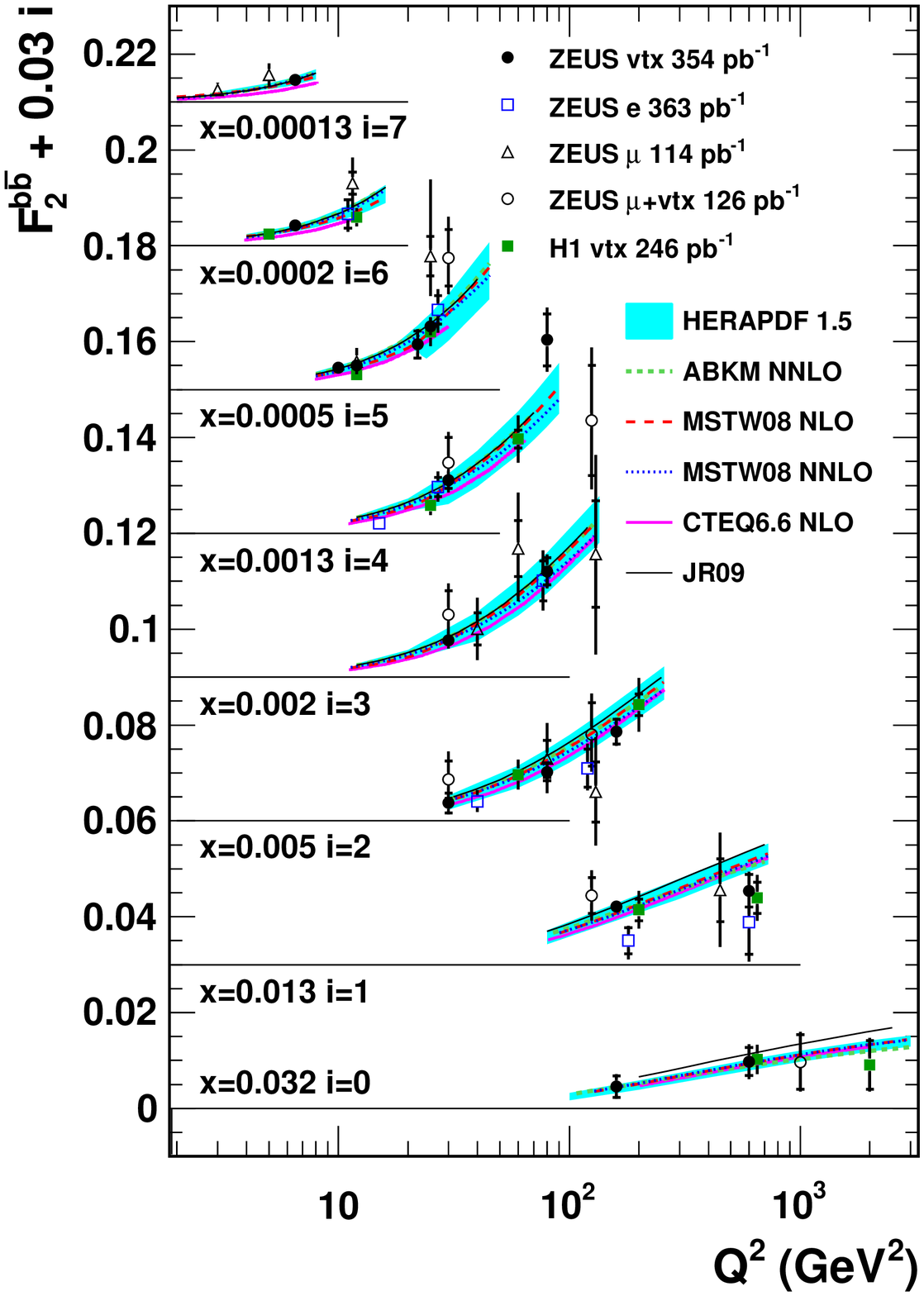}
\caption{{\em Left:} Combined \redc~\pcite{HERAcharmcomb} (filled circles) 
as a function of $x$ for fixed values of $Q^2$. 
The error bars represent the total uncertainty including uncorrelated, 
correlated and procedural uncertainties added in quadrature. 
The data are compared to the NLO predictions based on HERAPDF1.5~\cite{HERAPDF1.5} 
in the TR standard GMVFNS~\cite{TR}. 
The line represents the prediction using $\mct =1.4\gev$. 
The uncertainty band shows the full PDF uncertainty which is dominated 
by the variation of \mct~in the range $1.35 < \mct < 1.65\gev$. 
\emph{Right:} Measurements of \ftb~\pcite{zeusltt_hera2,h1ltt_hera2,zmu,zbmujet9600,zbe_hera2} 
(various symbols) as a function of $Q^2$ at fixed values of $x$.
The inner error bars are the statistical uncertainties, while 
the outer error bars are the statistical, systematic and extrapolation
uncertainties added in quadrature.
The data are compared to several NLO and NNLO predictions, 
including HERAPDF1.5~\cite{HERAPDF1.5} in the TR standard GMVFNS~\cite{TR}.
The uncertainty band shows the full PDF uncertainty which is dominated by the variation of \mbt.}
\label{fig:redxs_vs_theory} 
\end{figure}

Additionally, new statistically independent measurements of charm production
have been published (entries 18,19 and 22 in \Tab{cDIS}), in particular the 
ZEUS \dst measurement with \heraii data.
In \fig{redxs_vs_theory} the combined \redc~\cite{HERAcharmcomb} 
and individual measurements of \ftb (entries 3--5, 7--9, 11 in \Tab{bDIS})
are compared to NLO and NNLO QCD predictions.

The beauty measurements are all in good agreement with each other 
and the most precise 
data were obtained with inclusive lifetime
tagging.
The NLO QCD prediction in the GMVFNS approach based on HERAPDF1.5~\cite{HERAPDF1.5} 
is common for the two comparisons.
The good agreement between 
these predictions and the heavy-flavour data shows that the gluon 
density, which in HERAPDF1.5 is extracted from the scaling violations of \ft,
is adequate for the description of these gluon-induced production processes.
Other GMVFNS predictions were also compared to 
the combined charm reduced cross sections (not shown).
The best description of the data was provided by predictions including partial 
$\calo(\alpha_s^3)$ corrections, while predictions 
including $\calo(\alpha_s^2)$
terms agreed well with the data and predictions including $\calo(\alpha_s)$ have shown the largest deviations~\cite{HERAcharmcomb}.
The theoretical uncertainty for \redc and \ftb increases at low $Q^2$ 
and is dominated by the \mct~variation.
This indicates that the low-$Q^2$ data are sensitive to
the value of the heavy-quark mass used in the calculation, which
was exploited to extract the optimal \mct~values for different 
GMVFNS schemes as well as to measure the running heavy-quark masses 
(see \Sect{cbF2}).

\begin{figure}[tb!]
\centering
\bmp{c}{0.37\linewidth}
{\includegraphics[trim=50 50 60 60,clip,%
width=0.98\linewidth]{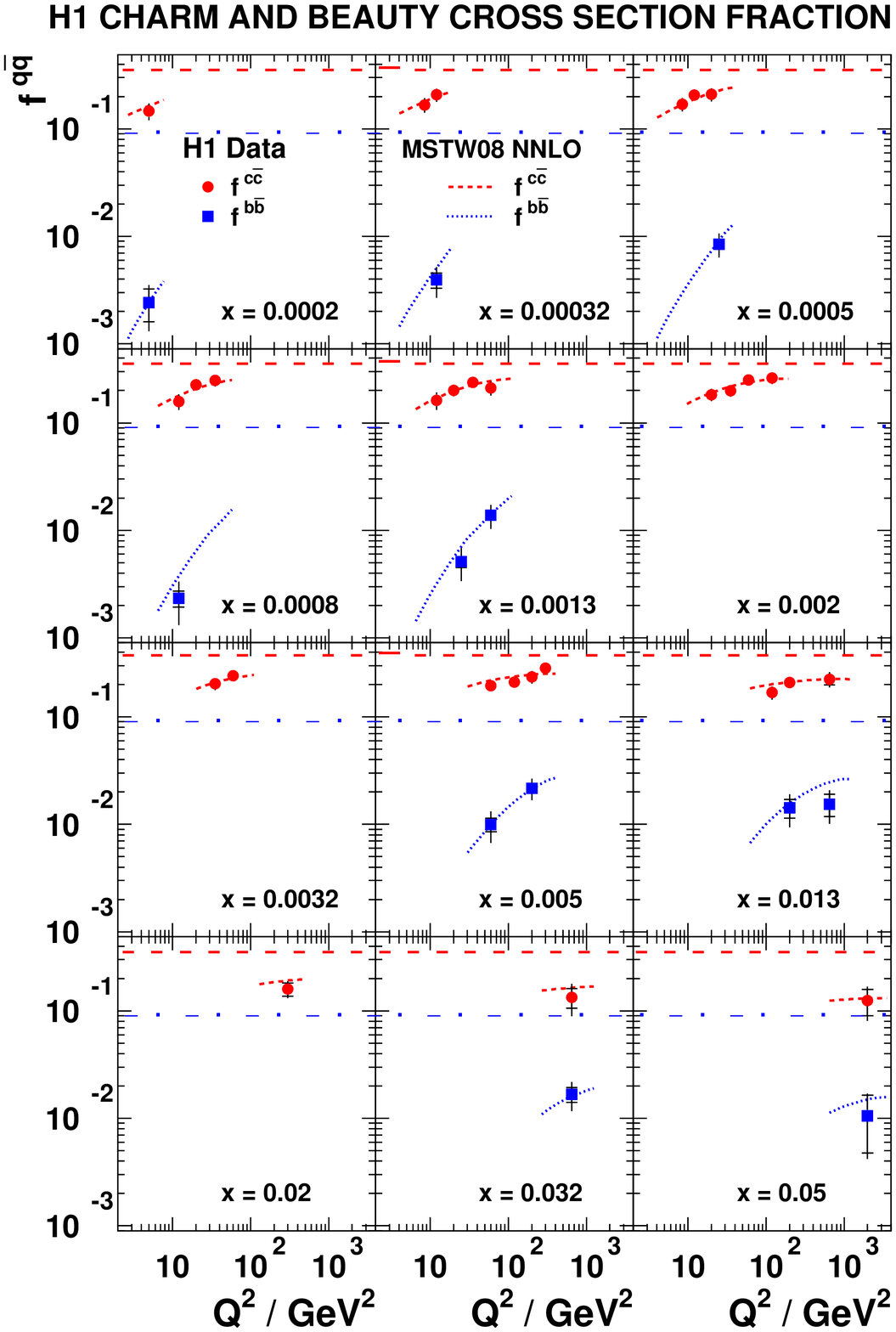}}
\color{black}
\emp%
\hspace{2mm}%
\bmp{c}{0.60\linewidth}
{\includegraphics[trim=10 30 50 70,clip,%
width=0.99\linewidth]{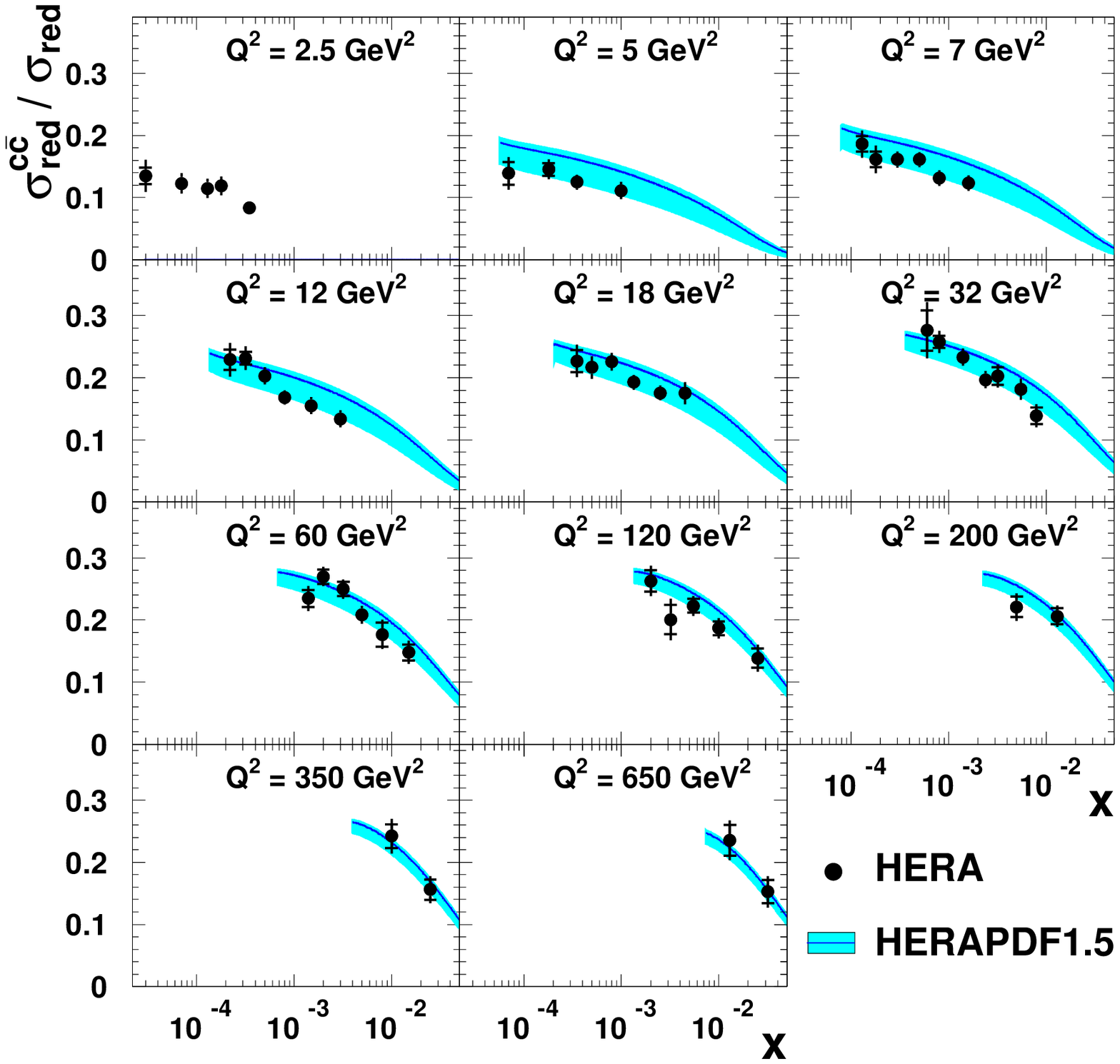}}
 \color{red}
 \put(-280,77){- - - - - - - - - - - - - - - - - - - - - - -}
 \put(-280,144){- - - - - - - - - - - - - - - - - - - - - - - - - - - - - - - - - - -}
 \put(-280,213){- - - - - - - - - - - - - - - - - - - - - - - - - - - - - - - - - - -}
 \put(-280,279){- - - - - - - - - - - - - - - - - - - - - - - - - - - - - - - - - - -}
\emp
\caption{
Fraction of charm and beauty contributions to the inclusive DIS cross section 
as a function of $x$ and $Q^2$~\cite{h1ltt_hera1,HERAcharmcomb,DIScomb}. 
Also shown (curves) is a GMVFNS prediction by 
MRST~\cite{MRST04} and HERAPDF1.5. The dashed and dash-dotted lines are the asymptotic
limits for charm and beauty from Eq.~(\ref{eq:cfrac}).
}
\label{fig:cbfrac}
\end{figure}

\fig{cbfrac} shows the fraction of the heavy-flavour 
component in the total inclusive DIS cross section:
$f^{q\bar{q}}=F_2^{q\bar{q}}/F_2$ and \redc/\redi.
%
%%%%%%%%%%%%%%%%%%%%%%%%%%%%%%%%%%%%%%%%%%%%%%%%%%%%%%%%%%%%%%%%%%%%%%%%%%%%%%%%%%%%%%%%%%%%%%%%%%%%%%%%%%%%%%%%%%%%%%%%
%
%---- H1 F2c + F2c from vertex, cb/total ratio
%
%
%
%////////////////////////////////////////////////////////////////////////////////////////////////////////
%
As expected, the heavy-flavour fractions increase
with increasing $Q^2$.
For $x \lsim 0.01$, the asymptotic limit is approached 
towards $Q^2 \sim 50\ m_c^2 \sim 100\gev^2$ for charm and 
$Q^2 \sim 50\ m_b^2 \sim 1000\gev^2$ for beauty.
The charm and beauty fractions in the high-$Q^2$ data come close to 
$4/11$ and $1/11$, respectively, stressing the importance of
the heavy-flavour component for the description of inclusive DIS. 
The observed suppression of the heavy-flavour fractions 
for $x \gsim 0.01$ originates from the rising importance of the valence-quark 
contribution to the inclusive DIS cross section in this kinematic domain. 

\subsection{Summary}

Large photon virtuality $Q^2$ provides an additional 
hard scale in the QCD calculations of heavy flavour production 
and allows probing the
parton dynamics inside the proton more directly than photoproduction.
The dominant contribution to the charm and beauty cross sections 
arises from photon-gluon fusion.
For $Q^2 \gg 4m_Q^2$, where the photon virtuality is the dominant hard scale, 
the cross-section behaviour is similar to the one of the inclusive 
cross section for deeply inelastic scattering. At high $Q^2$ and low $x$, 
the naively expected charm and beauty contributions of $4/11$ and $1/11$ are
asymptotically approached.   
For $Q^2 \ll 4m_Q^2$, where the quark mass is the dominant hard scale, 
the cross section behaves essentially like photoproduction, i.e. the 
photon can be approximated to be quasi-real.
NLO QCD predictions using zero-mass schemes (ZMVFNS) fail to describe the data 
in the vicinity of or below the so-called ``flavour threshold'' at
$Q^2 \sim m_Q^2$.
NLO QCD predictions
in the massive scheme (FFNS) give a good description of heavy flavour 
production at HERA over the complete accessible kinematic range. 
NLO predictions in variable flavour number schemes (GMVFNS) are only available
for inclusive quantitities, and perform about equally well. There is no
indication for the need of resummation of $\ln Q^2/m_Q^2$ terms at HERA
energies.
In particular for charm, the uncertainties from QCD corrections beyond NLO
and from the modelling of fragmentation are considerably larger than
the experimental uncertainties of the measured cross sections.
Improved QCD calculations would therefore be highly welcome.

\newpage

% Interpretation and impact on other experiments
\section{Measurement of QCD parameters, proton structure, and impact 
on LHC and other experiments}
\label{sect:QCD}

So far the emphasis was on direct cross-section measurements from the HERA 
data and on the comparison to and performance of different theoretical 
approximations for the perturbative QCD expansion. In this section 
the extraction of more fundamental QCD parameters and parametrisations 
will be discussed, which are of direct relevance to all high energy 
physics processes and to the Standard Model of particle physics in 
general. 

%%%%%%%%%%%%%%%%%%%%%%%%%%%%%%%%%%%%%%%%%%%%%%%%%%%%%%%%%%%%%%%%%%%%%%%%%%%%%%%%%%
%%%%%%%%%%%%%%%%%%%%% Fragmentation %%%%%%%%%%%%%%%%%%%%%%%%%%%%%%%%%%%%%%%%%%%%%%
%%%%%%%%%%%%%%%%%%%%%%%%%%%%%%%%%%%%%%%%%%%%%%%%%%%%%%%%%%%%%%%%%%%%%%%%%%%%%%%%%%

\subsection{Measurement of charm fragmentation functions and fragmentation
                fractions}

As outlined in Section \ref{sect:frag}, fragmentation fractions, i.e. the 
probability of a quark of a given flavour to form a specific final state 
hadron, and fragmentation functions, parametrising the fraction of the 
energy or momentum of the final state quark which will be taken by the 
final state hadron, are essential to relate theoretical QCD calculations at 
parton level to measurable hadronic final states.

Studies of the fragmentation process are based on a complete reconstruction 
of the final-state hadron. 
The statistics accessible at HERA for 
fully-reconstructed beauty hadrons 
is extremely low due to the moderate beauty-production cross section and 
small branching ratios. On the other hand, HERA is effectively a charm factory,
with about $10^8$ charm events recorded to tape.
Therefore, only the fragmentation of charm
quarks has been studied by H1 and ZEUS.
Charm fragmentation has been studied in both the DIS 
and photoproduction regimes.
A comparison between these results and \epem measurements
provides a so far unique test of the fragmentation universality in the
heavy-flavour sector for colour-neutral (electromagnetic) vs. coloured 
(strongly interacting) initial states.

%%%%%%%%%%%%%%%%%%%%%%%%%%%%%%%%%%%%%%%%%%%%%%%%%%%%%%%%%%%%%%%%%%%%%%%%%%%%%%%%%%
%%%%%%%%%%%%%%%%%%%%% Fragm function %%%%%%%%%%%%%%%%%%%%%%%%%%%%%%%%%%%%%%%%%%%%%
%%%%%%%%%%%%%%%%%%%%%%%%%%%%%%%%%%%%%%%%%%%%%%%%%%%%%%%%%%%%%%%%%%%%%%%%%%%%%%%%%%

\subsubsection{Charm fragmentation function}
\label{sect:cFragFunc}

The explicit reconstruction of a $D^*$ meson in the final state has 
the optimal signal sensitivity of all fully reconstructed charm final states
(cf. \Tab{r1} and \Tab{cDIS}).
Thus, it has been used for studies of the 
non-perturbative charm fragmentation function (Section \ref{sect:frag}).
Since the momentum of the charm quark is not measured in the detector,
the fragmentation function is not a directly accessible quantity.
It can be approximated either by jets to which 
a reconstructed $D^*$ meson is associated (for high-$p_T$ events)
or by the overall energy flow in the event hemisphere around the $D^*$ 
(for production close to the kinematic threshold).
Parameters of the fragmentation function were extracted from the data 
by fitting corresponding predictions to the measured normalised differential 
cross sections as a function of
$$z_{\mathrm{jet}}=\frac{(E+P)_D}{(E+P)_{\mathrm{jet}}}$$ and
$$z_{\mathrm{hem}}=\frac{(E+P)_D}{\sum_{\mathrm{hem}} (E+P)}$$ 
for the jet and hemisphere methods, respectively.
The tuning of the fragmentation parameters was done based on 
Monte Carlo simulations~\cite{PYTHIA,RAPGAP} with similar JETSET \cite{JETSET}
settings
or on NLO QCD calculations using the same ``heavy quark'' definition and 
similar schemes for the cancellation of collinear and infrared 
divergences\footnote{Due to the heavy quark masses most of these terms are
not really divergent. Nevertheless, events with ``similar'' topologies and
large but almost cancelling weights are produced in correlated groups.}   
in photoproduction and DIS~\cite{FMNR,hvqdis}.

\begin{figure}[!tb]
\centering
\bmp{c}{0.45\linewidth}
\includegraphics[width=0.95\linewidth]{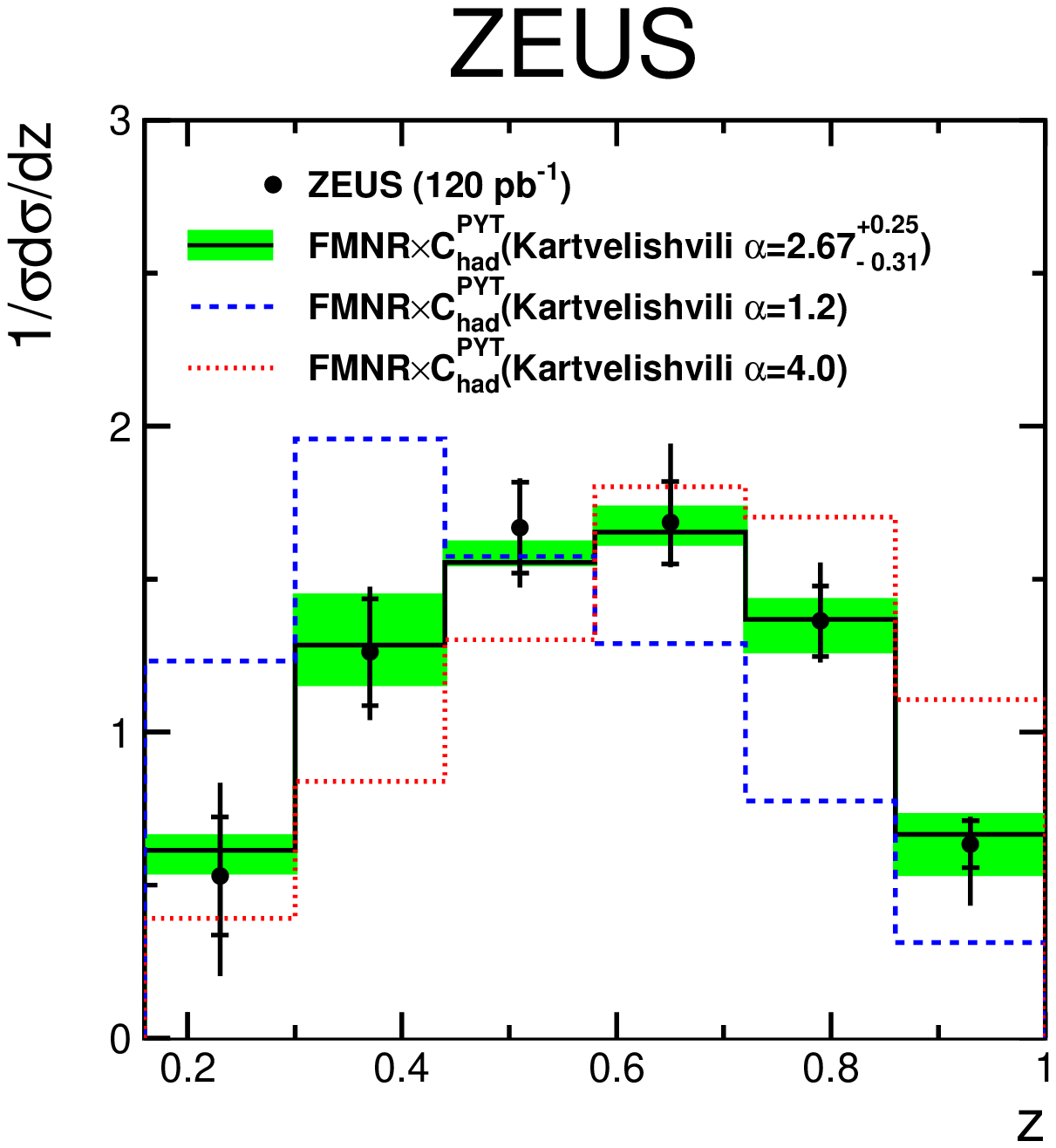}%
\put(-40,160){\Large (a)}
\emp%
\bmp{c}{0.35\linewidth}
\includegraphics[width=0.95\linewidth]{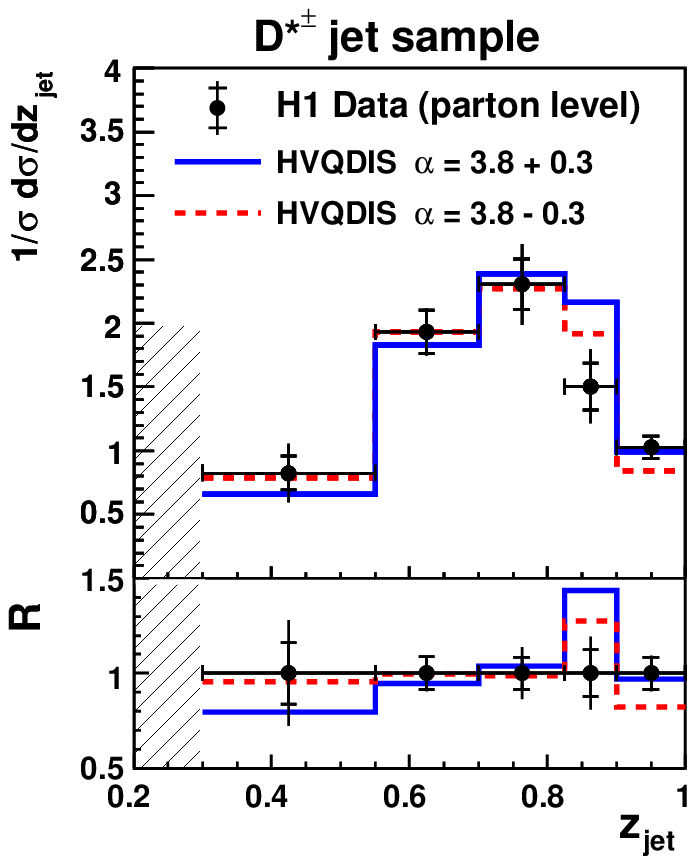}%
\put(-25,160){\Large (b)}
\emp\\
\bmp{c}{0.35\linewidth}
\includegraphics[width=\linewidth]{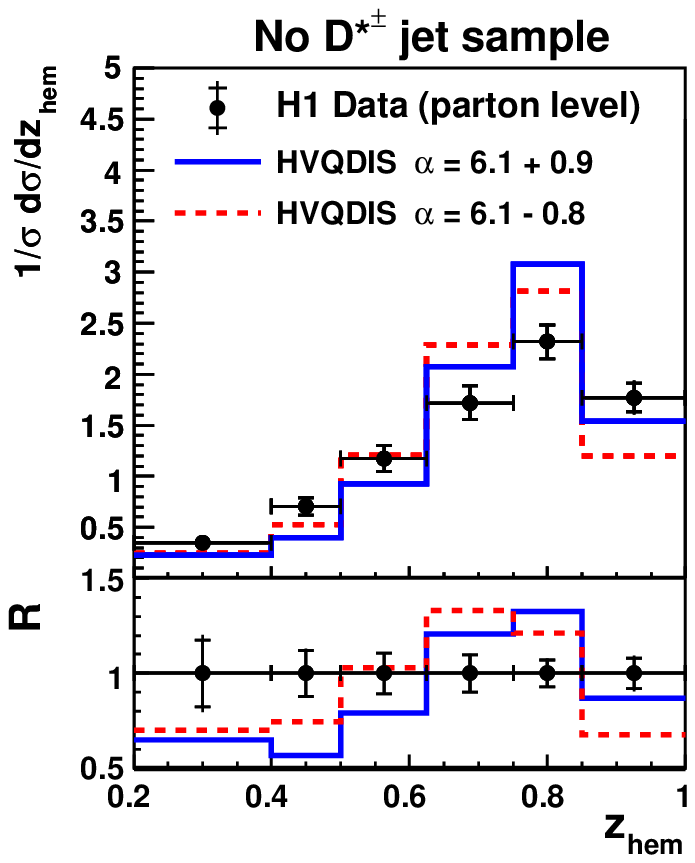}%
\put(-25,160){\Large (c)}
\emp%
\bmp{c}{0.35\linewidth}
\includegraphics[width=\linewidth]{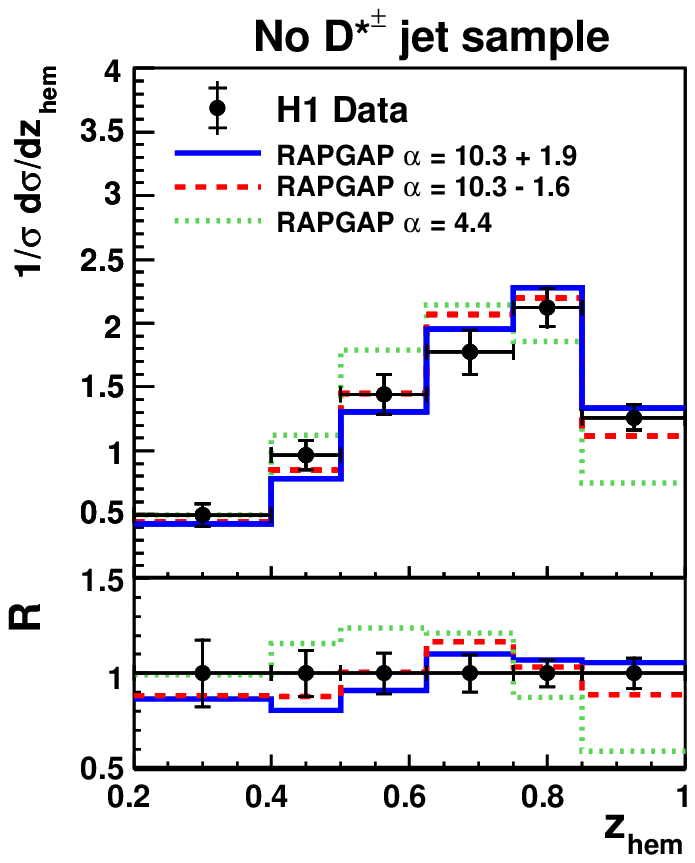}%
\put(-25,160){\Large (d)}
\emp
\caption{The normalised $D^*$ cross sections as a function of $z$
from the (a) ZEUS~\pcite{ZEUSfrag2009} and (b), (c), (d) H1~\pcite{H1frag2009} analyses.
The hard (b) and the threshold (c-d) regions are shown from the H1 measurement.
The statistical (inner error bars) and the statistical and systematic
uncertainties added in quadrature (outer error bars) are shown separately.
The data are compared to the NLO QCD predictions (a-c) from 
HVQDIS~\pcite{hvqdis} and FMNR~\pcite{FMNR} as well as
to the RAPGAP MC~\pcite{RAPGAP} (d) based on the string fragmentation model.
The parameter of the Kartvelishvili fragmentation function %~\cite{Kartvelishvili:1977pi} 
was tuned to the data in each case.
}
\label{fig:ffunc_data}
\end{figure}

The H1~\cite{H1frag2009} and ZEUS~\cite{ZEUSfrag2009} measurements were done 
in the DIS and photoproduction
regimes, respectively, utilising \herai data sets.
The H1 experiment investigated both the high $p_T$ and the threshold regions,
in order to cover the largest possible phase space, 
while ZEUS restricted the measurement to the high-$p_T$ regime, in order
to reach small $z$ values without strongly biasing the distributions, and 
in order to minimise perturbative fragmentation factorisation effects within 
the data set.
This led to H1 selecting jets with $E^{*}_T > 3\gev$ in the 
$\gamma^{*}p$  rest-frame and ZEUS cutting on $E_T > 9\gev$ for jets in the 
laboratory frame.
\Fig{ffunc_data} shows the measured normalised differential cross sections 
as well as predictions after tuning of the fragmentation model.
%
%%%%%%%%%%%%%%%%%%%%%%%%%%%%%%%%%%%%%%%%%%%%%%%%%%%%%%%%%%%%%%%%%%%%%%%%%
%%%%%%%%%%%%%%%%%%%%%%%%%%%%%%%%%%%%%%%%%%%%%%%%%%%%%%%%%%%%%%%%%%%%%%%%%
%
The extracted parameters of the fragmentation functions for the HVQDIS and 
FMNR NLO calculations, which should also be applicable to the conceptually
similar MNR \cite{MNR} calculations in hadroproduction, are presented in 
\tab{ffunc_data}.
%
%%%%%%%%%%%%%%%%%%%%%%%%%%%%%%%%%%%%%%%%%%%%%%%%%%%%%%%%%%%%%%%%%%%%%%%%%
\begin{table}[tb]
	\begin{center}
	\renewcommand{\arraystretch}{1.15}
	\begin{tabular}{|c|c|c|c|c|}
	\hline
	 & \multicolumn{3}{|c|}{H1~\pcite{H1frag2009}} & ZEUS~\pcite{ZEUSfrag2009}\\
	\hline
	kinematics & threshold & \multicolumn{2}{|c|}{high $p_T$} & high $p_T$\\
	$\hat{s}, \gev^2$ & $\sim 36$ & \multicolumn{2}{|c|}{$\sim 100$} & $\sim 350$ \\
	\hline
	$z$ method & hem & hem & jet & jet\\
	\hline
	Kartvelishvili ~\cite{Kartvelishvili:1977pi} & & & & \\
	$\alpha$ & $6.1^{+0.9}_{-0.8}$ & $3.3^{+0.4}_{-0.4}$ & $3.8^{+0.3}_{-0.3}$ & $2.7^{+0.2}_{-0.3}$ \\
	$(\chi^2/ndof)$ & $(37.6 / 4)$& $(4.4 / 4)$ & $(4.9 / 3)$ & n.a. \\
	\hline
	Peterson~\cite{Peterson:1982ak} & & & & \\
	$\varepsilon$ & $0.007^{+0.001}_{-0.001}$ & $0.068^{+0.015}_{-0.013}$ & $0.034^{+0.004}_{-0.004}$ & $0.079^{+0.013}_{-0.009}$ \\
	$(\chi^2/ndof)$ & $(38.6/4)$ & $(18.3 / 4)$ & $(23.3 / 3)$ & n.a. \\
	\hline
	\end{tabular}
	\caption{Parameters of fragmentation function extracted for the NLO
	QCD predictions by H1 and ZEUS. }
        \label{tab:ffunc_data}
	\end{center}
\end{table}
%%%%%%%%%%%%%%%%%%%%%%%%%%%%%%%%%%%%%%%%%%%%%%%%%%%%%%%%%%%%%%%%%%%%%%%%%
%
A few observations can be made from the distributions:
\begin{itemize}
\item
As expected, due to the high quark mass, a charm meson retains a large fraction
of the momentum of a charm quark. 
Therefore, the charm fragmentation 
is much harder than that of light hadrons. 

\item
The fragmentation of charm quarks to $D^*$ mesons 
near to the kinematic threshold is harder than in the region away
from the threshold (cf. \fig{ffunc_data} (b) and (c)).
This can be qualitatively understood as a consequence of the fact that the
phase space available for the production of additional particles is smaller 
near threshold. 
As a result, the fragmentation parameters extracted in the two
kinematic regions are significantly different.
This leads to the conclusion that the different kinematic regimes can not be 
described simultaneously within the framework of the independent
fragmentation function. This is to be expected, since the nonperturbative
phase space suppression is incompletely modeled in this approach.   

\item
NLO QCD calculations in conjunction with an independent fragmentation
fail to describe the data close to the kinematic threshold: 
$\chi^2/ndof\approx 38/4$ (see \fig{ffunc_data} (c) and \tab{ffunc_data}).
However, in the same phase space MC simulations can be tuned to 
provide a reasonable description of the $z$ distribution 
in the data: $\chi^2/ndof\approx 3/4$ (see \fig{ffunc_data} (d)).
This might be due to the proper treatment of phase space effects in the MC,
which are missing in the independent fragmentation approach used for the NLO
predictions.

\item
The Peterson fragmentation provides a much worse description of the data 
than the Kartvelishvili function. This has also been observed 
elsewhere~\cite{Seuster:2005tr}. 

\item
The jet and hemisphere methods in the region
where both are applicable, i.e. away from the threshold, yield
similar results for the Kartvelishvili parametrisation, while they remain 
somewhat different in the Peterson case. Thus again, the Kartvelishvili
parameterisation seems to be preferred.
\end{itemize}

For some recent measurements~\cite{HERAcharmcomb,h1dstar_hera2,zeusdpluslambda,
h1dstarhighQ2,zeusdstar_hera2,zeusdplus_hera2,HERAdstcomb}, 
these results have been used explicitly 
to model the fragmentation for the comparison of theoretical predictions to the 
charm HERA data (see \Sect{cbDIS}) and 
for the extrapolation to the full phase space in the context of the extraction
of the charm reduced cross sections (see \Sect{cbDIS} and \Sect{cbF2}).
This has shown that there are significant theory uncertainties due to
fragmentation.
A consistent phenomenological reanalysis of the H1 and ZEUS data is 
needed in order to resolve the differences observed in different
kinematic domains, which originate from neglecting perturbative evolution
and phase space effects, hopefully
resulting in an important reduction of the related theory uncertainties.
It is worth mentioning that the complete \heraii dataset, which has not yet
been analysed in this context, is available in principle for this purpose.

\begin{table}[tb]
	\begin{center}
	\renewcommand{\arraystretch}{1.15}
	\begin{tabular}{|c|c|c|c|c|}
	\hline
	 & \multicolumn{3}{|c|}{H1~\pcite{H1frag2009}} & ZEUS~\pcite{ZEUSfrag2009}\\
	\hline
	kinematics & threshold & \multicolumn{2}{|c|}{high $p_T$} & high $p_T$\\
	$\hat{s}, \gev^2$ & $\sim 36$ & \multicolumn{2}{|c|}{$\sim 100$} & $\sim 350$ \\
	\hline
	$z$ method & hem & hem & jet & jet\\
	\hline
	Kartvelishvili & & & & \\
	$\alpha$ & $7.5^{+1.3}_{-1.2}$ & $3.3^{+0.4}_{-0.4}$ & $3.1^{+0.3}_{-0.3}$ & n.a. \\
	$(\chi^2/ndof)$ & $(37.6 / 4)$& $(4.4 / 4)$ & $(4.9 / 3)$ & n.a. \\
	\hline
	Peterson & & & & \\
	$\varepsilon$ & $0.010^{+0.003}_{-0.003}$ & $0.049^{+0.012}_{-0.010}$ & $0.061^{+0.011}_{-0.009}$ & 
	$0.062^{+0.011}_{-0.008}$ \\
	$(\chi^2/ndof)$ & $(38.6/4)$ & $(18.3 / 4)$ & $(23.3 / 3)$ & n.a. \\
	\hline
	\end{tabular}
	\caption{Parameters of fragmentation function extracted by H1 and ZEUS for the 
	fragmentation model in PYTHIA with other parameter settings set to the default values. 
	Note, that also a set of fragmentation parameters was extracted by H1~\pcite{H1frag2009} 
	for the ALEPH PYTHIA tune~\pcite{Schael:2004ux}.}
        \label{tab:ffuncMC_data}
	\end{center}
\end{table}

Table \ref{tab:ffuncMC_data} shows the equivalent results extracted from 
LO+PS MCs using the ``default'' JETSET settings as used e.g. by the PYTHIA
and RAPGAP MCs. In this case a perturbative evolution of the fragmentation 
function
is partially included through the parton showering, and phase space corrections
are applied. Despite the poor $\chi^2$, the Peterson parameters extracted from 
the intermediate and high $p_T$ jet samples now agree with each other, as well 
as with the corresponding default parameter 0.05 extracted from $e^+e^-$ 
collisions
\cite{ZEUSfrag2009}. This confirms the universality of the nonperturbative 
part of fragmentation. Kartvelishvili parameters are unfortunately not 
available for all data sets and can hence not be compared.
Near threshold, even the MC model does not yield the same fragmentation 
parameters, and the $\chi^2$ is generally bad. 
This indicates that still not all threshold effects might have
been fully accounted for. 

%%%%%%%%%%%%%%%%%%%%%%%%%%%%%%%%%%%%%%%%%%%%%%%%%%%%%%%%%%%%%%%%%%%%%%%%%%%%%%%%%%%
%%%%%%%%%%%%%%%%%%%%%%% Fragmentation fractions %%%%%%%%%%%%%%%%%%%%%%%%%%%%%%%%%%%
%%%%%%%%%%%%%%%%%%%%%%%%%%%%%%%%%%%%%%%%%%%%%%%%%%%%%%%%%%%%%%%%%%%%%%%%%%%%%%%%%%%

\subsubsection{Charm fragmentation fractions and ratios}
\label{sect:cFragFrac}

The fractions of $c$ quarks hadronising into a particular charm hadron,
$f(c \to D,\Lambda_c)$, have been measured by H1 and ZEUS in the 
DIS~\cite{h1dmesons,zeusdmesons,zeusdpluslambda} and
photoproduction~\cite{ZEUSfrag2005,ZEUSfrag2013} regimes.
The measurements were done for $D^+$, $D^0$, $D^{*+}$, $D_s^+$
and $\Lambda_c$ based on full reconstruction of the charm-hadron decays.
The fragmentation fractions were extracted from integrated visible
cross sections.
The typical fiducial phase space of the charm hadrons was defined by
$p_T(D,\Lambda_c) > 3\gev$ and $\eta(D,\Lambda_c)<1.6$.
The fragmentation fractions were extracted with the additional constraint
that the sum of the fractions for all weakly-decaying 
open-charm hadrons (i.e. the ground states from the point of view of 
strong and electromagentic interactions) has to be equal to unity.
This was done by a constrained fit in H1~\cite{h1dmesons} and 
by an advanced procedure called equivalent phase space treatment 
in ZEUS~\cite{zeusdmesons}.
In addition to direct production, 
such experimentally measured fragmentation fractions 
include also all possible decay chains of excited charm hadrons.

\begin{figure}[!tb]
\centering
\includegraphics[width=0.75\linewidth]{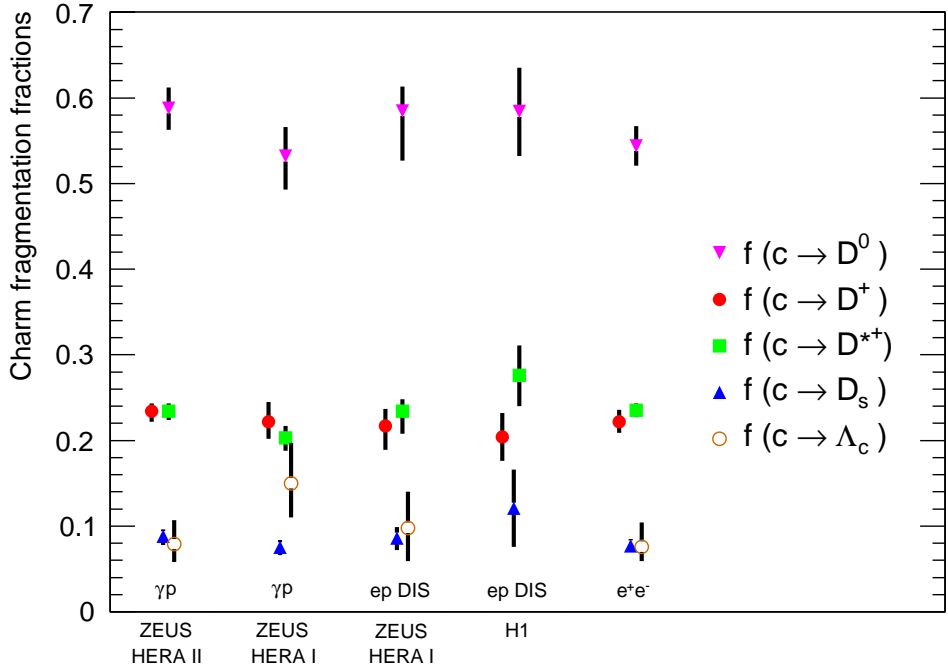}%
\caption{Fractions of charm quarks hadronising into a particular 
charm hadron~\pcite{ZEUSfrag2013}. 
Measurements from HERA are 
compared to the combined \epem data.
Different hadron species are shown with different marker types.
}
\label{fig:Fragmfract}
\end{figure}

\Fig{Fragmfract} shows a compilation of all available charm fragmentation 
fraction measurements. 
The HERA data are compared to an average of \epem 
measurements~\cite{Charmfrac, ATLAS:2011fea}.
To allow a direct comparison, all measurements have been 
corrected~\cite{CharmfragComb} 
to the decay branching fractions from PDG 2010~\cite{pdg2010}.
%
%%%%%%%%%%%%%%%%%%%%%%%%%%%%%%%%%%%%%%%%%%%%%%%%%%%%%%%%%%%%%%%%%%%%%%%%%
%%%%%%%%%%%%%%%%%%%%%%%%%%%%%%%%%%%%%%%%%%%%%%%%%%%%%%%%%%%%%%%%%%%%%%%%%
%
The HERA data reach very high precision, benefiting from
a partial cancellation of some systematic uncertainties in the ratio. 
In particular, the recent ZEUS measurement~\cite{ZEUSfrag2013} is based
on the full \heraii data sample and made use of the finalised tracking
with lifetime tagging for $D^0$, $D^+$ and $D_s^+$.
This allowed to reduce both statistical and systematic uncertainties.
The ultimate precision achieved with ZEUS \heraii data alone is fully 
competitive with the precision of the \epem average from several experiments.
All data from DIS, photoproduction and \epem collisions
are in agreement within the high accuracy of the data.
This demonstrates that the charm fragmentation fractions are 
independent of the production mechanism, and therefore supports
the hypothesis of universality of heavy-quark fragmentation.
The agreement between the fragmentation fractions has been checked 
quantitatively in the context of a combination~\cite{CharmfragComb}.

%%%%%%%%%%%%%%%%%%%%%%%%%%%%%%%%%%%%%%%%%%%%%%%%%%%%%%%%%%%%%%%%%%%%%%%%%
\begin{figure}[!tb]
\centering
\includegraphics[width=0.75\linewidth]{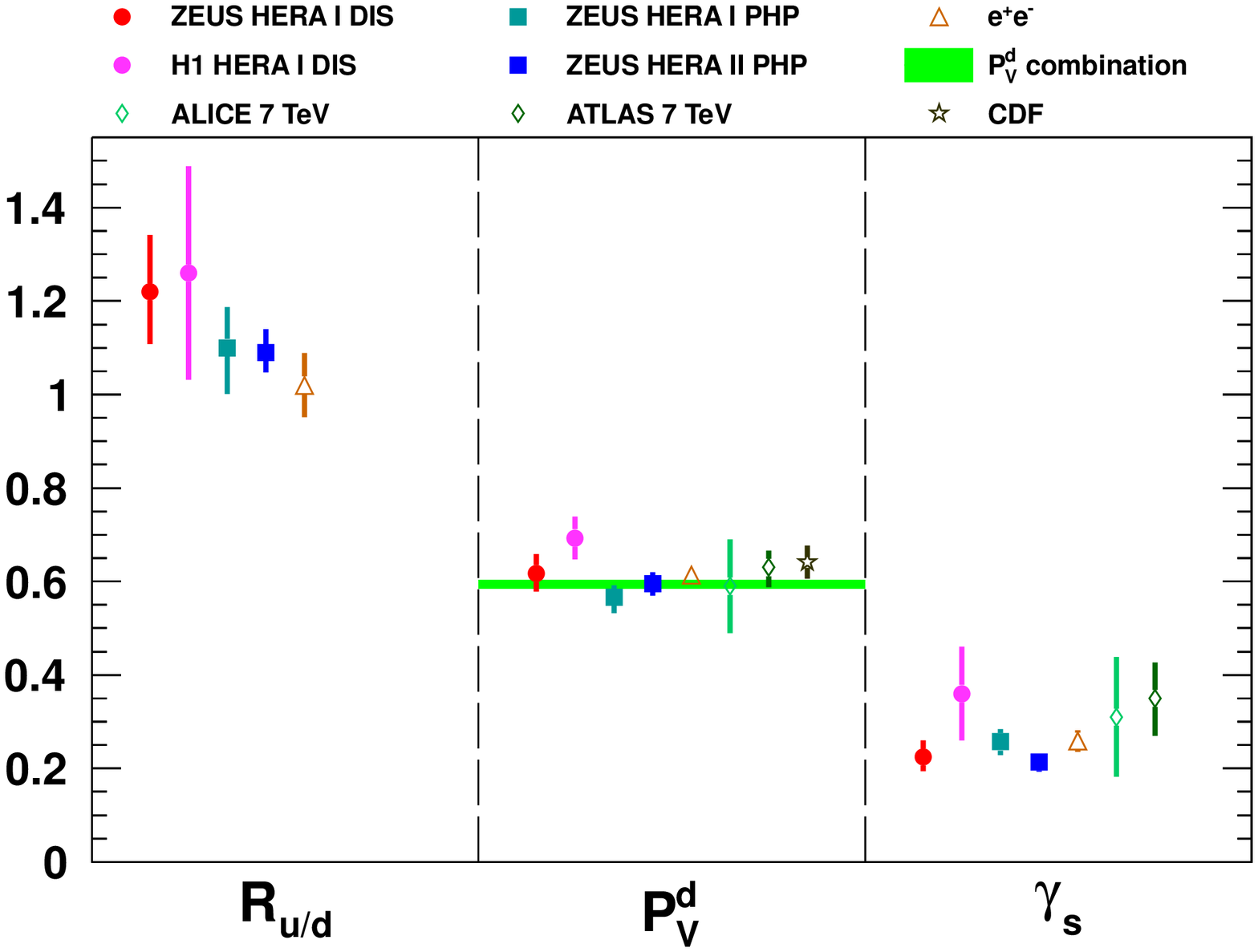}%
\caption{Fragmentation ratios \Rud, \Pdv, \gs 
measured at HERA and elsewhere~\pcite{h1dmesons,zeusdmesons,zeusdpluslambda,%
ZEUSfrag2005,ZEUSfrag2013,Charmfrac,CDFPdv,ALICEPdv,ALICEgammas,ATLAS:2011fea}.
Different measurements are shown with different marker types.
The error bars indicate the statistical and systematic uncertainties
added in quadrature.
The branching-ratio uncertainties are not shown due to the
high degree of correlation between experiments
and can be found in the original papers.
The filled band shows the result of the \Pdv combination~\pcite{David:2007iv}.
}
\label{fig:RPg}
\end{figure}
%%%%%%%%%%%%%%%%%%%%%%%%%%%%%%%%%%%%%%%%%%%%%%%%%%%%%%%%%%%%%%%%%%%%%%%%%
%
In addition to the fragmentation fractions, various charm 
fragmentation ratios were extracted:
the ratio of the neutral to charged $D$-meson production rates, \Rud,
the fraction of the charged D mesons produced in a vector state, \Pdv,
and the strangeness-suppression factor, \gs. 
\Fig{RPg} shows a comparison of HERA measurements~\cite{ZEUSfrag2013,
ZEUSfrag2005,h1dmesons,zeusdmesons,zeusdpluslambda} 
with results obtained in
\epem collisions~(numbers quoted in~\cite{zeusdmesons} using average 
from~\cite{Charmfrac})
and hadroproduction by CDF~\cite{CDFPdv}, ALICE~\cite{ALICEPdv, ALICEgammas} 
and ATLAS~\cite{ATLAS:2011fea}.
Also shown is a global average of \Pdv results from \epem, photo- and 
hadroproduction~\cite{David:2007iv}%
\footnote{The paper as well as other recent 
measurements~\pcite{CDFPdv,ALICEPdv,ATLAS:2011fea} report $P_V$ values,
which correspond to the fraction of charged \emph{and neutral} D mesons
produced in the vector state.
However the measurements rely on isospin symmetry assumption, 
which makes $P_V$ identical to \Pdv.},
which also 
includes some \epem, $ep$ and CDF data shown separately.
Note, that the uncertainty of the average is driven by the
$\pi^- \rnge A$ result from the WA92 experiment~\cite{Adinolfi:1999ih},
which is quoted with statistical uncertainty only, 
i.e. treating all systematic uncertainties, including the
branching-ratio uncertainty, as correlated between $D^+$ 
and $D^{*+}$.
The fragmentation ratios extracted from HERA, \epem and hadroproduction 
data agree within experimental uncertainties.
The ultimate precision achieved with the full \heraii data set
is competitive with the most precise measurements in other experiments.
Various simple theory expectations can be tested against the data.
The \Rud measurements are slightly above, but still in agreement within 
uncertainties, with the isospin invariance expectation of unity.
The \Pdv measurements are smaller than the naive spin-counting expectation
$0.75$ and the string fragmentation prediction 
$0.66$~\cite{Buchanan:1987ua, Pei:1996kq}.

Excited charm mesons have also been studied with the ZEUS detector
using the \herai~\cite{ZEUSDmes2009} and \heraii~\cite{ZEUSDmes2013} datasets.
Some parameters of the orbitally-excited charm states \donecz with $J^P=1^+$ and
\dtwocz with $J^P=2^+$ as well as charm-strange state \dsone
were measured.
The masses and widths were found to be in good agreement
between the two measurements and with the PDG average.
The helicity parameters $h$ for \donez and \dsone were found to be 
in agreement with \epem measurements.
The measured \donez parameter was found to prefer 
a mixture of $S$ and $D$ waves in the decay to $D^{*+}\pi^-$, 
although it is also consistent with a pure $D$ wave.
In addition, fragmentation fractions and ratios of branching 
ratios were extracted. 
For some parameters HERA can provide important or even
so far unique information.
For example, the fragmentation fractions for the studied 
excited mesons are so far very poorly experimentally determined.
The \donec and \donec fragmentation fractions were measured 
for the first time~\cite{ZEUSDmes2013}:
$$f(c \to D_1^+) = 4.6 \pm 1.8 \mathrm{(stat.)} 
{}^{+2.0}_{-0.3} \mathrm{(syst.)}\%,$$
$$f(c \to D_2^{*+}) = 3.2 \pm 0.8 \mathrm{(stat.)} 
{}^{+0.5}_{-0.2} \mathrm{(syst.)}\%.$$

%%%%%%%%%%%%%%%%%%%%%%%%%%%%%%%%%%%%%%%%%%%%%%%%%%%%%%%%%%%%%%%%%%%%%%%%%%%%%%%%%%
%%%%%%%%%%%%%%%%%%%%% Proton structure %%%%%%%%%%%%%%%%%%%%%%%%%%%%%%%%%%%%%%%%%%%
%%%%%%%%%%%%%%%%%%%%%%%%%%%%%%%%%%%%%%%%%%%%%%%%%%%%%%%%%%%%%%%%%%%%%%%%%%%%%%%%%%

\subsection{Measurement of parton density functions}
\label{sect:cbF2}

The gluon PDF at low- and medium-$x$ values is mostly constrained by 
the scaling violations of the inclusive structure function $F_2$.
In contrast, heavy-quark production at HERA provides a direct probe of the gluon
momentum distribution in the proton through the $\gamma^{*}g \to c\bar{c}$ process.
Such direct measurements are complementary to the indirect approach.

Already the very early charm measurements were used to directly extract 
the gluon PDF, as was done by the H1 collaboration in~\cite{H1gluon}.
The gluon densities extracted from the charm data were found to be in agreement 
with the result of a QCD analysis of inclusive $F_2$ measurements, although
the charm measurement was limited by statistics.

\begin{figure}[p]
\bmp{c}{\linewidth}
\center
\includegraphics[width=.45\linewidth]{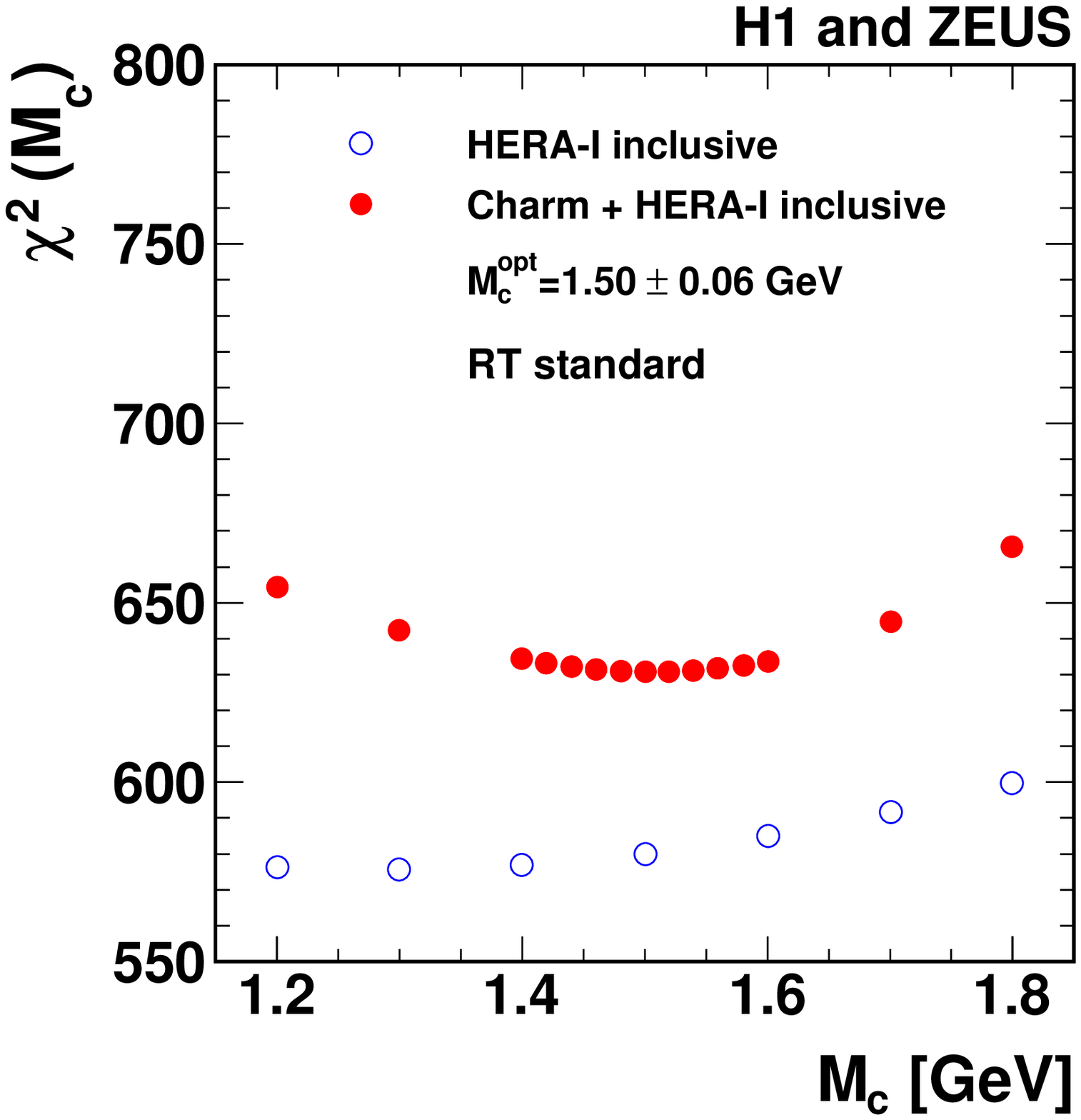}
\put(-60,175){\Large (a)}
\includegraphics[width=.45\linewidth]{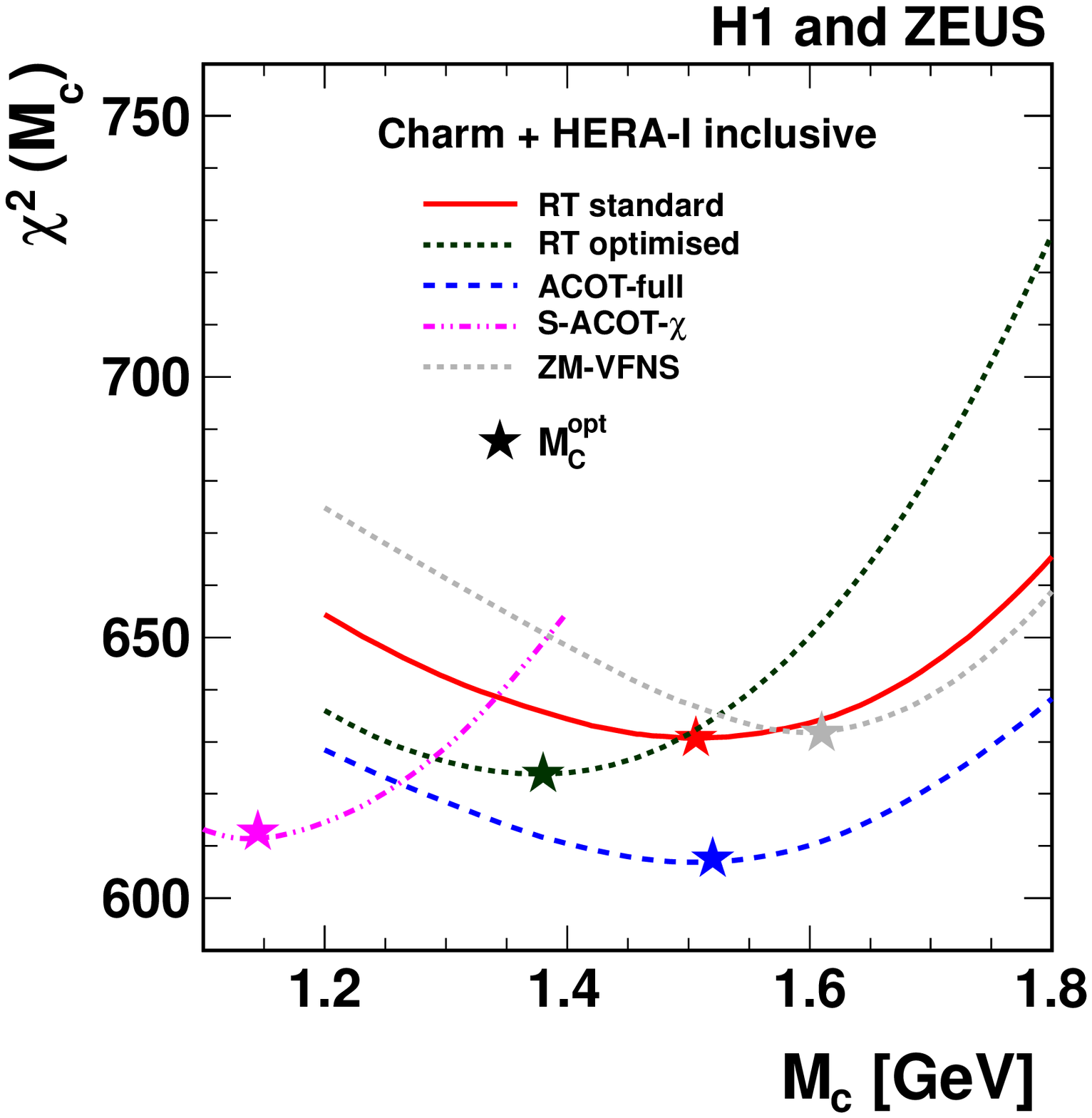}
\put(-60,175){\Large (b)}
\caption{(a) The values of $\chi^2(\mct)$ for the PDF fit to the combined HERA DIS 
data~\pcite{HERAcharmcomb} in the RT standard scheme~\pcite{TR}. 
The open symbols indicate the results of the fit to inclusive 
DIS data only. The results of the fit including the combined charm data are shown by 
filled symbols.
(b) The values of $\chi^2(\mct)$ for the PDF fit to the combined HERA inclusive DIS 
 and charm measurements~\pcite{HERAcharmcomb}. 
 Different heavy flavour schemes are used in the fit and presented 
 by lines with different styles. 
 The values of \mcto\ for each scheme are indicated by the stars.}
 \label{fig:charm_vfn_scan} 
\emp\\[15pt]
\bmp{c}{\linewidth}
\center
\includegraphics[width=.45\linewidth]{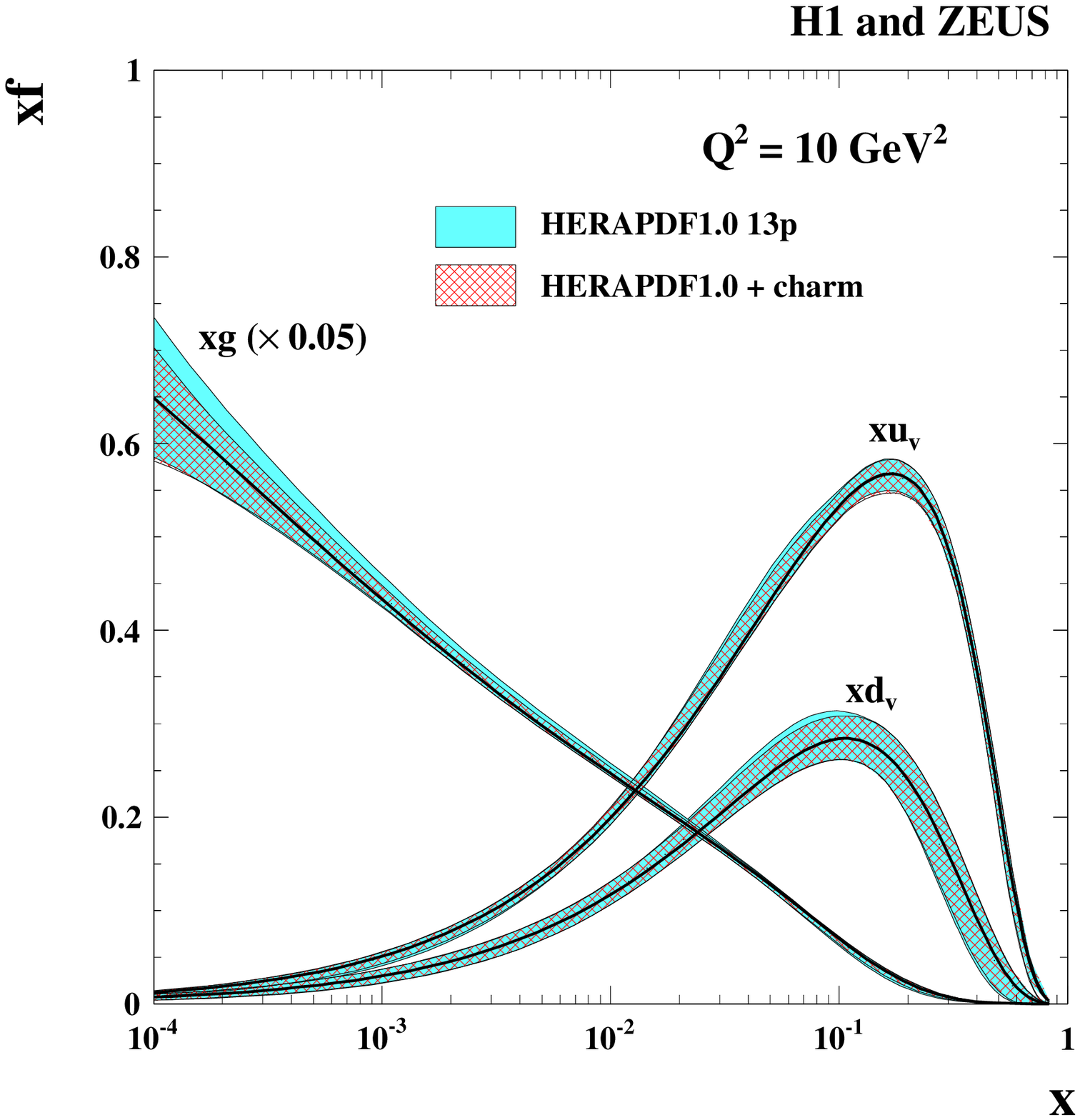}
\put(-60,175){\Large (a)}
\includegraphics[width=.45\linewidth]{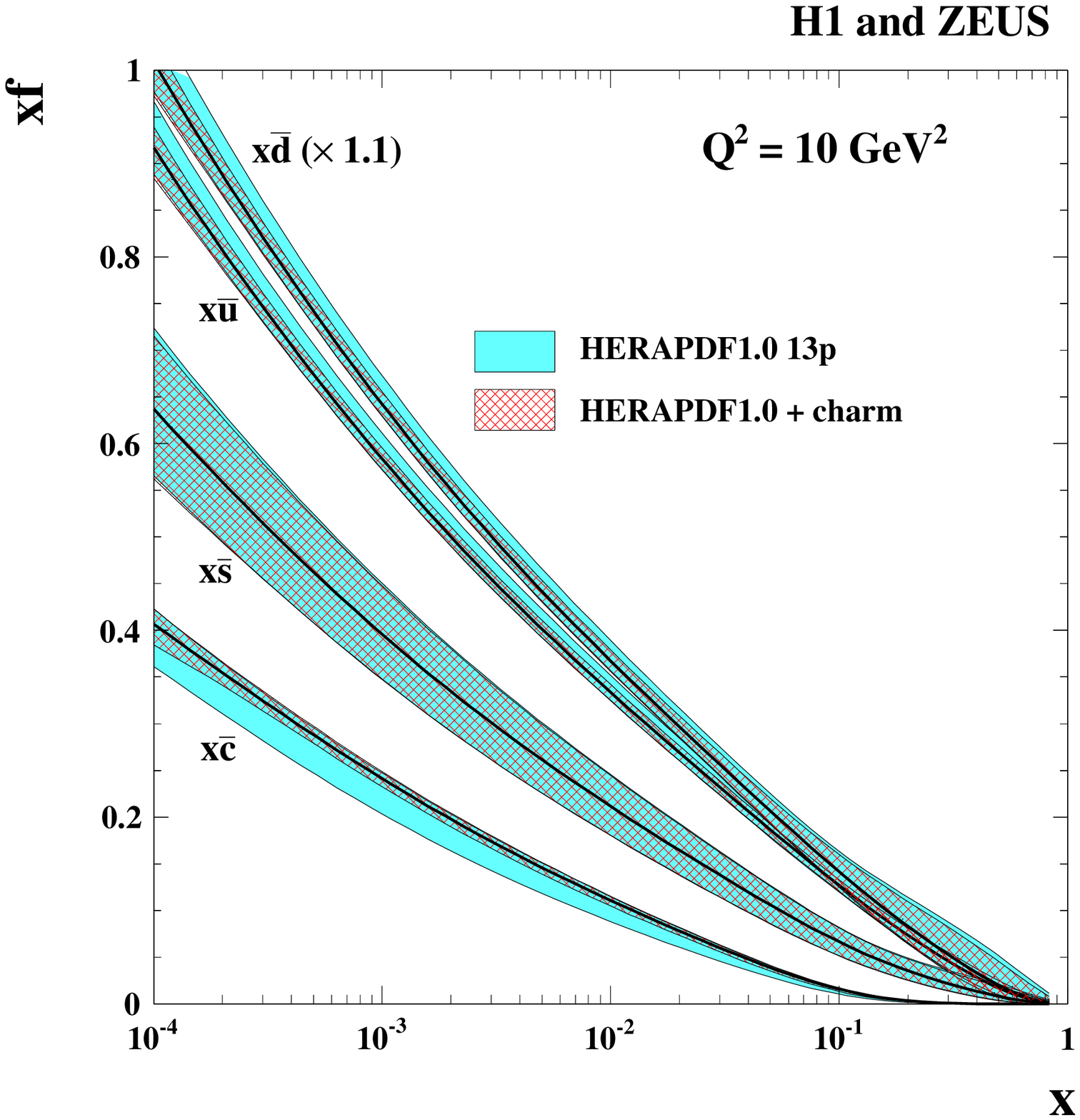}
\put(-60,175){\Large (b)}
\caption{Parton density functions~\pcite{HERAcharmcomb} $x\cdot f(x,Q^2)$ 
with $f=g,u_v,d_v,\overline{u},\overline{d},\overline{s},\overline{c}$ for 
(a) valence quarks and gluon and for 
(b) sea anti-quarks 
obtained from the combined QCD analysis of the inclusive DIS data 
and \redc (dark shaded bands) in the RT optimised scheme 
as a function of $x$ at $Q^2=10\gev^2$. 
Note that, somewhat confusingly but following common practice, 
here the variable $x$ refers to $x_b$ in Eq.~\ref{eq:factorization},
rather than to Bjorken $x$. 
For comparison the results of the QCD analysis of the inclusive DIS data only 
are also shown (light shaded bands). 
The gluon distribution function is scaled by a factor $0.05$ 
and the $x\overline{d}$ distribution function is scaled by a factor $1.1$ 
for better visibility.
The total PDF uncertainties include fit, model and parametrisation uncertainties.
}
\label{fig:charm_impact_on_pdfs} 
\emp
\end{figure}

The recent combined charm DIS data~\cite{HERAcharmcomb} were also used 
in a QCD analysis~\cite{HERAcharmcomb} together with the combined inclusive
\herai DIS cross sections~\cite{DIScomb}.
Only the data with $Q^2 > 3.5\gev^2$ were used in the analysis 
to assure applicability of pQCD calculations.
The analysis was performed at NLO using the 
HERAFitter package~\cite{H1comb,DIScomb,herafitter} and closely followed the 
HERAPDF1.0 prescription~\cite{DIScomb}. 
Various implementations of the NLO GMVFNS approach were used and 
the role of the value of the charm quark mass parameter (see \Sect{variable}),
\mct, was studied.
For each heavy-flavour scheme a number of PDF fits was performed 
to scan $\chi^2$ of the PDF fit as a function of \mct.
From the scan the optimal value, \mcto, of the charm-quark mass parameter
in a given scheme was determined by the minimum of the $\chi^2$
and the corresponding fit uncertainty%
\footnote{This minimisation uncertainty is usually referred to as the 
``experimental''
uncertainty in the HERAPDF context~\cite{DIScomb,HERAcharmcomb}.
However, it can absorb some other sources of uncertainties, 
e.g. variations of PDF parametrisation.
Therefore, the more general term ``fit uncertainty'' is used here.}
was evaluated from the 
$\Delta \chi^2 =1$ variation. 
The procedure is illustrated in \fig{charm_vfn_scan}(a)
which shows the fit to the inclusive DIS data alone and together with \redc.
The inclusive DIS cross sections alone only weakly constrain \mcto,
indicated by the shallowness of the $\chi^2(\mct)$ distribution.
The charm DIS cross sections provide the required constraint to extract \mcto. 
Additionally, for each GMVFNS approach the model and parametrisation 
assumptions in the fits were varied one-by-one and the corresponding $\chi^2$
scan as a function of \mct\ was repeated.
The difference between \mcto\ obtained with the default assumptions
and the result of each variation was taken as the corresponding source of
uncertainty.
The dominant contribution to the uncertainty was found to come from 
the variation of the minimum $Q^2$ value for inclusive DIS data used in the fit.

\Fig{charm_vfn_scan}(b) shows the $\chi^2$ distributions as a function of 
\mcto\ obtained from fits to the inclusive \herai data and the combined \redc 
for all variable-flavour-number schemes considered.
All schemes yield similar minimal $\chi^2$ values, 
however at quite different values of \mcto.
The resulting values of \mcto\ are given in \Tab{charm_vfn_scan} together 
with the evaluated uncertainties, the minimal total $\chi^2$ values and 
the $\chi^2$ contribution from the charm data.
The ACOT-full scheme provides the best global description of the 
inclusive and charm data together, while the RT optimised scheme yields 
the best description of the charm data alone.
The fits in the S-ACOT-$\chi$ scheme result in a very low value of \mcto\ compared
to other approaches. 
Since this scheme only includes a leading-order 
approximation of heavy-flavour production at the order considered here
(see Table \ref{tab:theories}), 
effectively no
distinction is made between pole or running mass.
%
%%%%%%%%%%%%%%%%%%%%%%%%%%%%%%%%%%%%%%%%%%%%%%%%%%%%%%%%%%%%%%%%%%%%%%%%%
\begin{table}[tb]
	\begin{center}
	\renewcommand{\arraystretch}{1.15}
	  \begin{tabular}[h]{|c|c|c|c|}
    \hline
   Scheme  & \mcto\,& $\chi^2/n_{\rm dof}$ & $\chi^2/{n_{\rm dp}}$ \\
                  &  [GeV]  &    $\sigma^{NC,CC}_{\rm red}$+\redc          & \redc   \\
  \hline
  \hline
      RT standard    & $1.50 \pm 0.06_{\rm{fit}} \pm 0.06_{\rm{mod \oplus param \oplus \alpha_s}}$ & $630.7/626$& $49.0/47$\\
      RT optimised   & $1.38 \pm 0.05_{\rm{fit}} \pm 0.03_{\rm{mod \oplus param \oplus \alpha_s}}$ & $623.8/626$& $45.8/47$\\
      ACOT-full         & $1.52 \pm 0.05_{\rm{fit}} \pm 0.12_{\rm{mod \oplus param \oplus \alpha_s}}$ & $607.3/626$& $53.3/47$\\
      S-ACOT-$\chi$ & $1.15 \pm 0.04_{\rm{fit}} \pm 0.01_{\rm{mod \oplus param \oplus \alpha_s}}$ & $613.3/626$& $50.3/47$\\
      ZMVFNS          & $1.60 \pm 0.05_{\rm{fit}} \pm 0.03_{\rm{mod \oplus param \oplus \alpha_s}}$ & $631.7/626$& $55.3/47$\\ 
  \hline
   \end{tabular}
	\caption{The values of the charm mass parameter \mcto\ as determined from 
	the \mct\ scans in different heavy flavour schemes~\pcite{HERAcharmcomb}. 
	The uncertainties of the minimisation procedure are denoted as ``fit'', 
	the model, parametrisation and $\alpha_s$ uncertainties were added in quadrature
	and are represented by ``$\mathrm{mod \oplus param \oplus \alpha_s}$''.
	The corresponding global and partial $\chi^2$ are presented per degrees of freedom,
	$n_{\mathrm {dof}}$, and per number of data points, $n_{\mathrm {dp}}$, respectively.}
        \label{tab:charm_vfn_scan}
	\end{center}
\end{table}
%%%%%%%%%%%%%%%%%%%%%%%%%%%%%%%%%%%%%%%%%%%%%%%%%%%%%%%%%%%%%%%%%%%%%%%%%
%
All NLO VFNS predictions using corresponding $\mcto$ values for each scheme 
provide a similarly good description of the \redc data~\cite{HERAcharmcomb}.

\Fig{charm_impact_on_pdfs} shows the PDFs extracted from 
the fit to the inclusive DIS data alone and together with the \redc data 
in the RT optimised VFNS\footnote{
Similar observations were made with other schemes.}.
A comparison of the extracted PDF uncertainties yields the following conclusions
about the impact of the \redc data~\cite{HERAcharmcomb}:
\begin{itemize}
 \item the uncertainty on the gluon PDF was reduced, mostly due to 
 a reduction of the parametrisation uncertainty due to the additional constraints
 that the charm data introduce due to the BGF process;
 
 \item the uncertainty on the charm-quark PDF is considerably reduced due to 
 the constrained range of \mct. The \mct\ variation was set to 
 $1.35 < \mct < 1.65\gev$ for the fit to the inclusive data only and 
 was defined by the evaluated total uncertainties as given 
 in~\Tab{charm_vfn_scan} for the fit including the charm data;
 
 \item the uncertainty on the up-quark sea PDF was correspondingly reduced,
 because the inclusive data constrain the sum of up- and charm-quark sea;
 
 \item the uncertainty on the down-quark sea was also reduced 
 because it was constrained to be equal to the up-quark sea at low $x$;

 \item the uncertainties on the valence-quark and strange-quark sea PDFs
 were almost unaffected;

 \item the central PDFs were not altered significantly and were found to be 
 within the uncertainties of the PDFs based on inclusive data only. This 
 reflects the good description of the charm data by the default PDFs
(section \ref{sect:F2cc}).
\end{itemize}

By now, the combined charm reduced cross sections~\cite{HERAcharmcomb} 
have been used in QCD analyses by various PDF-fitting
groups~\cite{Alekhin:2012vu,Alekhin:2013nda, CTEQmass,Gao:2013xoa,Jimenez-Delgado:2014twa,Ball:2014uwa,MMHT}. 
They are an important ingredient to constrain the proton flavour composition 
(see also next section) and to stabilise its gluon content. 
The latter is 
especially important for Higgs production at the LHC, for which the 
dominant process is gluon-gluon fusion via an intermediate
top-quark loop. 
Measuring this process precisely, in combination with a 
precise knowledge of the gluon content of the proton, allows the extraction
of a precise measurement of the Higgs-top Yukawa coupling. 
 
Instead focusing on the low-$x$ range, the HERA charm and beauty data have 
recently been used in conjunction with charm and beauty data from LHCb to
constrain the gluon distribution down to $x \sim 5 \times 10^{-6}$
\cite{LHCb}. This region is particularly relevant for the prediction of 
cross sections for processes occurring in cosmic ray interactions.

%%%%%%%%%%%%%%%%%%%%%%%%%%%%%%%%%%%%%%%%%%%%%%%%%%%%%%%%%%%%%%%%%%%%%%%%%%%%%%%%%%
%%%%%%%%%%%%%%%%%%%%% Charm mass %%%%%%%%%%%%%%%%%%%%%%%%%%%%%%%%%%%%%%%%%%%
%%%%%%%%%%%%%%%%%%%%%%%%%%%%%%%%%%%%%%%%%%%%%%%%%%%%%%%%%%%%%%%%%%%%%%%%%%%%%%%%%%

\subsection{Proton flavour composition and W/Z/H production at LHC}
\label{sect:WZ}

In the previous section it was outlined how the inclusion of charm data
into GMVNFS PDF fits, 
and in particular the constraint on the charm quark mass parameter derived
from these data, imposes constraints on the gluon content (relevant e.g. 
for Higgs production) and on the flavour composition of the 
quarks in the proton. This, in turn, affects theoretical predictions for 
processes which are sensitive to this flavour composition, such as the
production for W and Z bosons at LHC. 

\begin{figure}[htbp]
\begin{center}
\begin{minipage}[t]{8 cm}
\epsfig{file=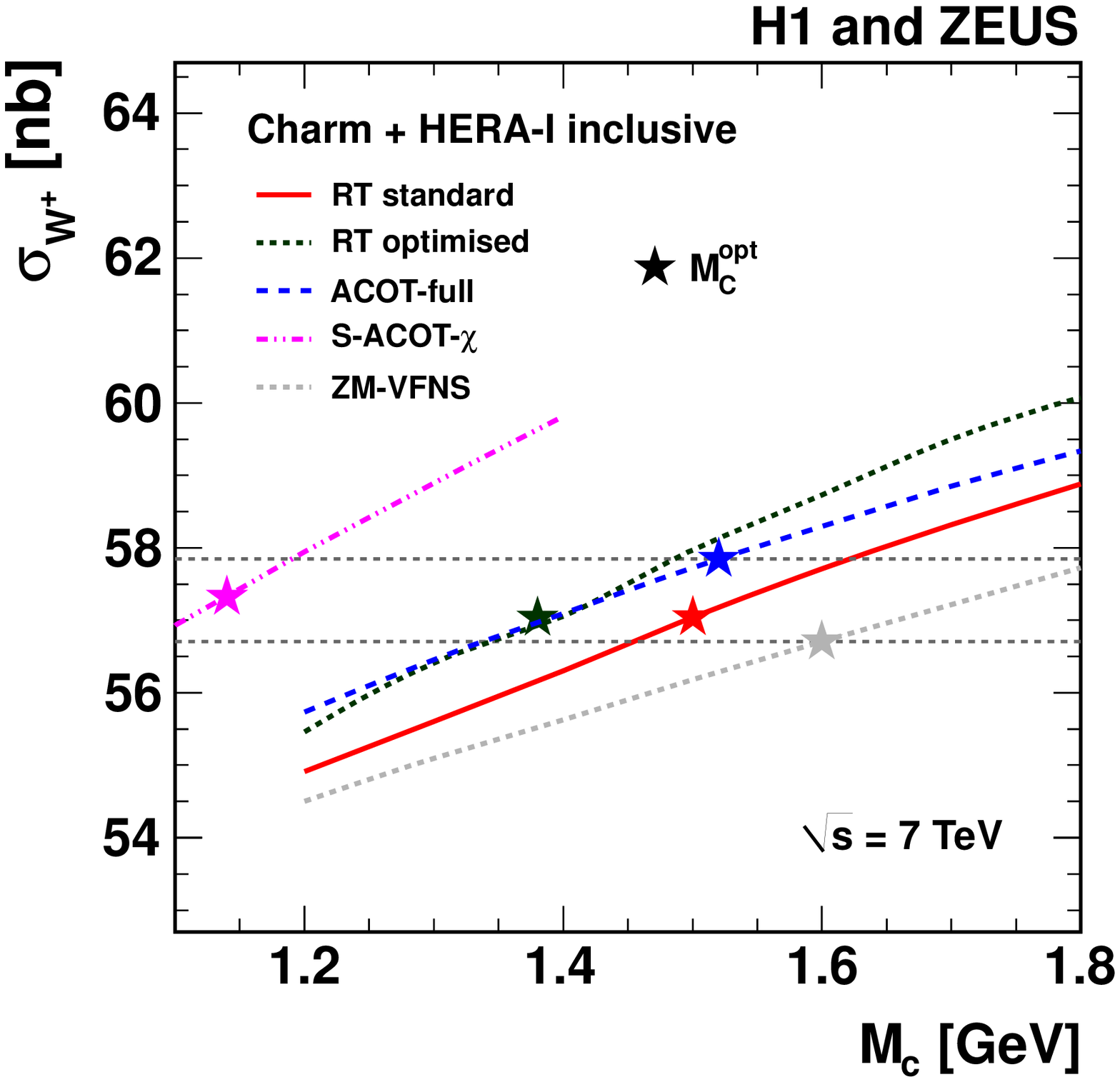,width=0.9\textwidth}
\put(6.5,15.4){\large (a)}
\end{minipage}
\begin{minipage}[t]{8 cm}
\epsfig{file=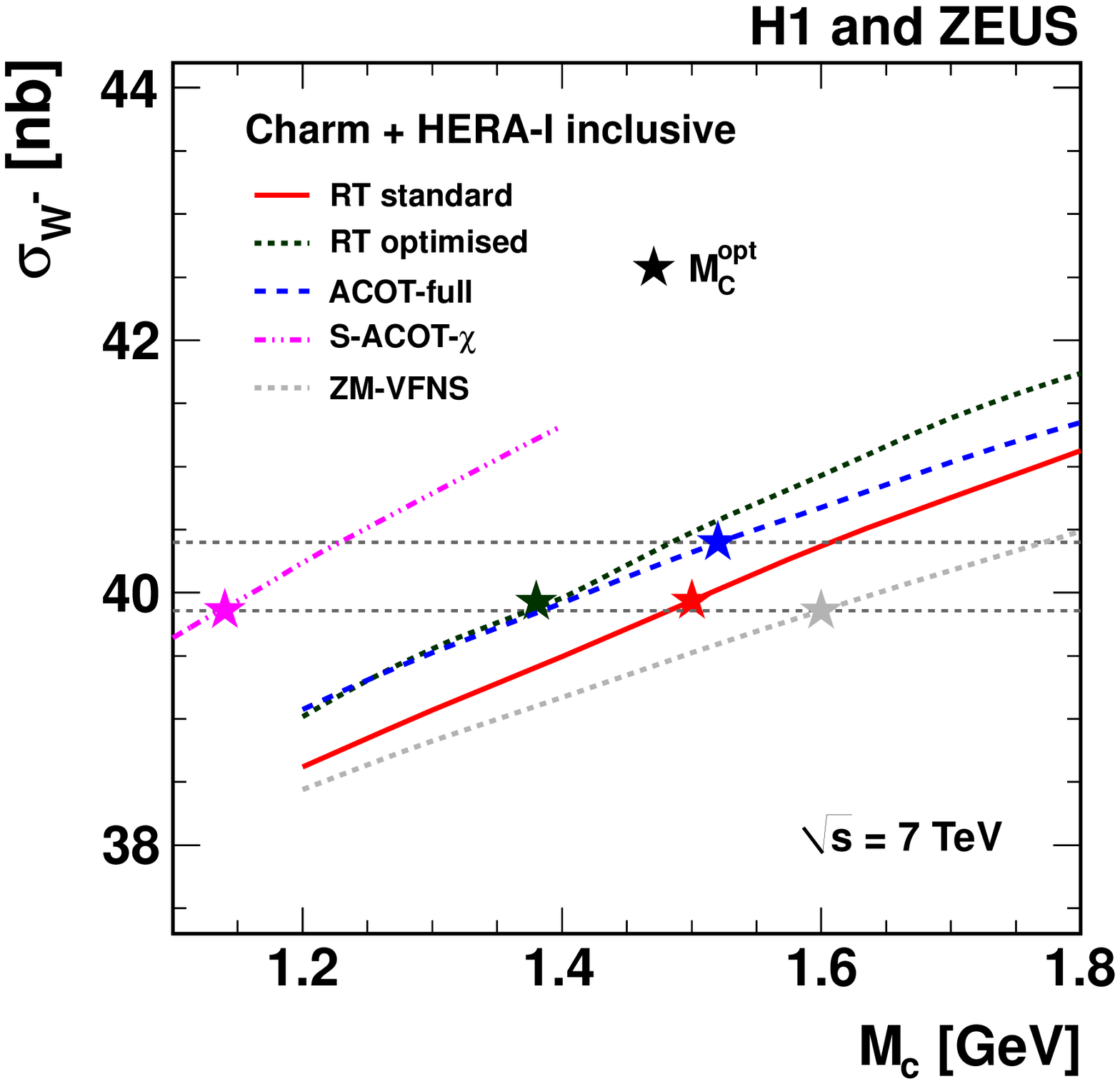,width=0.9\textwidth}
\put(14.5,15.4){\large (b)}
\end{minipage}
\begin{minipage}[t]{8 cm}
\epsfig{file=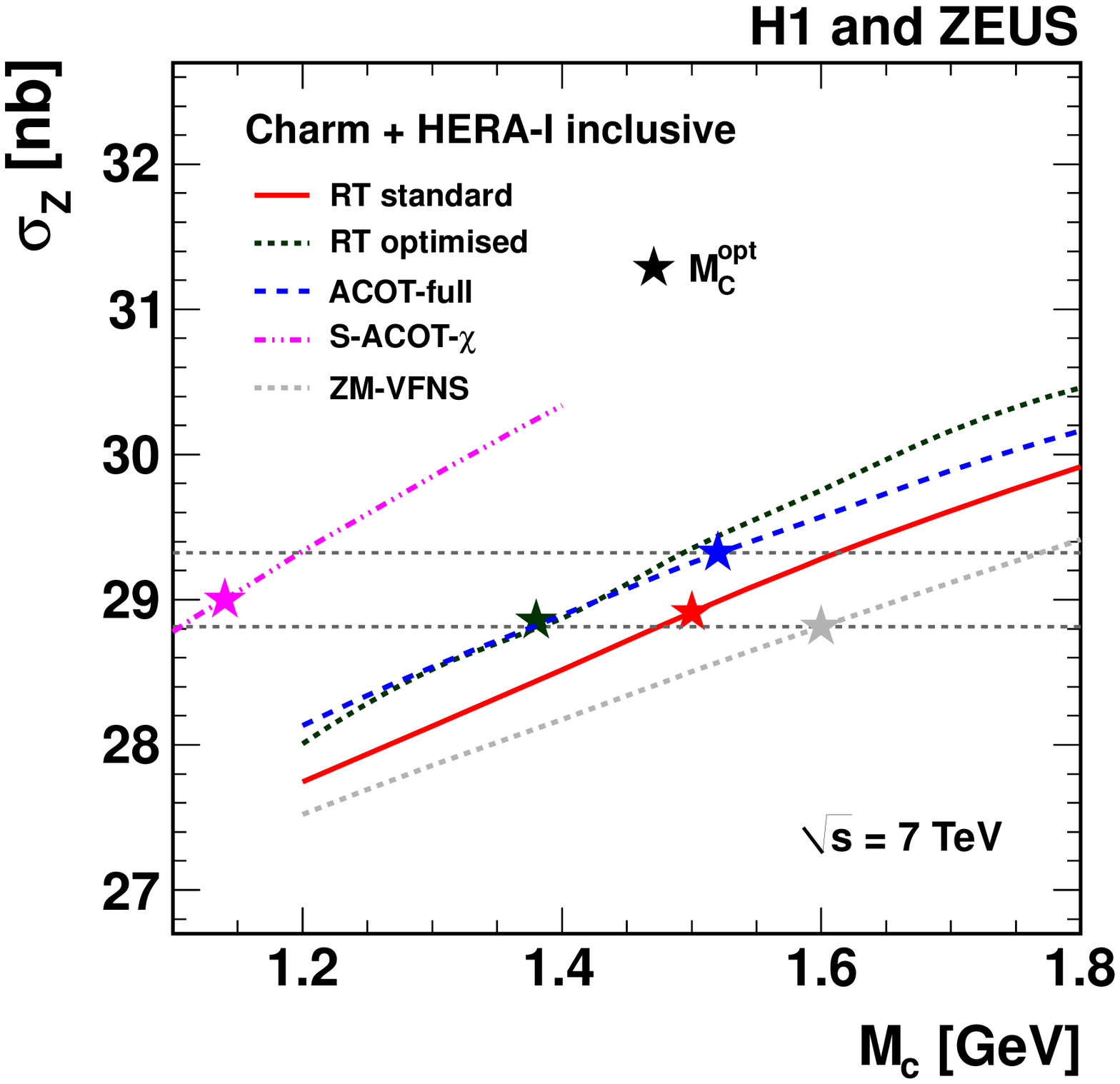,width=0.9\textwidth}
\put(10.,6.6){\large (c)}
\end{minipage}
\begin{minipage}[t]{17 cm}
\caption{NLO predictions for (a) $W^{+}$, (b) $W^{-}$ and (c) $Z$ production cross sections 
at the LHC 
for $\sqrt{s}=7\tev$ as a function of $\mct$ used in the corresponding PDF fit
\protect{\cite{HERAcharmcomb}}. 
The different lines represent predictions for different implementations of the VFNS. 
The predictions obtained with PDFs 
evaluated with the $\mcto$ values for each scheme are indicated by the stars. 
The horizontal dashed lines show the resulting spread of the predictions 
when choosing $M_c=\mcto$.}
\label{fig:wp_cross_section} 
\end{minipage}
\end{center}
\end{figure}

\Fig{wp_cross_section} shows NLO predictions for $W$ and $Z$ 
production at LHC for PDFs extracted in different heavy flavour schemes as 
discussed in \Sect{cbF2}, as a function of the charm quark mass 
parameter $M_c$ used in the PDF fit.  
For fixed $M_c$, these predictions differ by about $7\%$. The dependence on 
$M_c$ is opposite to what one would naively expect 
(see also \fig{charm_impact_on_pdfs}). A higher charm mass leads
to less charm in the proton (fewer gluons split) but to a higher gluon 
density. This in turn increases the amount of $u$ and $d$ sea quarks in the proton,
even more so since the total sea is constrained by the inclusive proton
structure functions. The larger number of $u$ and $d$ quarks overcompensates 
the smaller number of $c$ quarks and leads to an increase of the $W$ and 
$Z$ cross sections as shown in \fig{wp_cross_section}. The fit in 
\fig{charm_impact_on_pdfs} actually
led to a smaller charm mass than the default, therefore the effect on the 
PDF was opposite.

The stars in \fig{wp_cross_section} indicate the cross section 
predictions 
for the optimal mass for each heavy flavour scheme, as extracted from the 
charm data in \fig{charm_vfn_scan}(b). All predictions then coincide
to within $2\%$, independent of the heavy flavour scheme used. This demonstrates
that using the optimal mass for each scheme which best fits the HERA charm 
data stabilises the flavour composition in the proton, and leads to a reduction
of this contribution to the cross section uncertainty by about a factor 3.
To minimise the uncertaities arising from the charm and beauty masses, 
for GMVFNS schemes it is thus strongly recommended to use the optimal mass
parameters as derived from the heavy-flavour structure-function data rather 
than a mass obtained from external considerations.

A similar analysis for beauty, which remains to be done, will in addition 
yield experimental constraints on the $b$ PDF in the framework of 
5-flavour PDFs for LHC,
which are so far constrained by theory only. This in turn will be relevant
e.g. for a future measurement of the Higgs-$b$ Yukawa coupling from associated
Higgs-$b\bar b$ production.

\subsection{Measurements of the charm-quark mass and its running}
\label{sect:cmass}

The sensitivity of the HERA reduced charm cross sections to the charm-quark mass,
already partially studied in \Sect{cbF2}, can be used to measure 
the charm quark mass appearing in perturbative QCD, whose value depends on 
the renormalisation scheme within which it is being evaluated. The two mass 
definitions which are most commonly used are the pole mass and the 
$\overline{MS}$ running mass (\Sect{mass}). Since the 
$\overline{MS}$ mass is perturbatively better defined, recent charm mass 
measurements concentrate on this renormalization scheme.
The FFNS scheme (\Sect{massive}) is most suited for this 
evaluation, since it fully accounts 
for mass effects without any additional free parameters. FFNS calculations 
of the reduced cross section in this scheme exist at NLO 
and partial NNLO \cite{abkm09msbar,ABKMMSbar}. All results quoted in the 
following are obtained 
from these calculations unless otherwise quoted.

The first determination of the $\overline{MS}$ charm-quark mass \cite{cmass1}, 
from a subset of \dst charm data from the H1 collaboration, in which
also the details of the theoretical framework are given, obtained
\begin{equation}
m_c(m_c) = 1.27 \pm 0.04(\rm{fit}) \ ^{+0.06}_{-0.01}(\rm{scale})\gev
\end{equation}
at NLO, and 
\begin{equation}
m_c(m_c) = 1.36 \pm 0.04(\rm{fit}) \ ^{+0.04}_{-0.00}(\rm{scale}) \pm 0.1(\rm{theory})\gev
\end{equation}
at partial NNLO, where the last term reflects a very conservative estimate 
of the evaluation of the uncertainties of the NNLO approximation.

\begin{figure}[htbp]
\center
\epsfig{file=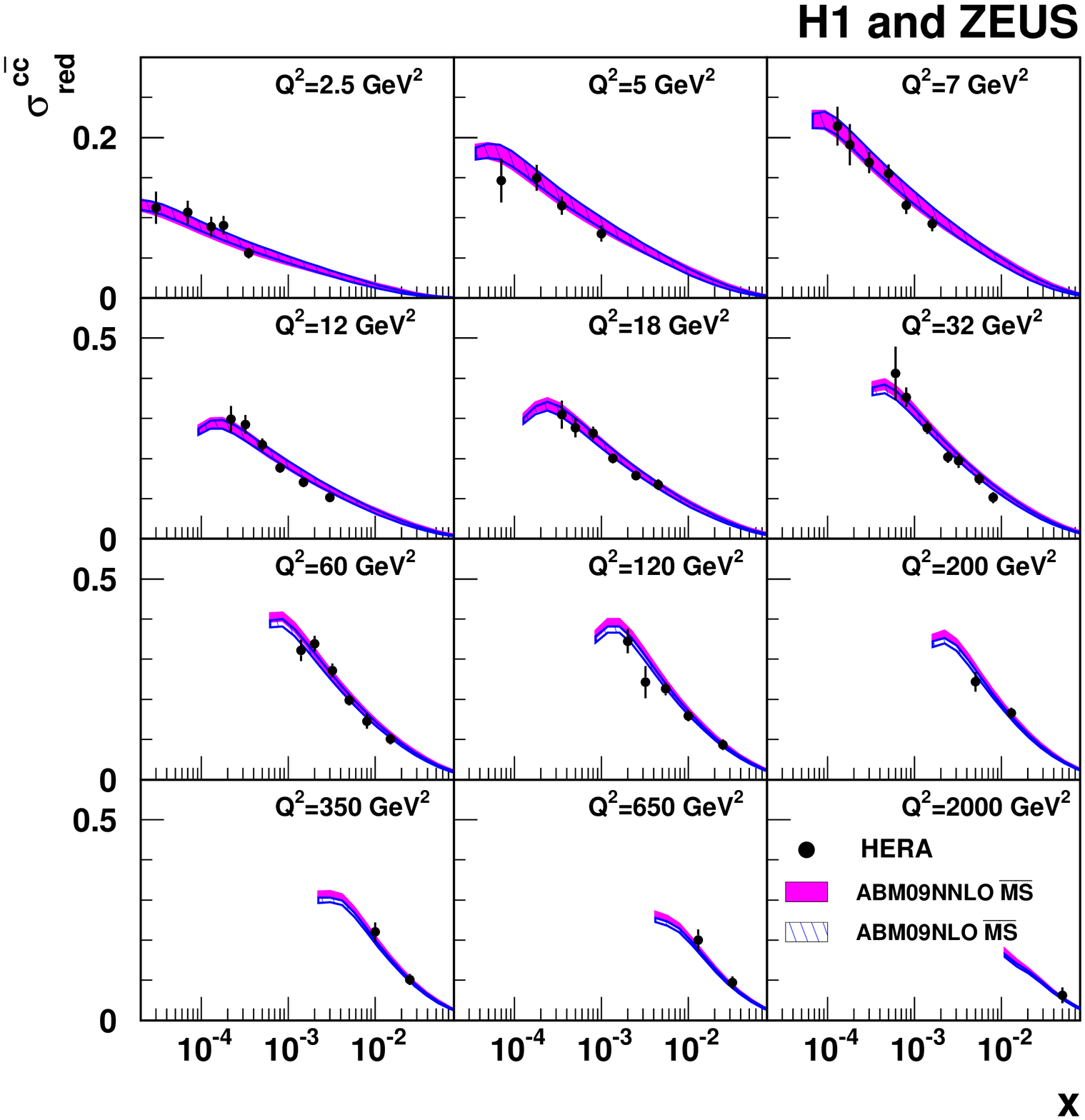,width=.6\textwidth}
\setlength{\unitlength}{1cm}
\caption{Combined reduced cross sections \cite{HERAcharmcomb} filled circles  as a function of $x$ 
for fixed values of $Q^2$. The error bars represent the total uncertainty including uncorrelated, correlated 
and procedural uncertainties added in quadrature. The data are compared to predictions of the ABM group at 
NLO (hashed band) and NNLO (shaded band) in FFNS using the $\overline{\rm{MS}}$ definition for the charm quark mass.}
\label{fig:combined_abkm} 
\end{figure}
                
\Fig{combined_abkm} shows the comparison of predictions of the ABKM 
group \cite{abkm}, using the above mass values as central values, to the 
combined HERA charm data \cite{HERAcharmcomb} discussed in 
\Sect{F2cc}, which have smaller
uncertainties than the data used for the initial measurement. 
Very good agreement is observed for both NLO and partial NNLO.
\begin{figure}[htbp]
\center
\epsfig{file=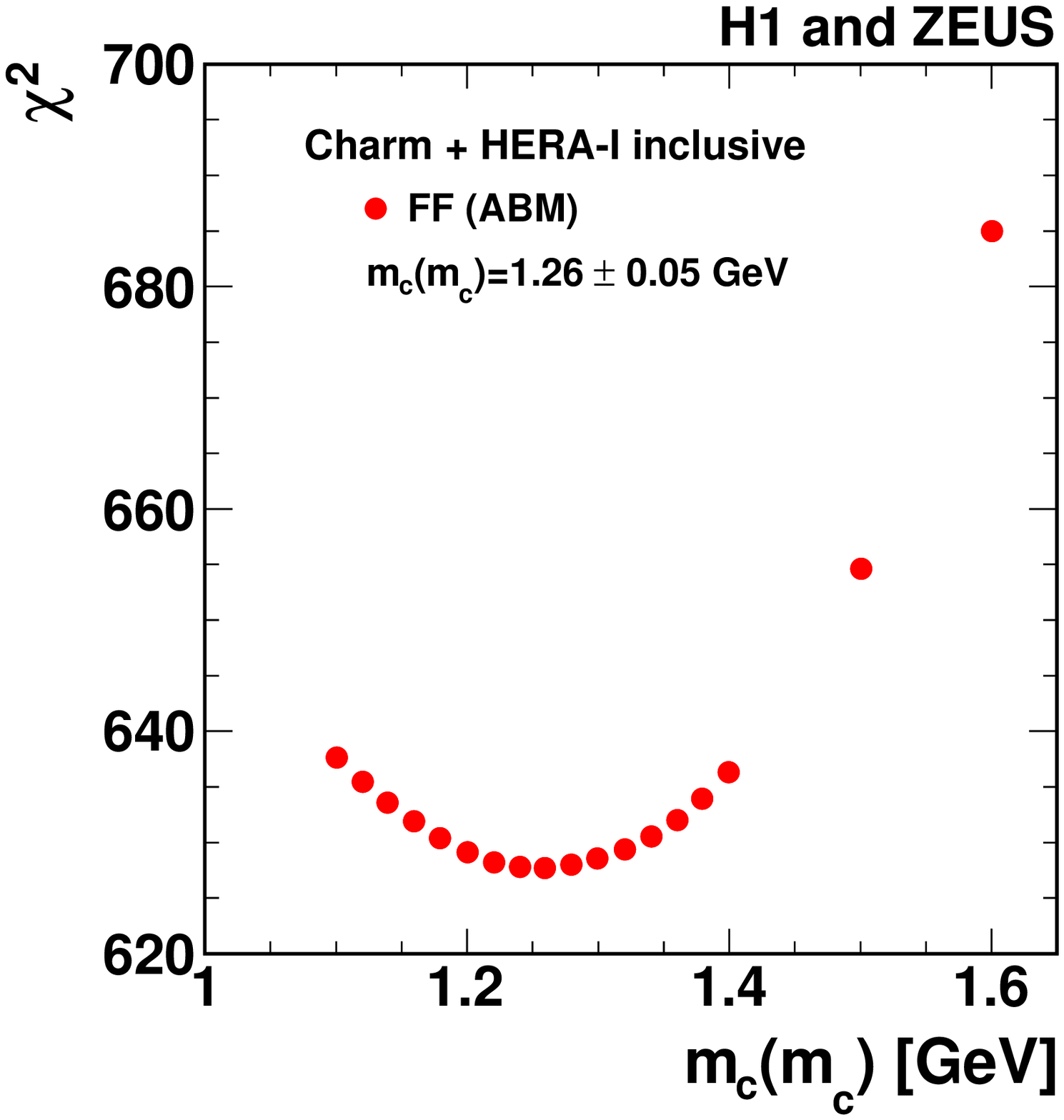,width=0.4\textwidth}
\hspace{1cm}
\epsfig{file=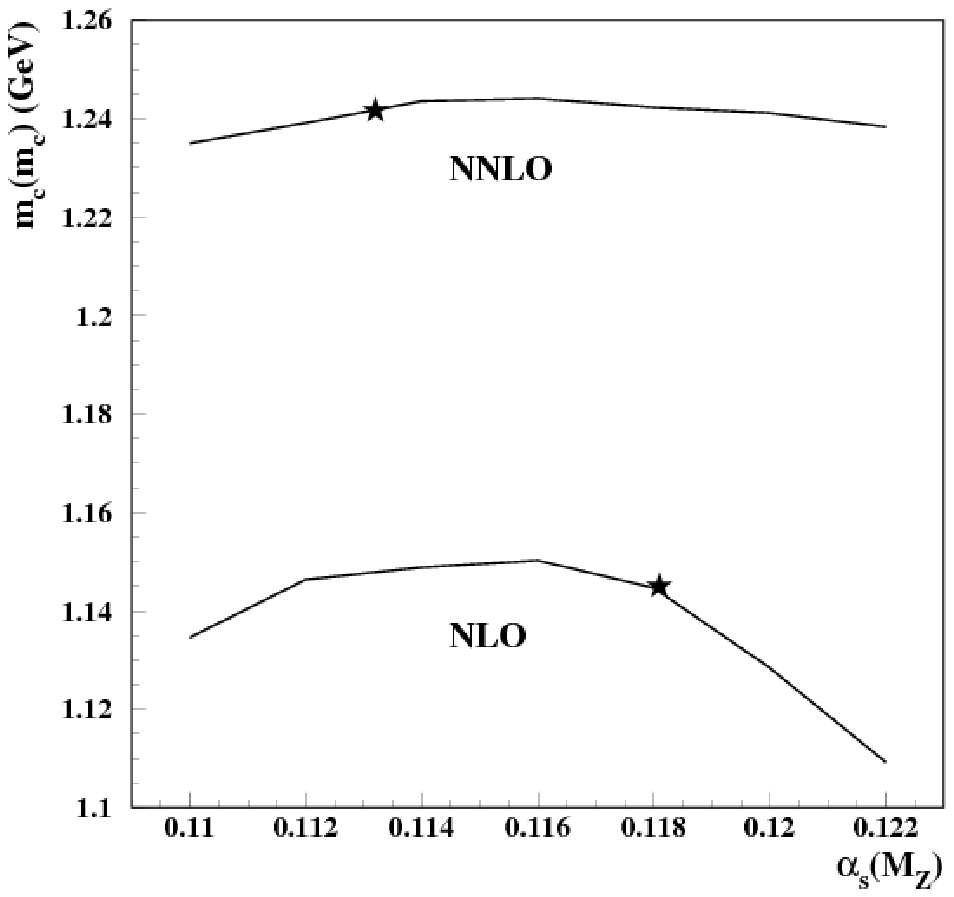,width=0.38\textwidth}
\caption{(left) The values of $\chi^2$ for the PDF fit \cite{HERAcharmcomb} 
to the combined HERA DIS data including charm measurements 
as a function of the running charm quark mass $m_c(m_c)$. The FFNS ABM scheme is used, where the charm quark 
mass is defined in the $\overline{\rm{MS}}$ scheme.
(right) The values of mc(mc) obtained in the NLO and NNLO variants of the ABM 
analysis \cite{mcalphas} with the value of $\alpha_s(M_Z)$ fixed. The 
position of the star displays the result with the value of $\alpha_s(M_Z)$ 
fitted. }
\label{fig:abm_scan} 
\end{figure}
The H1 and ZEUS collaborations have used a fit to these 
data (\fig{abm_scan}(left)), using the 
kinematic region $Q^2>3.5\gev^2$, to obtain the NLO 
measurement \cite{HERAcharmcomb}
\begin{equation}
m_c(m_c) = 1.26 \pm 0.05(\rm{exp}) \pm 0.03(\rm{mod}) \pm 0.02(\rm{param}) \pm 0.02(\alpha_s)\gev.
\end{equation}
This result has a slightly more elaborate evaluation of the uncertainties
related to the data extrapolation as well as other model and parametrisation
uncertainties, while it does not include uncertainties on the normalisation 
of the cross section predictions due to QCD scale variations. 

The same data where then used by the ABM group and collaborators \cite{Alekhin:2012vu} 
to reobtain similar evaluations, 
\begin{equation}
m_c(m_c) = 1.15 \pm 0.04(\rm{exp}) \ ^{+0.04}_{-0.00}(\rm{scale})\gev
\end{equation}
at NLO, and 
\begin{equation}
m_c(m_c) = 1.24 \pm 0.03(\rm{exp}) \ ^{+0.03}_{-0.02}(\rm{scale})\ ^{+0.00}_{-0.07}(\rm{theory})\gev
\end{equation}
at partial NNLO. The smaller central NLO value and the smaller uncertainty 
are mainly due to the fact that the charm data from the lowest $Q^2$ bin were 
included. 
A correlated measurement of $m_c(m_c)$ and the strong coupling constant
\cite{mcalphas} 
was also obtained (Fig. \ref{fig:abm_scan}(right)). In particular for 
the NLO case, the correlation between $m_c(m_c)$ and $\alpha_s$ is
non-negligible.

They were also used by the CTEQ group \cite{CTEQmass} to derive 
the $\overline{MS}$ mass in the context of the S-ACOT-$\chi$ VFNS, using 
charm matrix elements to one-loop order in the massive part of the calculation.
The result
\begin{equation}
m_c(m_c) = 1.19^{+0.08}_{-0.15}\gev
\end{equation}
exhibits a larger uncertainty than the previous extractions due to the 
additional uncertainty from the variation of the (single) free parameter
of this VFNS scheme, and due to conversions between the pole and 
running masses in the extraction process.

All these results from a predominantly space-like perturbative process are 
consistent with each other and with the world average \cite{PDG12}
\begin{equation}
m_c(m_c) = 1.275 \pm 0.025\gev
\end{equation}
obtained from lattice QCD and time-like processes.
This is a highly nontrivial triumph of QCD.
Some of the above measurements are now included in the latest world 
average \cite{PDG2014}, and further improvements on both the experimental
and theoretical sides have the potential to further improve the corresponding 
precision.   

\begin{figure}[h]
\center
\vspace{-0.2cm}
\epsfig{file=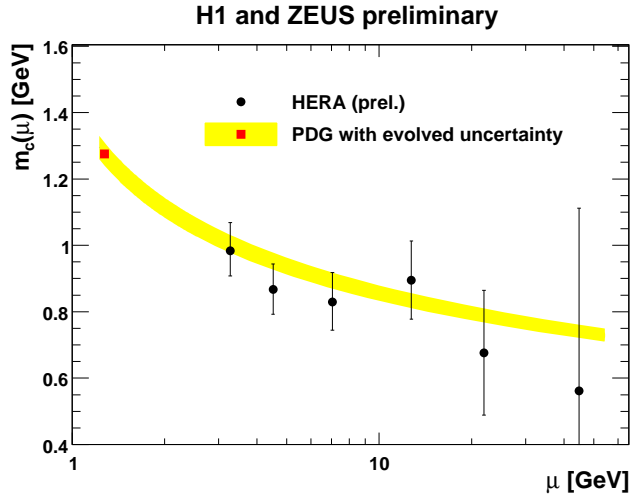,width=0.5\textwidth}
\caption{Measured charm mass $m_c(\mu)$ in the $\overline{MS}$ 
running mass scheme
as a function of the scale $\mu$ as defined in the text
(black points).
The red point at scale $m_c$ is the PDG world average \cite{PDG12} and the band 
is its expected running \cite{RUNdec}.  
}
\label{fig:mcrun} 
\end{figure}

In a recent preliminary result \cite{charmrun} the same data have again been 
used to determine the actual scale dependence (`running') of the charm-quark 
mass in the  $\overline{MS}$ scheme, according to Eq. (\ref{eq:run2}).
For this purpose, the charm data were subdivided into 6 different $Q^2$ ranges,
for which the charm mass was extracted separately at the scale $<Q^2>+4m_c^2$,
where $<Q^2>$ is the average of each range. The result is shown in 
Fig. \ref{fig:mcrun}. 
This is the first explicit measurement of the scale dependence of the charm 
quark mass.

%%%%%%%%%%%%%%%%%%%%%%%%%%%%%%%%%%%%%%%%%%%%%%%%%%%%%%%%%%%%%%%%%%%%%%%%%%%%%%%%%%
%%%%%%%%%%%%%%%%%%%%% beauty mass %%%%%%%%%%%%%%%%%%%%%%%%%%%%%%%%%%%%%%%%%%%
%%%%%%%%%%%%%%%%%%%%%%%%%%%%%%%%%%%%%%%%%%%%%%%%%%%%%%%%%%%%%%%%%%%%%%%%%%%%%%%%%%

\subsection{Measurement of the beauty-quark mass and its running}
\label{sect:bmass}

\begin{figure}[htbp]
\center
\epsfig{file=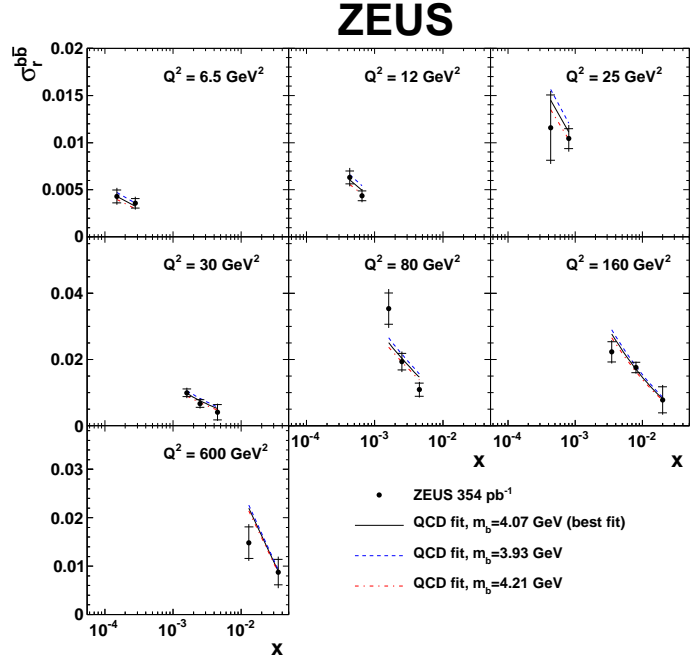,width=0.5\textwidth}
\caption{Reduced beauty cross section (filled symbols) as a function of $x$ 
for seven different values of $Q^2$ \cite{zeusltt_hera2}. Also shown are the 
results of a QCD fit for different values of the $\overline{MS}$ running mass
$m_b(m_b)$.
}
\label{fig:mbfit} 
\end{figure}

\begin{figure}[htbp]
\center
\vspace{-0.3cm}
\epsfig{file=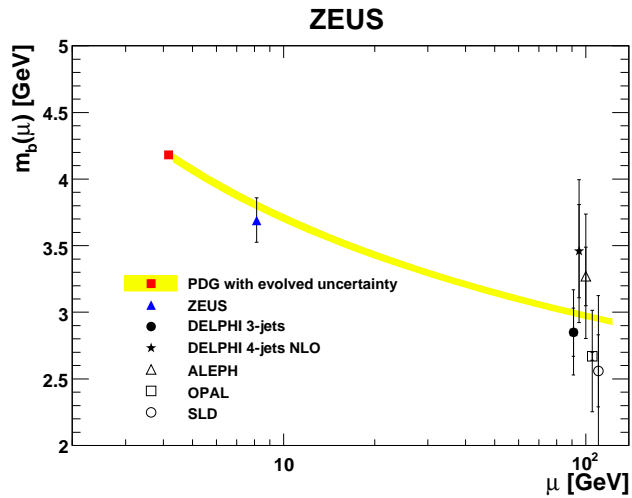,width=0.5\textwidth}
\caption{Measured beauty mass $m_b(\mu)$ in the $\overline{MS}$ 
running mass scheme
as a function of the scale $\mu$ from HERA \cite{zeusltt_hera2} and 
LEP \cite{LEPbmass} data.
The red point at scale $m_b$ is the PDG world average \cite{PDG12} and the band 
is its expected running \cite{RUNdec}.  
}
\label{fig:brun} 
\end{figure}

Using the same approach as outlined above for charm, the ZEUS collaboration 
has used a fit (Fig. \ref{fig:mbfit}) to the beauty reduced-cross-section 
data \cite{zeusltt_hera2} 
to extract the value of the beauty-quark $\overline{MS}$ running mass at NLO,
\begin{equation}
m_b(m_b) = 4.07 \pm 0.14(\rm{fit}) \ ^{+0.01}_{-0.07}(\rm{mod}) \ ^{+0.02}_{-0.00}(\rm{param}) 
\ ^{+0.08}_{-0.05}(\rm{theo})\gev,
\end{equation}
where the theoretical uncertainty is dominated by the scale variation 
uncertainty. This is the first such extraction from HERA data, and 
agrees well with the world average \cite{PDG12}
\begin{equation}
m_b(m_b) = 4.18 \pm 0.03\gev.
\end{equation}
\Fig{brun} shows this result, translated to the scale $4m_b^2$,
compared to the PDG value and its expected running and to values extracted 
from LEP data at the scale $M_Z$. The expected running of the $\overline{MS}$
beauty-quark mass is confirmed. This is a nontrivial test of the basics of 
QCD. 

\subsection{Summary}

Heavy flavour physics at HERA yields many results which are of interest
for particle physics in general.
The usage of HERA as a ``charm factory'' generates world-class information 
on charm fragmentation functions and fragmentation fractions
and allows tests of the fragmentation universality.
The constraints on PDFs from charm data, and to a lesser extent also from 
beauty data, help to reduce uncertainties for important cross sections 
at LHC, such as heavy flavour, W/Z and Higgs production. 
Constraints on the latter are
important for the measurement of the Higgs Yukawa couplings.
More directly, the charm and beauty DIS data have been used to extract
well defined measurements of the charm- and beauty-quark masses, which enter
the world average. The running of the charm-quark mass has been measured for 
the first time ever. By comparing with LEP data, the running of the 
beauty quark mass has also been confirmed. 
In general, the good agreement of QCD predictions with
the data support the applicability of the HERA results to all particle 
physics applications for which they might be directly or indirectly relevant.

\newpage

% Summary
\section{Summary and outlook}
\label{sect:conclusions}

Charm and beauty production at HERA are a great laboratory to test 
the theory of heavy flavour production in the framework of perturbative 
QCD and to measure some of its parameters. The occurrence of different
possibilities to treat the heavy quark masses in the PDF, matrix 
element and fragmentation parts of the calculation introduces a significant
level of complexity into the corresponding QCD calculations, in addition to 
the usual scheme and scale choices. Confronting such different choices 
with data can be helpful to understand the effects of different ways to 
truncate the perturbative series and to evaluate their impact on the 
measurement of fundamental parameters, both at HERA and at other colliders.

HERA was the first and so far only high energy $ep$ collider.
The heavy flavour results discussed in this review were obtained with 
the H1 and ZEUS detectors which were well suited for the detection 
of heavy flavoured particles. Adding the luminosities from the two 
collider experiments, a total luminosity of about $1\fbi$ was collected.

The availability of many different charm and beauty tagging methods 
allows results to be obtained through several different final states with 
different systematics. In addition to the statistical benefit from combining 
different samples, such combinations also profit from cross calibrations 
of the systematics from different methods and experiments.

Due to the high top mass, the only top final state which might have been 
detectable at HERA is single top production with non-Standard-Model couplings.
No signal is seen, and the coupling limits derived are competitive. 

The charm (beauty) quark masses provide semi-hard (hard) QCD scales which 
allow the succesful application of perturbative calculations 
over the complete phase space.
However, these masses also compete with other, often even harder 
perturbative scales.
Total cross sections for charm photoproduction and the total cross section 
for beauty production (including photoproduction and deeply inelastic
scattering) are reasonably described by perturbative 
calculations at next-to-leading order (NLO).
Single-differential cross sections already provide a good handle 
to test the applicability of different QCD approximations, although the 
theoretical uncertainties are mostly much larger than the experimental ones. 
The theory predictions agree with the data up to the highest accessible
transverse momenta or photon vitualities, 
showing no indications that final state resummation
corrections are needed for massive calculations in the HERA kinematical domain.
Double-differential cross sections, in particular those including jets, 
reveal a partial failure
of the massive scheme NLO predictions for kinematic 
observables which would need final states with four or more partons 
in the calculation.
Although statistics and therefore precision is higher for photoproduction,
qualitatively very similar conclusions are obtained 
for photoproduction and deeply inelastic scattering (DIS). 
The NLO calculations in the massless scheme, 
where available, do mostly not provide a
better description for the observables, and clearly fail for some 
DIS observables.
The LO+PS MCs PYTHIA and HERWIG, which are often used for acceptance 
corrections, are able to describe all topologies reasonably, often 
even very well. The CASCADE $k_t$-factorisation MC performs somewhat 
less well on average.

In DIS, the large photon virtuality $Q^2$ provides an additional 
hard scale in the QCD calculations of heavy flavour production 
and allows probing the
parton dynamics inside the proton more directly than in photoproduction.
The dominant contribution to the charm and beauty cross sections 
arises from photon-gluon fusion.
For $Q^2 \gg 4m_Q^2$, where the photon virtuality is the dominant hard scale, 
the cross-section behaviour is similar to the one of the inclusive 
cross section for deeply inelastic scattering. At high $Q^2$ and low $x$, 
the naively expected charm and beauty contributions of $4/11$ and $1/11$ are
asymptotically approached.   
NLO QCD predictions
in the massive scheme (FFNS) give a good description of heavy flavour 
production at HERA in DIS over the complete accessible kinematic range. 
NLO predictions in variable-flavour-number schemes (GMVFNS) are only available
for inclusive quantities, and perform about equally well.

In particular for charm, the uncertainties from QCD corrections beyond NLO
and from the modelling of fragmentation are considerably larger than
the experimental uncertainties of the measured cross sections.
Improved QCD calculations would therefore be highly welcome.

Heavy flavour physics at HERA yields many results which are of interest
for particle physics in general.
The usage of HERA as a ``charm factory'' generates world-class information 
on charm fragmentation functions and fragmentation fractions
and allows tests of the fragmentation universality.
The constraints on proton parton distribution functions (PDFs)
from charm and beauty data help to reduce 
uncertainties for important cross sections 
at the LHC, such as heavy flavour, W/Z and Higgs production. 
Constraints on the latter are
important for the measurement of the Higgs Yukawa couplings.
More directly, the charm and beauty DIS data have been used to extract
well defined measurements of the charm- and beauty-quark masses, which enter
the world average. The running of the charm-quark mass has been measured for 
the first time ever, and the running of the beauty quark 
mass has been confirmed. 

In general, the good agreement of QCD predictions with
the HERA data support the applicability of the QCD results derived from
these data to all particle 
physics applications for which they might be directly or indirectly relevant.
Some of the most important HERA heavy flavour results have been obtained 
during the last 2-3 years.
Even 8 years after the end of data taking the potential of the HERA heavy 
flavour data has still not been fully used in all cases, so there is 
room for significant further improvements, in particular also on the 
theory side, hoping e.g. for differential NNLO calculations in $ep$ collisions,
similar to those which have recently started to appear for the $pp$ case.

\section*{Acknowledgements} 

This review is a partial summary of the work of perhaps a 
thousand technicians, engineers and physicists for more than two decades.
It is our great pleasure to thank the many collegues who have contributed to 
the results presented here. 
We thank O. Kuprash and L. Schalow for technical contributions to this 
review.

\newpage

% Bibliography

\end{document}